%% file: sbika.tex
\documentclass[12pt, oneside]{book}
\usepackage{a4, amssymb, exscale}
\usepackage{makeidx, graphicx, multicol}

\newenvironment{eelist}[1]{
\begin{list}{}{
\setlength{\topsep}{0.5ex plus0.2ex minus0.1ex} 
\setlength{\leftmargin}{#1}
\setlength{\itemsep}{0.2ex plus0.2ex}
\setlength{\parsep}{0ex plus0.2ex} }}
{\end{list}}

\newenvironment{evlist}[2]{
\begin{list}{}{
\setlength{\topsep}{0.5ex plus0.2ex minus0.1ex} 
\setlength{\leftmargin}{#1}
\setlength{\itemsep}{#2 plus0.2ex}
\setlength{\parsep}{0ex plus0.2ex} }}
{\end{list}}

\newlength{\extextwidth}
\newenvironment{exframe}{\begin{minipage}{\extextwidth}{}}
{\end{minipage}}

\newcommand{\exparskip}{\par\vspace{1.2ex plus0.5ex minus0.5ex}}
\newcommand{\exsparskip}{\par\vspace{0.6ex plus0.3ex minus0.2ex}}
\newcommand{\tdom}[1]{\langle#1\rangle}
\newcommand{\sdom}[1]{#1^\triangleright}
\newcommand{\scod}[1]{#1_\triangleleft}
\newcommand{\adom}[2]{\sdom{#1}#2}
\newcommand{\domsdom}[1]{\tdom{\sdom{#1}}}
\newcommand{\xx}[1]{\mathbf{#1}}
\newcommand{\xxx}[1]{#1'}
\newcommand{\ftyped}[1]{\mathcal{T}(#1)}
\newcommand{\ffam}[1]{\mathcal{F}(#1)}
\newcommand{\utimes}[1]{\otimes #1}
\newcommand{\nonbot}[1]{#1^\natural}
\newcommand{\altbot}{\bot}
\newcommand{\ass}[2]{#2^{#1}} 
\newcommand{\opass}[2]{\mathop{{#2}^{#1}}} 
\newcommand{\rest}[2]{#1_{|#2}} 
\newcommand{\param}[2]{#1^{#2}}
\newcommand{\parax}[3]{#1^{#2}_{#3}}
\newcommand{\ipar}{\flat} 
\newcommand{\comp}{\circ}
\newcommand{\fcomp}{\circ}
\newcommand{\maximal}[1]{#1^{\uparrow}}
\newcommand{\fpi}[1]{#1^{\diamond}}

\newcommand{\Bool}{\mathbb{B}}
\newcommand{\Real}{\mathbb{R}}
\newcommand{\Nat}{\mathbb{N}}
\newcommand{\Int}{\mathbb{Z}}
\newcommand{\proj}{\mathrm{p}}
\newcommand{\True}{\mathrm{T}}
\newcommand{\False}{\mathrm{F}}
\newcommand{\id}{\mathrm{id}}
\newcommand{\dom}{\mathrm{dom}}
\newcommand\Hom{\mathop{\rm Hom}}
\newcommand\Idirected{\mathcal{I}_{\mathrm{d}}}
\newcommand\compact{\mathrm{K}}

\newcommand{\definition}[1]{\textit{#1}}
\newcommand{\proof}{{\textit{Proof}\enspace}}

\newcommand{\eop}{\ \vbox{\hrule
                       \hbox{\vrule
                             \hskip 6pt
                             \vrule height 6pt width 0pt
                             \vrule}%
                       \hrule}%
                     \vspace{\medskipamount}
                }

\newcommand{\oneto}[2]{#1 = 1,\, \ldots ,\, #2}
\newcommand{\svector}[2]{#1_1, \ldots , #1_{#2}}
\newcommand{\lvector}[2]{#1_1,\, \ldots , \, #1_{#2}}
\newcommand{\llist}[2]{#1_1 \cdots #1_{#2}}
\newcommand{\Oneptset}{\mathbb{I}}
\newcommand{\BOneptset}{\mathbb{T}}
\newcommand{\onept}{\varepsilon}

\newcommand{\SynInt}{\underline{Int}}
\newcommand{\sqle}{\sqsubseteq}

\newcommand{\lub}{\bigsqcup}
\newcommand{\directed}[1]{\mathrm{d}(#1)}
\newcommand{\total}[2]{#2^{#1}}
\newcommand\Bot[1]{\mathsf{bot}\left(#1\right)}

\newcommand{\blob}{\hskip 1pt \vbox{\hrule
                       \hbox{\vrule
                             \hskip 7pt
                             \vrule height 5pt width 0pt
                             \vrule}%
                       \hrule}%
           \hskip 1pt }

\newsavebox{\ttt}
\sbox{\ttt}{}
\newcommand{\startsection}[1]
    {\section{#1}
    \sbox{\ttt}{\thesection\ \ \textsc{#1}}
    \thispagestyle{plain}
}

\newcommand{\startchapter}[1]
    {\chapter{#1}
    \sbox{\ttt}{}}

\newtheorem{lemma}{Lemma}[section]
\newtheorem{proposition}{Proposition}[section]

\makeindex

\begin{document}

\author{Chris Preston}
\title{Specifying Data Objects with Initial Algebras}
\date{}
\maketitle

\frontmatter

\include{sbikapre}

\tableofcontents

\mainmatter
\include{sbikaint}
\include{sbika2}
\include{sbika3}
\include{sbika4}
\include{sbika5}
\include{sbika6}

\backmatter
\sbox{\ttt}{\textsc{Bibliography}}
\thispagestyle{plain}
\addcontentsline{toc}{chapter}{Bibliography}

\include{sbikabib}

\end{document}

%% file: sbikapre.tex
This study presents a systematic approach to specifying data objects with the help
of initial algebras. 
The primary aim is to describe the set-up to be found in modern functional programming 
languages such as \textit{Haskell} and \textit{ML}, although it can also 
be applied to more general situations.

The `initial algebra semantics' philosophy has been propagated by the ADJ group
consisting of J.A.\ Goguen, J.W.\ Thatcher, E.G.\ Wagner and J.B.\ Wright, for example 
in \cite{ADJ1}, and is now well-established. The approach presented here can be seen
as pushing this philosophy a stage further and consists of taking the following four steps. 

\begin{evlist}{25pt}{0.8ex}
\item[(1)] Data types are specified by a signature and the `fully-defined'
data objects are then described by the carrier sets in an initial algebra. 
\item[(2)] The initial algebra is extended to include `undefined' and 
`partially defined' data objects, leading to what is known as a bottomed algebra.
The correct bottomed algebra depends on the language being considered. However,
it can always be defined to be an initial object in a class of
bottomed algebras determined by a simple structure which we call a head type.
This set-up can deal with both lazy and strict languages as well as anything in between.
\item[(3)] 
The third step is based on the observation that 
there is a unique family of partial orders defined on the carrier sets 
of the initial bottomed algebra such that each partial order can
reasonably be interpreted as meaning `being less-defined than'. This leads to what are
called ordered algebras; these have partially ordered carrier sets and monotone
operators. The initial bottomed algebra turns out to also be an initial object
in an appropriate class of ordered algebras.
\item[(4)] 
The final step involves what is called the initial (or ideal) completion of a partially 
ordered set. This is used to  complete the ordered algebra to end up with a
continuous algebra
having complete partially ordered carrier sets and continuous operators.
The resulting continuous algebra is again an initial object in the appropriate
class of continuous algebras.
\end{evlist}
The bottomed, ordered and continuous algebras which occur here are uniquely determined
up to isomorphism by the signature and the head type, both of which
are essentially finite structures in any practical case.

The account includes a treatment of polymorphism, although to simplify things the
approach is a bit more restrictive than that to be found in languages such as 
\textit{Haskell} or \textit{ML}. Moreover, the polymorphic types here do not 
involve functional types, which means that data types such as lists whose components are 
functions are not allowed.

I have tried to keep the account self-contained, and have thus included 
all the standard results (together with their proofs) which are needed from universal 
algebra and the theory of partially ordered sets. It is not assumed that the reader knows 
anything about functional programming, but some experience
of this topic would, of course, not be amiss. 

This account is an expanded version of part of 
\textit{Computing with Equations} \cite{preston}, some notes I wrote about the
semantics of functional programming languages.

\vspace{2cm}
Chris Preston\\
Bielefeld\\ 
March 2003


%% file: sbikaint.tex
\startchapter{Introduction}
\label{intro}

In all modern functional programming languages the programmer can introduce new 
data types by what amounts to specifying a signature. Consider the following
simple example of how a signature can be represented in most such languages:
\begin{eelist}{120pt}
\item $\mathtt{nat\ ::=\ Zero\ |\ Succ\ nat}$
\item $\mathtt{pair\ ::=\ Pair\ nat\ nat}$
\item $\mathtt{list\ ::=\ Nil\ |\ Cons\ nat\ list}$
\end{eelist}
This declares three new types with the names $\mathtt{nat}$, $\mathtt{pair}$ and 
$\mathtt{list}$ together with five operator names $\mathtt{Zero}$, $\mathtt{Succ}$,
$\mathtt{Pair}$, $\mathtt{Nil}$ and $\mathtt{Cons}$, each with its functionality
(i.e., how many arguments it takes, the type of each argument and the type of the result, 
for instance $\mathtt{Cons}$ takes two arguments, the first of type
$\mathtt{nat}$, the second of type $\mathtt{list}$, with
$\mathtt{list}$ the type of the result).

Now the reason for giving a signature is because there is then a corresponding
class of algebras. In the example an algebra is any collection of three sets
$X_{\mathtt{nat}}$, $X_{\mathtt{pair}}$ and $X_{\mathtt{list}}$ together with five
mappings $p_{\mathtt{Zero}} : \Oneptset \to X_{\mathtt{nat}}$,
$p_{\mathtt{Succ}} : X_{\mathtt{nat}} \to X_{\mathtt{nat}}$,
$p_{\mathtt{Pair}} : X_{\mathtt{nat}} \times X_{\mathtt{nat}} \to X_{\mathtt{pair}}$,
$p_{\mathtt{Nil}} : \Oneptset \to X_{\mathtt{list}}$ and
$p_{\mathtt{Cons}} : X_{\mathtt{nat}} \times X_{\mathtt{list}} \to X_{\mathtt{list}}$,
where $\Oneptset$ is a set with one element (arising as an empty product) and where it 
should be clear how the domain and codomain of each mapping is determined by the 
functionality laid down in the signature. Such an algebra will be denoted for short
just by $(X,p)$.

Each algebra can be regarded as a realization of the signature. The sets 
$X_{\mathtt{nat}}$, $X_{\mathtt{pair}}$ and $X_{\mathtt{list}}$ are respectively sets of
data objects for the types $\mathtt{nat}$, $\mathtt{pair}$ and 
$\mathtt{list}$, the mappings $p_{\mathtt{Zero}}$ and $p_{\mathtt{Nil}}$ define
constants (namely the single elements in their images) and the mappings 
$p_{\mathtt{Succ}}$, $p_{\mathtt{Pair}}$ and 
$p_{\mathtt{Cons}}$ construct new data objects out of objects which have already been 
defined. 

The problem is now to decide which algebra should be considered as the `correct'
realization of the signature. Of course, in the example the names $\mathtt{nat}$, 
$\mathtt{pair}$ and $\mathtt{list}$ suggest perhaps singling out the algebra with 
$X_{\mathtt{nat}} = \Nat$, $X_{\mathtt{pair}} = \Nat^2$ and $X_{\mathtt{list}} = \Nat^*$  
(the set of all lists of elements from $\Nat$) and with $p_{\mathtt{Zero}}(\onept) = 0$,
$p_{\mathtt{Succ}}(n) = n+1$, $p_{\mathtt{Pair}}(m,n) = (m,n)$,
$p_{\mathtt{Nil}}(\onept) = \onept$ and $p_{\mathtt{Cons}}(n,s) = n \triangleleft s$, where
$\onept$ denotes both the single element in $\Oneptset$ and the empty list, and
$n \triangleleft s$ is the operation of adding the element $n$ to the beginning of the
list $s$. But there are clearly many other algebras which look nothing like
this particular choice, for instance with each occurrence of $\Nat$ replaced by $\Real$,
or with each of the sets $X_{\mathtt{nat}}$, $X_{\mathtt{pair}}$ and $X_{\mathtt{list}}$ 
consisting of just a single element. 

However, there is really only one natural choice for the `correct' realization
and this is to take a so-called initial algebra. An algebra $(X,p)$ is initial if for 
each algebra
$(X',p')$ there is a unique homomorphism from $(X,p)$ to $(X',p')$, where a homomorphism is 
a structure-preserving family of mappings. In the example it would consist of
three mappings $\pi_{\mathtt{nat}} : X_{\mathtt{nat}} \to X'_{\mathtt{nat}}$,
$\pi_{\mathtt{pair}} : X_{\mathtt{pair}} \to X'_{\mathtt{pair}}$ and
$\pi_{\mathtt{list}} : X_{\mathtt{list}} \to X'_{\mathtt{list}}$
satisfying five equations, one for each of the operator names with, for instance,
the equation for $\mathtt{Cons}$ requiring that
\[\pi_{\mathtt{list}}(p_{\mathtt{Cons}}(x,s))
= p'_{\mathtt{Cons}}(\pi_{\mathtt{nat}}(x),\pi_{\mathtt{list}}(s))\]
should hold for all $x \in X_{\mathtt{nat}}$, $s \in X_{\mathtt{list}}$.

It turns out that an initial algebra exists for each signature 
(and the algebra singled out above is initial). 
Moreover, any two initial algebras are isomorphic and so an initial algebra is uniquely 
determined, up to isomorphism, by the signature. 
Furthermore, there is an explicit class of initial algebras,
the term algebras, which are used in all functional programming languages to represent
data objects. In a term algebra $(E,\blob)$ the elements in the sets $E_b$ are terms
(or expressions), usually written in prefix form. For instance, in the example 
$\mathtt{Zero}$ and $\mathtt{Succ\ Succ\ Zero}$ are elements of
$E_{\mathtt{nat}}$, $\mathtt{Pair\ Succ\ Zero\ Zero}$ an element of
$E_{\mathtt{pair}}$ and 
$\mathtt{Nil}$ and $\mathtt{Cons\ Zero\ Cons\ Succ\ Zero\ Nil}$ elements of
$E_{\mathtt{list}}$. 
(If the reader prefers to use braces to increase the legibility then the terms above 
with more then one element become
$\mathtt{Succ\ (Succ\ Zero)}$,
$\mathtt{Pair\ (Succ\ Zero)\ Zero}$ 
and $\mathtt{Cons\ Zero\ (Cons\ (Succ\ Zero)\ Nil)}$.)

Now since term algebras are initial, there is a unique isomorphism from a term algebra
$(E,\blob)$ to any other initial algebra, and in particular to an initial algebra 
$(X,p)$ being used as the `correct' realization of the signature. This means that
each data object then has a unique representation as a term; for instance, in the
algebra singled out above, the data object $3 \in \Nat = X_{\mathtt{nat}}$
is uniquely represented by the term $\mathtt{Succ\ Succ\ Succ\ Zero} \in E_{\mathtt{nat}}$.
The results about algebras, and in particular about initial algebras, which are needed to 
make all this precise are developed in Chapter~\ref{uni_alg}.
 
The process outlined above is, however, only the first of several steps which are required
to completely specify the data objects, and we now consider the second step.
In any programming language data objects are manipulated by algorithms, and in any
non-trivial language it is an unavoidable fact that algorithms need not terminate. It is 
thus necessary to introduce an `undefined' element for each type in order to represent 
this state of affairs. Moreover, depending on the language, it may also be necessary
to have `partially defined' data objects, for example an element of type
$\mathtt{pair}$ in which the first component of a pair is defined but not the second,
or an element of type $\mathtt{list}$ in which only some of the components in a list 
are defined. To deal with this situation bottomed algebras will be introduced. These
are algebras containing for each type a special bottom element to denote an 
`undefined' element of the type. (The bottom element will always be denoted by the symbol
$\bot$, usually with a subscript to indicate which type is involved.)

The simplest way to obtain a bottomed algebra is to start with an ordinary algebra
$(X,p)$, add an `undefined' element $\bot_b$ to $X_b$ for each type $b$, and then extend 
each operator so that it produces an `undefined' value as soon as one of its arguments 
is `undefined'. This bottomed algebra is called the flat bottomed extension of $(X,p)$, 
and by definition it does not contain any `partially defined' objects. 

The other extreme is an initial bottomed algebra, where a bottomed algebra $(X,p)$ is 
initial if for each bottomed algebra $(X',p')$ there is a unique bottomed homomorphism 
from $(X,p)$ to $(X',p')$, and where a bottomed homomorphism is a homomorphism which maps 
bottom elements to bottom elements. As with ordinary algebras, there exist initial 
bottomed algebras and they are unique up to isomorphism.  Moreover, in some sense they 
contain all possible `partially defined' objects. One way to obtain an initial bottomed
algebra is as a term algebra with the bottom elements added as constant terms.
For the example this means that an initial bottomed algebra is given by the 
term algebra for the extended signature represented by
\begin{eelist}{100pt}
\item $\mathtt{nat\ ::=\ Zero\ |\ Succ\ nat\ |\ }\bot_{\mathtt{nat}}$
\item $\mathtt{pair\ ::=\ Pair\ nat\ nat\ |\ }\bot_{\mathtt{pair}}$
\item $\mathtt{list\ ::=\ Nil\ |\ Cons\ nat\ list\ |\ }\bot_{\mathtt{list}}$
\end{eelist}
The `correct' choice of a bottomed algebra depends on the application in hand.
The semantics of most modern functional programming languages (such as \textit{ML} or 
\textit{Haskell}) require an initial bottomed algebra, whereas for a language such as 
\textit{Lisp} the flat bottomed extension is the appropriate choice.

In Chapter~\ref{bot_algs} we develop a framework for dealing with bottomed algebras.
There is a condition, called regularity, which plays an important role here,
and requires that each element in the set $X_b \setminus \{\bot_b\}$ have a unique
representation of the form $p_k(\svector{x}{n})$. In particular, regularity is needed
in order to define the `pattern matching' and `case' operators which occur in all
typed functional programming languages. We introduce the concept of a head type to deal 
with the various kinds of bottomed algebras which could arise when specifying the semantics
of a programming language. To each head type there is a corresponding class of bottomed 
algebras and this class possesses an initial element which is regular and is,
as usual, unique up to isomorphism. This means that such an initial algebra is
essentially uniquely determined by the signature and the head type.
Moreover, the head types arising in practice are finite structures. This set-up includes
both flat bottomed extensions and initial bottomed algebras as special cases.

The third step in our programme of completely specifying data objects arises from the
observation that if $(X,p)$ is a regular bottomed algebra then for each type $b$
there is a unique partial order $\sqle_b$ on the set $X_b$ such that $x \sqle_b x'$
can reasonably be interpreted as meaning that $x$ is less-defined than $x'$.
In particular, this should mean that $\bot_b \sqle x$ for each $x \in X_b$ and that
each of the mappings $p_k$ be monotone (using the corresponding product order 
on the domain of $p_k$).

These partial orders are essential when considering algorithms which 
manipulate data objects as dynamical systems. If an algorithm
is designed to compute an element of type $b$ then at each point of time there is an 
element of $X_b$ which describes the present state of the computation.
For any reasonable algorithm this state will be monotone as a function of 
time and should converge in some sense to the expected `answer'. 
Now it is possible that an algorithm fails to terminate, not because it fails 
to produce an answer but because the answer is an infinite structure which cannot be
computed completely in finite time. Typical examples here are algorithms 
which produce infinite lists, for instance a list consisting of all the prime numbers. 
In order to deal with this situation it is natural to complete the partially ordered
set $(X_b,\sqle_b)$ for each type $b$ to include all data objects which can arise as
limits of finite computations. There is an appropriate completion which fits in with
this interpretation and is called either the initial or the ideal completion of the
partially ordered set.

The existence of the partial orders and the completion of the partially ordered sets are 
dealt with in Chapter~\ref{ord_cont_algs}. As preparation for this some general results 
about partial orders are first presented in Chapter~\ref{domains}.

There is an important aspect of specifying data objects which
has not yet been discussed, and this is  polymorphism. 
In the signature being used as an example there is a type $\mathtt{list}$ for lists 
having components of type $\mathtt{nat}$.
Now if lists having components of, say, type $\mathtt{pair}$ need 
to be implemented then the signature would have to be extended by adding a new type, say 
$\mathtt{listp}$, together with two new operator names, say $\mathtt{Nilp}$ and
$\mathtt{Consp}$, whose functionalities are given by 
\begin{eelist}{110pt}
\item $\mathtt{listp\ ::=\ Nilp\ |\ Consp\ pair\ listp}$
\end{eelist}
Moreover, the signature has to be extended in essentially the same way with a new list 
type and two new operator names for each type for which lists are required. This is 
clearly not very satisfactory. What is  needed is the possibility of defining once and 
for all lists of an arbitrary type, and this is a feature of all modern functional 
programming languages. Our example signature could, for instance, be replaced by something
like the following:
\begin{eelist}{110pt}
\item $\mathtt{nat\ ::=\ Zero\ |\ Succ\ nat}$
\item $\mathtt{pair\ a\ b\ ::=\ Pair\ a\ b}$
\item $\mathtt{list\ c\ ::=\ Nil\ |\ Cons\ c\ (list\ c)}$
\end{eelist}
The names $\mathtt{a}$, $\mathtt{b}$ and $\mathtt{c}$ are called type variables
and each can by replaced by a type to obtain a compound type. For example,
there are the compound types $\mathtt{pair\ nat\ nat}$ and $\mathtt{list\ nat}$
corresponding to the types $\mathtt{pair}$ and $\mathtt{list}$ in the original signature.
Arbitrarily complicated compound types can be generated, for example
$\mathtt{pair\ (list\ (list\ nat))\ (pair\ nat\ nat)}$, which is a type whose objects
are pairs with first component a list whose components are lists with components of type
$\mathtt{nat}$, and with second component a pair with both components of type
$\mathtt{nat}$.  
Signatures involving type variables and the corresponding algebras are treated in
Chapter~\ref{poly}.

In order to give a foretaste of one of the main results in Chapter~\ref{uni_alg} 
and to make some of the concepts introduced above a bit more precise
we end the Introduction by considering the following very simple signature
\begin{eelist}{110pt}
\item $\mathtt{nat\ ::=\ Zero\ |\ Succ\ nat}$
\end{eelist}
In this case an algebra consists of just a set $X_{\mathtt{nat}}$ together with two 
mappings $p_{\mathtt{Zero}} : \Oneptset \to X_{\mathtt{nat}}$ and
$p_{\mathtt{Succ}} : X_{\mathtt{nat}} \to X_{\mathtt{nat}}$, and is called
a\index{natural number algebra}\index{algebra!natural number}
\definition{natural number algebra}. It is convenient to represent such an
algebra in the form $(X_{\mathtt{nat}},p_{\mathtt{Nil}}(\onept),p_{\mathtt{Cons}})$, 
thus natural number algebras are exactly triples
$(X,e,p)$ consisting of a non-empty set $X$, an element $e \in X$ and a mapping 
$p : X \to X$. 
Of course, one such algebra is $(\Nat,0,\mathsf{s})$, where
$\mathsf{s}(n) = n + 1$ for each $n \in \Nat$ (and hence the name natural number
algebra). It can now be asked what is special about 
the algebra $(\Nat,0,\mathsf{s})$. 

To answer this question homomorphisms must be considered: 
If $(X,e,p)$ and $(X',e',p')$ are natural number algebras then a homomorphism
is here a mapping $\pi : X \to X'$ such that
$\pi(e) = e'$ and $p'\comp\pi = \pi\comp p$ (i.e., $p'(\pi(x)) = \pi(p(x))$ for all 
$x \in X$). Such a homomorphism $\pi$ is an isomorphism (i.e., 
there exists a homomorphism $\pi'$ from 
$(X',e',p')$ to $(X,e,p)$ with $\pi \comp \pi' = \id_{X'}$ and $\pi' \comp \pi = \id_X$) 
if and only if $\pi$ is a bijection, and then $\pi'$ is just the set-theoretic inverse 
$\pi^{-1}$ of $\pi$. In this case $(X,e,p)$ are $(X',e',p')$ said to be 
\index{isomorphic}\definition{isomorphic}.

The identity mapping is clearly a homomorphism and the composition of two homomorphisms is 
again a homomorphism; being isomorphic thus defines an equivalence relation on the class 
of all natural number algebras. Therefore if the equivalence class containing 
$(\Nat,0,\mathsf{s})$ can be identified in some reasonable way then the question asked 
above can be considered to have been answered satisfactorily. Two simple characterisations 
of this equivalence class are given below.

In the same way as before, a natural number algebra $(X,e,p)$ is said to be 
\definition{initial}\index{natural number algebra!initial} if for each such 
algebra $(X',e',p')$ there is a unique homomorphism from $(X,e,p)$ to $(X',e',p')$. Then
in particular $(\Nat,0,\mathsf{s})$ is initial, since given $(X',e',p')$, a homomorphism 
from
$(\Nat,0,\mathsf{s})$ to $(X',e',p')$ can be defined by induction, and its uniqueness also 
follows by induction. It follows that an algebra is isomorphic to $(\Nat,0,\mathsf{s})$ if 
and only if it is initial. In other words, the equivalence class containing 
$(\Nat,0,\mathsf{s})$ consists of exactly all the initial natural number algebras. This was 
the first characterisation.

Here is the second characterisation: A natural number algebra $(X,e,p)$ will be called a 
\definition{Peano triple}\index{Peano triple} if the following three conditions are 
satisfied:
\begin{evlist}{25pt}{0.5ex}
\item[(1)] The mapping $p$ is injective.
\item[(2)] $p(x) \ne e$ for all $x \in X$.
\item[(3)] The only subset $X'$ of $X$ containing $e$ with $p(x) \in X'$ for all 
$x \in X'$ is the set $X$ itself.
\end{evlist}
Then $(\Nat,0,\mathsf{s})$ is a Peano triple. (This is one of the possible 
formulations of the Peano axioms. In particular, the statement that $(\Nat,0,\mathsf{s})$ 
satisfies (3) is nothing but the principle of mathematical induction.) The reader is left 
to show that an algebra is isomorphic to $(\Nat,0,\mathsf{s})$ if and only if it is itself 
a Peano triple. In other words, the equivalence class containing $(\Nat,0,\mathsf{s})$ also 
consists of exactly all the Peano triples.

Of course, a corollary of these two characterisations is that a natural number algebra is 
initial if and only if it is a Peano triple. This is a special case of an important result 
presented in Section~\ref{init_algs} which states that an algebra is initial if and only 
if (in the terminology employed there) it is unambiguous and minimal. For a general algebra
unambiguity corresponds to conditions (1) and (2) in the definition of a Peano triple, 
while minimality corresponds to condition (3). This characterisation is sometimes 
expressed by saying that initial algebras are exactly those for which there is 
\textit{no confusion} (unambiguity) and \textit{no junk} (minimality).
The analysis of natural number algebras given here is essentially that to be found in 
Dedekind's book \textit{Was sind und was sollen die Zahlen?} first published in 1888.


%% file: sbika2.tex
\startchapter{Some universal algebra}
\label{uni_alg}

The material presented in this chapter is all part of standard 
universal algebra. The classical field of universal algebra 
deals with the case of signatures having a single type, and so
the only information required about each operator name is the
number of arguments it takes. In this form the main problems were
stated in Whitehead \cite{whitehead} and solved in Birkhoff \cite{birkhoff35};
standard texts are Cohn \cite{cohn} and Gr\"atzer \cite{graetzer}.
The generalisation of the theory to\index{multi-sorted algebra}
\index{algebra!multi-sorted}\definition{multi-sorted algebras} (i.e., to the algebras as 
they occur here) was made by Higgins \cite{higgins} and Birkhoff and 
Lipson \cite{birkhoff_lipson}. 
Birkhoff and Lipson (who speak of 
\definition{heterogeneous algebras})
\index{heterogenous algebra}\index{algebra!heterogenous}showed that essentially the 
whole of the classical theory carries over to the more general case.  

Some of the first uses of multi-sorted algebras in computer science can be
found in Maibaum \cite{maibaum} and Morris \cite{morris_fl}.
Their systematic use has been propagated by the 
ADJ group consisting of J.A.\ Goguen, J.W.\ Thatcher, E.G.\ Wagner 
and J.B.\ Wright, for example in the papers Goguen, Thatcher, Wagner and 
Wright \cite{ADJ1} and Goguen, Thatcher and Wagner \cite{ADJ2}. 
The emphasis here is very much on initial algebras in their various forms. The choice of 
material in this chapter is determined entirely by what will be needed later. The reader 
interested in a more balanced account of modern universal algebra should consult the books
mentioned above.

\startsection{Sets of various kinds}
\label{sets}  

Before beginning in the next section with universal algebra proper the simple notions of a 
typing and of a bottomed set will be introduced here. Moreover, at the end of the 
section we say something about initial objects.

The first task, however, is to fix some notation. The empty set will be denoted by
$\varnothing$ and the set $\{\varnothing\}$ by $\Oneptset$; thus $\Oneptset$ is the
`standard' set containing exactly one element. However, to avoid confusion
the single element in $\Oneptset$ will be denoted by $\onept$ (rather than by
$\varnothing$). The set $\{\True,\False\}$ of boolean values will be denoted by $\Bool$.
The set $\Nat$ of natural numbers is the set $\{0,1,2,\ldots\,\}$ (so  $0$ is considered 
to be a natural number). For each $n \in \Nat$ let $[n] = \{1,2,\ldots,n\}$; in particular
$[0] = \varnothing$.

The words \definition{function}\index{function}
and \definition{mapping}\index{mapping} are considered to be synonyms. 
Let $f : X \to Y$ be a mapping; then $X$ is called the \definition{domain}\index{domain}
of $f$ and will be denoted by $\dom(f)$ and $Y$ is called the 
\definition{codomain}\index{codomain}.
For each $A \subset X$ put
\[ f(A) = \{ y \in Y : y = f(x)\ \textrm{for some}\ x \in A \}\]
and for each $B \subset Y$ put $f^{-1}(B) = \{ x \in X : f(x) \in B \}$. The subset 
$f(X)$ of the codomain $Y$ is called the \definition{image}\index{image} of $f$ and will 
be denoted by 
$\Im(f)$. If $f : X \to Y$ is a mapping and $A \subset X$ then $\rest{f}{A}$ will be used
to denote the \definition{restriction}\index{restriction} of $f$ to $A$, thus 
$\rest{f}{A} : A \to Y$ is the mapping
given by $\rest{f}{A}(x) = f(x)$ for all $x \in A$.
If $f : X \to Y$ and $g : Y \to Z$ are mappings then their composition will be 
denoted by $g\comp f$, i.e., $g\comp f : X \to Z$ is the mapping defined by 
$(g\comp f)(x) = g(f(x))$ for all $x \in X$.

If $X$ and $Y$ are sets then $\total{X}{Y}$ will be used to denote the set of all mappings 
from $X$ to $Y$. In particular, $\total{\varnothing}{Y} = \Oneptset$ for each set $Y$.
If $\alpha : X \to Y$ and $J$ is a further set then $\alpha^J$ will be used to 
denote the induced mapping from $X^J$ to $Y^J$ defined by $\alpha^J(v) = \alpha \comp v$
for all $v \in X^J$.

Let $S$ be a set and $\mathcal{C}$ be some class of objects (such as the class of all 
sets). Then by an \definition{$S$-family}\index{family} of objects from $\mathcal{C}$ is 
just
meant a mapping $\alpha : S \to \mathcal{C}$. Such a family is said to be 
\definition{finite}\index{family!finite} if the set $S$ is finite. The usual convention 
for families
will be followed in that the value of $\alpha$ applied to the argument $s$ will be denoted
by $\alpha_s$ rather than by $\alpha(s)$. If $\alpha : S \to \mathcal{C}$ is an $S$-family
and $A \subset S$ then the restriction of $\alpha$ to $A$ will be denoted (as for mappings)
by $\rest{\alpha}{A}$, thus $\rest{\alpha}{A} : A \to \mathcal{C}$ is the $A$-family
with $(\rest{\alpha}{A})_a = \alpha_a$ for all $a \in A$.

It is useful to introduce some special notation for families of sets, i.e.,
for families $X : S \to \mathsf{Sets}$ with $\mathsf{Sets}$ the class of all sets.
If $X$ and $Y$ are $S$-families of sets then $Y \subset X$ will mean 
that $Y_s \subset X_s$ for each $s \in S$. This will also be indicated by saying that 
$X$ \definition{contains} $Y$ or that $Y$ \definition{is contained in} $X$.
Moreover, $X \cap Y$ and $X \cup Y$ are the $S$-families of sets defined
component-wise, i.e., $(X\cap Y)_s = X_s \cap Y_s$ and $(X\cup Y)_s = X_s \cup Y_s$ for
each $s \in S$, and $\varnothing$ will be used to denote the
$S$-family of sets with $\varnothing_s = \varnothing$ for each $s \in S$.

If $X$ is a finite $S$-family of sets then $\utimes{X}$ will be used to 
denote the \definition{cartesian product}\index{cartesian product}\index{product!cartesian}
 of the sets in the family $X$: This is 
defined to be the set of all mappings $v : S \to \bigcup_{s \in S} X_s$ such that 
$v(s) \in X_s$ for each $s \in S$. (Of course, this definition makes sense when $S$
is infinite, but we will only need finite products.) Note that $\utimes{X} = \Oneptset$ 
if $S = \varnothing$; moreover, if $Y \subset X$ then $\utimes{Y}$ can clearly be 
regarded as a subset of $\utimes{X}$.

Let $n \in \Nat$ and for each $\oneto{j}{n}$ let $X_j$ be a set; then
there is the $[n]$-family $X$ and thus the cartesian product $\utimes{X}$. This set 
will of course be denoted by $X_1 \times \cdots \times X_n$; it is the set of all mappings
$\varrho$ from $[n]$ to $\bigcup_{j=1}^n X_j$ such that $\varrho(j) \in X_j$ for each $j$.
In particular, if $n = 0$ then $X_1 \times \cdots \times X_n = \Oneptset$. For each $j$ 
let $x_j \in X_j$; then as usual $(\svector{x}{n})$ denotes the the element $\varrho$ of 
$X_1 \times \cdots \times X_n$ such that $\varrho(j) = x_j$ for each $j$; each 
element of $X_1 \times \cdots \times X_n$ has a unique representation of this form.

The simplest case of a cartesian product is when the sets $\lvector{X}{n}$ are all the 
same: Let $X$ be a set; then for each $n \in \Nat$ the $n$-fold cartesian product of $X$ 
with itself will be denoted by $X^n$, thus $X^n$ is the set of all mappings from $[n]$ to 
$X$. In particular, $X^0 = \Oneptset$; moreover, it is convenient to identify $X^1$ with 
$X$ in the obvious way. 

For each set $X$ the set $\bigcup_{n \ge 0} X^n$ will be denoted by $X^*$, which should 
be thought of as the set of all finite lists of elements from $X$. Note that since $X^1$ 
is being identified with $X$, it follows that $X$ is a subset of $X^*$. (In other words, 
each element $x$ of $X$ is identified with the list whose single component is equal to $x$.)
If $m \ge 1$ then the element $(\svector{x}{m})$ of $X^m \subset X^*$ will always be 
denoted simply by $\llist{x}{m}$.

If $S$ is a set then by an $S$-family of mappings is meant a mapping 
$\varphi : S \to \mathsf{Maps}$ with $\mathsf{Maps}$ the class of all mappings
between sets. Let $\varphi : S \to \mathsf{Maps}$ be such an $S$-family of mappings. Then
for each $s \in S$ there exist sets $X_s$ and $Y_s$ such that 
$\varphi_s : X_s \to Y_s$ and in this case we write $\varphi : X \to Y$.
More precisely, the statement that $\varphi : X \to Y$ is an 
$S$-family of mappings means that $X$ and $Y$ are $S$-families of sets and
$\varphi$ is an $S$-family of mappings with $\varphi_s : X_s \to Y_s$ for each
$s \in S$. If $Z$ is a further $S$-family of sets and $\psi : Y \to Z$
a further $S$-family of mappings then the $S$-family of composed mappings
will be denoted by $\psi\fcomp\varphi$, thus $\psi\fcomp\varphi : X \to Z$ is the
$S$-family of mappings with $(\psi\fcomp\varphi)_s =  \psi_s\comp\varphi_s$ for each
$s \in S$. 

Now let $S$ be a finite set, $X$ and $Y$ be $S$-families of sets and 
$\varphi : X \to Y$ be an $S$-family of mappings. Then there is a mapping 
$\utimes{\varphi} : \utimes{X} \to \utimes{Y}$ defined by
\[\utimes{\varphi} (v)(s) = \varphi_s (v(s))\]
for each $v \in \utimes{X}$, $s \in S$. Let $n \in \Nat$ and for each $\oneto{j}{n}$ let 
$\varphi_j : X_j \to Y_j$ be a mapping; 
then $\utimes{\varphi}$ is just the mapping from $X_1 \times \cdots \times X_n$ to 
$Y_1 \times \cdots \times Y_n$ defined for each 
$(\svector{x}{n}) \in X_1 \times \cdots \times X_n$ by
\[\utimes{\varphi}(\svector{x}{n}) = (\varphi_1(x_1),\ldots,\varphi_n(x_n))\; .\]

\newpage

\begin{lemma}\label{lemma_sets_1}
Let $S$ be a finite set.

(1)\enskip Let $X$ be an $S$-family of sets and $\id : X \to X$ be the $S$-family of 
identity mappings (i.e., with $\id_s : X_s \to X_s$ the identity mapping for each 
$s \in S$). Then $\utimes{\id} : \utimes{X} \to \utimes{X}$ is also the identity 
mapping. 

(2)\enskip Let $X$, $Y$ and $Z$ be $S$-families of sets and let 
$\varphi : X \to Y$ and $\psi : Y \to Z$ be $S$-families of mappings. Then 
$\utimes{(\psi\fcomp\varphi)} = \utimes{\psi} \comp \utimes{\varphi}$.

(3)\enskip Let $X$ and $Y$ be $S$-families of sets and let $\varphi : X \to Y$ be 
an $S$-family of mappings. If the mapping $\varphi_s : X_s \to Y_s$ is injective 
(resp.\ surjective) for each $s \in S$ then 
the mapping $\utimes{\varphi} : \utimes{X} \to \utimes{Y}$ is also injective 
(resp.\ surjective). In particular, if $\varphi_s : X_s \to Y_s$ is a bijection for 
each $s \in S$ then the mapping $\utimes{\varphi}$ is a bijection, and in this case 
$(\utimes{\varphi})^{-1} = \utimes{\varphi^{-1}}$, where $\varphi^{-1}$ is the
$S$-family of mappings with $(\varphi^{-1})_s = (\varphi_s)^{-1}$ for each $s \in S$.
\end{lemma}

\proof (1)\enskip This is clear.  

(2)\enskip Let $v \in \utimes{X}$ and $s \in S$; then
\begin{eqnarray*}
 \utimes{(\psi\fcomp\varphi)}(v)(s)
 &=& (\psi_s \comp\varphi_s )(v(s)) = \psi_s (\varphi_s (v(s)))\\
&=& \psi_s (\utimes{\varphi} (v))(s)
 = \utimes{\psi} (\utimes{\varphi}(v))(s)
 = (\utimes{\psi} \comp \utimes{\varphi})(v)(s)
\end{eqnarray*}
i.e., $\utimes{(\psi\fcomp\varphi)} = \utimes{\psi}\comp\utimes{\varphi}$.

(3)\enskip If $\varphi_s$ is injective for each $s$ then there is an $S$-family of mappings 
$\psi : Y \to X$ such that $\psi_s\comp\varphi_s$ is the identity mapping on $X_s$
for each $s \in S$. Then by (1) $\utimes{(\psi\fcomp\varphi)}$ is the identity mapping 
on $\utimes{X}$ and by (2) 
$\utimes{\psi} \comp\utimes{\varphi} = \utimes{(\psi\fcomp\varphi)}$. Hence 
$\utimes{\varphi}$ is injective. The other case is almost identical: If $\varphi_s$ is 
surjective for each $s$ then there exists an $S$-family of mappings $\psi : Y \to X$ 
such that $\varphi_s\comp\psi_s$ is the identity mapping on $Y_s$ for each $s \in S$, and 
as in the first part it then follows that $\utimes{\varphi} \comp\utimes{\psi}$ is 
the identity mapping on $\utimes{Y}$. Finally, if $\varphi_s$ is a bijection for each 
$s \in S$ then $\utimes{\varphi^{-1}}\comp\utimes{\varphi}$ is the identity 
mapping on $\utimes{X}$ and $\utimes{\varphi}\comp\utimes{\varphi^{-1}}$ is 
the identity mapping on $\utimes{Y}$, and thus 
$(\utimes{\varphi})^{-1} = \utimes{\varphi^{-1}}$. \eop

We now come to the notion of a typing. This takes into account the situation met with
in most modern programming languages in which each name occurring in a program is assigned,
either explicitly or implicitly, a type. The kind of object a name can refer to is then 
determined by its type. If $S$ is a set then a mapping $\gamma : I \to S$ will be called an 
\definition{$S$-typing}\index{typing}. The set $S$ should here be thought of as a set of 
`types', 
$I$ as a set of `names' and $\gamma$ as specifying the type of objects to which the `names' 
can be assigned. If the set $I$ is finite then $\gamma$ is called a 
\definition{finite}\index{typing!finite}
$S$-typing and the class of all finite $S$-typings will be denoted by $\ftyped{S}$.
In accordance with our previous notation the unique $S$-typing $\gamma$ with 
$\dom(\gamma) = \varnothing$ will be denoted by $\onept$. Note that there is an obvious
one-to-one correspondence between $\Oneptset$-typings and sets, since for each set $I$
there is a unique mapping from $I$ to $\Oneptset$.
In particular, $\ftyped{\Oneptset}$ can be considered as the class $\mathsf{FSets}$ of all
finite sets.

If $\gamma : I \to S$ is an $S$-typing then the pair $(I,\gamma)$ will be referred to as 
an \definition{$S$-typed set}\index{typed set}\index{set!typed}. Of course, there is no 
real difference between 
$S$-typings and $S$-typed sets. We will work mainly with $S$-typings because this tends
to result in simpler notation.

Let $\alpha : S \to \mathcal{C}$ be an $S$-family and $\gamma : I \to S$ be an $S$-typing. 
Then there is an $I$-family $\alpha \comp \gamma : I \to \mathcal{C}$, (which means that 
$(\alpha \comp \gamma)_\eta = \alpha_{\gamma(\eta)}$ for each $\eta \in I$). If $X$ is an 
$S$-family of sets and $\gamma : I \to S$ a finite $S$-typing then the product 
$\utimes{(X \comp \gamma)}$ will be denoted by $\ass{\gamma}{X}$; thus 
$\ass{\gamma}{X}$ is the set of all 
\definition{typed}\index{typed mapping}\index{mapping!typed} 
mappings from $I$ to 
$\bigcup_{\eta \in I} X_{\gamma(\eta)}$, i.e., the set of mappings 
$v : I \to \bigcup_{\eta\in I} X_{\gamma(\eta)}$ such that $v(\eta) \in X_{\gamma(\eta)}$ 
for each $\eta \in I$. The elements of $\ass{\gamma}{X}$ are called 
\definition{assignments}\index{assignment}. An assignment $v \in \ass{\gamma}{X}$ thus 
assigns to each 
`name' $\eta \in I$ an element $v(\eta) \in X_{\gamma(\eta)}$ of the appropriate type. 
Note that $\ass{\onept}{X} = \Oneptset$; moreover, if $Y \subset X$ then 
$\ass{\gamma}{Y}$ can clearly be regarded as a subset of $\ass{\gamma}{X}$.

The elements of $S^*$ will be considered as finite $S$-typings (and so $S^*$ 
will be regarded as a subset of $\ftyped{S}$): The list $\llist{s}{n}$
is identified with the mapping from $[n]$ to $S$ which assigns to each $j$ 
the type $s_j$. (The list $\onept$ with no components is then the $S$-typing 
$\onept$.) Let $X_S$ be a family of sets and $\sigma = \llist{s}{n} \in S^*$; then, 
considering $\sigma$ as an $S$-typed set, $\ass{\sigma}{X} = Y_1 \times \cdots \times Y_n$ 
with $Y_j = X_{s_j}$ for each $j$, which leads to the usual notation 
$X_{s_1} \times \cdots \times X_{s_n}$ for the set $\ass{\sigma}{X}$.

\begin{lemma}\label{lemma_sets_2}
Let $X$, $Y$ and $Z$ be $S$-families of sets and let 
$\varphi : X \to Y$ and $\psi : Y \to Z$ be $S$-families of mappings. Then 
$(\psi\fcomp\varphi)\comp\gamma = (\psi\comp\gamma)\fcomp(\varphi\comp\gamma)$
for each $S$-typing $\gamma$.
\end{lemma}

\proof This holds since for each $\eta \in \dom(\gamma)$
\[((\psi\fcomp\varphi)\comp\gamma)_\eta = (\psi\fcomp\varphi)_{\gamma(\eta)} 
= \psi_{\gamma(\eta)}\comp\varphi_{\gamma(\eta)} 
= (\psi\comp\gamma)_\eta \comp (\varphi\comp\gamma)_\eta
= ((\psi\comp\gamma)\fcomp(\varphi\comp\gamma))_\eta\;.\ \eop\]

Now let $X$ and $Y$ be $S$-families of sets and let $\varphi : X \to Y$ be an $S$-family of 
mappings. Then for each finite $S$-typing $\gamma$ the mapping 
$\utimes{(\varphi\comp\gamma)}$ will denoted by $\ass{\gamma}{\varphi}$. Thus
$\ass{\gamma}{\varphi} : \ass{\gamma}{X} \to \ass{\gamma}{Y}$ is the mapping given by
\[\ass{\gamma}{\varphi}(v)(\eta) = \varphi_{\gamma(\eta)} (v(\eta))\]
for each $v \in \ass{\gamma}{X}$, $\eta \in \dom(\gamma)$. Note that if 
$\sigma = \llist{s}{n} \in S^*$ then $\ass{\sigma}{\varphi}$ is just the mapping from 
$\ass{\sigma}{X} = X_{s_1} \times \cdots \times X_{s_n}$ to 
$\ass{\sigma}{Y} = Y_{s_1} \times \cdots \times Y_{s_n}$ given by
\[\ass{\sigma}{\varphi}(\svector{x}{n}) = (\varphi_{s_1}(x_1),
                                    \ldots,\varphi_{s_n}(x_n))\]
for each $(\svector{x}{n}) \in X_{s_1} \times \cdots \times X_{s_n}$.

\begin{lemma}\label{lemma_sets_3}
(1)\enskip Let $X$ be an $S$-family of sets and $\id : X \to X$ the $S$-family of 
identity mappings. Then $\ass{\gamma}{\id} : \ass{\gamma}{X} \to \ass{\gamma}{X}$ is 
also the identity mapping for each finite $S$-typing $\gamma$.

(2)\enskip Let $X$, $Y$ and $Z$ be $S$-families of sets and let 
$\varphi : X \to Y$ and $\psi : Y \to Z$ be $S$-families of mappings. Then 
$\ass{\gamma}{(\psi\fcomp\varphi)} = \ass{\gamma}{\psi}\comp\ass{\gamma}{\varphi}$
for each finite $S$-typing $\gamma$.

(3)\enskip Let $X$ and $Y$ be $S$-families of sets and let $\varphi : X \to Y$ be 
an $S$-family of mappings. If the mapping $\varphi_s : X_s \to Y_s$ is injective 
(resp.\ surjective) for each $s \in S$ then for each finite $S$-typing $\gamma$
the mapping $\ass{\gamma}{\varphi} : \ass{\gamma}{X} \to \ass{\gamma}{Y}$ is also injective 
(resp.\ surjective). In particular, if $\varphi_s : X_s \to Y_s$ is a bijection for 
each $s \in S$ then the mapping $\ass{\gamma}{\varphi}$ is a bijection, and in this case 
$(\ass{\gamma}{\varphi})^{-1} = \ass{\gamma}{(\varphi^{-1})}$.
\end{lemma}

\proof (1) is clear; moreover, by Lemma~\ref{lemma_sets_1}~(2)
and Lemma~\ref{lemma_sets_2} 
\[\ass{\gamma}{(\psi\fcomp\varphi)} = \utimes{((\psi\fcomp\varphi)\comp\gamma)} 
= \utimes{((\psi\comp\gamma)\fcomp(\varphi\comp\gamma))} 
= \utimes{(\psi\comp\gamma)} \comp \utimes{(\varphi\comp\gamma)} 
= \ass{\gamma}{\psi}\comp\ass{\gamma}{\varphi}\;,\]
which is (2), and (3) follows exactly as in Lemma~\ref{lemma_sets_1}~(3). \eop

The notion of a $\mathcal{C}$-typing still makes sense when $\mathcal{C}$ is a class.
In particular, $\ftyped{\mathcal{C}}$ will be used to denote the class of all
finite $\mathcal{C}$-typings. Note, however, that 
a $\mathcal{C}$-typing $\gamma : I \to \mathcal{C}$ is exactly an $I$-family of objects
from $\mathcal{C}$, and so $\ftyped{\mathcal{C}}$ is also the class of all finite 
families of objects from $\mathcal{C}$.

We now come to the second simple notion, that of a bottomed set, which will play an 
important role. A \definition{bottomed set}\index{bottomed set}\index{set!bottomed}
 is a pair $(X,\bot)$ consisting of a set $X$ 
and a distinguished \definition{bottom element}\index{bottom element}
 $\bot \in X$. The bottom element $\bot$ 
should be thought of as representing an `undefined' or `completely unknown' value which, 
if it were to be `more defined' or `become more known', would take on the value of one of 
the remaining elements of $X$.

If $(X,\bot)$ and $(X',\bot')$ are bottomed sets then by a mapping 
$f : (X,\bot) \to (X',\bot')$ is just meant a mapping  $f : X \to X'$. Such a mapping is 
said to be \definition{bottomed}\index{bottomed mapping}\index{mapping!bottomed}
(or \definition{strict})\index{strict mapping}\index{mapping!strict}
if $f(\bot) = \bot'$. It should 
be noted, however, that the mappings between bottomed sets which will occur here are 
typically not bottomed. A bottomed mapping $f : X \to X'$ will be called 
\definition{proper}\index{proper bottomed mapping}\index{bottomed mapping!proper}
if $f(x) \ne \bot'$ for all $x \in X \setminus \{\bot\}$.

If $(X,\bot)$ and $(X',\bot')$ are bottomed sets then $(X,\bot) \subset (X',\bot')$ will 
mean that $X \subset X'$ and $\bot = \bot'$.

If $(X,\bot)$ is a bottomed set then it is usual to write just $X$ instead of $(X,\bot)$ 
and to assume that the bottom element $\bot$ can be inferred from the context. The set 
$X$ will be referred to as the \definition{underlying set}\index{underlying set}
\index{set!underlying} if it is necessary to 
distinguish it from the bottomed set $X$. If $X$ is a bottomed set then the set 
$X \setminus \{\bot\}$ will be denoted by $\nonbot{X}$.

Note that if $X$ and $X'$ are bottomed sets then $X \subset X'$ means that
$X \subset X'$ (as sets) and that $X$ and $X'$ have a common bottom element.

Let $X$ be a finite $S$-family of bottomed sets with $\bot_s$ the bottom 
element of $X_s$ for each $s \in S$. Then the product set $\utimes{X}$ (defined in terms of 
the $S$-family of sets $X$) will be considered as a bottomed set by stipulating that 
its bottom element $\bot$ be the mapping given by $\bot(s) = \bot_s$ for each $s \in S$. 
A special case of this is 
when $n \ge 2$ and $X_j$ is a bottomed set with bottom element $\bot_j$ for each 
$\oneto{j}{n}$; then the product $X_1 \times \cdots \times X_n$ is considered as a 
bottomed set with bottom element $(\svector{\bot}{n})$. 
Note that this definition implies in particular that $\Oneptset$ is considered to be
a bottomed set (with of course $\onept$ as bottom element).

If $S$ is an arbitrary set, $X$ an $S$-family of bottomed sets and $\gamma$ a 
finite $S$-typing then the bottomed set $\utimes(X\fcomp \gamma)$ will be denoted by
$\ass{\gamma}{X}$. Thus the bottomed set $\ass{\gamma}{X}$ is just the set
$\ass{\gamma}{X}$ together with the bottom element $\bot$ given by 
$\bot(\eta) = \bot_{\gamma(\eta)}$ for each $\eta \in \dom(\gamma)$. 

Almost all the constructions to be made in these notes result in what are called
\definition{initial objects} and there is one trivial property of such objects
which is worth noting, namely that if they exist in a given category then they are 
isomorphic. Isolating this fact here helps to avoid repeating the same kind of argument 
for each special case. Now since the concept of being initial involves a category, we must 
first say what this is: A \definition{category}\index{category} $\mathsf{C}$ consists of
\begin{evlist}{25pt}{0.8ex}
\item[---] a class of elements $\mathcal{C}$ called the \definition{objects}\index{object}
of the category,
\item[---] for each $X,\, Y \in \mathcal{C}$ a set $\Hom(X,Y)$, whose elements 
are called \definition{morphisms}\index{morphism} with \definition{domain}\index{domain} 
$X$ and \definition{codomain}\index{codomain} $Y$, 
\item[---] for each $X,\,Y,\,Z \in \mathcal{C}$ a mapping $(f,g) \mapsto g \comp f$ from 
$\Hom(X,Y) \times \Hom(Y,Z)$ to $\Hom(X,Z)$,
\end{evlist}
such that the following two conditions hold:
\begin{evlist}{25pt}{0.8ex}
\item[---] \textit{(Associativity)} If $f \in \Hom(W,X)$, $g \in \Hom(X,Y)$ and 
$h \in \Hom(Y,Z)$ then 
\[(h\comp g)\comp f = h \comp (g\comp f)\;.\]
\item[---] \textit{(Identity)} For each $X \in \mathcal{C}$ there exists a morphism 
$\id_X \in \Hom(X,X)$ such that $f \comp \id_X = f$ and $\id_X \comp g = g$
for all $f \in \Hom(X,Y)$, $g \in \Hom(Y,X)$ and all $Y \in \mathcal{C}$.
\end{evlist}
For each $X \in \mathcal{C}$ the morphism $\id_X$ is unique: If $\id'_X \in \Hom(X,X)$ is 
a further morphism with this property then 
$\id'_X = \id'_X \comp \id_X = \id_X \comp \id'_X = \id_X$.

In the categories we will be dealing with the composition of morphisms $\comp$ is 
always some kind of composition of mappings (or families of mappings). This means that
the associativity of $\comp$ is a trivial consequence of the fact that the
composition of mappings is associative. Similarly, the identity morphisms will always be 
identity mappings (or families of identity mappings), and so they will trivially
satisfy the defining condition. 

The simplest example of a category is the the category of sets, i.e., the category with 
$\mathcal{C} = \mathsf{Sets}$, with $\Hom(X,Y)$ the set of all mappings from $X$ to $Y$ 
for each $X,\, Y \in \mathcal{C}$, and with $\comp$ the usual composition of mappings.
This category will be denoted (along with its objects) by  $\mathsf{Sets}$. 

A second example is the category of bottomed sets, i.e., the category with $\mathcal{C}$ 
the class of all bottomed sets, with $\Hom(X,Y)$ the set of all mappings from $X$ to $Y$ 
for each $X,\, Y \in \mathcal{C}$, and again with $\comp$ the usual composition of mappings.
This category will be denoted (along with its objects) by  $\mathsf{BSets}$. 

Note that in both these categories there is a mapping 
$\utimes : \ftyped{\mathcal{C}} \to \mathcal{C}$ which gives the product of each finite
family of objects.
Moreover, if $S$ is an arbitrary set, $X$ an $S$-family of objects from $\mathcal{C}$ 
and $\gamma$ a finite $S$-typing then in both
$\mathsf{Sets}$ and $\mathsf{BSets}$ the object $\utimes (X \fcomp \gamma)$ is being
denoted by $\ass{\gamma}{X}$.

Let $\mathsf{C}$ be a category;
a morphism $f \in \Hom(X,Y)$ is called an \definition{isomorphism}\index{isomorphism} 
if there exists
a morphism $g \in \Hom(Y,X)$ such that $g \comp f = \id_X$ and $f \comp g = \id_Y$.
In this case the inverse $g$ is uniquely determined by $f$. Objects 
$X,\, Y \in \mathcal{C}$ are said to be \definition{isomorphic}\index{isomorphic} if there 
exists an
isomorphism $f \in \Hom(X,Y)$. It is easy to see that being isomorphic defines an 
equivalence relation on the class $\mathcal{C}$.

An object $X \in \mathcal{C}$ is said to be \definition{initial}\index{initial object}
\index{object!initial} if for each object
$Y \in \mathcal{C}$ there exists a unique morphism $f : X \to Y$, i.e., if the set
$\Hom(X,Y)$ consists of a single element for each $Y \in \mathcal{C}$. In particular,
it must then be the case that $\id_X$ is the only element in $\Hom(X,X)$.
Initial objects need not exist in a category $\mathsf{C}$: For instance, they do not exist
in the category $\mathsf{BSets}$ (since morphisms are not necessarily strict mappings).
Moreover, they can exist but be rather trivial: In the category $\mathsf{Sets}$
the empty set $\varnothing$ is the only initial object.
However, for the categories involving various kinds of algebras which we will be dealing
with there are non-trivial initial objects, and the following result shows they
are then unique up to isomorphism:

\begin{proposition}\label{prop_sets_1}
Initial objects $X,\, Y \in \mathcal{C}$ are isomorphic. Conversely, if $X$ is an initial
object and $Y$ is isomorphic to $X$ then $Y$ is also initial.
\end{proposition}

\proof Suppose first that $X$ and $Y$ are initial objects. In 
particular, $\id_X$ is the only element in $\Hom(X,X)$ and $\id_Y$ is the only element in 
$\Hom(Y,Y)$. Moreover, there exists a unique morphism $f : X \to Y$ and a unique morphism 
$f : Y \to X$. Thus $g \comp f \in \Hom(X,X)$ and so $g \comp f = \id_X$, and in the same
way $f \comp g = \id_Y$. Hence $f$ is an isomorphism and $X$ and $Y$ are isomorphic.

Conversely, let $X \in \mathcal{C}$ be an initial object and $Y$ be isomorphic to $X$.
Since $X$ and $Y$ are isomorphic there exist morphisms $f : X \to Y$, $g : Y \to X$ with 
$g \comp f = \id_X$ and $f \comp g = \id_Y$. Let $Z \in \mathcal{C}$; since $X$ is initial
there exists a (unique) morphism $h : X \to Z$ and then $h \comp g \in \Hom(Y,Z)$.
Consider $h_1,\, h_2 \in \Hom(Y,Z)$; then 
$h_1\comp g$ and $h_2\comp g$ are both elements of $\Hom(X,Z)$ and hence 
$h_1 \comp g = h_2 \comp g$, since
$X$ is initial. It thus follows that
\[h_1 = h_1 \comp (g\comp f) = (h_1 \comp g) \comp f
= (h_2 \comp g) \comp f = h_2 \comp (g \comp f) = h_2\;,\]
which shows there exists a unique morphism in $\Hom(Y,Z)$ for each object $Z$. \eop

The statement in Proposition~\ref{prop_sets_1} is an example of what Lang would refer to as
`abstract nonsense'. However, it should be borne in mind each time an initial object
arises in what follows. Here is further piece of `abstract nonsense' which will be 
applied a couple of times:

\begin{proposition}\label{prop_sets_2}
Let $P$ be a property of the objects $\mathcal{C}$ satisfying the following three 
conditions:
\begin{evlist}{20pt}{0.5ex}
\item[(1)] There exists an object having property $P$.
\item[(2)] Every object having property $P$ is initial.
\item[(3)] Any object isomorphic to an object having
property $P$ also has property $P$.
\end{evlist}
Then each initial object has property $P$, and so having property $P$ is equivalent to 
being initial. 
\end{proposition}

\proof Let $X$ be an initial object; by (1) there exists an object $Y$ having property 
$P$, and by (2) and Proposition~\ref{prop_sets_1} $X$ and $Y$ are isomorphic; thus by (3) 
$X$ has property $P$. \eop

A category $\mathsf{C}'$ is a \definition{subcategory}\index{subcategory} of a category 
$\mathsf{C}$ if 
\begin{evlist}{20pt}{0.5ex}
\item[---] the objects of $\mathsf{C}'$ are a subclass of the objects of 
$\mathsf{C}$, 
\item[---] for all objects $X,\, Y$ of $\mathsf{C}'$ the set of morphisms $\Hom'(X,Y)$
in $\mathsf{C}'$ is a subset of the set of morphisms $\Hom(X,Y)$ in $\mathsf{C}$,
\item[---] the composition of morphisms in $\mathsf{C}'$ is the restriction of the
composition of morphisms in $\mathsf{C}$.
\end{evlist}
$\mathsf{C}'$ is called a \definition{full subcategory}\index{full subcategory}
\index{subcategory!full} of $\mathsf{C}$
if $\Hom'(X,Y) = \Hom(X,Y)$ for all objects $X,\, Y$ of $\mathsf{C}'$. 
Given $\mathsf{C}$, such a full subcategory is determined by specifying a subclass
of the objects of $\mathsf{C}$.

Let us emphasise that categories play a very superficial role here, and are needed
simply to formalise what it means to be initial. Readers not familiar with this kind
of stuff are recommended to look in Mac Lane and Birkhoff's Algebra text 
\cite{maclane_birkhoff}.

\newpage

\startsection{Algebras and homomorphisms}
\label{algs_homs}

The structure which plays a fundamental role in all of what follows is that of an algebra 
associated with a signature. What is usually called a signature we will call an
\index{enumerated signature}\index{signature!enumerated}\definition{enumerated signature}.
This is a triple $\Lambda = (B,K,\Theta)$, where $B$ and $K$ are non-empty sets and
$\Theta : K \to B^* \times B$ is a mapping. The set $B$ should here be thought of as a 
set of \index{type}\definition{types}, $K$ can be regarded as a set of 
\index{operator name}\index{name!operator}\definition{operator names}, and for each 
$k \in K$ the pair $\Theta(k)$ specifies the type of the domain and the codomain of the 
operator named by $k$. If $k \in K$ with $\Theta(k) = (\llist{b}{n},b)$ 
then we say that $k$ \definition{has type} $\llist{b}{n} \to b$.

If $\Lambda = (B,K,\Theta)$ is an enumerated signature then a 
\definition{$\Lambda$-algebra}\index{algebra} is any pair $(X,p)$ consisting of a 
$B$-family of sets $X$ and a $K$-family of mappings $p$ such that $p_k$ is a mapping from 
$X_{b_1} \times \cdots \times X_{b_n}$ to $X_b$ whenever $k \in K$ is of type 
$\llist{b}{n} \to b$. For each $b \in B$ the set $X_b$ should be thought of as a set of 
elements of type $b$ and for each $k \in K$ the mapping $p_k$ can be thought of as the 
operator corresponding to the operator name $k$.

A simple enumerated signature $\Lambda = (B,K,\Theta)$ with a corresponding 
`natural' $\Lambda$-algebra $(X,p)$ are given in Example~\thesection.1 on the 
following page. The usual way of representing such a signature is then illustrated in 
Example~\thesection.2.

Recall that $\ftyped{B}$ denotes the class of all finite $B$-typings and that
$B^*$ is considered to be a subset of $\ftyped{B}$ in that
$\sigma = \llist{b}{n} \in B^*$ is identified with the mapping from $[n]$ to $B$ which 
assigns $j$ the type $b_j$ for each $\oneto{j}{n}$. Our definition of a signature is 
obtained by replacing $B^*$ with $\ftyped{B}$ in the definition of an 
enumerated signature. This means that a \definition{signature}\index{signature} is a triple 
$\Lambda = (B,K,\Theta)$ consisting of non-empty sets $B$ and $K$ and a mapping
$\Theta : K \to \ftyped{B} \times B$. 

The two components of $\Theta$ will be denoted by $\sdom{\Theta}$ and $\scod{\Theta}$, 
thus $\sdom{\Theta} : K \to \ftyped{B}$, $\scod{\Theta} : K \to B$, and 
$\Theta(k) = (\sdom{\Theta}(k),\scod{\Theta}(k))$ for each $k \in K$. However, we almost 
always just write $\sdom{k}$ instead of $\sdom{\Theta}(k)$ and $\scod{k}$ instead of 
$\scod{\Theta}(k)$. Moreover, for each $k \in K$ the set $\dom(\sdom{k})$ will be 
denoted by $\domsdom{k}$, so $\sdom{k} : \domsdom{k} \to B$ and $\adom{k}{\eta}$ will be 
used as an alternative notation for $\sdom{k}(\eta)$ for each $\eta \in \domsdom{k}$.

If $\Lambda = (B,K,\Theta)$ is a signature then a 
\definition{$\Lambda$-algebra}\index{algebra} is any pair $(X,p)$ consisting of a 
$B$-family of sets $X$ and a 
$K$-family of mappings $p$ such that $p_k$ is a mapping from $\ass{\sdom{k}}{X}$ to 
$X_{\scod{k}}$ for each $k \in K$. For enumerated signatures this gives the same 
definition as above, since if $\sigma = \llist{b}{n} \in B^*$ then 
$\ass{\sigma}{X} = X_{b_1} \times \cdots \times X_{b_n}$.

The reason for working with the more general definition is that in the long run it turns 
out to be more natural. However, the reader not wanting to believe this can simply assume 
that all the signatures occurring are enumerated. In this case 
$\sdom{k}$ is always an element $\llist{b}{n}$ of $B^*$ and 
$\ass{\sdom{k}}{X}$ should just be thought of as a compact way of denoting the
product $X_{b_1} \times \cdots \times X_{b_n}$. Moreover, $\domsdom{k}$ is then the set 
$[n]$ and $\sdom{k}\eta = b_\eta$ for each $\eta \in [n]$.

It is always possible to replace a general signature with an `equivalent' enumerated 
signature. (This really just amounts to fixing an enumeration of the elements in the set
$\domsdom{k}$ for each $k \in K$.) 

\bigskip
\fbox{\begin{exframe}
\textit{Example \thesection.1\enspace} 
$\SynInt$ will always be used to denote the subset of the set
$\{\mathtt{0},\mathtt{1},\mathtt{2},\mathtt{3},\mathtt{4},\mathtt{5},\mathtt{6},
                                  \mathtt{7},\mathtt{8},\mathtt{9},\mathtt{-}\}^*$
containing for each integer $n$ its standard representation
$\underline{n}$ as a string of characters. The mapping 
$n \mapsto \underline{n}$ thus maps $\Int$ bijectively onto $\SynInt$.
\exparskip
Define an enumerated signature $\Lambda = (B,K,\Theta)$ by letting
\begin{eelist}{20pt}
\item $B = \{\mathtt{bool},\mathtt{nat},\mathtt{int},\mathtt{pair},\mathtt{list}\}$,
\item $K = \{\mathtt{True},\mathtt{False},\mathtt{Zero},\mathtt{Succ},\mathtt{Pair},
                      \mathtt{Nil},\mathtt{Cons}\} \cup \SynInt$,
\end{eelist}
and with $\Theta : K \to B^* \times B$ defined by
\begin{eelist}{20pt}
\item $\Theta(\mathtt{True}) = \Theta(\mathtt{False}) = (\onept,\mathtt{bool})$, 
\item $\Theta(\mathtt{Zero}) = (\onept,\mathtt{nat})$, \enskip
$\Theta(\mathtt{Succ}) = (\mathtt{nat},\mathtt{nat})$, 
\item $\Theta(\mathtt{Pair}) = (\mathtt{int}\ \mathtt{int},\mathtt{int})$,
\item $\Theta(\mathtt{Nil}) = (\onept,\mathtt{list})$, \enskip
$\Theta(\mathtt{Cons}) = (\mathtt{int}\ \mathtt{list},\mathtt{list})$, 
\item $\Theta(\underline{n}) = (\onept,\mathtt{int})$ for each $n \in \Int$.
\end{eelist}
This means that $\mathtt{True}$ and $\mathtt{False}$ are of type 
$\onept \to \mathtt{bool}$, $\mathtt{Zero}$ is of type $\onept \to \mathtt{nat}$,
$\mathtt{Succ}$ of type $\mathtt{nat} \to \mathtt{nat}$, 
$\mathtt{Pair}$ of type $\mathtt{int\ int} \to \mathtt{pair}$, 
$\mathtt{Nil}$ of type $\onept \to \mathtt{list}$, 
$\mathtt{Cons}$ of type $\mathtt{int\ list} \to \mathtt{list}$,
and $\underline{n}$ is of type 
$\onept \to \mathtt{int}$ for each $n \in \Int$.
\exparskip
Now define a $\Lambda$-algebra $(X, p)$ 
with a $B$-family of sets $X$ and a $K$-family of mappings $p$ by letting
\begin{eelist}{20pt}
\item $X_{\mathtt{bool}} = \Bool$,\enskip 
$X_{\mathtt{nat}} = \Nat$, \enskip $X_{\mathtt{int}} = \Int$, 
$X_{\mathtt{pair}} = \Int^2$, \enskip $X_{\mathtt{list}} = \Int^*$, \enskip
\item $p_{\mathtt{True}} : \Oneptset \to X_{\mathtt{bool}}$ 
      with $p_{\mathtt{True}}(\onept) = \True$, 
\item $p_{\mathtt{False}} : \Oneptset \to X_{\mathtt{bool}}$ 
      with $p_{\mathtt{False}}(\onept) = \False$, 
\item $p_{\mathtt{Zero}} : \Oneptset \to X_{\mathtt{nat}}$ 
      with $p_{\mathtt{Zero}}(\onept) = 0$, 
\item $p_{\mathtt{Succ}} : X_{\mathtt{nat}} \to X_{\mathtt{nat}}$ 
      with $p_{\mathtt{Succ}}(n) = n + 1$, 
\item $p_{\underline{n}} : \Oneptset \to X_{\mathtt{int}}$ 
      with $p_{\underline{n}}(\onept) = n$ for each $n \in \Int$,
\item $p_{\mathtt{Pair}} : X_{\mathtt{int}} \times X_{\mathtt{int}} \to X_{\mathtt{pair}}$
      with $p_{\mathtt{Pair}}(m,n) = (m,n)$, 
\item $p_{\mathtt{Nil}} : \Oneptset \to X_{\mathtt{list}}$
      with $p_{\mathtt{Nil}}(\onept) = \onept$, 
\item $p_{\mathtt{Cons}} : X_{\mathtt{int}} \times X_{\mathtt{list}} \to X_{\mathtt{list}}$
      with $p_{\mathtt{Cons}}(m,s) = m \triangleleft s$,
\end{eelist}
where $m \triangleleft s$ is the element of $\Int^*$ obtained 
by adding $m$ to the beginning of the list $s$, i.e.,
\[ m \triangleleft s  = \left\{
  \begin{array}{cl}
                    m\,\llist{m}{n}  &\ \textrm{if}\ s = \llist{m}{n}\  \textrm{with}
                            \ n \ge 1, \\
                      m   &\  \textrm{if}\  s = \onept.
    \end{array} \right. \]
\end{exframe}}

\fbox{\begin{exframe}
\textit{Example \thesection.2\enspace} 
An enumerated signature $\Lambda = (B,K,\Theta)$ 
with $B$ and $K$ finite can (and in most functional programming languages will) 
be represented in a form similar to the following,
where $\lvector{b}{n}$ is some enumeration of the elements in the set $B$,
$k_{j1},\,\ldots,\,k_{jm_j}$ an enumeration of the elements of 
$K_{b_j}$ for each $j$ and $\Theta(k_{j\ell}) = (\gamma_{j\ell},b_j)$
for each $\ell,\,j$:
\begin{eqnarray*}
 b_1\ &\mathtt{::=}&\ k_{11}\ \gamma_{11}\ |\ \cdots 
                  \ \mathtt{|}\ k_{1m_1}\ \gamma_{1m_1}\\
b_2\ &\mathtt{::=}&
    \ k_{21}\ \gamma_{21}\ \mathtt{|}\ \cdots\ \mathtt{|}\ k_{2m_2}\ \gamma_{2m_2} \\
          &&\qquad\qquad\vdots \\
 b_n\ &\mathtt{::=}&\ k_{n1}\ \gamma_{n1}
    \ \mathtt{|}\ \cdots\ \mathtt{|}\ k_{nm_n}\ \gamma_{nm_n}
\end{eqnarray*}
The enumerated signature $\Lambda = (B,K,\Theta)$ introduced in Example~\thesection.1 
can thus be represented in the form
\begin{eelist}{100pt}
\item $\mathtt{bool\ ::=\ True\ |\ False}$
\item $\mathtt{nat\ ::=\ Zero\ |\ Succ\ nat}$
\item $\mathtt{int\ ::=\ } \cdots\, \mathtt{\ -2\ |\ -1\ |\ 0\ |\ 1\ |\ 2\ }\, \cdots$
\item $\mathtt{pair\ ::=\ Pair\ int\ int}$
\item $\mathtt{list\ ::=\ Nil\ |\ Cons\ int\ list}$
\end{eelist}
Of course, there is a problem here with the type
$\mathtt{int}$, since $K_{\mathtt{int}}$ is infinite, but in all
real programming languages this type is `built-in' and so it does not need to
be included in the part of the signature specified by the programmer.
\end{exframe}}

\bigskip

For the remainder of the chapter let $\Lambda = (B,K,\Theta)$ be a 
signature. For each $b \in B$ put $K_b = \{ k \in K : \scod{k} = b \}$.
The sets $K_b$, $b \in B$, thus form a partition of the set $K$.
(For instance, in the signature $\Lambda$ in Example~\thesection.1 
$K_{\mathtt{bool}} = \{\mathtt{True},\mathtt{False}\}$,
$K_{\mathtt{nat}} = \{\mathtt{Zero},\mathtt{Succ}\}$, 
$K_{\mathtt{int}} = \SynInt$, 
$K_{\mathtt{pair}} = \{\mathtt{Pair}\}$ and
$K_{\mathtt{list}} = \{\mathtt{Nil},\mathtt{Cons}\}$.)

The next task is to explain what are the structure-preserving mappings 
between algebras. Let $(X,p)$ and $(Y,q)$ be $\Lambda$-algebras and let 
$\pi : X \to Y$ be a $B$-family of mappings, i.e., $\pi_b : X_b \to Y_b$ for each 
$b \in B$. Then the family $\pi$ is called a \definition{homomorphism}\index{homomorphism}
from $(X,p)$ to $(Y,q)$ if 
\[ q_k \comp \ass{\sdom{k}}{\pi} = \pi_{\scod{k}} \comp p_k\]
for each $k \in K$. This fact will also be expressed by saying that
$\pi : (X,p) \to (Y,q)$ is a homomorphism.

If $\Lambda$ is enumerated then $\pi : (X,p) \to (Y,q)$ being a homomorphism
means that if $k \in K$ is of type $\llist{b}{n} \to b$ then
\[ q_k(\pi_{b_1}(x_1),\ldots,\pi_{b_n}(x_n)) = \pi_b(p_k(\svector{x}{n}))\]
must hold for all $(\svector{x}{n}) \in X_{b_1} \times \cdots \times X_{b_n}$,
this condition being interpreted as $q_k(\onept) = \pi_b(p_k(\onept))$ 
when $k$ is of type $\onept \to b$.

\begin{proposition}\label{prop_algs_homs_1}
(1)\enskip The $B$-family of identity mappings $\id : X \to X$ defines a homomorphism from a
$\Lambda$-algebra $(X,p)$ to itself. 

(2)\enskip If $\pi : (X,p) \to (Y,q)$ and 
$\varrho : (Y,q) \to (Z,r)$ are homomorphisms then the composition $\varrho\fcomp\pi$ 
is a homomorphism from $(X,p)$ to $(Z,r)$.
\end{proposition}

\proof (1)\enskip This follows immediately from Lemma~\ref{lemma_sets_3}~(1). 

(2)\enskip Let $k \in K$; then by Lemma~\ref{lemma_sets_3}~(2)
\[ r_k \comp \ass{\sdom{k}}{(\varrho\fcomp\pi)} 
  = r_k \comp\ass{\sdom{k}}{\varrho}\comp \ass{\sdom{k}}{\pi}
  = \varrho_{\scod{k}} \comp q_k \comp\ass{\sdom{k}}{\pi} 
   = \varrho_{\scod{k}} \comp\pi_{\scod{k}} \comp p_k 
  = (\varrho\fcomp\pi)_{\scod{k}} \comp p_k\]
and hence $\varrho\fcomp\pi$ is a homomorphism from $(X,p)$ to $(Z,r)$. \eop

Proposition~\ref{prop_algs_homs_1} implies that there is a category whose objects are
$\Lambda$-algebras and whose morphisms are homomorphisms between $\Lambda$-algebras.
With the terminology introduced at the end of Section~\ref{sets}, a homomorphism 
$\pi : (X,p) \to (Y,q)$ is an \definition{isomorphism}\index{isomorphism} if 
there exists a homomorphism $\varrho : (Y,q) \to (X,p)$ such that $\varrho\fcomp\pi$ and 
$\pi\fcomp\varrho$ are the families of identity mappings (on $X$ and $Y$ respectively). 
If this is the case then for each $b \in B$ the mapping $\pi_b : X_b \to Y_b$ must be a 
bijection and $\varrho_b$ must be the set-theoretic inverse $\pi_b^{-1}$ of $\pi_b$ (and 
so in particular $\varrho$ is uniquely determined by $\pi$). The converse also holds:

\begin{proposition}\label{prop_algs_homs_2}
If $\pi : (X,p) \to (Y,q)$ is a homomorphism such that for each $b \in B$ the mapping
$\pi_b : X_b \to Y_b$ is a bijection then $\pi$ is an isomorphism. 
\end{proposition}

\proof This amounts to showing that the family $\pi^{-1}$ of set-theoretic inverses is 
also a homomorphism. Let $k \in K$. Then 
$q_k\comp\ass{\sdom{k}}{\pi} = \pi_{\scod{k}}\comp p_k$, and therefore by 
Lemma~\ref{lemma_sets_3}~(3) it follows that 
$p_k\comp\ass{\sdom{k}}{(\pi^{-1})} = p_k\comp (\ass{\sdom{k}}{\pi})^{-1} 
 = \pi_{\scod{k}}^{-1}\comp q_k$, which 
implies that $\pi^{-1}$ is a homomorphism. \eop

We are going to divide up the signatures into two kinds. The set 
\[B \setminus \Im(\scod{\Theta}) = \{ b \in B : K_b = \varnothing \}\]
will be called the\index{parameter set}\index{set!parameter}
\definition{parameter set of $\Lambda$}, and is always denoted in what 
follows by $A$. Thus $A$ is a proper subset of $B$, since $B$ and $K$ are both non-empty.
The signature $\Lambda$ will be called 
\definition{closed}\index{closed signature}\index{signature!closed} if $A = \varnothing$
(which is the case for the signature in Example~\thesection.1)
and \definition{open}\index{open signature}\index{signature!open} if $A \ne \varnothing$. 

Open signatures are typically involved when dealing with what goes under the name of
polymorphism.
Consider the signature in Example~\thesection.1. This contains a type $\mathtt{list}$ 
with 
associated operator names $\mathtt{Nil}$ of type $\onept \to \mathtt{list}$ and
$\mathtt{Cons}$ of type $\mathtt{int\ list} \to \mathtt{list}$. In the $\Lambda$-algebra 
$(X,p)$ defined in the example $X_{\mathtt{list}} = \Int^*$ with
$p_{\mathtt{Nil}} : \Oneptset \to \Int^*$ given by $p_{\mathtt{Nil}}(\onept) = \onept$ 
and $p_{\mathtt{Cons}} : \Int \times \Int^* \to \Int^*$ given by
$p_{\mathtt{Cons}}(m,s) = m \triangleleft s$. Now if lists of some other type $t$ need 
to be implemented then the signature would have to be extended by adding a new type, say 
$\mathtt{tlist}$, together with two new operator names, one of type 
$\onept \to \mathtt{tlist}$ and the other of type $t\ \mathtt{tlist} \to \mathtt{tlist}$.
Moreover, the signature has to be extended in essentially the same way with a new list 
type and two new operator names for each type for which lists are required. This is 
clearly not very satisfactory. What is  needed is the possibility of defining once and 
for all lists of an arbitrary type. The reader should look at Example~\thesection.3 
on the next 
page but one in order to get some inkling of how this might be done using an open 
signature. This topic (i.e., polymorphism) will be dealt with systematically in 
Chapter~\ref{poly}.

The definitions and results in this chapter do not distinguish explicitly between open 
and closed signatures. However, many of the results are only really relevant when applied 
to the `right' kind of signature.

The following simple construction will play an important role in Chapter~\ref{poly}:
Let $F$ be a non-empty set and for each $i \in F$ let 
$\Lambda_i = (B_i,K_i,\Theta_i)$ be a signature. Then 
$\Lambda$ is said to be the \definition{disjoint union}
\index{disjoint union of signatures}\index{signatures!disjoint union of}of the signatures 
$\Lambda_i$, $i \in F$, if the following conditions hold:
\begin{itemize}
\item[(1)] $B_i \cap B_i = \varnothing$ and $K_i \cap K_i = \varnothing$
whenever $i \ne j$.
\item[(2)] $B = \bigcup_{i\in F} B_i$ and $K = \bigcup_{i\in F} K_i$.
\item[(3)] $\Theta_i(k) = \Theta(k)$ for all $k \in K_i$, $i \in F$.
\end{itemize}
In particular, if $A_i$ is the parameter set of $\Lambda_i$ for each $i \in F$ then 
$A = \bigcup_{i\in F} A_i$ is the parameter set of $\Lambda$.
If $\Lambda$ is the disjoint union of the signatures $\Lambda_i$, $i \in F$, and 
$(X^i,p^i)$ is a $\Lambda_i$-algebra for each $i \in F$ then a $\Lambda$-algebra $(X,p)$ 
can be defined by putting $X_b = X^i_b$ for each 
$b \in B_i$ and $p_k = p^i_k$ for each $k \in K_i$. $(X,p)$ will be
called the \definition{sum}\index{sum of algebras}\index{algebras!sum of}
of the $\Lambda_i$-algebras $(X^i,p^i)$, $i \in F$, and will
be denoted by $\oplus_{i\in F} (X^i,p^i)$. The converse also holds:

\begin{lemma}\label{lemma_algs_homs_1}
Suppose $\Lambda$ is the disjoint union of the signatures $\Lambda_i$, $i \in F$.
Let $(X,p)$ be a $\Lambda$-algebra and for each $i \in F$ put
$X^i = \rest{X}{B_i}$ and $p^i = \rest{p}{K_i}$. Then $(X^i,p^i)$ is a 
$\Lambda_i$-algebra and $(X,p) = \oplus_{i\in F} (X^i,p^i)$.
\end{lemma}

\proof Straightforward. \eop

\begin{lemma}\label{lemma_algs_homs_2}
Suppose $\Lambda$ is the disjoint union of the signatures $\Lambda_i$, $i \in F$, and 
for each $i \in F$ let $(X^i,p^i)$, $(Y^i,q^i)$ be $\Lambda_i$-algebras
and $\pi^i : (X^i,p^i) \to (Y^i,q^i)$ be a homomorphism. Then
$\pi : \oplus_{i\in F} (X^i,p^i) \to \oplus_{i\in F}(Y^i,q^i)$ is a homomorphism, where
$\pi$ is given by $\pi_b = \pi^i_b$ for each $b \in B_i$.
\end{lemma}

\proof Straightforward. \eop

A signature $\Lambda' = (B',K',\Theta')$ is said to be an 
\definition{extension}\index{extension of a signature}\index{signature!extension of} of 
$\Lambda$ if $B \subset B'$, $K \subset K'$ with $\Theta = \rest{\Theta'}{K}$. Thus if 
$k \in K$ then $\sdom{k}$ and $\scod{k}$ have the same meaning in both $\Lambda$ and 
$\Lambda'$. Note that $\Lambda$ is trivially an extension of itself.

Let $\Lambda'$ be an extension of $\Lambda$. A $\Lambda'$-algebra $(Y,q)$ is 
then called an
\definition{extension}\index{extension of an algebra}\index{algebra!extension of} 
of a $\Lambda$-algebra $(X,p)$ if 
$X_b \subset Y_b$ for each $b \in B$ and $p_k$ is the restriction of $q_k$ to 
$\ass{\sdom{k}}{X}$ for each $k \in K$. (The case $\Lambda' = \Lambda$ is certainly not 
excluded here, and in fact it will occur frequently in what follows.)
Note that if $(Y,q)$ is a $\Lambda'$-algebra then $(\rest{Y}{B},\rest{q}{K})$ is a
$\Lambda$-algebra. Moreover, if $\pi$ is a homomorphism from  $(Y,q)$ to 
$(Z,r)$ then the family $\rest{\pi}{B}$ is a homomorphism from 
$(\rest{Y}{B},\rest{q}{K})$ to $(\rest{Z}{B},\rest{r}{K})$. 

Let us now look at the special case of a 
\index{single-sorted signature}\index{signature!single-sorted}\definition{single-sorted} 
signature, i.e., a 
signature of the form $(\Oneptset, K,\Theta)$. In this case there is no choice 
for the second component of $\Theta$ (since there is only one mapping possible from $K$ to 
$\Oneptset$) and so a single-sorted signature can be regarded as being a pair 
$(K,\vartheta)$ consisting of a set $K$ and a mapping $\vartheta : K \to \mathsf{FSets}$, 
recalling that $\ftyped{\Oneptset}$ can be identified with the class $\mathsf{FSets}$ of 
all finite sets.

Let $\Lambda = (K,\vartheta)$ be a single-sorted signature; then (identifying a 
$\Oneptset$-family $Z$ with the object $Z_\onept$) a $\Lambda$-algebra 
is here a pair $(X,p)$ consisting of a set $X$ and a $K$-family of mappings $p$ with 
$p_k : \total{\vartheta(k)}{X} \to X$ for each $k \in K$.

Consider now the even more  special case of an enumerated single-sorted signature
$\Lambda = (K,\vartheta)$. Then, since $\Oneptset^*$ can clearly be identified with the set 
of natural numbers $\Nat$, $\vartheta$ can here be regarded as a mapping from $K$ to 
$\Nat$. If $(X,p)$ is a  $\Lambda$-algebra then $p_k$ is a mapping from the cartesian 
product $X^{\vartheta(k)}$ to $X$, so $\vartheta(k)$ is just the number of arguments 
taken by the operator $p_k$.

If $\Lambda = (B,K,\Theta)$ is an arbitrary signature then there is a single-sorted 
signature $\Lambda^o = (K,\vartheta)$ with $\vartheta : K \to \mathsf{FSets}$ 
defined by $\vartheta(k) = \domsdom{k} = \dom(\sdom{k})$ for each $k \in K$. This means 
that $\Lambda^o$ 
is obtained from $\Lambda$ by no longer distinguishing between the various types. Now let 
$(Y,p)$ be a $\Lambda^o$-algebra, so $Y$ is a set and 
$p_k : \total{\vartheta(k)}{X} \to X$ for each $k \in K$; for each $b \in B$ put 
$X_b = Y$. Then $\ass{\sdom{k}}{X} = \total{\vartheta(k)}{Y}$ for each $k \in K$, since 
any mapping from $\vartheta(k) = \domsdom{k}$ to 
$\bigcup_{\eta \in \domsdom{k}} X_{\adom{k}{\eta}} = Y$ is automatically 
typed, which implies that $(X,p)$ is a $\Lambda$-algebra. This almost trivial method of 
obtaining $\Lambda$-algebras turns out to be surprisingly useful.

\bigskip
\fbox{\begin{exframe}
\textit{Example \thesection.3\enspace} 
Consider the signature $\Lambda = (B,K,\Theta)$ with
\begin{eelist}{20pt}
\item  $B = \{\mathtt{bool},\mathtt{atom},\mathtt{int},\mathtt{pair},\mathtt{list},
            \mathtt{lp}, \mathtt{x},\mathtt{y},\mathtt{z}\}$,
\item $K = \{\mathtt{True},\mathtt{False},\mathtt{Atom},\mathtt{Pair},
                 \mathtt{Nil},\mathtt{Cons},\mathtt{L},\mathtt{P}\}\, \cup\, \SynInt$,
\end{eelist}
and with $\Theta : K \to B^* \times B$ defined by
\begin{eelist}{20pt}
\item $\Theta(\mathtt{True}) = \Theta(\mathtt{False}) = (\onept,\mathtt{bool})$,
\item $\Theta(\mathtt{Atom}) = (\onept,\mathtt{atom})$,
\item $\Theta(\mathtt{Pair}) = (\mathtt{x}\ \mathtt{y},\mathtt{pair})$, 
\item $\Theta(\mathtt{Nil}) = (\onept,\mathtt{list})$, \enskip
      $\Theta(\mathtt{Cons}) = (\mathtt{z}\ \mathtt{list},\mathtt{list})$,
\item $\Theta(\mathtt{L}) = (\mathtt{list},\mathtt{lp})$, \enskip
      $\Theta(\mathtt{P}) = (\mathtt{pair},\mathtt{lp})$,
\item $\Theta(\underline{n}) = (\onept,\mathtt{int})$ for each $n \in \Int$.
\end{eelist}
\exparskip
The parameter set is here the set $A = \{\mathtt{x},\mathtt{y},\mathtt{z}\}$.
Using the conventions introduced  in Example~\thesection.2, this signature can be 
represented as
\begin{eelist}{100pt}
\item $\mathtt{bool\ ::=\ True\ |\ False}$
\item $\mathtt{atom\ ::=\ Atom}$
\item $\mathtt{int\ ::=\ } \cdots\, \mathtt{\ -2\ |\ -1\ |\ 0\ |\ 1\ |\ 2\ }\, \cdots$
\item $\mathtt{pair\ ::=\ Pair\ x\ y}$
\item $\mathtt{list\ ::=\ Nil\ |\ Cons\ z\ list}$
\item $\mathtt{lp\ ::=\ L\ list\ |\ P\ pair}$
\end{eelist}
\exparskip
Let $V$ be an $A$-family of sets. A $\Lambda$-algebra $(X, p)$ with $\rest{X}{A} = V$ can 
then be defined by putting
\begin{eelist}{20pt}
\item $X_{\mathtt{bool}} = \Bool$, \enskip 
      $X_{\mathtt{atom}} = \Oneptset$, \enskip $X_{\mathtt{int}} = \Int$, \enskip 
      $X_{\mathtt{pair}} = V_{\mathtt{x}} \times V_{\mathtt{y}}$, \enskip
      $X_{\mathtt{list}} = V_{\mathtt{z}}^*$, 
\item $X_{\mathtt{lp}} = (V_{\mathtt{x}} \times V_{\mathtt{y}}) \cup V_{\mathtt{z}}^*$ 
      \enskip (this union being considered to be disjoint), 
\item $X_{\mathtt{x}} = V_{\mathtt{x}}$, $X_{\mathtt{y}} = V_{\mathtt{y}}$, 
      $X_{\mathtt{z}} = V_{\mathtt{z}}$, 
\item $p_{\mathtt{True}} : \Oneptset \to X_{\mathtt{bool}}$ 
      with $p_{\mathtt{True}}(\onept) = \True$, 
\item $p_{\mathtt{False}} : \Oneptset \to X_{\mathtt{bool}}$ 
      with $p_{\mathtt{False}}(\onept) = \False$, 
\item $p_{\mathtt{Atom}} : \Oneptset \to X_{\mathtt{atom}}$ 
      with $p_{\mathtt{Atom}}(\onept) = \onept$, 
\item $p_{\underline{n}} : \Oneptset \to \Int$
      with $p_{\underline{n}}(\onept) = n$ for each $n \in \Int$,
\item $p_{\mathtt{Pair}} : V_{\mathtt{x}} \times V_{\mathtt{y}} \to X_{\mathtt{pair}}$
      with $p_{\mathtt{Pair}}(x,y) = (x,y)$, 
\item $p_{\mathtt{Nil}} : \Oneptset \to X_{\mathtt{list}}$ 
      with $p_{\mathtt{Nil}}(\onept) = \onept$, 
\item $p_{\mathtt{Cons}} : V_{\mathtt{z}} \times X_{\mathtt{list}} \to X_{\mathtt{list}}$
      with $p_{\mathtt{Cons}}(z,s) = m \triangleleft s$,
\item $p_{\mathtt{L}} : X_{\mathtt{list}} \to X_{\mathtt{lp}}$
      with $p_{\mathtt{L}}(s) = s$,
\item $p_{\mathtt{P}} : X_{\mathtt{pair}} \to X_{\mathtt{lp}}$ 
      with $p_{\mathtt{P}}(p) = p$.
\end{eelist}
\exparskip
In Chapter~\ref{poly} an augmented form of the representation of the signature $\Lambda$ 
will be introduced, providing the types $\mathtt{pair}$, $\mathtt{list}$ and 
$\mathtt{lp}$ with the parameters from the set $A$ on which they depend.
\end{exframe}}

\newpage

\startsection{Invariant families and minimal algebras}
\label{inv_fams}

For the whole of the section let $(X,p)$ be a $\Lambda$-algebra. 

Let $Y \subset X$ (i.e., $Y$ is a $B$-family of sets with $Y_b \subset X_b$ for each 
$b \in B$). The family $Y$ is said to be
\index{invariant family}\index{family!invariant}\definition{invariant in $(X,p)$}, 
or just \definition{invariant}, if $p_k(\ass{\sdom{k}}{Y}) \subset Y_{\scod{k}}$ 
for all $k \in K$. In particular, the family $X$ is itself trivially invariant. 

A related notion is that of a  subalgebra: A $\Lambda$-algebra $(Y,q)$ is a 
\definition{subalgebra}\index{subalgebra} of $(X,p)$ if $Y \subset X$ and $q_k$ is the 
restriction 
of $p_k$ to $\ass{\sdom{k}}{Y}$ for each $k \in K$. In this case the family $Y$ is clearly 
invariant. Conversely, let $Y$ be any invariant family and for each $k \in K$ 
let $q_k$ denote the restriction of $p_k$ to $\ass{\sdom{k}}{Y}$, so $q_k$ can be considered
as a mapping 
from $\ass{\sdom{k}}{Y}$ to $Y_{\scod{k}}$. Then $(Y,q)$ is a subalgebra of $(X,p)$. 
This means 
there is a one-to-one correspondence between invariant families and subalgebras of 
$(X,p)$. If $Y$ is an invariant family then the corresponding subalgebra $(Y,q)$ 
is called the \index{associated subalgebra}\definition{subalgebra associated with $Y$}.

\begin{lemma}\label{lemma_inv_fams_1}
Let $\pi : (X,p) \to (Y,q)$ be a homomorphism.

(1)\enskip If the family $\breve{X}$ is invariant in $(X,p)$ then the family $\breve{Y}$ 
given by $\breve{Y}_b = \pi_b(\breve{X}_b)$ for each $b \in B$ is invariant in $(Y,q)$.

(2)\enskip If the family $\breve{Y}$ is invariant in $(Y,q)$ then the family $\breve{X}$ 
given by $\breve{X}_b = \pi^{-1}_b(\breve{Y}_b)$ for each $b \in B$ is invariant in 
$(X,p)$. 
\end{lemma}

\proof (1)\enskip Let $k \in K$; then by Lemma~\ref{lemma_sets_3}~(3) 
$\ass{\sdom{k}}{\pi}(\ass{\sdom{k}}{\breve{X}}) = \ass{\sdom{k}}{\breve{Y}}$ and hence 
\[q_k(\ass{\sdom{k}}{\breve{Y}}) 
=  q_k(\ass{\sdom{k}}{\pi}(\ass{\sdom{k}}{\breve{X}})) 
= \pi_{\scod{k}}(p_k(\ass{\sdom{k}}{\breve{X}})) 
\subset \pi_{\scod{k}}(\breve{X}_{\scod{k}}) = \breve{Y}_{\scod{k}}\;.\]
The family $\breve{Y}$ is thus invariant in $(Y,q)$.

(2)\enskip Let $k \in K$; then by Lemma~\ref{lemma_sets_3}~(3) 
$\ass{\sdom{k}}{\pi} (\ass{\sdom{k}}{\breve{X}}) \subset \ass{\sdom{k}}{\breve{Y}}$, 
since $\pi_b(\breve{X}_b) \subset \breve{Y}_b$ for each $b \in B$. Hence
\[\pi_{\scod{k}}(p_k(\ass{\sdom{k}}{\breve{X}})) 
= q_k(\ass{\sdom{k}}{\pi}(\ass{\sdom{k}}{\breve{X}})) \subset q_k(\ass{\sdom{k}}{\breve{Y}})
\subset \breve{Y}_{\scod{k}}\]
and therefore
$p_k(\ass{\sdom{k}}{\breve{X}}) \subset \pi_{\scod{k}}^{-1}(\breve{Y}_{\scod{k}})
= \breve{X}_{\scod{k}}$.
This implies that the family $\breve{X}$ is invariant in $(X,p)$. \eop

Lemma~\ref{lemma_inv_fams_1} says that both the image and the pre-image of a subalgebra under a 
homomorphism are again subalgebras. In what follows let $U$ be a $B$-family of sets with
$U \subset X$.

\begin{lemma}\label{lemma_inv_fams_2} 
If $Y$ is an invariant family in $(X,p)$ containing $U$ and
\[\breve{Y}_b = U_b \cup \bigcup_{k \in K_b} p_k(\ass{\sdom{k}}{Y})\]
for each $b \in B$ then the family $\breve{Y}$ 
is invariant in $(X,p)$ and $U \subset \breve{Y} \subset Y$. 
\end{lemma}

\proof By definition $U \subset \breve{Y}$, and $\breve{Y} \subset Y$ holds because
$Y$ is invariant. Moreover, $\breve{Y}$ is invariant, since then
$p_k(\ass{\sdom{k}}{\breve{Y}}) \subset p_k(\ass{\sdom{k}}{Y})\subset \breve{Y}_{\scod{k}}$ 
for all $k \in K$. \eop

\begin{lemma}\label{lemma_inv_fams_3} 
There is a minimal invariant family containing $U$ (i.e., there is an invariant family 
$\hat{X}$ with $U \subset \hat{X}$ such that if $Y$ is any invariant family containing 
$U$ then $\hat{X} \subset Y$). 
\end{lemma}

\proof As already noted, the family $X$ is itself invariant, and it contains of course
$U$. Moreover, it is easy to see that an arbitrary intersection of invariant families 
is again invariant. (More precisely, if $X^{t}$ is an invariant family for each $t \in T$ 
and $\grave{X}_b = \bigcap_{t \in T} X_b^{t}$ for each $b \in B$ then $\grave{X}$ is also 
invariant.) The intersection of all the invariant families containing $U$ is thus the 
required minimal family. In fact, this minimal family $\hat{X}$ can be given somewhat 
more explicitly: For each $n \in \Nat$ define a family $\hat{X}^{n} \subset X$
by putting $\hat{X}^{0} = U$ and for each $n \in \Nat$, $b \in B$ letting
$\hat{X}^{n+1}_b = \hat{X}^{n}_b \cup \bigcup_{k \in K_b} \Im(\hat{p}^{n}_k)$,
where $\hat{p}^{n}_k$ is the restriction of $p_k$ to 
$\ass{\sdom{k}}{(\hat{X}^{n})} $. Then it is straightforward to check that 
$\hat{X}_b = \bigcup_{n \in \Nat} \hat{X}^{n}_b$ for each $b \in B$. This shows that 
$\hat{X}_b$ consists exactly of those elements of $X_b$ which can be `constructed' in a 
finite number of steps out of elements from the family $U$ and elements which have
already been `constructed'. \eop

\begin{lemma}\label{lemma_inv_fams_4}  
If $\hat{X}$ is the minimal invariant family containing $U$ then 
\[\hat{X}_b = U_b \cup \bigcup_{k \in K_b} p_k(\ass{\sdom{k}}{\hat{X}})\]
for each $b \in B$.
\end{lemma}

\proof This follows immediately from Lemma~\ref{lemma_inv_fams_2}. \eop

If $\hat{X}$ is the minimal invariant family containing $U$ then the associated subalgebra 
will be referred to as the\index{minimal subalgebra}\index{subalgebra!minimal}
\definition{minimal subalgebra of $(X,p)$ containing $U$}.

The $\Lambda$-algebra $(X,p)$ is now said to be \definition{$U$-minimal} if $X$ is the 
only invariant family containing $U$. Note that the minimal subalgebra of 
$(X,p)$ containing $U$ is always a $U$-minimal $\Lambda$-algebra. 

\begin{proposition}\label{prop_inv_fams_1}
If $(X,p)$ is $U$-minimal then for each $b \in B$
\[U_b \cup \bigcup_{k \in K_b} \Im(p_k) = X_b\;.\] 
\end{proposition}

\proof This follows immediately from Lemma~\ref{lemma_inv_fams_4}. \eop

Consider the special case when $U = \varnothing$: The minimal invariant family containing 
$\varnothing$ is of course just the minimal invariant family, and the associated subalgebra 
will be referred to as the \definition{minimal subalgebra of $(X,p)$}. Moreover, $(X,p)$ 
is said to be \definition{minimal} if $X$ is the only invariant family, thus
$(X,p)$ is minimal if and only if it is $\varnothing$-minimal.

If $(X,p)$ is minimal then by Proposition~\ref{prop_inv_fams_1}
$\bigcup\limits_{k \in K_b} \Im(p_k) = X_b$ for each $b \in B$.

As in Section~\ref{algs_homs} let $A$ be the parameter set of $\Lambda$ (so 
$A = \{ b \in B : K_b = \varnothing \}$). Note that if $(X,p)$ is $U$-minimal then by 
Proposition~\ref{prop_inv_fams_1} $\rest{X}{A} = \rest{U}{A}$. In particular, if $(X,p)$ is 
minimal then $\rest{X}{A} = \varnothing$. This indicates that minimality is
usually not an appropriate requirement when the signature is open (i.e., when 
$A \ne \varnothing$).

The converse of Proposition~\ref{prop_inv_fams_1} does not hold in general. However, the 
condition occurring there can be combined with a second condition to give a useful 
sufficient criterion for being minimal. A $B$-family of mappings $\#$ with 
$\#_b : X_b \to \Nat$ for each $b \in B$ will be called a 
\definition{grading}\index{grading for an algebra} for $(X,p)$  if for each $k \in K$
\[ \#_{\adom{k}{\eta}} (v(\eta)) < \#_{\scod{k}}(p_k(v))\]
for all $\eta \in \domsdom{k}$. If there exists a grading then $(X,p)$ is said to be 
\index{graded algebra}\index{algebra!graded}\definition{graded}.
If $\Lambda$ is enumerated then a family $\#$ being a grading means that
\[\#_{b_j}(x_j) <  \#_b(p_k(\svector{x}{n}))\]
must hold for each $\oneto{j}{n}$ for each 
$(\svector{x}{n}) \in X_{b_1} \times \cdots \times X_{b_n}$ whenever $k \in K$ is of type 
$\llist{b}{n} \to b$. 

\begin{lemma}\label{lemma_inv_fams_5}  
If $(X,p)$ is graded and for each $b \in B$ 
\[U_b \cup \bigcup_{k \in K_b} \Im(p_k) = X_b\]
then $X$ is $U$-minimal. 
\end{lemma}

\proof Let $\#$ be a grading for $(X,p)$, let $\hat{X}$ be the minimal invariant 
family containing $U$ and suppose $\hat{X} \ne X$. There thus exists $b \in B$ and 
$x \in X_b \setminus \hat{X}_b$ such that $\#_b(x) \le \#_{b'}(x')$ whenever
$x' \in X_{b'} \setminus \hat{X}_{b'}$ for some $b' \in B$. Then $x \in \Im(p_k)$
some $k \in K_b$, since $U_b \subset \hat{X}_b$, and so there
exists $v \in \ass{\sdom{k}}{X}$ with $x = p_k(v)$. But it then follows that 
$\#_{\adom{k}{\eta}}(v(\eta)) < \#_{\scod{k}}(x)$ and hence that 
$v(\eta) \in \hat{X}_{\adom{k}{\eta}}$ for each 
$\eta \in \domsdom{k}$ (by the minimality of $\#_b(x)$). 
However, this implies $x \in \hat{X}_b$, 
since the family $\hat{X}$ is invariant, which is a contradiction. \eop

\begin{proposition}\label{prop_inv_fams_2} 
If $(X,p)$ is graded then it is $U$-minimal if and only if 
\[U_b \cup \bigcup_{k \in K_b} \Im(p_k) = X_b\] 
for each $b \in B$.
\end{proposition}

\proof This follows immediately from Proposition~\ref{prop_inv_fams_1} and 
Lemma~\ref{lemma_inv_fams_5}. \eop

There is an obvious grading for the $\Lambda$-algebra $(X,p)$ defined in 
Example~\ref{algs_homs}.1, and it is thus easy to check that this algebra is minimal. 

\begin{proposition}\label{prop_inv_fams_3} 
(1)\enskip If $(X,p)$ is $U$-minimal and $\pi$ and $\varrho$ are homomorphisms from 
$(X,p)$ to a $\Lambda$-algebra $(Y,q)$ with $\pi_b(x) = \varrho_b(x)$ for all $x \in U_b$, 
$b \in B$, then $\pi = \varrho$. In particular, if $(X,p)$ is minimal then there exists at 
most one homomorphism from $(X,p)$ to a $\Lambda$-algebra $(Y,q)$. 

(2)\enskip If $(X,p)$ is $U$-minimal then any homomorphism $\pi$ from a $\Lambda$-algebra 
$(Y,q)$ to $(X,p)$ with $U_b \subset \pi_b(Y_b)$ for each $b \in B$ is surjective (i.e., 
$\pi_b$ is surjective for each $b \in B$). In particular, if $(X,p)$ is minimal then any 
homomorphism $\pi : (Y,q) \to (X,p)$ is surjective.
\end{proposition}

\proof (1)\enskip For each $b \in B$ let 
$\breve{X}_b = \{ x \in X_b : \pi_b(x) = \varrho_b(x) \}$; thus $U \subset \breve{X}$. 
Consider $k \in K$ and let $v \in \ass{\sdom{k}}{\breve{X}}$; then 
$v(\eta) \in \breve{X}_{\adom{k}{\eta}}$ for each $\eta \in \domsdom{k}$ and hence
\[ \ass{\sdom{k}}{\pi}(v)(\eta) = \pi_{\adom{k}{\eta}} (v(\eta))
         = \varrho_{\adom{k}{\eta}} (v(\eta)) = \ass{\sdom{k}}{\varrho}(v)(\eta)\]
which implies that $\ass{\sdom{k}}{\pi}(v) = \ass{\sdom{k}}{\varrho}(v)$. Therefore
\[ \pi_{\scod{k}}(p_k(v)) = q_k(\ass{\sdom{k}}{\pi}(v)) 
= q_k(\ass{\sdom{k}}{\varrho}(v)) = \varrho_{\scod{k}}(p_k(v))\; ,\]
i.e., $p_k(v) \in \breve{X}_{\scod{k}}$. This shows that the family $\breve{X}$ is 
invariant and thus $\breve{X} = X$, since $(X,p)$ is $U$-minimal. In other words, 
$\pi = \varrho$. 

(2)\enskip This follows immediately from Lemma~\ref{lemma_inv_fams_1}~(1). \eop

\newpage

\startsection{Initial algebras}
\label{init_algs}

Using the terminology introduced in Section~\ref{sets}, 
a $\Lambda$-algebra $(X,p)$ is said to be 
\definition{initial}\index{initial algebra}\index{algebra!initial}
if for each $\Lambda$-algebra $(Y,q)$ there exists a unique 
homomorphism from $(X,p)$ to $(Y,q)$. Let us note here that for open signatures there 
is a more appropriate notion which will be dealt with in Section~\ref{bound_algs}.

Initial algebras will be characterised as those that are minimal and possess a further 
property, here called regularity. The existence of an initial $\Lambda$-algebra will be 
established by showing that a minimal regular $\Lambda$-algebra exists.

Again let $A = \{ b \in B : K_b = \varnothing \}$ be the parameter set of $\Lambda$.
A $\Lambda$-algebra $(X,p)$ is said to be 
\definition{regular}\index{regular algebra}\index{algebra!regular} if for each 
$b \in B\setminus A$ and each $x \in X_b$ 
there exists a unique $k \in K_b$ and a unique element $v \in \ass{\sdom{k}}{X}$ such 
that $p_k(v) = x$. Thus $(X,p)$ is regular if and only if the mapping $p_k$ is injective 
for each $k \in K$ and for each $b \in B\setminus A$ the sets $\Im(p_k)$, $k \in K_b$, 
form a partition of $X_b$.

In particular, the $\Lambda$-algebra $(X,p)$ in Example~\ref{algs_homs}.1 is clearly 
regular. 

It is useful to introduce a further condition, which turns out to be  equivalent to being 
regular for minimal algebras. A $\Lambda$-algebra $(X,p)$ is said to be 
\definition{unambiguous}\index{unambiguous algebra}\index{algebra!unambiguous}
 if the mapping $p_k$ is injective for each $k \in K$ and for 
each $b \in B$ the sets $\Im(p_k)$, $k \in K_b$, are disjoint subsets of $X_b$. 
(Of course, in the definition of being unambiguous it would make no difference
if $B$ were replaced by $B \setminus A$, since $K_a = \varnothing$ for each $a \in A$.)

\begin{proposition}\label{prop_init_algs_1}
There exists an initial $\Lambda$-algebra. Moreover, the following are equivalent
for a $\Lambda$-algebra $(X,p)$:
\begin{evlist}{15pt}{0.5ex}
\item[(1)] $(X,p)$ is initial.
\item[(2)] $(X,p)$ is minimal and regular.
\item[(3)] $(X,p)$ is minimal and unambiguous.
\end{evlist}
\end{proposition}

\proof This occupies the rest of the section. \eop

The characterisation of initial algebras in Proposition~\ref{prop_init_algs_1}
is sometimes referred to as stating that initial algebras are exactly
those for which there is \textit{no junk} and  
\textit{no confusion}  (i.e., those which are minimal and unambiguous).

\textit{Remark:} Recall that if $(X,p)$ is minimal then by 
Proposition~\ref{prop_inv_fams_2} 
$\rest{X}{A} = \varnothing$, and so Proposition~\ref{prop_algs_homs_1} implies that also
$\rest{X}{A} = \varnothing$ for each initial $\Lambda$-algebra $(X,p)$. (This indicates 
why being initial is usually not an appropriate requirement when the signature is open.) 
Note further that this property of minimal algebras means that, as far as 
Proposition~\ref{prop_init_algs_1} is concerned, $B\setminus A$ could be replaced by $B$ 
in the definition of being regular. However, many of the $\Lambda$-algebras which will be 
considered are not minimal and then the definition of regularity made above is the 
`correct' one.

Before going any further consider again the natural number algebras discussed in the
Introduction. The equivalence of initial natural number algebras and Peano triples is 
easily seen to be just a special case of Proposition~\ref{prop_init_algs_1}.

The equivalence of (2) and (3) in Proposition~\ref{prop_init_algs_1} is dealt with by the 
following simple lemma:

\begin{lemma}\label{lemma_init_algs_1}
A minimal $\Lambda$-algebra is unambiguous if and only if it is regular.
\end{lemma}

\proof It is clear that any regular $\Lambda$-algebra is unambiguous.
Conversely, an unambiguous $\Lambda$-algebra $(X,p)$ will be regular if
$\bigcup_{k \in K_b} \Im(p_k) = X_b$ for each $b \in B \setminus A$, and this holds for 
minimal $\Lambda$-algebras by Proposition~\ref{prop_inv_fams_1}. \eop

The main steps in the proof of Proposition~\ref{prop_init_algs_1} are first to show that 
there exists a minimal regular $\Lambda$-algebra and then to show that any such 
$\Lambda$-algebra is initial. 

In order to get started an unambiguous $\Lambda$-algebra is needed, and the following 
trivial observation is useful here: Let $\Lambda^o = (K,\vartheta)$ be the 
single-sorted signature defined at the end of Section~\ref{algs_homs} (with the mapping
$\vartheta : K \to \mathsf{FSets}$ given by $\vartheta(k) = \domsdom{k}$ for each 
$k \in K$, recalling that $\domsdom{k} = \dom(\sdom{k})$). Let $(Y,p)$ be a 
$\Lambda^o$-algebra and $(X,p)$ be the 
$\Lambda$-algebra with $X_b = Y$ for each $b \in B$.

\begin{lemma}\label{lemma_init_algs_2}
If the $\Lambda^o$-algebra $(Y,p)$ is unambiguous then so is the $\Lambda$-algebra 
$(X,p)$. 
\end{lemma}

\proof This is clear. \eop

\begin{lemma}\label{lemma_init_algs_3}
There exists an unambiguous $\Lambda$-algebra. 
\end{lemma}

\proof By Lemma~\ref{lemma_init_algs_2} it is enough to show that if 
$\Lambda' = (K,\vartheta)$ 
is a single-sorted signature then there exists a unambiguous $\Lambda'$-algebra. The 
construction given below is just one of many possibilities of defining such an algebra.

Let $M = M_o \cup K$, where $M_o$ is the set of all pairs of the form $(k,\eta)$ with
$k \in K$ and $\eta \in \vartheta(k)$, and let $X$ be the set of all non-empty
finite subsets of $M^*$. Now if $k \in K$ with $\vartheta(k) = \varnothing$ then define
$p_k : \Oneptset \to X$ by letting $p_k(\onept) = \{k\}$ (where here $k$ is the 
list consisting of the single component $k$), and if  
$k \in K$ with $\vartheta(k) \ne \varnothing$ then define a mapping
$p_k : \total{\vartheta(k)}{X} \to X$ by letting
\[ p_k(v) 
    = \{k\} \cup \bigcup_{\eta \in \vartheta(k)} 
\{ (k,\eta) \triangleleft s : s \in v(\eta) \} \]
for each $v \in X^{\vartheta(k)}$, where $\triangleleft : M \times M^* \to M^*$ is the 
(infix) 
operation of adding an element to the beginning of a list. This gives a $\Lambda'$-algebra
$(X,p)$. But it is easily checked that $\Im(p_{k_1})$ and $\Im(p_{k_2})$ are 
disjoint subsets of $X$ whenever $k_1 \ne k_2$, and also that $p_k$ is injective 
for each $k \in K$. Hence $(X,p)$ is unambiguous. \eop

\begin{lemma}\label{lemma_init_algs_4} 
There exists a minimal regular $\Lambda$-algebra. 
\end{lemma}

\proof By Lemma~\ref{lemma_init_algs_3} there exists an unambiguous $\Lambda$-algebra 
$(X,p)$. But then any subalgebra of $(X,p)$ is also unambiguous. In particular, the minimal 
sub\-algebra is minimal and unambiguous, and hence by Lemma~\ref{lemma_init_algs_1} it 
is minimal and regular. \eop

\begin{lemma}\label{lemma_init_algs_5} 
Let $(X,p)$ be a minimal regular $\Lambda$-algebra. Then there exists a unique 
$B$-family $\#$ with $\#_b : X_b \to \Nat$ for each $b \in B$ with
$\#_{\scod{k}}(p_k(\onept)) = 0$ if $k \in K$ with $\sdom{k} = \onept$ and such that 
\[ \#_{\scod{k}}(p_k(v)) 
 = 1 + \max \{ \#_{\adom{k}{\eta}} (v(\eta)) : \eta \in  \domsdom{k}\}\]
for all $v \in \ass{\sdom{k}}{X}$ whenever $k \in K$ with 
$\sdom{k} \ne \onept$. 
\end{lemma}

\proof Let us emphasise again that $\rest{X}{A} = \varnothing$, since $(X,p)$ is minimal.
The family $\#$ will be obtained as the limit of a sequence 
$\{\#^m\}_{m \ge 0}$, where $\#^m$ is a $B$-family of mappings with $\#^m_b : X_b \to \Nat$
for each $b \in B$. First define $\#^0_b = 0$ for each $b \in B$. Next suppose that the 
family $\#^m$ has already been defined for some $m \in \Nat$. Then, since $(X,p)$ is 
regular and $\rest{X}{A} = \varnothing$, there is a unique family of mappings $\#^{m+1}$ 
such that $\#^{m+1}_{\scod{k}}(p_k(\onept)) = 0$ if $k \in K$ with $\sdom{k} = \onept$ 
and such that
\[ \#^{m+1}_{\scod{k}}(p_k(v)) 
 = 1 + \max \{ \#^m_{\adom{k}{\eta}} (v(\eta)) : \eta \in \domsdom{k} \}\]
for all $v \in \ass{\sdom{k}}{X}$ whenever $k \in K$ with $\sdom{k} \ne \onept$. 
By induction this defines the family $\#^m$ for each $m \in \Nat$.

Now $\#^m \le \#^{m+1}$ holds for each $m \in \Nat$ (i.e., $\#^m_b(x) \le \#^{m+1}_b(x)$ 
for all $x \in X_b$, $b \in B$): This follows by induction on $m$, since 
$\#^0 \le \#^1$ holds by definition and if $\#^m \le \#^{m+1}$ for some 
$m \in \Nat$ and $k \in K$ with $\sdom{k} \ne \onept$ then
\begin{eqnarray*}
 \#^{m+1}_{\scod{k}}(p_k(v))
  &=& 1 + \max \{ \#^m_{\adom{k}{\eta}} (v(\eta)) : \eta \in \domsdom{k} \} \\
 &\le& 1 + \max \{ \#^{m+1}_{\adom{k}{\eta}} (v(\eta)) : \eta \in \domsdom{k} \} 
= \#^{m+2}_{\scod{k}}(p_k(v)) 
\end{eqnarray*}
for all $v \in \ass{\sdom{k}}{X}$, which implies that $\#^{m+1} \le \#^{m+2}$. Moreover, 
the sequence $\{\#^m_b(x)\}_{m \ge 0}$ is bounded for each $x \in X_b$, $b \in B$: Let 
$X'_b$ denote the set of those elements $x \in X_b$ for which this is the case. Then it is 
easily checked that the $B$-family $X'$ is invariant, and hence $X' = X$, since 
$(X,p)$ is minimal. 

Let $x \in X_b$; then by the above $\{\#^m_b(x)\}_{m \ge 0}$ is a bounded increasing 
sequence from $\Nat$, and so there exists an element $\#_b(x) \in \Nat$ such that 
$\#^m_b(x) = \#_b(x)$ for all but finitely many $m$. This defines a mapping
$\#_b : X_b \to \Nat$ for each $b \in B$, and it immediately follows that the family 
$\#$ has the required property.
It remains to show the uniqueness, so suppose $\#'$ is another $B$-family of mappings with 
this property. For each $b \in B$ let $X'_b = \{ x \in X_b : \#'_b(x) = \#_b(x) \}$;
then the family $X'$ is clearly invariant and hence $X' = X$, since $(X,p)$ is 
minimal. \eop

It can now be shown that a minimal regular $\Lambda$-algebra $(X,p)$ is initial:
Let $(Y,q)$ be any $\Lambda$-algebra; then it is enough to just construct a homomorphism 
$\pi : (X,p) \to (Y,q)$, since Proposition~\ref{prop_inv_fams_1}~(1)
implies that this homomorphism is unique.

Let $\#$ be the family of mappings given by Lemma~\ref{lemma_init_algs_5} (with 
$\#_b : X_b \to \Nat$ for 
each $b \in B$) and for each $b \in B$, $m \in \Nat$ let
$X^m_b = \{ x \in X_b : \#_b(x) = m \}$. Define $\pi$ by defining $\pi_b$ on $X^m_b$ for 
each $b \in B$ using induction on $m$. Let $x \in X^0_b$; then, since $(X,p)$ is regular
and $\rest{X}{A} = \varnothing$,
there exists a unique $k \in K_b$ and a unique element $v \in \ass{\sdom{k}}{X}$ such that 
$p_k(v) = x$, and here $\sdom{k} = \onept$ and so $x = p_k(\onept)$. Thus put 
$\pi_b(x) = q_k(\onept)$, which defines $\pi_b$ on $X^0_b$ for each $b \in B$. Now let 
$m > 0$ and suppose $\pi_{b'}$ has already been defined on $X^n_{b'}$ for all $n < m$ and 
all ${b'} \in B$. Let $x \in X^m_b$; again using regularity and the fact that
$\rest{X}{A} = \varnothing$, there exists a unique 
$k \in K_b$ and a unique $v \in \ass{\sdom{k}}{X}$ such that $x = p_k(v)$. In this case
$v \in \ass{\sdom{k}}{X}$ with $\sdom{k} \ne \onept$, and by the defining 
property of $\#$ it follows that $\#_{\adom{k}{\eta}} (v(\eta)) < m$ for each 
$\eta \in \domsdom{k}$, which means $\pi_{\adom{k}{\eta}} (v(\eta))$ is already 
defined for each $\eta \in \domsdom{k}$ and hence that 
$\ass{\sdom{k}}{\pi}(v)$ is already defined (i.e., $\ass{\sdom{k}}{\pi}(v)$ is the element 
$v'$ of $\ass{\sdom{k}}{Y}$ given by 
$v'(\eta) =  \pi_{\adom{k}{\eta}}(v(\eta))$ for each $\eta \in \domsdom{k}$). Thus here 
put $\pi_b(x) = q_k(\ass{\sdom{k}}{\pi}(v))$. In this way $\pi_b$ is defined on 
$X^m_b$ for each $b \in B$ and each $m \in \Nat$, and the family $\pi$ is a homomorphism 
more-or-less by construction. 

This shows that any minimal regular $\Lambda$-algebra is initial which, together
with Lemma~\ref{lemma_init_algs_4}, also implies that an initial $\Lambda$-algebra exists. 
In order to show that an initial $\Lambda$-algebra is minimal and regular the following
fact is needed:

\begin{lemma}\label{lemma_init_algs_6} 
A $\Lambda$-algebra isomorphic to a minimal unambiguous $\Lambda$-algebra is itself 
minimal and unambiguous. Thus by Lemma~\ref{lemma_init_algs_1} a $\Lambda$-algebra 
isomorphic to a minimal regular $\Lambda$-algebra is itself minimal and regular.
\end{lemma}

\proof Let $(Y,q)$ a minimal unambiguous $\Lambda$-algebra and $\pi : (X,p) \to (Y,q)$ be 
an isomorphism. If $X'$ is an invariant family in $(X,p)$ then 
Lemma~\ref{lemma_inv_fams_1}~(1) implies that $\pi_b(X'_b) = Y_b$ for each $b \in B$, since 
$Y$ is the only invariant family in $(Y,q)$. Thus 
$X'_b = \pi^{-1}_b(\pi_b(X'_b)) = \pi^{-1}_b(Y_b) = X_b$ for each $b \in B$, i.e., 
$X' = X$, and this shows that $(X,p)$ is minimal. Now if $k \in K_b$ and 
$x \in \Im(p_k)$ then by the definition of a homomorphism $\pi_b(x) \in \Im(q_k)$. It 
immediately follows that if $k_1,\, k_2 \in K_b$ with $k_1 \ne k_2$ then 
$\Im(p_{k_1}) \cap \Im(p_{k_2}) = \varnothing$. Finally, 
$q_k \comp \ass{\sdom{k}}{\pi} = \pi_{\scod{k}} \comp p_k$ and, moreover, $q_k$ is 
injective, $\pi_{\scod{k}}$ is a bijection and by Lemma~\ref{lemma_sets_3}~(3) 
$\ass{\sdom{k}}{\pi}$ is also a bijection. Thus $p_k$ is injective for each $k \in K$. \eop

Now consider the property $P$ of $\Lambda$-algebras of being minimal and regular.
Then it has already been shown that there exists a $\Lambda$-algebra having property $P$
and that every $\Lambda$-algebra having property $P$ is initial. Moreover, 
Lemma~\ref{lemma_init_algs_4} shows that any $\Lambda$-algebra isomorphic to a 
$\Lambda$-algebra 
having property $P$ has property $P$. Thus by Proposition~\ref{prop_sets_2} every initial 
$\Lambda$-algebra has property $P$, and this completes the proof of 
Proposition~\ref{prop_init_algs_1}. \eop

Proposition~\ref{prop_init_algs_1} implies that the $\Lambda$-algebra $(X,p)$ in 
Example~\ref{algs_homs}.1 is initial.

A type $b \in B\setminus A$ will be said to be 
\definition{primitive}\index{primitive type}\index{type!primitive} if 
$\sdom{k} = \onept$ for each $k \in K_b$. (In the signature $\Lambda$ in 
Example~\ref{algs_homs}.1 the primitive types are therefore $\mathtt{bool}$ and 
$\mathtt{int}$.) Note that if 
$(X,p)$ is an initial $\Lambda$-algebra and $b \in B\setminus A$ is a primitive type 
then the mapping $k \mapsto p_k(\onept)$ maps $K_b$ bijectively onto $X_b$. (For the 
$\Lambda$-algebra $(X,p)$ in Example~\ref{algs_homs}.1 this gives the obvious bijections
between $K_{\mathtt{bool}} = \{\mathtt{True},\mathtt{False}\}$ and
$X_{\mathtt{bool}} = \Bool = \{\True,\False\}$ and between $K_{\mathtt{int}} = \SynInt$ and 
$X_{\mathtt{int}} = \Int$.)

\begin{lemma}\label{lemma_init_algs_7}  
The minimal subalgebra of an unambiguous $\Lambda$-algebra is initial. 
\end{lemma}

\proof This follows from Proposition~\ref{prop_init_algs_1} and 
Lemma~\ref{lemma_init_algs_1}, since, as 
was already noted, any subalgebra of an unambiguous algebra is unambiguous. \eop

Let $\Lambda' = (B',K,\Theta')$ be a signature which is an extension of $\Lambda$. The 
following is a useful extension of Proposition~\ref{prop_init_algs_1}:

\begin{proposition}\label{prop_init_algs_2} 
Let $(X,p)$ be an initial $\Lambda$-algebra. Then there exists an initial 
$\Lambda'$-algebra which is an extension of $(X,p)$. 
\end{proposition}

\proof By Proposition~\ref{prop_init_algs_1} there exists an initial $\Lambda'$-algebra 
$(Z,r)$ and the pair $(Y,q) = (\rest{Z}{B},\rest{r}{K})$ is then a $\Lambda$-algebra; let 
$(\hat{Y},\hat{q})$ be the minimal subalgebra of $(Y,q)$. Now 
by Proposition~\ref{prop_init_algs_1} and Lemma~\ref{lemma_init_algs_1} $(Z,r)$ is 
unambiguous and thus $(Y,q)$ is also unambiguous. Hence by Lemma~\ref{lemma_init_algs_7}
$(\hat{Y},\hat{q})$ is an initial $\Lambda$-algebra, and by construction
$(Z,r)$ is an extension of $(\hat{Y},\hat{q})$. The unique isomorphism 
from $(X,p)$ to $(\hat{Y},\hat{q})$ can therefore be used to construct an 
initial $\Lambda'$-algebra which is an extension of $(X,p)$. \eop

We end the section by noting that if $\Lambda = (K,\vartheta)$ is a single-sorted signature 
then the unique homomorphism from an initial $\Lambda$-algebra to a $\Lambda$-algebra 
$(Y,q)$ exists even when $Y$ is a class. This observation forms the basis for several 
of the constructions which will occur in later chapters.

If $\mathcal{C}$ is a class and $J$ a set then $\mathcal{C}^J$ will be used to denote the 
class of all mappings from $J$ to $\mathcal{C}$; in particular, 
$\mathcal{C}^\varnothing = \Oneptset$. If $X$ is a set and $\pi : X \to \mathcal{C}$ is a 
mapping then $\pi^J$ will be used to denote the mapping from $X^J$ to $\mathcal{C}^J$ 
defined by
\[\pi^J(u) (\eta) = \pi(u(\eta))\]
for all $u \in X^J$, $\eta \in J$.

Let $\Lambda = (K,\vartheta)$ be a single-sorted signature, so 
$\vartheta : K \to \mathsf{FSets}$ and a $\Lambda$-algebra is here a pair $(X,p)$ 
consisting of a set $X$ and a $K$-family of mappings $p$ with 
$p_k : \total{\vartheta(k)}{X} \to X$ for each $k \in K$. Let $\mathcal{C}$ be a class and 
let $\varphi$ be a $K$-family of mappings with
$\varphi_k : \mathcal{C}^{\vartheta(k)} \to \mathcal{C}$ for each $k \in K$.
Then $(\mathcal{C},\varphi)$ would be a $\Lambda$-algebra, except that
$\mathcal{C}$ is in general not a set. However, if $(X,p)$ is a $\Lambda$-algebra
then a mapping $\pi : X \to \mathcal{C}$ will still be called a 
\definition{homomorphism}\index{homomorphism} from $(X,p)$ to $(\mathcal{C},\varphi)$ if 
\[ \pi\comp p_k = \varphi_k \comp \pi^{\vartheta(k)}\]
for all $k \in K$. 

\begin{proposition}\label{prop_init_algs_3} 
Let $(X,p)$ be an initial $\Lambda$-algebra; then there exists a unique homomorphism 
$\pi : (X,p) \to (\mathcal{C},\varphi)$. 
\end{proposition}

\proof The construction of a homomorphism from $(X,p)$ to a $\Lambda$-algebra given in 
the proof of Proposition~\ref{prop_init_algs_1} takes place essentially in $(X,p)$. This 
construction 
can thus also be used to give a homomorphism from $(X,p)$ to $(\mathcal{C},\varphi)$. 
Moreover, the uniqueness follows exactly as in the proof of 
Proposition~\ref{prop_init_algs_1}
from the fact that $(X,p)$ is minimal. \eop

\startsection{Free algebras}
\label{free_algs}

The concept of being initial will now be generalised, leading to what are known as free 
algebras. The existence of free algebras will be established by looking at a new class of 
initial algebras, which also plays an important role in the following section.

Let $U$ be a $B$-family of sets, which is considered to be fixed in what follows. 
A $\Lambda$-algebra $(X,p)$ is said to be 
\definition{$U$-based}\index{based algebra}\index{algebra!based} if $U \subset X$.
If $(X,p)$ and $(X',p')$ are $U$-based $\Lambda$-algebras then a homomorphism 
$\pi : (X,p) \to (X',p')$ is said to 
\definition{fix $U$}\index{homomorphism!fixing a family} if 
$\pi_b(x) = x$ for all $x \in U_b$, $b \in B$.

\begin{proposition}\label{prop_free_algs_1}
(1)\enskip If $(X,p)$ is a $U$-based $\Lambda$-algebra then
the $B$-family of identity mappings $\id : X \to X$ defines a homomorphism 
from $(X,p)$ to itself fixing $U$.

(2)\enskip If $(X,p)$, $(Y,q)$ and $(Z,r)$ are $U$-based $\Lambda$-algebras and
$\pi : (X,p) \to (Y,q)$ and $\varrho : (Y,q) \to (Z,r)$ are homomorphisms fixing $U$ 
then the composition $\varrho\fcomp\pi$ 
is a homomorphism from $(X,p)$ to $(Z,r)$ fixing $U$.
\end{proposition}

\proof This follows immediately from Proposition~\ref{prop_algs_homs_1}. \eop

Proposition~\ref{prop_free_algs_1} implies there is a category whose objects are $U$-based 
$\Lambda$-algebras and whose morphisms are homomorphisms fixing $U$. 
A $\Lambda$-algebra $(X,p)$ is said to be 
\definition{$U$-initial}\index{initial algebra}\index{algebra!initial} if it is an initial
object in this category, i.e., if it is $U$-based
and if for each $U$-based $\Lambda$-algebra $(X',p')$ there exists a unique homomorphism 
from $(X,p)$ to $(X',p')$ fixing $U$. 

It will turn out that there exist $U$-initial $\Lambda$-algebras and that
these are exactly the $U$-free algebras:
A $\Lambda$-algebra $(X,p)$ is said to be 
\definition{$U$-free}\index{free algebra}\index{algebra!free} if it is $U$-based and 
for each $\Lambda$-algebra $(Y,q)$ and each 
$B$-family of mappings $v : U \to Y$ there exists a unique homomorphism 
$\pi^v : (X,p) \to (Y,q)$ such that $\pi^v_b (\eta) = v(\eta)$ for all $\eta \in U_b$, 
$b \in B$.

\begin{proposition}\label{prop_free_algs_2} 
There exists a $U$-free $\Lambda$-algebra. Moreover, 
a $\Lambda$-algebra is $U$-free if and only if it is $U$-initial.
\end{proposition}

\proof The proof starts by showing that there exists a $U$-initial $\Lambda$-algebra,
and such an algebra will be obtained from an initial 
$\Lambda_U$-algebra, where $\Lambda_U$ is the following extension of the signature 
$\Lambda$: For each $\eta \in U_b$, $b \in B$ let $\diamond^b_\eta$ be some element not in 
$K$ and 
such that $\diamond^b_{\eta_1} \ne \diamond^b_{\eta_2}$ whenever $\eta_1 \ne \eta_2$ and,
moreover, such that $\diamond^{b_1}_{\eta_1} \ne \diamond^{b_2}_{\eta_2}$ whenever 
$b_1 \ne b_2$. Put 
\[ K_U = K \cup \{ \diamond^b_\eta : \eta \in U_b,\, b \in B \}\] 
and define a signature $\Lambda_U = (B,K_U,\Theta_U)$ with
$\Theta_U : K_U \to \ftyped{B} \times B$ given by
\[ \Theta_U(k) = \left\{
  \begin{array}{cl}
         \Theta(k) &\ \textrm{if}\ k \in K,\\
        (\onept,b) &\ \textrm{if}\ k = \diamond^b_\eta\  \textrm{for some}\ \eta \in U_b\;.
    \end{array} \right. \]

A $U$-based $\Lambda$-algebra $(X,p)$ can be extended to a $\Lambda_U$-algebra 
$(X,p')$, called the
\index{associated algebra}\definition{$\Lambda_U$-algebra associated with $(X,p)$},
by putting $p'_k = p_k$ for each $k \in K$ and defining 
$p'_{\diamond^b_\eta} : \Oneptset \to X_b$ to be the mapping
with $p'_{\diamond^b_\eta}(\onept) = \eta$ for each $\eta \in U_b$, $b \in B$.

\begin{lemma}\label{lemma_free_algs_1}
Let $(X,p)$ and $(Y,q)$ be $U$-based $\Lambda$-algebras, let $(X,p')$ 
and $(Y,q')$ be the associated $\Lambda_U$-algebras and let $\pi : X \to Y$ 
be a family of mappings. Then $\pi$ is a homomorphism from $(X,p)$ to $(Y,q)$ 
fixing $U$ if and only if it is a homomorphism from $(X,p')$ to $(Y,q')$.
\end{lemma}

\proof This is clear. \eop

\begin{lemma}\label{lemma_free_algs_2}
There exists a $U$-based $\Lambda$-algebra whose associated $\Lambda_U$-algebra 
is an initial $\Lambda_U$-algebra. 
\end{lemma}

\proof By Proposition~\ref{prop_init_algs_1} there exists an initial $\Lambda_U$-algebra, 
and it is easy to see that there then exists an initial $\Lambda_U$-algebra $(Y,q')$ with 
$Y \cap U = \varnothing$. For each $b \in B$ put 
$V_b = \{ q'_{\diamond^b_\eta}(\onept) : \eta \in U_b \}$; then $V_b \subset Y_b$ and, 
since $(Y,q')$ is regular, there is a unique bijection $\varrho_b : V_b \to U_b$ such that
$\varrho_b(q'_{\diamond^b_\eta}(\onept)) = \eta$ for each $\eta \in U_b$. Now put 
$X_b = U_b \cup (Y_b \setminus V_b)$ and extend $\varrho_b$ to a bijection 
$\varrho_b : Y_b \to X_b$ by letting $\varrho_b(y) = y$ for all $y \in Y_b \setminus V_b$.
Also for each $k \in K_U$ let $p'_k : \ass{\sdom{k}}{X} \to X_{\scod{k}}$ be  
given by $p'_k =  \varrho_{\scod{k}} \comp q'_k\comp\ass{\sdom{k}}{(\varrho^{-1})}$. This 
results in a $\Lambda_U$-algebra $(X,p')$, and 
\[ \varrho_{\scod{k}}\comp q'_k 
  = \varrho_{\scod{k}} \comp q'_k\comp \ass{\sdom{k}}{(\varrho^{-1})} 
\comp \ass{\sdom{k}}{\varrho} 
      = p'_k\comp\ass{\sdom{k}}{\varrho}\]
for each $k \in K_U$,
which means that $\varrho$ is a homomorphism, and therefore an isomorphism, from 
$(Y,q')$ to $(X,p')$. Hence $(X,p')$ is itself an initial 
$\Lambda_U$-algebra. But by definition $(X,p) = (X,\rest{p'}{K})$ is a $U$-based 
$\Lambda$-algebra and clearly $(X,p')$ is the $\Lambda_U$-algebra associated with $(X,p)$,
since $p'_{\diamond^b_\eta}(\onept) = \varrho_b(q'_{\diamond^b_\eta}(\onept)) = \eta$ for 
each $\eta \in U_b$ and each $b \in B$. \eop

\begin{lemma}\label{lemma_free_algs_3} 
A $U$-based $\Lambda$-algebra is $U$-initial if and only if the associated 
$\Lambda_U$-algebra is initial.
\end{lemma}

\proof Let $(X,p)$ be a $U$-based $\Lambda$-algebra and suppose first that the associated 
$\Lambda_U$-algebra $(X,p')$ is initial. Let $(Y,q)$ be an arbitrary $U$-based 
$\Lambda$-algebra and let $(Y,q')$ be the associated $\Lambda_U$-algebra. There is thus a 
unique homomorphism $\pi : (X,p') \to (Y,q')$, which by Lemma~\ref{lemma_free_algs_1} is 
the unique homomorphism from $(X,p)$ to $(Y,q)$ fixing $U$. Thus $(X,p)$ is a $U$-initial 
$\Lambda$-algebra.

Suppose conversely $(X,p)$ is a $U$-initial $\Lambda$-algebra. By 
Lemma~\ref{lemma_free_algs_2} there exists a  $U$-based $\Lambda$-algebra $(Y,q)$ whose 
associated $\Lambda_U$-algebra $(Y,q')$ is initial. Let $\pi$ be the unique homomorphism 
from $(X,p)$ to $(Y,q)$ fixing $U$ and let $\pi'$ be the unique $\Lambda_U$-homomorphism 
from $(Y,q')$ to $(X,p')$. Then by Proposition~\ref{prop_algs_homs_1} and 
Lemma~\ref{lemma_free_algs_1} it 
follows that $\pi'\fcomp\pi$ is a homomorphism fixing $U$ from $(X,p)$ to itself and so 
$\pi'\fcomp\pi_b$ must be the family of identity mappings on $X$. In the same way 
$\pi\fcomp\pi'$ is a homomorphism from $(Y,q')$ to itself and so $\pi\fcomp\pi'$ is 
the family of identity mappings on $Y$. This implies that $\pi'$ is an 
isomorphism and hence $(X,p')$, being isomorphic to the initial $\Lambda_U$-algebra 
$(Y,q')$, is itself initial. \eop

In particular Lemma~\ref{lemma_free_algs_3}, together with Lemma~\ref{lemma_free_algs_2}, 
implies that a $U$-initial $\Lambda$-algebra exists. It will next be shown that $U$-initial 
$\Lambda$-algebra is $U$-free.

\begin{lemma}\label{lemma_free_algs_4} 
A $U$-initial $\Lambda$-algebra $(X,p)$ is $U$-free.
\end{lemma}

\proof By Lemma~\ref{lemma_free_algs_3} the associated $\Lambda_U$-algebra $(X,p')$ is 
initial. Let $(Y,q)$ be any $\Lambda$-algebra and let $\varrho : U \to Y$ be a $B$-family 
of mappings. For each $k \in K$ let $q'_k = q_k$ and for 
$\eta \in U_b$ let $q'_{\diamond^b_\eta} : \Oneptset \to Y_b$ be given by 
$q'_{\diamond^b_\eta}(\onept) = \varrho_b(\eta)$; thus $(Y,q')$ is a 
$\Lambda_U$-algebra and so there exists a unique homomorphism $\pi$ from 
$(X,p')$ to $(Y,q')$. But then $\pi$ is also a homomorphism from
$(X,p)$ to $(Y,q)$ with
\[ \pi_b(\eta) = \pi_b(p_{\diamond^b_\eta}(\onept)) = q'_{\diamond^b_\eta}(\onept)
  = \varrho_b(\eta)\]
for each $\eta \in U_b$, $b \in B$. Moreover, $\pi$ is the unique such homomorphism 
since, conversely, any homomorphism $\pi' : (X,p) \to (Y,q)$ with 
$\pi'_b(\eta) = \varrho_b(\eta)$ for all $\eta \in U_b$, $b \in B$ is also a homomorphism 
from $(X,p')$ to $(Y,q')$. Hence $(X,p)$ is $U$-free. \eop

This completes the proof of Proposition~\ref{prop_free_algs_2},
since clearly a $U$-free $\Lambda$-algebra is $U$-initial. \eop

There is a characterisation of $U$-initial $\Lambda$-algebras, and thus of
$U$-free $\Lambda$-algebras, which corresponds to the second part of 
Proposition~\ref{prop_init_algs_1}. Again let $A$ be the parameter set of $\Lambda$.

A $U$-based $\Lambda$-algebra $(X,p)$ is said to be 
\definition{$U$-regular}\index{regular algebra}\index{algebra!regular} if 
$\Im(p_k) \cap U_b = \varnothing$ for each $k \in K_b$ and if for each $b \in B \setminus A$
and each $x \in X_b \setminus U_b$ there exists a unique $k \in K_b$ and a unique element 
$v \in \ass{\sdom{k}}{X}$ such that $p_k(v) = x$. Thus $(X,p)$ is $U$-regular if and only 
if the mapping $p_k$ is injective for each $k \in K$ and for each $b \in B\setminus A$ 
the sets $\Im(p_k)$, $k \in K_b$, form a partition of $X_b \setminus U_b$.
Similarly, a $U$-based $\Lambda$-algebra $(X,p)$ is said to be 
\definition{$U$-unambiguous}\index{unambiguous algebra}\index{algebra!unambiguous}
if the mapping $p_k$ is injective for each $k \in K$ and for each $b \in B\setminus A$ 
the sets $\Im(p_k)$, $k \in K_b$, are disjoint subsets of $X_b \setminus U_b$.

\begin{proposition}\label{prop_free_algs_3} 
 The following are equivalent for a $\Lambda$-algebra $(X,p)$:
\begin{evlist}{15pt}{0.5ex}
\item[(1)] $(X,p)$ is $U$-free.
\item[(2)] $(X,p)$ is $U$-initial.
\item[(3)] $(X,p)$ is $U$-minimal and $U$-regular.
\item[(4)] $(X,p)$ is $U$-minimal and $U$-unambiguous.
\end{evlist}
\end{proposition}

\proof By Proposition~\ref{prop_init_algs_1} and Lemma~\ref{lemma_free_algs_3} it follows 
that a $U$-based $\Lambda$-algebra 
$(X,p)$ is $U$-initial if and only if the associated $\Lambda_U$-algebra $(X,p')$ is 
minimal and regular. But clearly $(X,p')$ is regular if and only if $(X.p)$ is
$U$-regular and it is minimal if and only if $(X,p)$ is $U$-minimal
(since a family is invariant in $(X,p')$ if and only if it contains $U$ 
and is invariant in $(X,p)$). Thus a $\Lambda$-algebra is $U$-initial if and only if
it $U$-minimal and $U$-regular. This, together with Proposition~\ref{prop_free_algs_2},
shows the equivalence of (1), (2) and (3). The equivalence of (3) and (4) then follows from
Proposition~\ref{prop_inv_fams_1}. \eop

If $(X,p)$ is $U$-minimal then by Proposition~\ref{prop_inv_fams_1} 
$\rest{X}{A} = \rest{U}{A}$ and therefore by Proposition~\ref{prop_free_algs_3} this also 
holds for $U$-free and $U$-initial $\Lambda$-algebras.

\begin{lemma}\label{lemma_free_algs_5} 
Let $(X,p)$ be a $U$-free $\Lambda$-algebra which is an extension of an initial 
$\Lambda$-algebra $(Z,r)$. Then $U \cap Z = \varnothing$.
\end{lemma}

\proof By Proposition~\ref{prop_free_algs_3} $\Im(p_k) \subset X_b \setminus U_b$ for all 
$k \in K_b$, 
$b \in B$. But clearly $\Im(r_k) \subset \Im(p_k)$ for each $k \in K$, and so by 
Proposition~\ref{prop_inv_fams_1} $Z_b \subset X_b \setminus U_b$ for all $b \in B$,
i.e., $U \cap Z = \varnothing$. \eop 

The following is a variation on Proposition~\ref{prop_free_algs_2}:

\begin{proposition}\label{prop_free_algs_4}  
Let $(Z,r)$ be an initial $\Lambda$-algebra with $U \cap Z = \varnothing$. Then there  
exists a $U$-free $\Lambda$-algebra extending $(Z,r)$. Moreover, if $(X,p)$ and 
$(X',p')$ are any two such $U$-free $\Lambda$-algebras and $\pi$ is the unique 
isomorphism from $(X,p)$ to $(X',p')$ such that $\pi_b(\eta) = \eta$ for each 
$\eta \in I_b$, $b \in B$ then $\pi_b(z) = z$ for all $z \in Z_b$, $b \in B$.
\end{proposition}

\proof By Proposition~\ref{prop_free_algs_2} there exists a $U$-free $\Lambda$-algebra 
$(\bar{X},\bar{p})$;
consider the minimal subalgebra $(\bar{Z},\bar{r})$ of $(\bar{X},\bar{p})$. By 
Proposition~\ref{prop_free_algs_3} $(\bar{Z},\bar{r})$ is unambiguous and therefore by 
Proposition~\ref{prop_init_algs_1} and Lemma~\ref{lemma_init_algs_1} $(\bar{Z},\bar{r})$ 
is an initial $\Lambda$-algebra. 
Moreover, by Lemma~\ref{lemma_free_algs_5} $(\bar{Z},\bar{r})$ is disjoint from $U$ and by 
construction 
$(\bar{X},\bar{p})$ is an extension of $(\bar{Z},\bar{r})$. The unique isomorphism from 
$(Z,r)$ to $(\bar{Z},\bar{r})$ can now be used to construct a $U$-free $\Lambda$-algebra 
$(X,p)$ which is an extension of $(Z,r)$ (and note that here the assumption is needed that
$U \cap Z = \varnothing$). Finally, consider any $\Lambda$-algebra $(X',p')$ which is 
an extension of $(Z,r)$ and let $\pi$ be any homomorphism from $(X,p)$ to $(X',p')$. Then 
$\pi_b(z) = z$ for all $z \in Z_b$, $b \in B$, since the family $Z'$ defined by 
$Z'_b = \{ z \in Z_b : \pi_b(z) = z \}$ for each $b \in B$ is invariant in $(Z,r)$.
\eop

Let us close the discussion of free algebras with a result which corresponds to 
Proposition~\ref{prop_init_algs_3}. Let
$\Lambda = (K,\vartheta)$ be a single-sorted signature, let 
$\mathcal{C}$ be a class and let $\varphi$ be a $K$-family of mappings with
$\varphi_k : \mathcal{C}^{\vartheta(k)} \to \mathcal{C}$ for each $k \in K$. Recall that if 
$(X,p)$ is a $\Lambda$-algebra then a mapping $\pi : X \to \mathcal{C}$ is still called a 
homomorphism from $(X,p)$ to $(\mathcal{C},\varphi)$ if 
$\pi(p_k(v)) = \varphi_k(\pi^{\vartheta(k)}(v))$ for all $v \in X^{\vartheta(k)}$, 
$k \in K$. 

\begin{proposition}\label{prop_free_algs_5}   
Let $I$ be a set and $\lambda : I \to \mathcal{C}$ be a mapping. If $(X,p)$ is an 
$I$-free $\Lambda$-algebra then there exists a unique homomorphism 
$\pi : (X,p) \to (\mathcal{C},\varphi)$ such that $\pi(\eta) = \lambda(\eta)$ for 
each $\eta \in I$. 
\end{proposition}

\proof By Lemma~\ref{lemma_free_algs_3} the $\Lambda_I$-algebra $(X,p')$ associated with 
$(X,p)$ is initial. For $k \in K$ put $\varphi'_k = \varphi_k$ and for each $\eta \in I$  
define $\varphi'_{\diamond^\onept_\eta} : \Oneptset \to \mathcal{C}$ by 
$\varphi'_{\diamond^\onept_\eta}(\onept) = \lambda(\eta)$. 
Hence by Proposition~\ref{prop_init_algs_3} there exists a homomorphism 
$\pi : (X,p') \to (\mathcal{C},\varphi')$. But then $\pi$ is also a homomorphism from 
$(X,p)$ to $(\mathcal{C},\varphi)$ with 
\[\pi(\eta) = \pi(p'_{\diamond^\onept_\eta}(\onept)) 
   = \varphi'_{\diamond^\onept_\eta}(\onept) = \lambda(\eta)\]
for each $\eta \in I$. The uniqueness follows from the fact given in
Proposition~\ref{prop_free_algs_3}~(3) that $X$ is the only invariant set in $(X,p)$ 
containing $I$. \eop

\startsection{Bound algebras}
\label{bound_algs}

The title of this section should not be seen as having anything to do with the title of 
the previous section. It should be considered rather as an abbreviation of `algebras
bound to an $A$-family of sets'. 

As before $A$ denotes the parameter set 
$B \setminus \Im(\scod{\Theta}) = \{ b \in B : K_b = \varnothing \}$ of $\Lambda$. 
For open signatures it is usually the case that an $A$-family of sets $V$ 
is given and the interest is then only in $\Lambda$-algebras $(X,p)$ with 
$\rest{X}{A} = V$. This is the situation which will be dealt with here.

Let $V$ be an $A$-family of sets, which is considered to be fixed in what follows. 
A $\Lambda$-algebra $(X,p)$ is said to be 
\definition{bound to $V$}\index{bound algebra}\index{algebra!bound to a family} 
if $\rest{X}{A} = V$. 
Of course, if $\Lambda$ is closed (i.e., if $A = \varnothing$) then there is only one 
$A$-family of sets and any $\Lambda$-algebra is bound to it.
If $(X,p)$ and $(X',p')$ are $\Lambda$-algebras bound to $V$ 
then a homomorphism $\pi : (X,p) \to (X',p')$ is said to 
\definition{fix}\index{homomorphism!fixing a family} $V$ if $\pi_a(x) = x$ for 
each $x \in V_a$, $a \in A$. 
Again, if $\Lambda$ is closed then this imposes no requirement on a homomorphism.

\begin{proposition}\label{prop_bound_algs_1}
(1)\enskip If $(X,p)$ is a $\Lambda$-algebra bound to $V$ then
the $B$-family of identity mappings $\id : X \to X$ defines a homomorphism 
from $(X,p)$ to itself fixing $V$.

(2)\enskip If $(X,p)$, $(Y,q)$ and $(Z,r)$ are $\Lambda$-algebras bound to $V$ and
$\pi : (X,p) \to (Y,q)$ and $\varrho : (Y,q) \to (Z,r)$ are homomorphisms fixing $V$ 
then the composition $\varrho\fcomp\pi$ 
is a homomorphism from $(X,p)$ to $(Z,r)$ fixing $V$.
\end{proposition}

\proof This follows immediately from Proposition~\ref{prop_algs_homs_1}. \eop

Proposition~\ref{prop_bound_algs_1} implies there is a category whose objects are 
$\Lambda$-algebras bound to $V$ and whose morphisms are homomorphisms fixing $V$. 
A $\Lambda$-algebra $(X,p)$ is called 
\definition{$V$-initial}\index{initial algebra}\index{algebra!initial} if it is an 
initial object in this category, i.e., if it is bound to $V$ and 
for each $\Lambda$-algebra $(X',p')$ bound to $V$ there exists a unique homomorphism
$\pi : (X,p) \to (X',p')$  fixing $V$.

\begin{proposition}\label{prop_bound_algs_2} 
There exists a $V$-initial $\Lambda$-algebra.
\end{proposition}

\proof 
For the rest of the section let $\bar{V}$ be the 
$B$-family of sets with $\rest{\bar{V}}{A} = V$ and 
$\rest{\bar{V}}{B\setminus A} = \varnothing$; this $B$-family  will be called the
\index{trivial extension}\index{extension!trivial}\definition{trivial extension of $V$ to $B$}. 
Note that any $\Lambda$-algebra 
bound to $V$ is then $\bar{V}$-based. To show the existence of a $V$-initial 
$\Lambda$-algebra part of the following result is needed (the remainder being required 
for the proof of Proposition~\ref{prop_bound_algs_3} below). 

\begin{lemma}\label{lemma_bound_algs_1}   
An initial $\bar{V}$-based $\Lambda$-algebra is $V$-initial. Moreover,
a $\Lambda$-algebra is $V$-initial if and only if it is $\bar{V}$-initial.
\end{lemma}

\proof Proposition~\ref{prop_free_algs_3} and Proposition~\ref{prop_inv_fams_2} imply 
that a $\bar{V}$-initial 
$\Lambda$-algebra is bound to $V$, and thus is a $V$-initial $\Lambda$-algebra (since 
each $\Lambda$-algebra bound to $V$ is 
$\bar{V}$-based, and fixing $\bar{V}$ is of course the same as fixing $V$). This also shows that
a $\Lambda$-algebra bound to $V$ and which is a $\bar{V}$-initial $\Lambda$-algebra 
is $V$-initial. Conversely, let $(X,p)$ be $V$-initial,
and let $(X',p')$ be a $\bar{V}$-initial $\Lambda$-algebra (whose existence is 
guaranteed by Proposition~\ref{prop_free_algs_3}). By the first part of the proof 
$(X',p')$ is then  $V$-initial. Therefore $(X,p)$ and $(X',p')$ are 
isomorphic as $\Lambda$-algebras bound to $V$, and so they are also isomorphic as 
$\bar{V}$-based $\Lambda$-algebras. This 
implies then that $(X,p)$ is itself $\bar{V}$-initial. \eop

The existence of a $V$-initial $\Lambda$-algebra now follows from 
Proposition~\ref{prop_free_algs_3} and the first statement in 
Lemma~\ref{lemma_bound_algs_1}. \eop

For the situation being considered here there is a result which corresponds to 
Proposition~\ref{prop_init_algs_2}. A $\Lambda$-algebra $(X,p)$ bound to $V$ is said to be  
\definition{$V$-minimal}\index{minimal algebra}\index{algebra!minimal} 
if $X$ is the only invariant family $Y$ in $(X,p)$ such 
that $\rest{Y}{A} = V$.

\begin{lemma}\label{lemma_bound_algs_2}
A $V$-minimal $\Lambda$-algebra is $\bar{V}$-minimal. Conversely, a $\bar{V}$-minimal 
$\Lambda$-algebra bound to $V$ is $V$-minimal. 
\end{lemma}

\proof Let $(X,p)$ be a $\Lambda$-algebra bound to $V$. Then
an invariant family $Y$ satisfies $\rest{Y}{A} = V$ if and only if it
contains $\bar{V}$. This implies that $(X,p)$ is $V$-minimal if and only if it is 
$\bar{V}$-minimal.
The result thus follows because a $V$-minimal $\Lambda$-algebra is bound to $V$
by definition. \eop

\begin{proposition}\label{prop_bound_algs_3} 
 The following are equivalent for a $\Lambda$-algebra $(X,p)$:
\begin{evlist}{15pt}{0.5ex}
\item[(1)] $(X,p)$ is $V$-initial.
\item[(2)] $(X,p)$ is $V$-minimal and regular.
\item[(3)] $(X,p)$ is $V$-minimal and unambiguous.
\end{evlist}
\end{proposition}

\proof Let $(X,p)$ be a $\Lambda$-algebra bound to $V$. Then by 
Lemma~\ref{lemma_bound_algs_2}
$(X,p)$ is $V$-minimal if and only if it is $\bar{V}$-minimal. Furthermore, (since 
$\bar{V}_b = \varnothing$ for each $b \in B\setminus A$) $(X,p)$ is regular if and only if 
it is $\bar{V}$-regular and unambiguous if and only if it is $\bar{V}$-unambiguous. The result 
thus follows from Proposition~\ref{prop_free_algs_3} and the second statement in 
Lemma~\ref{lemma_bound_algs_1}. \eop

\begin{proposition}\label{prop_bound_algs_4}  
If $(X,p)$ is a $V$-minimal $\Lambda$-algebra then $X_b = \bigcup_{k \in K_b} \Im(p_k)$ 
for each $b \in B\setminus A$. Moreover, a graded $\Lambda$-algebra $(X,p)$ is $V$-minimal
if and only if $X_b  = \bigcup_{k \in K_b} \Im(p_k)$ for each $b \in B\setminus A$.
\end{proposition}

\proof This follows from Propositions \ref{prop_inv_fams_2} and \ref{prop_inv_fams_3} and 
Lemma~\ref{lemma_bound_algs_2}. \eop

There is an obvious grading for the $\Lambda$-algebra $(X,p)$ defined in
Example~\ref{algs_homs}.3, and it is thus easy to check that this algebra is $V$-minimal. 
It then follows that $(X,p)$ is $V$-initial, since it is also clearly unambiguous.

Note that if $(X,p)$ is a $V$-initial $\Lambda$-algebra and $b \in B\setminus A$ is a 
primitive type then by Proposition~\ref{prop_bound_algs_3} the mapping 
$k \mapsto p_k(\onept)$ maps $K_b$
bijectively onto $X_b$. (For the $\Lambda$-algebra $(X,p)$ in Example~\ref{algs_homs}.3 
with primitive 
types $\mathtt{atom}$, $\mathtt{bool}$ and $\mathtt{int}$ and this gives the obvious 
bijections between $K_{\mathtt{atom}} = \{\mathtt{Atom}\}$ and
$X_{\mathtt{atom}} = \Oneptset$, between 
$K_{\mathtt{bool}} = \{\mathtt{True},\mathtt{False}\}$ and
$X_{\mathtt{bool}} = \Bool = \{\True,\False\}$ and between $K_{\mathtt{int}} = \SynInt$ and 
$X_{\mathtt{int}} = \Int$.)

It turns out that a $V$-initial $\Lambda$-algebra has a seemingly stronger property
than that of just being $V$-initial: A $\Lambda$-algebra $(X,p)$ will be called 
\definition{intrinsically free}\index{intrinsically free algebra}\index{algebra!intrinsically free} 
if for each $\Lambda$-algebra $(Y,q)$ and each $A$-family 
of mappings $\varrho : \rest{X}{A} \to \rest{Y}{A}$ there exists a unique 
homomorphism $\pi : (X,p) \to (Y,q)$ such that $\rest{\pi}{A} = \varrho$.

\begin{proposition}\label{prop_bound_algs_42}  
A $\Lambda$-algebra $(X,p)$ bound to $V$ is $V$-initial if and only if it is
intrinsically free.
\end{proposition}

\proof 
An intrinsically free $\Lambda$-algebra bound to $V$ is clearly $V$-initial and the converse
follows immediately from Lemma~\ref{lemma_bound_algs_1} and Lemma~\ref{lemma_free_algs_4}.
\eop

\begin{proposition}\label{prop_bound_algs_5}  
Let $(X,p)$ be a $\Lambda$-algebra and let $U$ be an $A$-family of sets with
$U \subset \rest{X}{A}$. Then there exists a 
unique $U$-minimal subalgebra $(\hat{X},\hat{p})$ of $(X,p)$.
\end{proposition}

\proof Let $(\hat{X},\hat{p})$ be the minimal subalgebra of $(X,p)$ containing $\bar{U}$
(the trivial extension of $U$ to $B$).
Then by Lemma~\ref{lemma_inv_fams_4} $\rest{\hat{X}}{A} = U$, since $K_a = \varnothing$ 
for each $a \in A$, i.e., $(\hat{X},\hat{p})$ is bound to $U$. Moreover,
$(\hat{X},\hat{p})$ is $U$-minimal, since if $X'$ is an invariant family in
$(\hat{X},\hat{p})$ with $\rest{X'}{A} = \rest{\hat{X}}{A}$ then $X'$ is an invariant 
family in $(X,p)$ containing $\bar{U}$ and hence $X' = \hat{X}$. 
Finally, if $(Y,q)$ is any $U$-minimal subalgebra of $(X,p)$ bound to $U$
then the $B$-family $X' = \hat{X} \cap Y$ is invariant in
$(X,p)$ and contains $\bar{U}$. Thus $Y = X' = \hat{X}$, which gives the 
uniqueness of $(\hat{X},\hat{p})$. \eop

\begin{lemma}\label{lemma_bound_algs_3}
A $\Lambda$-algebra is initial if and only if it is $\varnothing$-initial.
\end{lemma}

\proof First note that a homomorphism between $\Lambda$-algebras bound to $\varnothing$
trivially fixes $\varnothing$. Moreover, as already mentioned above, an initial 
$\Lambda$-algebra is bound to $\varnothing$, and therefore it is $\varnothing$-initial.
Conversely, let $(X,p)$ be $\varnothing$-initial and consider any $\Lambda$-algebra 
$(Y,q)$. Now the family $Y'$ with $\rest{Y'}{B\setminus A} = \rest{Y}{B\setminus A}$
and $\rest{Y'}{A} = \varnothing$ is invariant in $(Y,q)$ and the associated
subalgebra $(Y',q')$ is bound to $\varnothing$, there thus exists a unique
homomorphism $\pi : (X,p) \to (Y',q')$. But $\pi$ is then the unique 
homomorphism from $(X,p)$ to $(Y,q)$ (it being unique because any  
homomorphism from $(X,p)$ to $(Y,q)$ is also a homomorphism from $(X,p)$ to 
$(Y',q')$). Hence $(X,p)$ is initial. \eop


\startsection{Term algebras}
\label{term_algs}

This section introduces what are called term algebras. These provide the simplest explicit 
examples of initial algebras, and they can be used as the basic components of a `real' 
programming language. The main result
(Proposition~\ref{prop_term_algs_1}) is really just a version of the classical result of
{\L}ukasiewicz \cite{lukasiewicz} concerning prefix or (left Polish) 
notation. Until further notice assume that $\Lambda$ is an enumerated signature;
the general case is dealt with at the end of the section.

Let $Z$ be a set; then the concatenation of two lists $\ell,\, \ell' \in Z^*$ will be 
denoted by $\ell\,\ell'$. Thus $\ell\,\onept = \onept\,\ell = \ell$ and if 
$\ell = \llist{z}{m}$, $\ell' = \llist{z'}{n}$ with $m,\, n \ge 1$ then
\[\ell \,\ell'  = \llist{z}{m}\,\llist{z'}{n}\; .\] 
Concatenation is clearly associative, and hence if $p \ge 1$ and 
$\lvector{\ell}{p} \in Z^*$ then \textit{the} concatenation of $\lvector{\ell}{p}$ can be 
denoted simply by $\llist{\ell}{p}$. This notation is clearly compatible with the 
notation being employed for the elements of $Z^*$ (in the sense that 
$\ell = \llist{z}{n} \in Z^*$ can be considered as the concatenation of the $n$ one 
element lists $\lvector{z}{n}$). If $z \in Z$ and $\ell \in Z^*$ then in particular there 
is the list $z\,\ell$ obtained by adding $z$ to the beginning of the list $\ell$
(a list which is also being denoted by $z \triangleleft \ell$).

Let $\Omega$ be a set and $\Gamma : K \to \Omega$ be a mapping. A $\Lambda$-algebra 
$(Y,q)$ can then be obtained as follows: For each $b \in B$ put $Y_b = \Omega^*$; 
if $k \in K$ is of type $\llist{b}{n} \to b$ (with $n \ge 0$) then let
$q_k : Y_{b_1} \times \cdots \times Y_{b_n} \to Y_b$ be the mapping defined by
\[ q_k(\svector{y}{n}) = \Gamma(k)\,\llist{y}{n}\]
for each $(\svector{y}{n}) \in Y_{b_1} \times \cdots \times Y_{b_n}$,
i.e., $q_k(\svector{y}{n})$ is the list obtained by concatenating the element $\Gamma(k)$ 
and the lists $\lvector{y}{n}$. In particular, if $k \in K$ is of type $\onept \to b$ then
$q_k(\onept)$ is just the list consisting of the single component $\Gamma(k)$. 
(This is really just an instance of the construction described at the end of 
Section~\ref{algs_homs}.)

Now let $(E,\blob)$ be the minimal subalgebra of $(Y,q)$, i.e., $E$ is the 
minimal invariant family and $\blob_k$ is the corresponding restriction of $q_k$ for each 
$k \in K$. This minimal $\Lambda$-algebra $(E,\blob)$ is called the 
\definition{term $\Lambda$-algebra specified by $\Gamma$}\index{term algebra} and 
$\Gamma$ is then referred to as a
\index{term algebra specifier}\index{specifier!term algebra}\definition{term algebra specifier}. 
By Proposition~\ref{prop_inv_fams_2} it is clear that each element of $E_b$ contains at 
least one component, i.e., $\onept \notin E_b$ for each $b \in B$. 

The simplest example of this construction is with $\Omega = K$ and with $\Gamma : K \to K$ 
the identity mapping: The term $\Lambda$-algebra specified by this mapping is called the
\index{standard term algebra}\index{term algebra!standard}\definition{standard term $\Lambda$-algebra}.

It turns out that the standard term $\Lambda$-algebra is initial. This is a special case 
of Proposition~\ref{prop_term_algs_1} below.

Let $\Gamma : K \to \Omega$ be a term algebra specifier and $(E,\blob)$ be the term 
$\Lambda$-algebra specified by $\Gamma$. For each $k \in K$ let 
$\chi_k : \ass{\sdom{k}}{E} \times \Omega^* \to \Omega^*$ be defined by 
\[\chi_k(v,\alpha)  = \blob_k(v)\,\alpha\; .\] 

\begin{lemma}\label{lemma_term_algs_1}
Suppose $\Im(\chi_{k_1})$ and $\Im(\chi_{k_2})$ are disjoint subsets of $\Omega^*$ whenever 
$b \in B$ and $k_1,\, k_2 \in K_b$ with $k_1 \ne k_2$. Then $(E,\blob)$ is initial. 
\end{lemma}

\proof By Proposition~\ref{prop_init_algs_2} it is enough to show $(E,\blob)$ is 
unambiguous. But
\[\Im(\blob_k) = \chi_k(\ass{\sdom{k}}{E} \times \{\onept\}) \subset \Im(\chi_k)\]
for each $k \in K$; thus if $k_1,\, k_2 \in K_b$ with $k_1 \ne k_2$ then 
$\Im(\blob_{k_1})$ and $\Im(\blob_{k_2})$ are clearly disjoint. It therefore remains to 
show that $\blob_k$ is injective for each $k \in K$, and for this the following lemma is 
required:

\begin{lemma}\label{lemma_term_algs_2}
Suppose that the assumption in the statement of Lemma~\ref{lemma_term_algs_1} is satisfied 
and let $\chi_b : E_b \times \Omega^* \to \Omega^*$ be defined by 
$\chi_b(e,\alpha) = e\,\alpha$. Then for each $b \in B$ the mapping $\chi_b$ is injective.
\end{lemma}

\proof For each $b \in B$ define a subset $G_b$ of $\Omega^*$ by
\[G_b  = \{ e \in E_b : e\,\alpha \in E_b 
    \ \textrm{for some}\ \alpha \in \Omega^* \ \textrm{with}\ \alpha \ne \onept \}\]  
and put $G = \bigcup_{b \in B} G_b$. Then, since $\chi_b$ is 
injective if and only if $G_b = \varnothing$, it follows that $G \ne \varnothing$
if and only if $\chi_b$ is not injective for some $b \in B$. 

Suppose that $G \ne \varnothing$ and let $m = \min \{ |\alpha| : \alpha \in G \}$, where 
$|\alpha|$ denotes the number of components in the list $\alpha \in \Omega^*$. There thus 
exists $b \in B$, $e \in E_b$ and $\alpha \in \Omega^* \setminus \{\onept\}$ such that 
$|e| = m$ and $e' = e\,\alpha \in E_b$. Now by Proposition~\ref{prop_inv_fams_2} there 
exist $k,\,k' \in K_b$ and $v,\,v'$ such that 
$e = \blob_k(v)$ and $e' = \blob_{k'}(v')$, and this implies that
$\chi_k(v,\alpha) = e\,\alpha = e' = e'\,\onept = \chi_{k'}(v',\onept)$, which by 
assumption is only possible if $k' = k$. Let $k$ be of type $\llist{b}{n} \to b$ and put
$\Gamma(k) = \alpha'$. Thus if $v = (\svector{e}{n})$ and $v' = (\svector{e'}{n})$ then
$e = \alpha'\,\llist{e}{n}$ and $e\,\alpha = \alpha'\,\llist{e'}{n}$. There must therefore 
exist $1 \le j \le n$ with $e_j \ne e'_j$ (and so in particular $n > 0$). Let $i$ be the 
least index with $e_i \ne e'_i$ and let $\hat{e}$ be the shorter of the lists $e_i$ and 
$e'_i$. But then $\hat{e} \in G_{b_i}$ and $|\hat{e}| < |e|$, which by the minimality
of $|e|$ is not possible.

This shows that $G = \varnothing$ and hence that $\chi_b$ is injective for each $b \in B$.
\eop 

The proof of Lemma~\ref{lemma_term_algs_1} can now be completed. Thus let $k \in K$ be of 
type $\llist{b}{n}\to b$ and let $v,\,v' \in \ass{\sdom{k}}{E}$ with 
$\blob_k(v) = \blob_k(v')$. 
This means that if $v = (\svector{e}{n})$, $v' = (\svector{e'}{n})$ then 
$\Gamma(k)\,\llist{e}{n} = \Gamma(k)\,\llist{e'}{n}$. Consider $1 \le j \le n$ and assume 
$e_i = e'_i$ for each $i < j$. Then 
\[ \chi_{b_j}(e_j,e_{j+1} \cdots e_n) = e_j\,e_{j+1}\cdots e_n
 = e'_j\,e'_{j+1}\cdots e'_n = \chi_{b_j}(e'_j,e'_{j+1}\cdots e'_n)\]
and so by Lemma~\ref{lemma_term_algs_1} $e_j = e'_j$. Therefore by $n$ applications of 
Lemma~\ref{lemma_term_algs_2} it follows that $e_j = e'_j$ for each $\oneto{j}{n}$; i.e., 
$v = v'$. This shows that $\blob_k$ is injective. \eop

The term algebra specifier $\Gamma : K \to \Omega$ will be called
\definition{locally injective}\index{term algebra specifier!locally injective}
if for each $b \in B$ the restriction of $\Gamma$ to $K_b$ 
is injective. (Of course, since $\rest{K}{A} = \varnothing$, this is 
equivalent to requiring that the restriction of $\Gamma$ to $K_b$ be injective
for each $b \in B\setminus A$.)

\begin{proposition}\label{prop_term_algs_1}
The term $\Lambda$-algebra $(E,\blob)$ specified by a locally injective specifier 
$\Gamma$ is initial. 
\end{proposition}

\proof This follows immediately from the fact that $\Gamma(k)$ is the first component of 
each element of $\Im(\chi_k)$, and hence the hypothesis of Lemma~\ref{lemma_term_algs_1}
is satisfied. \eop

\begin{proposition}\label{prop_term_algs_2}
The standard term $\Lambda$-algebra is initial. 
\end{proposition}

\proof This follows from Proposition~\ref{prop_term_algs_1}, since the identity mapping 
used to specify the standard term $\Lambda$-algebra is trivially locally injective. \eop

If $(E,\blob)$ is initial then the family $E$ can be thought of, somewhat informally, 
as being defined by the following rules:
\begin{evlist}{25pt}{0.8ex}
\item[(1)] If $k \in K$ is of type $\onept \to b$ then the list consisting of the single 
component $\Gamma(k)$ is an element of $E_b$.
\item[(2)] If $k \in K$ is of type $\llist{b}{n} \to b$ with $n \ge 1$ and 
$e_j \in E_{b_j}$ for each $j$ then $\Gamma(k)\,\llist{e}{n}$ is an element of $E_b$.
\item[(3)] Each element of $E_b$ can be obtained in a unique way using finitely
many applications of rules (1) and (2). 
\end{evlist}
These rules are just meant to be a rather imprecise statement of 
Proposition~\ref{prop_init_algs_1}.
In the special case when $(E,\blob)$ is the standard term $\Lambda$-algebra then 
$E_b \subset K^*$ for each $b \in B$ and the family $E$ is defined by the following 
rules:
\begin{evlist}{25pt}{0.8ex}
\item[(1)] If $k \in K$ is of type $\onept \to b$ then the list consisting of the single 
component $k$ is an element of $E_b$.
\item[(2)] If $k \in K$ is of type $\llist{b}{n} \to b$ with $n \ge 1$ and 
$e_j \in E_{b_j}$ for each $j$ then $k\,e_1 \cdots e_n$ is an element of $E_b$.
\item[(3)] Each element of $E_b$ can be obtained in a unique way using finitely
many applications of rules (1) and (2). 
\end{evlist}

\bigskip
\fbox{\begin{exframe}
\textit{Example~\thesection.1} Consider the standard term $\Lambda$-algebra 
$(E, \blob)$ arising from the enumerated signature $\Lambda$ in Example~\ref{algs_homs}.1.
Then:
\begin{eelist}{10pt}
\item $E_{\mathtt{bool}}$ consists of the two elements $\mathtt{True}$ and 
      $\mathtt{False}$ of $K^*$.
\item $E_{\mathtt{nat}}$ consists of exactly the following elements of $K^*$:
      \[\mathtt{Zero},\, \mathtt{Succ\ Zero},\, \mathtt{Succ\ Succ\ Zero}, 
                               \, \mathtt{Succ\ Succ\ Succ\ Zero},\, \ldots\;.\]
\item $E_{\mathtt{int}}$ consists of all elements of $K^*$ having the form 
      $\underline{n}$ with $n \in \Int$.
\item $E_{\mathtt{pair}}$ consists of all elements of $K^*$ of the form 
      $\mathtt{Pair}\,\underline{m}\,\underline{n}$ with $m,\,n \in \Int$. (Each element 
      of $E_{\mathtt{pair}}$ thus has exactly three components.)
\item $E_{\mathtt{list}}$ consists of the element $\mathtt{Nil}$ plus all elements of $K^*$ 
      having the form $\mathtt{Cons}\,\underline{n}\,e$ with $n \in \Int$ and 
      $e \in E_{\mathtt{list}}$. For instance, 
      \[\mathtt{Cons\ 42\ Cons\ -128\ Cons\ 0\ Cons\ -21\ Nil}\]
      is an element of $E_{\mathtt{list}}$.
\end{eelist}
\end{exframe}}

\bigskip
It will now no longer be assumed that $\Lambda$ is enumerated. In order to apply the above 
results to $\Lambda$ the problem of replacing the general signature 
$\Lambda = (B,K,\Theta)$ with an `equivalent' enumerated signature 
$\Lambda' = (B,K,\Theta')$ must be considered. In order to carry this out fix for each 
$k \in K$ an enumeration of the elements in the set $\domsdom{k} = \dom(\sdom{k})$: More 
precisely, choose a bijective mapping $i_k$ from $[n_k]$ to $\domsdom{k}$, where 
$n_k$ is the cardinality of $\domsdom{k}$. There is then a mapping $k \mapsto k^\circ$
from $K$ to $B^*$ given by $k^\circ = \sdom{k}\comp i_k$ for each $k \in K$, i.e., with
\[k^\circ = \sdom{k}(i_k(1)) \cdots \sdom{k}(i_k(n_k))\;,\]
and the mapping $\Theta' : K \to B^* \times B$ with 
$\Theta' = (k^\circ,\scod{k})$. This defines an enumerated signature 
$\Lambda' = (B,K,\Theta')$, which will be called the
\definition{signature obtained from $\Lambda$ and the family of enumerations $i$}.

\begin{lemma}\label{lemma_term_algs_3}
Let $X$ be a family of sets and let $k \in K$ with $k^\circ = \llist{b}{n_k}$. Then 
the mapping $i_k^* : \ass{\sdom{k}}{X} \to X_{b_1} \times \cdots \times X_{b_{n_k}}$
given by 
\[ i_k^*(v) = (v(i_k(1)),\ldots,v(i_k(n_k)))\]
for each $v \in \ass{\sdom{k}}{X}$ is a bijection. 
\end{lemma}

\proof This is clear. \eop

\begin{proposition}\label{prop_term_algs_3}
Let $(X,p')$ be a $\Lambda'$-algebra and put $p = p' \fcomp i^*$.
Then $(X,p)$ is a $\Lambda$-algebra. Conversely, if $(X,p)$ is any 
$\Lambda$-algebra then there exists a unique $\Lambda'$-algebra $(X,p')$ such that 
$p = p' \fcomp i^*$. 
\end{proposition}

\proof This follows immediately from Lemma~\ref{lemma_term_algs_3}. \eop

If $(X,p')$ is a $\Lambda'$-algebra and $p = p'\fcomp i^*$ then
$(X,p)$ is called the
\index{associated algebra}\definition{$\Lambda$-algebra associated with $(X,p')$ and $i$}.

Proposition~\ref{prop_term_algs_3} says there is a one-to-one correspondence between 
$\Lambda$-algebras
and $\Lambda'$-algebras. Moreover, Proposition~\ref{prop_term_algs_4} below implies that a 
$\Lambda'$-algebra has a property (such as being minimal or initial) if and only if the 
corresponding $\Lambda$-algebra also has this property. In this sense the signatures 
$\Lambda'$ and $\Lambda$ can be regarded as being equivalent.

\begin{proposition}\label{prop_term_algs_4}
Let $(X,p')$ be a $\Lambda'$-algebra and let $(X,p)$ be the $\Lambda$-algebra associated 
with $(X,p')$ and $i$. Then:

(1)\enskip $(X,p)$ is minimal if and only if $(X,p')$ is a minimal $\Lambda'$-algebra. 

(2)\enskip $(X,p)$ is initial if and only if $(X,p')$ is an initial $\Lambda'$-algebra. 
\end{proposition}

\proof (1)\enskip This follows from the easily verified fact that a family $\grave{X}$ 
is invariant in $(X,p')$ if and only if it is invariant in $(X,p)$. 

(2)\enskip This follows from (1) and Proposition~\ref{prop_init_algs_2}. \eop

Now let $\Gamma : K \to \Omega$ be a mapping, which will still be referred to as a 
\index{term algebra specifier}\definition{term algebra specifier}. Then $\Gamma$ is also 
a term algebra specifier for 
the signature $\Lambda'$, so let $(E,\blob')$ be the term $\Lambda'$-algebra
specified by $\Gamma$. Put $\blob = \blob'\fcomp i^*$ (with $i^*$ given by 
Lemma~\ref{lemma_term_algs_3}). This means that $E_b \subset \Omega^*$ for each $b \in B$ 
and for each $k \in K$ the mapping $\blob_k : \ass{\sdom{k}}{E} \to E_b$ is given by
\[ \blob_k(v) =\Gamma(k)\,v(i_k(1)) \cdots v(i_k(n_k))\]
for each $v \in \ass{\sdom{k}}{E}$. Moreover, the 
family $E$ can be regarded as being defined by the following rules:
\begin{evlist}{25pt}{0.8ex}
\item[(1)] If $k \in K$ with $\sdom{k} = \onept$ then the list consisting of the 
single component $\Gamma(k)$ is an element of $E_{\scod{k}}$.
\item[(2)] If $k \in K$ with $\sdom{k} \ne \onept$ and $e_j \in E_{b_j}$ for 
$\oneto{j}{n_k}$, where $b_j = \sdom{k}(i_k(j))$, then 
$\Gamma(k)\,e_1 \cdots e_{n_k}$ is an element of $E_{\scod{k}}$.
\item[(3)] Each element of $E_b$ can be obtained in a unique way using finitely
many applications of rules (1) and (2). 
\end{evlist}
The algebra $(E,\blob)$ is called the \definition{term $\Lambda$-algebra specified 
by $\Gamma$ and the family of enumerations $i$}, or simply the 
\definition{standard term $\Lambda$-algebra defined by the family $i$}
\index{standard term algebra}\index{term algebra!standard}in the special 
case when $\Gamma : K \to K$ is the identity mapping.

\begin{lemma}\label{lemma_term_algs_4}
The $\Lambda$-algebra $(E,\blob)$ is initial if and only if $(E,\blob')$ is an 
initial $\Lambda'$-algebra. 
\end{lemma}

\proof This follows from Proposition~\ref{prop_term_algs_4}~(2), since by definition 
$(E,\blob)$ is the $\Lambda$-algebra associated with $(E,\blob')$ and $i$. \eop

\begin{proposition}\label{prop_term_algs_5}
If $\Gamma$ is locally injective then $(E,\blob)$ is an initial $\Lambda$-algebra. In 
particular, the standard term $\Lambda$-algebra defined by any family of enumerations is 
initial. 
\end{proposition}

\proof This follows immediately from Proposition~\ref{prop_term_algs_1}
and Lemma~\ref{lemma_term_algs_4}. \eop

Of course, if the signature $\Lambda$ is enumerated and the above constructions are 
applied with the family of identity enumerations $i$ then the end-result is that 
nothing happens, since in this case $\Lambda' = \Lambda$.

The final topic of this section considers the construction of term algebras when a 
further signature $\Lambda' = (B',K',\Theta')$ is given which is an extension 
of $\Lambda$. Suppose in what follows that $\Gamma' : K' \to \Omega'$ is a term algebra 
specifier which is an \definition{extension}\index{term algebra specifier!extension of}
of a term algebra specifier 
$\Gamma : K \to \Omega$ in the sense that $\Omega \subset \Omega'$ and 
$\Gamma(k) = \Gamma'(k)$ for each $k \in K$.

\begin{lemma}\label{lemma_term_algs_5}
Assume that the signatures $\Lambda$ and $\Lambda'$ are both enumerated. Then the term 
$\Lambda'$-algebra specified by $\Gamma'$ is an extension of the term $\Lambda$-algebra 
$(E,\blob)$ specified by $\Gamma$. 
\end{lemma}

\proof This follows from Lemma~\ref{lemma_term_algs_6}, because the $\Lambda'$-algebra 
corresponding to the $\Lambda$-algebra $(Y,q)$ in the definition of $(E,\blob)$ is an 
extension of $(Y,q)$. \eop

\begin{lemma}\label{lemma_term_algs_6}
Let $(Y,q)$ be a $\Lambda'$-algebra which is an extension of a $\Lambda$-algebra $(X,p)$, 
and let $\hat{X}$ be the minimal invariant family in $(X,p)$ and $\hat{Y}$ the minimal 
invariant family in $(Y,q)$. Then $\hat{X} \subset \rest{\hat{Y}}{B}$. 
\end{lemma}

\proof Let $\grave{Y}$ be any invariant family in $(Y,q)$ and put 
$Z = \rest{\grave{Y}}{B} \cap X$. Then the family $Z$ is invariant in 
$(X,p)$, since 
\[ p_k(\ass{\sdom{k}}{Z}) \subset q_k(\ass{\sdom{k}}{\grave{Y}} ) \cap X_{\scod{k}}
                        \subset \grave{Y}_{\scod{k}} \cap X_{\scod{k}} =  Z_{\scod{k}}\]
for each $k \in K$. Thus $\hat{X} \subset Z$ and hence $\hat{X} \subset \grave{Y}$. In 
particular, $\hat{X} \subset \rest{\hat{Y}}{B}$. \eop

Lemma~\ref{lemma_term_algs_6} implies that the standard term $\Lambda'$-algebra defined 
by the family of enumerations $i'$ is an extension of the standard term 
$\Lambda$-algebra defined by the family $i$.


%% file: sbika3.tex
\startchapter{Bottomed algebras}
\label{bot_algs}

With the preparations made in Chapter~\ref{uni_alg}, the next step in building a framework
for specifying data objects will now be introduced. This involves the treatment
of `undefined' and `partially defined' objects.

In any programming language data objects are manipulated by algorithms, and in any
non-trivial language it is an unavoidable fact that algorithms need not terminate. It is 
thus necessary to introduce an `undefined' element for each type in order to represent 
this state of affairs. Moreover, depending on the language, it may also be necessary
to have `partially defined' data objects, for example a pair whose
first component is defined but not the second, or a list in which only some of the 
components are defined. To deal with this situation bottomed algebras will be introduced. 
These are algebras containing for each type $b \in B$ a special bottom element 
$\bot_b$ to denote an `undefined' element of the type.

\startsection{Bottomed algebras and homomorphisms}
\label{bot_algs_homs}

Let $\Lambda = (B,K,\Theta)$ be a signature which is considered to be
fixed throughout the chapter; $A$ will always denote the parameter set of $\Lambda$,
i.e., $A = \{ b \in B : K_b = \varnothing \}$. A 
\definition{bottomed $\Lambda$-algebra}\index{bottomed algebra}\index{algebra!bottomed}
is any pair $(X,p)$ consisting of a 
$B$-family of bottomed sets $X$ and a $K$-family of
mappings $p$ such that $p_k$ is a mapping from $\ass{\sdom{k}}{X}$ to $X_{\scod{k}}$ for 
each $k \in K$. If $(X,p)$ is a bottomed $\Lambda$-algebra and $\breve{X}_b$ 
is the underlying set of $X_b$ for each $b \in B$, then $(\breve{X},p)$ is an 
`ordinary' $\Lambda$-algebra which will be called the 
\definition{underlying $\Lambda$-algebra}\index{underlying algebra}\index{algebra!underlying}
or just the \definition{underlying algebra}.
Following the convention made in Section~\ref{sets} $X_b$ will also be used to denote
the underlying set $\breve{X}_b$, which means that the underlying and the bottomed 
$\Lambda$-algebra will both be denoted by $(X,p)$.

If $(X,p)$ is a bottomed $\Lambda$-algebra then the bottom element of $X_b$ will 
(almost) always be denoted by $\bot_b$. Thus if $(Y,q)$ is a further bottomed 
$\Lambda$-algebra then the bottom element of $Y_b$ will also be denoted by $\bot_b$, 
although it is not to be assumed that the bottom elements of $X_b$ and $Y_b$ are the same.
Recall that if $Z$ is a bottomed set with bottom element $\bot$ then $\nonbot{Z}$ denotes 
the set $Z \setminus \{\bot\}$; thus if $(X,p)$ is a bottomed $\Lambda$-algebra 
then $\nonbot{X}_b$ is the set $X_b \setminus \{\bot_b\}$ for each $b \in B$.

If $(Y,q)$ is a $\Lambda$-algebra then a bottomed $\Lambda$-algebra $(X,p)$ will be
called a 
\definition{bottomed extension}\index{bottomed extension}\index{extension!bottomed}
of $(Y,q)$ if the underlying algebra
of $(X,p)$ is an extension of $(Y,q)$ and $\bot_b \notin Y_b$ for each $b \in B$.
The simplest way to obtain bottomed $\Lambda$-algebras is 
a special case of a such an extension: Let $(Y,q)$ 
be any $\Lambda$-algebra and for each $b \in B$ choose an element $\bot_b$ not in $Y_b$.
For each $b \in B$ put $Y^\altbot_b = Y_b \cup \{\bot_b\}$, and consider $Y^\altbot_b$ as 
a bottomed set with bottom element $\bot_b$. For each $k \in K$ 
let $q^\altbot_k : \ass{\sdom{k}}{(Y^\altbot)} \to Y^\altbot_{\scod{k}}$ be given by
\[ q^\altbot_k(v) = \left\{ \begin{array}{cl} 
                 q_k(v) &\ \textrm{if}\ v \in \ass{\sdom{k}}{Y},\\
                 \bot_{\scod{k}} &\ \textrm{otherwise}. 
\end{array}\right.\]
Then $(Y^\altbot,q^\altbot)$ is a bottomed extension  called the 
\definition{flat bottomed extension}\index{flat bottomed extension}\index{bottomed extension!flat}
of $(Y,q)$. 

A bottomed $\Lambda$-algebra $(X,p)$ should be thought of a describing all data objects,
both the basic defined objects and those in various degrees of not being defined. 
In particular, $(X,p)$ will be a bottomed extension of a $\Lambda$-algebra $(Y,q)$ 
describing the defined data objects. However, it simplifies things to only work with 
$(X,p)$ without continually having to worry about $(Y,q)$. On the other hand, it only makes 
sense to work with bottomed $\Lambda$-algebras which are bottomed extensions
of suitable $\Lambda$-algebras. There is a simple condition ensuring that this is
the case which will be introduced in the the following section (since it fits in better
with the material considered there).

If $(X,p)$ and $(Y,q)$ are bottomed $\Lambda$-algebras then a
\definition{bottomed homomorphism}\index{bottomed homomorphism}\index{homomorphism!bottomed}
$\pi : (X,p) \to (Y,q)$ is a homomorphism 
of the underlying algebras such that the mappings in the family $\pi$ are all bottomed, 
i.e., $\pi_b(\bot_b) = \bot_b$ for each $b \in B$. 

We consider a set-up including the situation typical for open signatures, 
in which an $A$-family of bottomed sets $V$ is given and the interest is then only in 
bottomed $\Lambda$-algebras $(X,p)$ with $\rest{X}{A} = V$. 

Let $V$ be an $A$-family of bottomed sets, which is considered to be fixed in what follows. 
A bottomed $\Lambda$-algebra $(X,p)$ is said to be 
\definition{bound to $V$}\index{bottomed algebra!bound to a family} if $\rest{X}{A} = V$. 
Of course, if $\Lambda$ is closed (i.e., if $A = \varnothing$) then there is only one 
$A$-family of bottomed sets and any bottomed $\Lambda$-algebra is bound to it.

Let $U$ be an $A$-family of sets and let $(Y,q)$ be a $\Lambda$-algebra fixing $U$. 
Then the flat bottomed extension $(Y^\altbot,q^\altbot)$ of $(Y,q)$ is a bottomed 
$\Lambda$-algebra fixing $U^\altbot$, where $U^\altbot_a = U_a \cup \{\bot_a\}$ (considered
as a bottom set with bottom element $\bot_a$) for each $a \in A$. Conversely, 
a simple way to obtain a bottomed $\Lambda$-algebra fixing $V$ is to construct a 
$\Lambda$-algebra $(Y,q)$ fixing $\nonbot{V}$; the flat 
bottomed extension of $(Y,q)$ is then a bottomed $\Lambda$-algebra fixing $V$. 
In Example~\thesection.1 this method is employed to obtain a bottomed 
$\Lambda$-algebra fixing a given $A$-family of bottomed sets with $\Lambda$ the 
signature in Example~\ref{algs_homs}.3. (The $\Lambda$-algebra $(Y,q)$ used in the 
example is essentially that occurring in Example~\ref{algs_homs}.3.)

\bigskip
\fbox{\begin{exframe}
\textit{Example \thesection.1} Let $\Lambda = (B,K,\Theta)$ be the signature
in Example~\ref{algs_homs}.3 (so $A = \{\mathtt{x},\mathtt{y},\mathtt{z}\}$) and let 
$V$ be an
$A$-family of bottomed sets. Then a bottomed $\Lambda$-algebra $(Y,q)$ bound to $V$ can 
be defined by letting
\begin{eelist}{20pt}
\item $Y_{\mathtt{bool}} = \Bool^\altbot = \Bool \cup\{\bot_{\mathtt{bool}}\}$, \enskip 
      $Y_{\mathtt{atom}} = \Oneptset^\altbot = \Oneptset \cup\{\bot_{\mathtt{atom}}\}$,
\item $Y_{\mathtt{int}} = \Int^\altbot = \Int \cup \{\bot_{\mathtt{int}}\}$, \enskip
      $Y_{\mathtt{pair}} 
      = (\nonbot{V}_{\mathtt{x}} \times \nonbot{V}_{\mathtt{y}}) 
      \cup \{\bot_{\mathtt{pair}}\}$, 
\item $Y_{\mathtt{list}} = (\nonbot{V}_{\mathtt{z}})^* \cup \{\bot_{\mathtt{list}}\}$, 
      \enskip
      $Y_{\mathtt{lp}} = (\nonbot{V}_{\mathtt{x}} \times \nonbot{V}_{\mathtt{y}}) 
                  \cup (\nonbot{V}_{\mathtt{z}})^* \cup \{\bot_{\mathtt{lp}}\}$, 
\item $Y_{\mathtt{x}} = V_{\mathtt{x}}$,\enskip $Y_{\mathtt{y}} = V_{\mathtt{y}}$,\enskip
      $Y_{\mathtt{z}} = V_{\mathtt{z}}$,
\end{eelist}
all these unions being considered to be disjoint and $Y_b$ being considered as a bottomed
set with bottom element $\bot_b$ for each $b \in B\setminus A$,
\begin{eelist}{20pt}
\item $q_{\mathtt{True}} : \Oneptset \to Y_{\mathtt{bool}}$ 
      with $q_{\mathtt{True}}(\onept) = \True$, 
\item $q_{\mathtt{False}} : \Oneptset \to Y_{\mathtt{bool}}$ 
      with $q_{\mathtt{False}}(\onept) = \False$, 
\item $q_{\mathtt{Atom}} : \Oneptset \to Y_{\mathtt{atom}}$ 
      with $q_{\mathtt{Atom}}(\onept) = \onept$, 
\item $q_{\underline{n}} : \Oneptset \to Y_{\mathtt{int}}$ 
      with $q_{\underline{n}}(\onept) = n$ for each $n \in \Int$,
\item $q_{\mathtt{Pair}} : Y_{\mathtt{x}} \times Y_{\mathtt{y}} \to Y_{\mathtt{pair}}$ with
      \[ q_{\mathtt{Pair}}(x,y) = \left\{ \begin{array}{cl}
            (x,y) & \ \textrm{if}\ x \in \nonbot{V}_{\mathtt{x}}\ \textrm{and}
                         \ y \in \nonbot{V}_{\mathtt{y}},\\
                    \bot_{\mathtt{pair}} &\ \textrm{otherwise},
      \end{array}\right.\]
\item $q_{\mathtt{Nil}} : \Oneptset \to Y_{\mathtt{list}}$ 
      with $q_{\mathtt{Nil}}(\onept) = \onept$, 
\item $q_{\mathtt{Cons}} : Y_{\mathtt{z}} \times Y_{\mathtt{list}} \to Y_{\mathtt{list}}$ 
      with 
      \[ q_{\mathtt{Cons}}(z,s) = \left\{ \begin{array}{cl}
            m \triangleleft s &\ \textrm{if}\ z \in \nonbot{V}_{\mathtt{z}}\ \textrm{and} 
                             \ s \in (\nonbot{V}_{\mathtt{z}})^*,\\
                \bot_{\mathtt{list}} &\ \textrm{otherwise},
      \end{array}\right.\]
\item $q_{\mathtt{L}} : Y_{\mathtt{list}} \to Y_{\mathtt{lp}}$ with
      $ q_{\mathtt{L}}(s) = \left\{ \begin{array}{cl}
                s &\ \textrm{if}\ s \in (\nonbot{V}_{\mathtt{z}})^*,\\
                    \bot_{\mathtt{lp}} &\ \textrm{otherwise}, 
      \end{array}\right.$
\item $q_{\mathtt{P}} : Y_{\mathtt{pair}} \to Y_{\mathtt{lp}}$ with 
      $ q_{\mathtt{P}}(p) = \left\{ \begin{array}{cl}
          p &\ \textrm{if}\ p \in \nonbot{V}_{\mathtt{x}} \times \nonbot{V}_{\mathtt{y}},\\
                    \bot_{\mathtt{lp}} &\ \textrm{otherwise}. 
      \end{array}\right.$
\end{eelist}
\end{exframe}}

\bigskip

If $(X,p)$ and $(X',p')$ are bottomed $\Lambda$-algebras bound to $V$ 
then a bottomed homomorphism $\pi : (X,p) \to (X',p')$ is said to 
\definition{fix}\index{bottomed homomorphism!fixing a family} $V$ 
if $\pi_a(x) = x$ for each $x \in V_a$, $a \in A$. Again, if $\Lambda$ is closed then
this imposes no requirement on a bottomed homomorphism.

\begin{proposition}\label{prop_bot_algs_homs_1}
(1)\enskip If $(X,p)$ is a bottomed $\Lambda$-algebra bound to $V$ then
the $B$-family of identity mappings $\id : X \to X$ defines a bottomed homomorphism 
from $(X,p)$ to itself fixing $V$.

(2)\enskip If $\pi : (X,p) \to (Y,q)$ and $\varrho : (Y,q) \to (Z,r)$ are bottomed 
homomorphisms fixing 
$V$ then the composition $\varrho\fcomp\pi$ is a bottomed homomorphism from $(X,p)$ to 
$(Z,r)$ fixing $V$.
\end{proposition}

\proof This follows immediately from Proposition~\ref{prop_algs_homs_1}. \eop

Proposition~\ref{prop_bot_algs_homs_1} implies that there is a category whose objects are 
bottomed $\Lambda$-algebras bound to $V$ with morphisms bottomed homomorphisms fixing $V$. 
A bottomed $\Lambda$-algebra $(X,p)$ is called 
\definition{$V$-initial}\index{initial bottomed algebra}\index{bottomed algebra!initial}
if it is an initial object in this category, i.e., if it is bound to $V$ and 
for each bottomed $\Lambda$-algebra $(X',p')$ bound to $V$ there exists a unique 
bottomed homomorphism $\pi : (X,p) \to (X',p')$  fixing $V$. 
If $\Lambda$ is closed then `initial' will be used instead of `$V$-initial'.
Thus in this special case a bottomed $\Lambda$-algebra $(X,p)$ is initial if for each 
bottomed $\Lambda$-algebra $(X',p')$ there exists a unique 
bottomed homomorphism $\pi : (X,p) \to (X',p')$.

\begin{proposition}\label{prop_bot_algs_homs_2} 
There exists a $V$-initial bottomed $\Lambda$-algebra.
\end{proposition}

\proof 
For the rest of the section let $U$ be the $B$-family of sets with 
$\rest{U}{A} = V$ and $U_b = \{\bot_b\}$ for all $b \in B \setminus A$; this $B$-family  
will be called the 
\definition{$\bot$-trivial extension of $V$ to $B$}\index{trivial extension}\index{extension!trivial}
(and note that $U$ is considered just as a family of sets, and not as a family of bottomed sets).
For each $U$-based $\Lambda$-algebra $(X,p)$ there is then an associated bottomed 
$\Lambda$-algebra, also denoted by $(X,p)$, obtained by stipulating that $\bot_b$ 
be the bottom element of $X_b$ for each $b \in B$ (with $\bot_a$ the bottom element of 
$V_a$ for each $a \in A$). Conversely, the underlying algebra of a bottomed 
$\Lambda$-algebra bound to $V$ is a $U$-based $\Lambda$-algebra. 
For the existence of a $V$-initial bottomed $\Lambda$-algebra part of the following 
result is needed (the remainder being required for the proof of Propositions 
\ref{prop_bot_algs_homs_3} and \ref{prop_bot_algs_homs_4} below).

\begin{lemma}\label{lemma_bot_algs_homs_1}   
The bottomed $\Lambda$-algebra associated with a $U$-initial 
$\Lambda$-algebra is $V$-initial. Moreover, a bottomed $\Lambda$-algebra is $V$-initial 
if and only if the underlying algebra is $U$-initial.
\end{lemma}

\proof This is essentially the same as the proof of Lemma~\ref{lemma_bound_algs_1}. \eop

The existence of a $V$-initial bottomed $\Lambda$-algebra now follows from the first
statement in Lemma~\ref{lemma_bot_algs_homs_1} and from 
Proposition~\ref{prop_free_algs_2}. \eop

For the present situation there is a result corresponding to 
Proposition~\ref{prop_bound_algs_3}
which gives an explicit characterisation of $V$-initial bottomed $\Lambda$-algebras.
Before preceeding, however, let us note that $V$-initial bottomed $\Lambda$-algebras
are rather special. For example, the flat bottomed extension of an initial 
$\Lambda$-algebra is needed to describe the semantics
of several programming languages, but it is almost never an initial bottomed 
$\Lambda$-algebra.
In the following sections we will thus be considering a much larger class of bottomed 
algebras. 

Let $(X,p)$ be a bottomed $\Lambda$-algebra. A family of sets $Y$ with 
$Y \subset X$ is said to be \definition{invariant}
in $(X,p)$ if it is invariant in the underlying $\Lambda$-algebra, and it is  
\definition{bottomed}\index{invariant bottomed family}\index{bottomed family!invariant} 
if $\bot_b \in Y_b$ for each $b \in B$. A bottomed $\Lambda$-algebra $(Y,q)$ is said to 
be a 
\definition{bottomed subalgebra}\index{bottomed subalgebra}\index{subalgebra!bottomed} 
of $(X,p)$ if $Y \subset X$ as bottomed 
sets (i.e., $Y \subset X$ as sets with $Y_b$ and $X_b$ having the same bottom element 
for each $b \in B$) and $q_k$ is the restriction of $p_k$ to $\ass{\sdom{k}}{Y}$ for each 
$k \in K$. In this case the family of sets $Y$ is clearly bottomed and invariant.
Conversely, let $Y$ be any invariant bottomed family and for each $k \in K$ 
let $q_k$ denote the restriction of $p_k$ to $\ass {\sdom{k}}{Y}$. Then $(Y,q)$ is 
a bottomed subalgebra of $(X,p)$, regarding $Y_b$ as a bottomed set with bottom 
element $\bot_b$ (the bottom element of $X_b$) for each $b \in B$. If $Y$ is an 
invariant bottomed family then the corresponding bottomed subalgebra $(Y,q)$ is 
called the 
\index{associated bottomed subalgebra}\definition{bottomed subalgebra associated with $Y$}.

A bottomed $\Lambda$-algebra $(X,p)$ bound to $V$ is said to be 
\definition{$V$-minimal}\index{minimal bottomed algebra}\index{bottomed algebra!minimal}
if $X$ is the only invariant bottomed family $\breve{X}$ in $(X,p)$ such that 
$\rest{\breve{X}}{A} = V$. If $\Lambda$ is closed then `minimal' will be used instead of 
`$V$-minimal'. Thus in this special case a bottomed $\Lambda$-algebra $(X,p)$ is minimal
$X$ is the only invariant bottomed family in $(X,p)$.

A bottomed $\Lambda$-algebra $(X,p)$ is said to be 
\definition{strictly regular}\index{strictly regular bottomed algebra}\index{bottomed algebra!strictly regular}
if the mapping $p_k$ is injective for each $k \in K$ and for each $b \in B\setminus A$ the sets 
$\Im(p_k)$, $k \in K_b$, form a partition of $\nonbot{X}_b$. Moreover, $(X,p)$ is said to 
be \definition{strictly unambiguous}\index{bottomed algebra!strictly unambiguous}
if the mapping $p_k$ is injective for each 
$k \in K$ and for each $b \in B$ the sets $\Im(p_k)$, $k \in K_b$, are disjoint subsets of 
$\nonbot{X}_b$. The qualification `strictly' is used here because 
we want to reserve `regular' and `unambiguous' for somewhat weaker properties to be 
introduced in Section~\ref{reg_bot_algs}.
(As in Section~\ref{init_algs}, in the definition of being strictly 
 unambiguous it would make no difference
if $B$ were replaced by $B \setminus A$, since $K_a = \varnothing$ for each $a \in A$.)

\begin{proposition}\label{prop_bot_algs_homs_3}  
The following are equivalent for a bottomed $\Lambda$-algebra $(X,p)$:

\begin{evlist}{15pt}{0.5ex}
\item[(1)] $(X,p)$ is $V$-initial.
\item[(2)] $(X,p)$ is $V$-minimal and strictly regular.
\item[(3)] $(X,p)$ is $V$-minimal and strictly unambiguous.
\end{evlist}
\end{proposition}

\proof The following fact will be needed:

\begin{lemma}\label{lemma_bot_algs_homs_2} 
A bottomed $\Lambda$-algebra $(X,p)$ bound to $V$ is $V$-minimal if and only if
its underlying algebra is $U$-minimal. 
\end{lemma}

\proof 
An invariant bottomed family ${\breve X}$ satisfies $\rest{\breve{X}}{A} = V$ if and only 
if it contains $U$. This implies that $(X,p)$ is $V$-minimal if and only if the underlying
algebra is $U$-minimal. \eop

Let $(X,p)$ be a bottomed $\Lambda$-algebra bound to $V$. Then by 
Lemma~\ref{lemma_bot_algs_homs_2}
$(X,p)$ is $V$-minimal if and only if the underlying algebra is $U$-minimal. Furthermore, 
(since $U_b = \{\bot_b\}$ for each $b \in B\setminus A$) $(X,p)$ is strictly regular if 
and only if the underlying algebra is $U$-regular and strictly unambiguous if and only if
the underlying algebra is $U$-unambiguous. Proposition~\ref{prop_bot_algs_homs_3} 
therefore follows from Proposition~\ref{prop_free_algs_3} and the second statement in 
Lemma~\ref{lemma_bot_algs_homs_1}. \eop

The result corresponding to Proposition~\ref{prop_bound_algs_42} also holds for bottomed
algebras: A  bottomed $\Lambda$-algebra $(X,p)$ will be called 
\definition{intrinsically free}\index{intrinsically free algebra}\index{algebra!intrinsically free} 
if for each bottomed $\Lambda$-algebra $(Y,q)$ and each 
$A$-family of bottomed mappings $\varrho : \rest{X}{A} \to \rest{Y}{A}$ there exists a unique 
bottomed homomorphism $\pi : (X,p) \to (Y,q)$ such that $\rest{\pi}{A} = \varrho$.

\begin{proposition}\label{prop_bot_algs_homs_4}  
A bottomed $\Lambda$-algebra bound to $V$ is $V$-initial if and only if it is
intrinsically free.
\end{proposition}

\proof 
An intrinsically free bottomed $\Lambda$-algebra bound to $V$ is clearly $V$-initial and the converse
follows immediately from Lemma~\ref{lemma_free_algs_4} and 
Lemma~\ref{lemma_bot_algs_homs_1}. \eop

\begin{lemma}\label{lemma_bot_algs_homs_3} 
Let $\pi,\,\varrho$ be bottomed homomorphisms from a $V$-minimal bottomed $\Lambda$-algebra 
$(X,p)$ to a bottomed $\Lambda$-algebra $(Y,q)$ with 
$\rest{\pi}{A} = \rest{\varrho}{A}$. Then $\pi = \varrho$.
In particular, if $(Y,q)$ is bound to $V$ then there exists at
most one bottomed homomorphism from $(X,p)$ to $(Y,q)$ fixing $V$. 
\end{lemma}

\proof This follows from Proposition~\ref{prop_inv_fams_3}~(1) and 
Lemma~\ref{lemma_bot_algs_homs_2}. \eop

In Section~\ref{inv_fams} a sufficient condition for a $\Lambda$-algebra to be minimal 
was given 
in terms of a grading. For a bottomed $\Lambda$-algebra $(X,p)$ the definition of a 
grading has to be relaxed a bit: A $B$-family of mappings $\#$ with $\#_b : X_b \to \Nat$ 
for each $b \in B$ will be called a 
\definition{bottomed grading}\index{bottomed grading}\index{grading!bottomed}
for $(X,p)$ if $\#_b(\bot_b) = 0$ for all $b \in B$ and
$\#_{\adom{k}{\eta}} (v(\eta)) < \#_{\scod{k}}(p_k(v))$
for all $\eta \in \domsdom{k}$ whenever $v \in \ass{\sdom{k}}{X}$ is 
such that $p_k(v) \ne \bot_{\scod{k}}$. If there exists a bottomed grading then $(X,p)$ is 
said to be
\index{graded bottomed algebra}\index{bottomed algebra!graded}\definition{graded}.

\begin{lemma}\label{lemma_bot_algs_homs_4}  
If $(X,p)$ is a graded bottomed $\Lambda$-algebra bound to $V$ and
\[X_b =  \{\bot_b\} \cup \bigcup_{k \in K_b} \Im(p_k)\]
for each $b \in B\setminus A$ then $(X,p)$ is $V$-minimal.
\end{lemma}

\proof Let $\#$ be a bottomed grading for $(X,p)$ and let $\hat{X}$ be the minimal 
invariant bottomed family with $\rest{\hat{X}}{A} = V$; suppose $\hat{X} \ne X$. 
There thus exists $b \in B \setminus A$ and $x \in X_b \setminus \hat{X}_b$ such that 
$\#_b(x) \le \#_{b'}(x')$ whenever $x' \in X_{b'} \setminus \hat{X}_{b'}$ for some 
$b' \in B$. In particular, $x \ne \bot_b$, since $\bot_b \in \hat{X}_b$, and hence 
$x \in \Im(p_k)$ for some $k \in K_b$. There therefore exists 
$v \in \ass{{\sdom{k}}}{X}$ with $x = p_k(v)$. But it then follows that 
$\#_{\adom{k}{\eta}}(v(\eta)) < \#_{\scod{k}}(x)$ and hence that 
$v(\eta) \in \hat{X}_{\adom{k}{\eta}}$ for each $\eta \in \domsdom{k}$ (by the 
minimality of $\#_b(x)$). 
However, this implies $x \in \hat{X}_b$, 
since the family $\hat{X}$ is invariant, which is a contradiction. \eop

\begin{proposition}\label{prop_bot_algs_homs_5}   
If $(X,p)$ is a $V$-minimal bottomed $\Lambda$-algebra then 
\[X_b = \{\bot_b\} \cup \bigcup_{k \in K_b} \Im(p_k)\]
for each $b \in B\setminus A$. Moreover, a graded bottomed $\Lambda$-algebra $(X,p)$ 
bound to $V$ is $V$-minimal if and only if this equality holds for each 
$b \in B\setminus A$.
\end{proposition}

\proof This follows from Lemmas \ref{lemma_bot_algs_homs_2} and 
\ref{lemma_bot_algs_homs_4} and Proposition~\ref{prop_inv_fams_1}. \eop

For the rest of the section consider the case when $\Lambda$ is the disjoint union of the 
signatures $\Lambda_i$, $i \in F$ (as defined at the end of Section~\ref{algs_homs}).
For each $i \in F$ let $(X^i,p^i)$ be a bottomed $\Lambda_i$-algebra. Then the sum
$\oplus_{i\in F} (X^i,p^i)$ is clearly a bottomed $\Lambda$-algebra. Moreover, if
$A_i$ is the parameter set of $\Lambda_i$ and $(X^i,p^i)$ is bound to the $A_i$-family 
of bottomed sets $V^i$ for each $i \in F$ then $\oplus_{i\in F} (X^i,p^i)$ is bound to
the $A$-family $V$ with $V_a = V^i_a$ for each $a \in A^i_a$.

\begin{proposition}\label{prop_bot_algs_homs_6}   
If $(X^i,p^i)$ is $V^i$-minimal for each $i \in F$ then 
$\oplus_{i\in F} (X^i,p^i)$ is $V$-minimal.
\end{proposition}

\proof Straightforward. \eop

\bigskip
\fbox{\begin{exframe}
\textit{Example \thesection.2} The following notation will be  employed here (and also 
later): If 
$Z$ is a set then $\Bot{Z}$ denotes a disjoint copy of $Z$; the element in $\Bot{Z}$ 
corresponding to the element $z \in Z$ will be denoted by $z^\altbot$ (so
$\Bot{Z} = \{ z^\altbot : z \in Z \}$).
\exparskip
Let $\Lambda$ be the signature in Example~\ref{algs_homs}.1 and consider the bottomed 
$\Lambda$-algebra $(Y,q)$ defined by
\begin{eelist}{20pt}
\item $Y_{\mathtt{bool}} = \Bool^\altbot = \Bool \cup \{\bot_{\mathtt{bool}}\}$, 
\item $Y_{\mathtt{nat}} = \Nat \cup \Bot{\Nat}$ with $0^\altbot = \bot_{\mathtt{nat}}$, 
\item $Y_{\mathtt{int}} = \Int \cup \{\bot_{\mathtt{int}}\}$, \quad
$Y_{\mathtt{pair}} = Y_{\mathtt{int}}^2 \cup \{\bot_{\mathtt{pair}}\}$,
\item $Y_{\mathtt{list}} 
        = Y_{\mathtt{int}}^* \cup \Bot{Y_{\mathtt{int}}^*}$
with $\onept^\altbot = \bot_{\mathtt{list}}$,
\end{eelist}
all these unions being considered to be disjoint and $Y_b$ being considered as a bottomed
set with bottom element $\bot_b$ for each $b \in B$,
\begin{eelist}{20pt}
\item $q_{\mathtt{True}} : \Oneptset \to Y_{\mathtt{bool}}$ with
$q_{\mathtt{True}}(\onept) = \True$, 
\item $q_{\mathtt{False}} : \Oneptset \to Y_{\mathtt{bool}}$ with
$q_{\mathtt{False}}(\onept) = \False$, 
\item $q_{\mathtt{Zero}} : \Oneptset \to Y_{\mathtt{nat}}$ with
$q_{\mathtt{Zero}}(\onept) = 0$,
\item $q_{\mathtt{Succ}} : Y_{\mathtt{nat}} \to Y_{\mathtt{nat}}$
with
\[ q_{\mathtt{Succ}}(x) = \left\{ \begin{array}{cl}
             n + 1 &\ \textrm{if}\ x = n\ \textrm{for some}\ n \in \Nat, \\
            (n + 1)^\altbot &\ \textrm{if}\ x = n^\altbot\  \textrm{for some}\ n \in \Nat,
\end{array}\right.\]
\item $q_{\underline{n}} : \Oneptset \to Y_{\mathtt{int}}$ with
$q_{\underline{n}}(\onept) = n$ for each $n \in \Int$,
\item $q_{\mathtt{Pair}} : Y_{\mathtt{int}} \times Y_{\mathtt{int}}
 \to Y_{\mathtt{pair}}$
with $q_{\mathtt{Pair}}(x_1,x_2) = (x_1,x_2)$, 
\item $q_{\mathtt{Nil}} : \Oneptset \to Y_{\mathtt{list}}$ with
$q_{\mathtt{Nil}}(\onept) = \onept$, 
\item $q_{\mathtt{Cons}} : Y_{\mathtt{int}} \times Y_{\mathtt{list}}
 \to Y_{\mathtt{list}}$ 
with 
\[ q_{\mathtt{Cons}}(x,z) = \left\{\begin{array}{cl} 
           x \triangleleft s &\ \textrm{if}\ z = s\ \textrm{for some}
                   \ s \in Y_{\mathtt{int}}^*,\\
        (x \triangleleft s)^\altbot &\ \textrm{if}\ z = s^\altbot\ \textrm{for some}
                 \ s \in Y_{\mathtt{int}}^*.
\end{array}\right.\]
\end{eelist}
The reader is left to check that $(Y,q)$ is an initial 
bottomed $\Lambda$-algebra. Of course, the underlying $\Lambda$-algebra
is an extension of the $\Lambda$-algebra $(X,p)$ introduced in Example~\ref{algs_homs}.1.
\exparskip
Note that an element of $Y_{\mathtt{list}}$ has either the form $\llist{x}{n}$ or 
the form $(\llist{x}{n})^\altbot$, where $n \ge 0$ and
$x_j \in \Int \cup \{\bot_{\mathtt{int}}\}$ for $\oneto{j}{n}$. The element $\llist{x}{n}$
describes a `real' list with $n$ components (although some or all of these components 
may be `undefined'). The element $(\llist{x}{n})^\altbot$, on the other hand, should be 
thought of as a `partial' list containing at least $n$ components, of which the first $n$ 
components are `known' to be $\lvector{x}{n}$.
\end{exframe}}

\bigskip

\bigskip
\fbox{\begin{exframe}
\textit{Example \thesection.3} Again let $\Lambda$ be the signature in 
Example~\ref{algs_homs}.3 and let $V$ 
be an $A$-family of bottomed sets. Then a bottomed $\Lambda$-algebra $(Y,q)$ bound to 
$V$ can be defined by letting
\begin{eelist}{20pt}
\item $Y_{\mathtt{bool}} = \Bool^\altbot = \Bool \cup\{\bot_{\mathtt{bool}}\}$, \enskip 
      $Y_{\mathtt{atom}} = \Oneptset^\altbot = \Oneptset \cup\{\bot_{\mathtt{atom}}\}$,
\item $Y_{\mathtt{int}} = \Int^\altbot = \Int \cup \{\bot_{\mathtt{int}}\}$, \enskip
      $Y_{\mathtt{pair}} 
      = (V_{\mathtt{x}} \times V_{\mathtt{y}}) \cup \{\bot_{\mathtt{pair}}\}$, 
\item $Y_{\mathtt{list}} = V_{\mathtt{z}}^* \cup \Bot{V_{\mathtt{z}}^*}$
with $\bot_{\mathtt{list}} = \onept^\altbot$.
\item $Y_{\mathtt{lp}} = Y_{\mathtt{pair}} 
                  \cup Y_{\mathtt{list}} \cup \{\bot_{\mathtt{lp}}\}$, 
\item $Y_{\mathtt{x}} = V_{\mathtt{x}}$,\enskip $Y_{\mathtt{y}} = V_{\mathtt{y}}$,\enskip
      $Y_{\mathtt{z}} = V_{\mathtt{z}}$,
\end{eelist}
all these unions being considered to be disjoint and $Y_b$ being considered as a bottomed
set with bottom element $\bot_b$ for each $b \in B\setminus A$,
\begin{eelist}{20pt}
\item $q_{\mathtt{True}} : \Oneptset \to Y_{\mathtt{bool}}$ 
      with $q_{\mathtt{True}}(\onept) = \True$, 
\item $q_{\mathtt{False}} : \Oneptset \to Y_{\mathtt{bool}}$ 
      with $q_{\mathtt{False}}(\onept) = \False$, 
\item $q_{\mathtt{Atom}} : \Oneptset \to Y_{\mathtt{atom}}$ 
      with $q_{\mathtt{Atom}}(\onept) = \onept$, 
\item $q_{\underline{n}} : \Oneptset \to Y_{\mathtt{int}}$ 
      with $q_{\underline{n}}(\onept) = n$ for each $n \in \Int$,
\item $q_{\mathtt{Pair}} : Y_{\mathtt{x}} \times Y_{\mathtt{y}} \to Y_{\mathtt{pair}}$ with
      $q_{\mathtt{Pair}}(x,y) = (x,y)$,
\item $q_{\mathtt{Nil}} : \Oneptset \to Y_{\mathtt{list}}$ 
      with $q_{\mathtt{Nil}}(\onept) = \onept$, 
\item $q_{\mathtt{Cons}} : Y_{\mathtt{z}} \times Y_{\mathtt{list}} \to Y_{\mathtt{list}}$ 
      with 
\[ q_{\mathtt{Cons}}(x,z) = \left\{\begin{array}{cl} 
           x \triangleleft s &\ \textrm{if}\ z = s\ \textrm{for some}
                   \ s \in  V_{\mathtt{z}}^*,\\
        (x \triangleleft s)^\altbot &\ \textrm{if}\ z = s^\altbot\ \textrm{for some}
                 \ s \in  V_{\mathtt{z}}^*.
\end{array}\right.\]
\item $q_{\mathtt{L}} : Y_{\mathtt{list}} \to Y_{\mathtt{lp}}$ with
      $ q_{\mathtt{L}}(s) = s$,
\item $q_{\mathtt{P}} : Y_{\mathtt{pair}} \to Y_{\mathtt{lp}}$ with 
      $ q_{\mathtt{P}}(p) = p$.
\end{eelist}
\exparskip
The reader is left to show, making use of Proposition~\ref{prop_bot_algs_homs_3}, that 
$(Y,q)$ is a $V$-initial bottomed $\Lambda$-algebra.
\end{exframe}}

\startsection{Regular bottomed algebras}
\label{reg_bot_algs}

A bottomed $\Lambda$-algebra $(X,p)$ is said to be 
\definition{regular}\index{regular bottomed algebra}\index{bottomed algebra!regular}
if for each $b \in B\setminus A$ and each $x \in \nonbot{X}_b$ there exists a unique 
$k \in K_b$ and a unique $v \in \ass{\sdom{k}}{X}$ such that $p_k(v) = x$. 
Regularity is essential if `case' or `pattern matching' operators (as they occur in all 
modern functional programming languages) are to be defined. It says that if 
$b \in B\setminus A$ and $x \in X_b$ is not completely undefined then $x$ can be 
constructed in a unique way by applying a constructor $p_k$ (with $k \in K_b$) to one of 
its arguments.

\begin{lemma}\label{lemma_reg_bot_algs_1}
A strictly regular bottomed $\Lambda$-algebra is regular.
Conversely, a regular bottomed $\Lambda$-algebra $(X,p)$ is strictly regular if and only
if $\bot_{\scod{k}} \notin \Im(p_k)$ for each $k \in K$.
\end{lemma}

\proof This is clear. \eop

By Proposition~\ref{prop_bot_algs_homs_3} each $V$-initial bottomed $\Lambda$-algebra is 
strictly regular and hence by Lemma~\ref{lemma_reg_bot_algs_1} it is regular. Moreover, 
if $(Y,q)$ is an initial $\Lambda$-algebra
then the flat bottomed extension $(Y^\altbot,q^\altbot)$ of $(Y,q)$ is a regular
bottomed $\Lambda$-algebra.

Let us now  look at the problem left over 
from the previous section of knowing when a bottomed $\Lambda$-algebra $(X,p)$ is a 
bottomed extension of a suitable $\Lambda$-algebra $(Y,q)$ describing the defined data 
objects. In practice it can 
be assumed that $(Y,q)$ is $U$-initial for some $A$-family of sets $U$ and then $(Y,q)$ 
must be the unique $U$-minimal subalgebra of the underlying algebra $(X,p)$. 
Note that here $U \subset \rest{\nonbot{X}}{A}$, since $\bot_a \notin Y_a$ for each
$a \in A$.

A simple condition which, together with being regular, solves this problem is the following:
A bottomed $\Lambda$-algebra $(X,p)$ will be called \definition{$\natural$-invariant} 
if the family $\nonbot{X}$ is invariant in the underlying algebra, i.e., if
$p_k(v) \ne \bot_{\scod{k}}$ whenever $v \in \ass{\sdom{k}}{X}$ is such that
$v(\eta) \ne \bot_{\adom{k}{\eta}}$ for all $\eta \in \domsdom{k}$.
In particular, a strictly regular bottomed $\Lambda$-algebra as well as
any flat bottomed extension is $\natural$-invariant.

\begin{proposition}\label{prop_reg_bot_algs_2}
Let $(X,p)$ be a regular $\natural$-invariant bottomed $\Lambda$-algebra and let $U$ be 
an $A$-family of sets with $U \subset \rest{\nonbot{X}}{A}$. Then the unique $U$-minimal 
subalgebra $(Y,q)$ of the underlying algebra $(X,p)$ is $U$-initial and $\bot_b \notin Y_b$
for each $b \in B$ (and so in particular $(X,p)$ is a bottomed extension of $(Y,q)$). 
\end{proposition}

\proof Let $\breve{U}$ be the trivial extension of $U$ to $B$ (i.e., the $B$-family of sets 
with $\rest{\breve{U}}{A} = U$ and $\rest{\breve{U}}{B\setminus A} = \varnothing$).
The proof of Proposition~\ref{prop_bound_algs_5} shows that $Y$ is the minimal invariant 
family in the underlying $\Lambda$-algebra $(X,p)$ 
containing $\breve{U}$. But $\nonbot{X}$ is also an invariant family in $(X,p)$ containing 
$\breve{U}$, and therefore $Y \subset \nonbot{X}$, i.e., 
$\bot_b \notin Y_b$ for each $b \in B$. Now, since $(X,p)$ is regular and
$Y \subset \nonbot{X}$, it follows that 
$q_k$ is injective for each $k \in K$ and that the sets $\Im(q_k)$, $k \in K_b$, are 
disjoint subsets of $Y_b$ for each $b \in B\setminus A$. 
Hence $(Y,q)$ is unambiguous, and so by Proposition~\ref{prop_bound_algs_3} it is 
$U$-initial. \eop

Note the following special case of Proposition~\ref{prop_reg_bot_algs_2}: If the signature
$\Lambda$ is closed (i.e., if $A = \varnothing$) and $(X,p)$ is a regular 
$\natural$-invariant bottomed $\Lambda$-algebra then the unique minimal 
subalgebra $(Y,q)$ of the underlying algebra $(X,p)$ is initial and $\bot_b \notin Y_b$
for each $b \in B$ (and so in particular $(X,p)$ is a bottomed extension of $(Y,q)$). 

Recall that if $X$ and $Y$ are bottomed sets then a bottomed mapping $f : X \to Y$ is 
said to be 
\definition{proper}\index{proper bottomed mapping}\index{bottomed mapping!proper} 
if $f(\nonbot{X}) \subset \nonbot{Y}$ (i.e., if 
$f(x) \ne \bot_Y$ for all $x \in X \setminus \{\bot_X\}$). A bottomed homomorphism 
is called \definition{proper} if it is a family of proper mappings.

\begin{lemma}\label{lemma_reg_bot_algs_2}
Let $(X,p)$, $(Y,q)$ be bottomed $\Lambda$-algebras with $(Y,q)$ $\natural$-invariant
and suppose there exists a proper bottomed homomorphism $\pi : (X,p) \to (Y,q)$. 
Then $(X,p)$ is also $\natural$-invariant.
\end{lemma}

\proof This follows from Lemma~\ref{lemma_inv_fams_1}~(2), since 
$\nonbot{X}_b = \pi_b^{-1}(\nonbot{Y}_b)$ for each $b \in B$. \eop

The rest of the section is taken up with the statement and proof of a somewhat technical
result which, however, plays a crucial role in the next section.
In what follows let $V$ be an $A$-family of bottomed sets, let $(H,\diamond)$ be a bottomed 
$\Lambda$-algebra and let $\sigma : V \to \rest{H}{A}$ be an $A$-family of proper 
bottomed mappings. A bottomed $\Lambda$-algebra $(X,p)$ bound to $V$ is said to be 
\definition{classified by $(H,\diamond)$ and $\sigma$} if there exists a proper bottomed 
homomorphism $\pi : (X,p) \to (H,\diamond)$ with $\rest{\pi}{A} = \sigma$.

\begin{proposition}\label{prop_reg_bot_algs_1}
There exists a $V$-minimal regular bottomed $\Lambda$-algebra which is classified by 
$(H,\diamond)$ and $\sigma$. Moreover, any such bottomed $\Lambda$-algebra $(X,p)$ is an 
initial object: For each bottomed $\Lambda$-algebra $(Y,q)$ bound to $V$ and classified 
by $(H,\diamond)$ and $\sigma$ there exists a unique bottomed homomorphism 
$\pi : (X,p) \to (Y,q)$ fixing $V$.
\end{proposition}

\proof This occupies the rest of the section. \eop

It is useful to define a bottomed $\Lambda$-algebra $(X,p)$ to be 
\definition{unambiguous}\index{unambiguous bottomed algebra}\index{bottomed algebra!unambiguous}
if for each $b \in B$ and each $x \in \nonbot{X}_b$ there exists 
at most one $k \in K_b$ with $x \in \Im(p_k)$ and, moreover, if there is such a $k$ then 
there exists at most one $v \in \ass{\sdom{k}}{X}$ with $x = p_k(v)$. In particular,
any regular bottomed $\Lambda$-algebra is unambiguous.

\begin{lemma}\label{lemma_reg_bot_algs_3}
A $V$-minimal bottomed $\Lambda$-algebra is unambiguous if and only if it is regular.
\end{lemma}

\proof This follows from the first statement in Proposition~\ref{prop_bot_algs_homs_5}. \eop

\begin{lemma}\label{lemma_reg_bot_algs_4}
There exists a $V$-minimal regular bottomed $\Lambda$-algebra which is classified by 
$(H,\diamond)$ and $\sigma$. 
\end{lemma}

\proof
By Proposition~\ref{prop_bot_algs_homs_2} there exists a $V$-initial bottomed 
$\Lambda$-algebra $(Z,r)$ and by Proposition~\ref{prop_bot_algs_homs_4} there then 
exists a unique bottomed homomorphism 
$\delta$ from $(Z,r)$ to $(H,\diamond)$ such that $\rest{\delta}{A} = \sigma$. Put
\[ Z'_b  = \{ z \in Z_b : \delta_b(z) \ne \bot_b \} \cup \{\bot_b\}\]
for each $b \in B$; hence $Z'$ is a $B$-family of bottomed sets with 
$\rest{Z'}{A} = V$. For each $k \in K$ define a mapping 
$r'_k : \ass{\sdom{k}}{Z} \to Z_{\scod{k}}$ by letting
\[ r'_k(v) = \left\{ \begin{array}{cl}
        r_k(v) &\ \textrm{if}\ r_k(v) \in Z'_{\scod{k}},\\
        \bot_{\scod{k}} &\ \textrm{otherwise}. 
\end{array}\right.\]
Then $(Z,r')$ is a bottomed $\Lambda$-algebra bound to $V$, and it is easy to see 
that $(Z,r')$ is unambiguous (since by Proposition~\ref{prop_bot_algs_homs_3} $(Z,r)$ is unambiguous 
and if $z \in \nonbot{Z}_b$ with $z = r'_k(v)$ then also 
$z = r_k(v)$). Moreover, $Z'$ is an invariant bottomed family in $(Z,r')$, since
$\Im(r'_k) \subset Z'_{\scod{k}}$ for each $k \in K$.

Let $X$ be the minimal invariant bottomed family in $(Z,r')$ with $\rest{X}{A} = V$, 
and let $(X,p)$ be the associated bottomed subalgebra (so 
$p_k : \ass{\sdom{k}}{X} \to X_{\scod{k}}$ is the restriction of $r'_k$ to 
$\ass{\sdom{k}}{X}$). 
Then $(X,p)$ is a bottomed $\Lambda$-algebra bound to $V$, and by construction it 
is $V$-minimal. But $(X,p)$, being a bottomed subalgebra of the unambiguous bottomed
$\Lambda$-algebra $(Z,r')$, is itself unambiguous, and therefore by Lemma~\ref{lemma_reg_bot_algs_3}
$(X,p)$ is regular, i.e., $(X,p)$ is a $V$-minimal regular bottomed $\Lambda$-algebra.
Note that $X \subset Z'$, since $Z'$ is an invariant bottomed family in $(Z,r')$
with $\rest{Z'}{A} = V$. 

For each $b \in B$ let $\pi_b : X_b \to H_b$ be the restriction of $\delta_b$ to $X_b$.
Then $\pi_b$ is a proper bottomed mapping, since if $x \in \nonbot{X}_b$ then
$x \in Z'_b \setminus \{\bot_b\}$ and so $\pi_b(x) = \delta_b(x) \ne \bot_b$.
Moreover, $\rest{\pi}{A} = \sigma$. The proof will therefore be 
completed by showing that $\pi$ is a homomorphism from $(X,p)$ to $(H,\diamond)$.
Thus consider $k \in K$ and $v \in \ass{\sdom{k}}{X}$.
If $r_k(v) \in Z'_{\scod{k}}$ then $r'_k(v) = r_k(v)$ and in this case
\[\pi_{\scod{k}}(p_k(v)) = \delta_{\scod{k}}(r'_k(v)) 
= \delta_{\scod{k}}(r_k(v)) 
= \diamond_k(\ass{\sdom{k}}{\delta}(v))
= \diamond_k(\ass{\sdom{k}}{\pi}(v))\;.\]
On the other hand, if $r_k(v) \notin Z'_{\scod{k}}$ then 
$\delta_{\scod{k}}(r_k(v)) = \bot_{\scod{k}}$ and $r'_k(v) = \bot_{\scod{k}}$, and hence 
also $\delta_{\scod{k}}(r'_k(v)) = \bot_{\scod{k}}$, and so it again follows that
\[\pi_{\scod{k}}(p_k(v)) = \delta_{\scod{k}}(r'_k(v)) = \bot_{\scod{k}} 
= \delta_{\scod{k}}(r_k(v)) = \diamond_k(\ass{\sdom{k}}{\delta}(v))
= \diamond_k(\ass{\sdom{k}}{\pi}(v))\;.\]
This shows that $\pi_{\scod{k}}(p_k(v)) = \diamond_k(\ass{\sdom{k}}{\pi}(v))$ for all 
$v \in \ass{\sdom{k}}{X}$, $k \in K$, i.e., $\pi$ is a homomorphism from $(X,p)$ to 
$(H,\diamond)$. \eop

For the moment let us just assume that $(X,p)$ is a $V$-minimal regular bottomed 
$\Lambda$-algebra.

\begin{lemma}\label{lemma_reg_bot_algs_5}
There exists a 
unique $B$-family of mappings $\#$ with $\#_b : X_b \to \Nat$ for each $b \in B$ such that 
$\#_a(x) = 0$ for each $x \in X_a$, $a \in A$, $\#_b(\bot_b) = 0$ for each $b \in B$, and 
for each $k \in K$
\[ \#_{\sdom{k}}(p_k(v)) 
 = 1 + \max \{ \#_{\adom{k}{\eta}}(v(\eta)) : \eta \in \domsdom{k} \}\]
for all $v \in \ass{\sdom{k}}{X}$ with $p_k(v) \ne \bot_{\scod{k}}$ (the maximum  
being taken to be $0$ if $\sdom{k} = \onept$). 
\end{lemma}

\proof This is very similar to the proof of Lemma~\ref{lemma_init_algs_5}. We give the proof in 
full, however, because Lemma~\ref{lemma_reg_bot_algs_5} is the key to the proof of 
Proposition~\ref{prop_reg_bot_algs_1}. The family $\#$ will be obtained as the limit of a sequence 
$\{\#^m\}_{m \ge 0}$, where $\#^m$ is a $B$-family of mappings with $\#^m_b : X_b \to \Nat$
for each $b \in B$. First define $\#^0_b = 0$ for each $b \in B$. Next suppose the 
family $\#^m$ has already been defined for some $m \in \Nat$. Then, since $(X,p)$ is 
a regular bottomed $\Lambda$-algebra, there exists a unique family of mappings $\#^{m+1}$ 
such that $\#^{m+1}_a(x) = 0$ for each $x \in X_a$, $a \in A$, $\#^{m+1}_b(\bot_b) = 0$ 
for each $b \in B$, and for each $k \in K$ 
\[ \#^{m+1}_{\sdom{k}}(p_k(v)) 
 = 1 + \max \{ \#^m_{\adom{k}{\eta}}(v(\eta)) : \eta \in \domsdom{k} \}\]
for all $v \in \ass{\sdom{k}}{X}$ with $p_k(v) \ne \bot_{\scod{k}}$ (the maximum  
being taken to be $0$ if $\sdom{k} = \onept$). 
By induction this defines the family $\#^m$ for each $m \in \Nat$.

Now $\#^m \le \#^{m+1}$ holds for each $m \in \Nat$: By definition 
$\#^m_a(x) = \#^{m+1}_a(x) = 0$ for each $x \in X_a$, $a \in A$, and 
$\#^m_b(\bot_b) = \#^{m+1}_b(\bot_b) = 0$ for each $b \in B$; also $\#^0 \le \#^1$ holds 
by definition. But if $\#^m \le \#^{m+1}$ for some  $m \in \Nat$ and $k \in K$ then
\begin{eqnarray*}
 \#^{m+1}_{\scod{k}}(p_k(v))
  &=& 1 + \max \{ \#^m_{\adom{k}{\eta}} (v(\eta)) : \eta \in \domsdom{k} \} \\
 &\le& 1 + \max \{ \#^{m+1}_{\adom{k}{\eta}} (v(\eta)) : \eta \in \domsdom{k} \} 
= \#^{m+2}_{\scod{k}}(p_k(v)) 
\end{eqnarray*}
for all $v \in \ass{\sdom{k}}{X}$ with $p_k(v) \ne \bot_{\scod{k}}$. This implies that
$\#^{m+1}_b(x) \le \#^{m+2}_b(x)$ for all $x \in \nonbot{X}_b$, $b \in B\setminus A$,
and hence that $\#^{m+1} \le \#^{m+2}$. Thus by induction $\#^m \le \#^{m+1}$ for each 
$m \in \Nat$.

Moreover, the sequence $\{\#^m_b(x)\}_{m \ge 0}$ is bounded for each $x \in X_b$, 
$b \in B$: Let 
$X'_b$ denote the set of those elements $x \in X_b$ for which this is the case. Then 
$X'$ is bottomed, $\rest{X'}{A} = V$, and it is easily checked that the $B$-family $X'$ 
is invariant, and hence $X' = X$, since $(X,p)$ is $V$-minimal. 

Let $x \in X_b$; then by the above $\{\#^m_b(x)\}_{m \ge 0}$ is a bounded increasing 
sequence from $\Nat$, and so there exists an element $\#_b(x) \in \Nat$ such that 
$\#^m_b(x) = \#_b(x)$ for all but finitely many $m$. This defines a mapping
$\#_b : X_b \to \Nat$ for each $b \in B$, and it immediately follows that the family 
$\#$ has the required property.
It remains to show the uniqueness, so suppose $\#'$ is another $B$-family of mappings with 
this property. For each $b \in B$ let $X'_b = \{ x \in X_b : \#'_b(x) = \#_b(x) \}$;
then $X'$ is clearly an invariant bottomed family with $\rest{X'}{A} = V$, and hence 
$X' = X$, since $(X,p)$ is $V$-minimal. \eop

In what follows again let $(Z,r)$ be a fixed $V$-initial bottomed $\Lambda$-algebra, and let 
$\lambda$ be the unique bottomed homomorphism from $(Z,r)$ to $(X,p)$ fixing $V$.

\begin{lemma}\label{lemma_reg_bot_algs_6}
There exists a unique family of bottomed mappings $\varrho : X \to Z$ fixing $V$ 
and such that for each $k \in K$
\[\varrho_{\scod{k}}(p_k(v)) = r_k(\ass{\sdom{k}}{\varrho}(v))\]
for all $v \in \ass{\sdom{k}}{X}$ with $p_k(v) \ne \bot_{\scod{k}}$. Moreover, 
$\lambda_b(\varrho_b(x)) = x$ for all $x \in X_b$, $b \in B$. 
\end{lemma}

\proof Let $\#$ be the family of mappings given by Lemma~\ref{lemma_reg_bot_algs_5}
and for each $b \in B$, 
$m \in \Nat$ let $X^m_b = \{ x \in X_b : \#_b(x) = m \}$. Define $\varrho_b$ on $X^m_b$ 
for each $b \in B$ using induction on $m$. If $x \in X^0_b$ then either $b \in A$ or 
$x = \bot_b$, in which case put $\varrho_b(x) = x$. Now let $m > 0$ and suppose 
$\varrho_c$ is already defined on $X^k_c$ for each $c \in B$ and for all 
$k < m$. Let $x \in X^m_b$; then in particular $b \in B\setminus A$ and 
$x \in \nonbot{X}_b$ and so there exists a unique $k \in K_b$ and a unique 
element $v \in \ass{\sdom{k}}{X}$ such that $x = p_k(v)$. Moreover, 
$v(\eta) \in X^{m_\eta}_{\adom{k}{\eta}}$ for some $m_\eta$ with $m_\eta < m$, which 
means that $\varrho_{\adom{k}{\eta}}(v(\eta))$ is already 
defined for each $\eta \in \domsdom{k}$. It thus makes sense to put
$\varrho_b(x) = r_b(\ass{\sdom{k}}{\varrho}(v))$ (where of course 
$\ass{\sdom{k}}{\varrho}(v)$ is the element $v' \in \ass{\sdom{k}}{Z}$ with 
$v'(\eta) = \varrho_{\adom{k}{\eta}}(v(\eta))$ for each $\eta \in \domsdom{k}$). In this 
way $\varrho_b$ is defined on $X^m_b$ for each $m \in \Nat$ 
and the family $\varrho$ has the required property by construction. The uniqueness of the 
family $\varrho$ also follows using induction on $m$.

Finally, for each $b \in B$ let $X'_b = \{ x \in X_b : \lambda_b(\varrho_b(x)) = x \}$;
then $\rest{X'}{A} = V$ and the family $X'$ is bottomed. Moreover, it is also invariant. 
(Let $k \in K$ and $v \in \ass{\sdom{k}}{X}$ with 
$v(\eta) \in X'_{\adom{k}{\eta}}$ for 
each $\eta \in \domsdom{k}$; put $x = p_k(v)$. If $x = \bot_{\scod{k}}$ then 
$x \in X'_{\scod{k}}$ holds immediately. On the other hand, if 
$x \in \nonbot{X}_{\scod{k}}$ then $p_k(v) \ne \bot_{\scod{k}}$; in this case
$\varrho_{\scod{k}}(x) = r_k(\ass{\sdom{k}}{\varrho}(v))$ and hence, since 
$\lambda_{\adom{k}{\eta}}(\varrho_\eta(v(\eta))) = v(\eta)$ for each $\eta \in \domsdom{k}$
and so by Lemma~\ref{lemma_sets_3}~(2) 
$\ass{\sdom{k}}{\lambda} (\ass{\sdom{k}}{\varrho}(v)) = v$, it follows that
\[\lambda_{\scod{k}}(\varrho_{\scod{k}}(x)) 
 = \lambda_{\scod{k}}(r_k(\ass{\sdom{k}}{\varrho}(v)))
 = p_k(\ass{\sdom{k}}{\lambda} (\ass{\sdom{k}}{\varrho}(v))) = p_k(v) = x\;.\]
Thus in both cases $x \in X'_{\scod{k}}$.) But $(X,p)$ is $V$-minimal and thus 
$X' = X$, i.e., $\lambda_b(\varrho_b(x)) = x$ for all $x \in X_b$, $b \in B$. \eop

Now let $(Y,q)$ be any bottomed
$\Lambda$-algebra and let $\tau : V \to \rest{Y}{A}$ be any family of bottomed mappings.
By Proposition~\ref{prop_bot_algs_homs_4} there exists a unique
bottomed homomorphism $\mu : (Z,r) \to (Y,q)$ such that $\rest{\mu}{A} = \tau$.
Let $M,\, N \subset Z$ be the families defined by  
$M_b = \{ z \in Z_b : \lambda_b(z) = \bot_b \}$ and $N_b = \{ z \in Z_b : \mu_b(z) = \bot_b \}$
for each $b \in B$.

\begin{lemma}\label{lemma_reg_bot_algs_7}
Let $\varrho : X \to Z$ be the family defined in Lemma~\ref{lemma_reg_bot_algs_6}. If $M \subset N$
then $\omega = \mu \fcomp \varrho$ is a bottomed homomorphism from $(X,p)$ to $(Y,q)$ 
with $\rest{\omega}{A} = \tau$. Moreover, $\omega$ is the unique such bottomed homomorphism.
\end{lemma}

\proof Let $k \in K$ and $v \in \ass{\sdom{k}}{X}$; then by Lemma~\ref{lemma_sets_3}~(2)
\[ q_k(\ass{\sdom{k}}{\omega}(v)) = q_k(\ass{\sdom{k}}{\mu} (\ass{\sdom{k}}{\varrho}(v)))
 = \mu_{\scod{k}}(r_k(\ass{\sdom{k}}{\varrho} (v)))\;.\]
If $p_k(v) \ne \bot_{\scod{k}}$ then 
$\omega_{\scod{k}}(p_k(v)) = \mu_{\scod{k}}(\varrho_{\scod{k}}(p_k(v))) 
= \mu_{\scod{k}}(r_k(\ass{\sdom{k}}{\varrho}(v)))$, and 
so in this case $\omega_{\scod{k}}(p_k(v)) = q_k(\ass{\sdom{k}}{\omega}(v))$. But if 
$p_k(v) = \bot_{\scod{k}}$ 
then $\lambda_{\adom{k}{\eta}}(\varrho_{\adom{k}{\eta}}(v(\eta))) = v(\eta)$ for each 
$\eta \in \domsdom{k}$ and so by Lemma~\ref{lemma_sets_3}~(2) 
$\ass{\sdom{k}}{\lambda} (\ass{\sdom{k}}{\varrho}(v)) = v$, thus
\[ \lambda_{\scod{k}}(r_k(\ass{\sdom{k}}{\varrho} (v))) 
= p_k(\ass{\sdom{k}}{\lambda} (\ass{\sdom{k}}{\varrho} (v)))
   = p_k(v) = \bot_{\scod{k}}\; ,\]
which means that $r_k(\ass{\sdom{k}}{\varrho}(v)) \in M_{\scod{k}}$.
Therefore
$\mu_{\scod{k}}(r_k(\ass{\sdom{k}}{\varrho} (v))) = \bot_{\scod{k}}$, since
$M_{\scod{k}} \subset N_{\scod{k}}$, and again
\[ q_k(\ass{\sdom{k}}{\omega}(v)) = \mu_{\scod{k}}(r_k(\ass{\sdom{k}}{\varrho}(v)))
               = \bot_{\scod{k}} = \omega_{\scod{k}}(\bot_{\scod{k}}) 
               = \omega_{\scod{k}}(p_k(v))\; .\]
This shows that $\omega$ is a homomorphism, and it is clear that $\omega$ is bottomed 
and that $\rest{\omega}{A} = \tau$. 
The uniqueness follows from Lemma~\ref{lemma_bot_algs_homs_3}. \eop

Suppose now that $(X,p)$ is a $V$-minimal regular bottomed $\Lambda$-algebra 
classified by $(H,\diamond)$ and $\sigma$ and that $(Y,q)$ is any bottomed 
$\Lambda$-algebra bound to $V$ and classified 
by $(H,\diamond)$ and $\sigma$. Suppose further that $\tau = \id$, and so
$\mu : (Z,r) \to (Y,q)$ is the unique bottomed homomorphism fixing $V$.

\begin{lemma}\label{lemma_reg_bot_algs_8}
With the above assumptions $M = N$.
\end{lemma}

\proof 
Again let $\delta : (Z,r) \to (H,\diamond)$ be the unique bottomed homomorphism such that 
$\rest{\delta}{A} = \sigma$ and let $L \subset Z$ be the family with
$L_b = \{ z \in Z_b : \delta_b(z) = \bot_b \}$ for each $b \in B$. 
Now by definition there exists a proper bottomed homomorphism 
$\pi : (X,p) \to (H,\diamond)$ such that $\rest{\pi}{A} = \sigma$, and
by Proposition~\ref{prop_bot_algs_homs_1}~(2) $\pi \fcomp \lambda$ is then a bottomed homomorphism 
from $(Z,r)$ to $(H,\diamond)$ with $\pi_a (\lambda_a(z)) =  \pi_a(z) = \sigma_a(z)$ for each 
$z \in Z_a$, $a \in A$. But $\delta$ is the unique such homomorphism and hence 
$\delta = \pi \fcomp \lambda$. In particular, $M = L$, since $\pi$ is a family of
proper bottomed mappings. In the same way $N = L$ 
and therefore $M = N$. \eop

\textit{Proof of Proposition~\ref{prop_reg_bot_algs_1}\enskip} This follows from 
Lemmas \ref{lemma_reg_bot_algs_4}, \ref{lemma_reg_bot_algs_7} 
and \ref{lemma_reg_bot_algs_8}. \eop

To end the section consider the case when $\Lambda$ is the disjoint union of the 
signatures $\Lambda_i$, $i \in F$ (as defined at the end of Section~\ref{algs_homs}).

\begin{proposition}\label{prop_reg_bot_algs_4}
For each $i \in F$ let $(X^i,p^i)$ be a regular bottomed $\Lambda_i$-algebra. Then
the sum $\oplus_{i\in F} (X^i,p^i)$ is regular.
\end{proposition}

\proof Straightforward. \eop

\newpage

\startsection{Algebras associated with a head type}
\label{head_type}

In this section we introduce the concept of a head type, which is
a simple kind of bottomed $\Lambda$-algebra. Head types will be
used to classify the bottomed $\Lambda$-algebras which typically arise
when dealing with functional programming languages.
The main result is Proposition~\ref{prop_head_type_1}; 
this is just a special case of 
Proposition~\ref{prop_reg_bot_algs_1}. In particular,
the set-up considered includes as special cases flat bottomed 
extensions and $V$-initial bottomed algebras. 

Let $\altbot$ and $\natural$ be distinct elements and regard
$\BOneptset = \{\altbot,\natural\}$ as a bottomed set with bottom element $\altbot$.
A bottomed $\Lambda$-algebra $(H,\diamond)$ will be called a 
\definition{head type}\index{head type} if 
$H_a = \BOneptset$ for each $a \in A$. Note that if $V$ is any $A$-family of
bottomed sets and $(H,\diamond)$ is a head type then there is a unique $A$-family 
of proper bottomed mappings $\sigma : V \to \rest{H}{A}$. 

If $(H,\diamond)$ is a head type then a bottomed $\Lambda$-algebra $(X,p)$ 
is said to be a bottomed $\Lambda$-algebra
\definition{associated with $(H,\diamond)$}\index{algebra!associated with head type}
or, more simply, to be 
an \definition{$(H,\diamond)$-algebra} if there exists a proper bottomed
homomorphism from $(X,p)$ to $(H,\diamond)$.

In the context of programming languages the choice of a head type can be seen as a design 
decision for the language being considered. We will usually assume that a head type 
$(H,\diamond)$ is given and then only be interested in $(H,\diamond)$-algebras. 

\begin{proposition}\label{prop_head_type_1}
Let $(H,\diamond)$ be a head type and $V$ be an $A$-family of bottomed sets. Then there 
exists a $V$-minimal regular $(H,\diamond)$-algebra. Moreover, any such 
$(H,\diamond)$-algebra is an initial object in the full subcategory of 
$(H,\diamond)$-algebras bound to $V$. 
\end{proposition}

\proof This is just a special case of Proposition~\ref{prop_reg_bot_algs_1},
since a bottomed $\Lambda$-algebra
bound to $V$ is an $(H,\diamond)$-algebra if and only if it is classified by $(H,\diamond)$
and the unique $A$-family of proper bottomed mappings $\sigma : V \to \rest{H}{A}$. \eop

\begin{lemma}\label{lemma_head_type_1}
A bottomed subalgebra $(Y,q)$ of an $(H,\diamond)$-algebra $(X,p)$ is itself an 
$(H,\diamond)$-algebra.
\end{lemma}

\proof If $\pi : (X,p) \to (H,\diamond)$ is a proper bottomed homomorphism and $\varrho_b$
is the restriction of $\pi_b$ to $Y_b$ for each $b \in B$ then $\varrho$ is a proper 
bottomed homomorphism from $(Y,q)$ to $(H,\diamond)$. \eop

\begin{proposition}\label{prop_head_type_2}
A bottomed $\Lambda$-algebra which is an $(H,\diamond)$-algebra for some 
$\natural$-invariant head type $(H,\diamond)$ is itself $\natural$-invariant. 
\end{proposition}

\proof This follows immediately from Lemma~\ref{lemma_reg_bot_algs_2}. \eop

A head type $(H,\diamond)$ will be called 
\definition{simple}\index{simple head type}\index{head type!simple} if $H_b = \BOneptset$
for each $b \in B$. Let $(H,\diamond)$ be a simple head type and $(X,p)$ be a bottomed 
$\Lambda$-algebra. Then there is only one $A$-family of proper bottomed mappings
$\varepsilon : X \to H$, namely with $\varepsilon_b(\bot_b) = \altbot$ and
$\varepsilon_b(x) = \natural$ for all $x \in \nonbot{X}_b$, $b \in B$. Thus $(X,p)$
is an $(H,\diamond)$-algebra if and only if this family $\varepsilon$ is a homomorphism 
from $(X,p)$ to $(H,\diamond)$.

If $(H,\diamond)$ is a simple head type then we usually just write $\diamond$ instead of 
$(H,\diamond)$.

Each flat bottomed extension of a $\Lambda$-algebra and each strictly regular
bottomed $\Lambda$-algebra is associated with a simple head 
type: Let $H_b = \BOneptset$ for each $b \in B$ and for each $k \in K$ let 
$\diamond^\altbot_k : \ass{\sdom{k}}{H} \to H_{\scod{k}}$ be the mapping defined by
\[ \diamond^\altbot_k(v) = \left\{ \begin{array}{cl}
    \natural  &\ \textrm{if}\ v = \natural^k,\\
    \altbot     &\ \textrm{otherwise}, 
\end{array}\right.\]
where $\natural^k$ is the element of the set $\ass{\sdom{k}}{H}$ defined by 
$\natural^k(\eta) = \natural$ for each $\eta \in \domsdom{k}$,
and let $\diamond^\natural_k : \ass{\sdom{k}}{H} \to H_{\scod{k}}$ be given by
$\diamond^\natural_k(v) = \natural$ for all $v \in \ass{\sdom{k}}{H}$. Then 
$\diamond^\altbot$ and $\diamond^\natural$ are both $\natural$-invariant head types
and the following holds:

\begin{proposition}\label{prop_head_type_3}
(1)\enskip A bottomed $\Lambda$-algebra is a $\diamond^\altbot$-algebra if and 
only if it is the flat bottomed extension of some $\Lambda$-algebra.

(2)\enskip A bottomed $\Lambda$-algebra $(X,p)$ is a $\diamond^\natural$-algebra if and 
only if $p_k(v) \ne \bot_{\scod{k}}$ for all $v \in \ass{\sdom{k}}{X}$, $k \in K$.
In particular, each strictly regular bottomed $\Lambda$-algebra is a
$\diamond^\natural$-algebra.

\end{proposition}

\proof (1)\enskip 
If $(X,p)$ is a $\diamond^\altbot$-algebra then by Proposition~\ref{prop_head_type_2}
$\nonbot{X}$ is invariant in $(X,p)$, and it is easy to check that $(X,p)$ is then 
the flat bottomed extension of the subalgebra associated with $\nonbot{X}$.
Conversely, the flat bottomed extension of a $\Lambda$-algebra is clearly a
$\diamond^\altbot$-algebra.

(2)\enskip This is clear. \eop

\bigskip
\fbox{\begin{exframe}
\textit{Example \thesection.1} The bottomed $\Lambda$-algebra $(Y,q)$ given in 
Example~\ref{bot_algs_homs}.2 is a $V$-minimal regular $\diamond^\altbot$-algebra.
\exparskip
The bottomed $\Lambda$-algebra $(Y,q)$ in Example~\ref{bot_algs_homs}.3 is a $V$-minimal 
regular $\diamond^\natural$-algebra.
\end{exframe}}

\bigskip
Besides $\diamond^\altbot$ and $\diamond^\natural$ there are two further simple head
types which should perhaps be mentioned: One is the `degenerate' head type 
$\diamond^{\flat}$, where $\diamond^{\flat}_k$ is the constant mapping with 
value $\altbot$ for each $k \in K$ (and so in particular $\diamond^{\flat}$ is not 
$\natural$-invariant). A bottomed $\Lambda$-algebra $(X,p)$ is clearly a
$\diamond^{\flat}$-algebra if and only if
$p_k(v) = \bot_{\scod{k}}$ for all $v \in \ass{\sdom{k}}{X}$, $k \in K$.
Of course, such algebras are of little practical use.

Now for the fourth head type: A type $b \in B$ is called a 
\definition{product type}\index{product type}\index{type!product} if 
$K_b$ consists of exactly one constructor name $k$ with
$\domsdom{k}$ containing at least two elements and $\adom{k}{\eta} \ne b$
for each $\eta \in \domsdom{k}$. (In the signature $\Lambda$ in Example~\ref{algs_homs}.3, 
for instance, $\mathtt{pair}$ is a product type.) Define a simple head type
$\diamond^{\Join}$ by letting
\[ \diamond^{\Join}_k = \left\{ \begin{array}{cl}
   \diamond^o_k &\ \textrm{if}
      \ k\ \textrm{is the single constructor for some product type}, \\
         \diamond^\natural_k &\ \textrm{otherwise}, 
\end{array}\right.\]
where if $k \in K$ with $\sdom{k} \ne \onept$ then 
$\diamond^o_k : \ass{\sdom{k}}{H} \to H_{\scod{k}}$ is given by
\[ \diamond^o_k(v) = \left\{ \begin{array}{cl}
     \natural  &\ \textrm{if}\ v(\eta) = \natural\ 
  \textrm{for at least one}\ \eta \in \domsdom{k}, \\
  \altbot &\ \textrm{otherwise}.
\end{array}\right.\]
Then $\diamond^{\Join}$ is a $\natural$-invariant head type which is sometimes used 
instead of 
$\diamond^\natural$. The reason for perhaps preferring $\diamond^{\Join}$ 
to $\diamond^\natural$ is that it provides more `natural' bottomed products.
This can be seen by looking at Example~\thesection.2 and comparing it with
Example~\ref{bot_algs_homs}.2.

It is worth noting the following two explicit cases of Proposition~\ref{prop_head_type_1}.
Let $V$ be an $A$-family of bottomed sets.

\begin{proposition}\label{prop_head_type_4}
The flat bottomed extension of any $\nonbot{V}$-initial $\Lambda$-algebra is a $V$-minimal
regular $\diamond^\altbot$-algebra. Furthermore, each $V$-initial bottomed 
$\Lambda$-algebra is a $V$-minimal regular $\diamond^\natural$-algebra. 
\end{proposition}

\proof This is left for the reader. \eop

\bigskip
\fbox{\begin{exframe}
\textit{Example \thesection.2} Let $\Lambda = (B,K,\Theta)$ be the signature
in Example~\ref{algs_homs}.3 (so $A = \{\mathtt{x},\mathtt{y},\mathtt{z}\}$) and let 
$V$ be an
$A$-family of bottomed sets. Define a bottomed $\Lambda$-algebra $(Y,q)$ bound to $V$ 
exactly as in Example~\ref{bot_algs_homs}.3 except that $Y_{\mathtt{pair}}$ is now 
defined to be
\begin{eelist}{100pt}
\item $V_{\mathtt{x}} \times V_{\mathtt{y}}$ 
      \ with $\bot_{\mathtt{pair}} = (\bot_{\mathtt{x}},\bot_{\mathtt{y}})$
\end{eelist}
(instead of being $V_{\mathtt{x}} \times V_{\mathtt{y}} \cup \{\bot_{\mathtt{pair}}\}$).
The reader is left to show that $(Y,q)$ is a $V$-minimal regular
$\diamond^{\Join}$-algebra.
\end{exframe}}

\bigskip

There is a condition, being $\natural$-stable, which plays an important role in 
Chapter~\ref{ord_cont_algs} and which is strongly related to a certain class of simple head 
types. A bottomed $\Lambda$-algebra $(Y,q)$ is called 
\definition{$\natural$-stable}\index{stable bottomed algebra}\index{bottomed algebra!stable}
if whenever $k \in K$ and $v_1,\,v_2 \in \ass{\sdom{k}}{Y}$ are such that 
$q_k(v_1) \in \nonbot{Y}_{\scod{k}}$
and $v_2(\eta) \in \nonbot{Y}_{\adom{k}{\eta}}$ for all $\eta \in \domsdom{k}$ with 
$v_1(\eta) \in \nonbot{Y}_{\adom{k}{\eta}}$ then also $q_k(v_2) \in \nonbot{Y}_{\scod{k}}$.
Expressed in terms of the contraposition, this says that if $q_k(v_2) = \bot_{\scod{k}}$ and 
$v_1$ is at least as bad as $v_2$ in the sense that $v_1(\eta) = \bot_{\adom{k}{\eta}}$ whenever
$v_2(\eta) = \bot_{\adom{k}{\eta}}$ then it must also be the case that $q_k(v_1) = \bot_{\scod{k}}$.
A head type is said to be 
\definition{$\natural$-stable}\index{stable head type}\index{head type!stable} if it is 
a $\natural$-stable bottomed $\Lambda$-algebra, and in particular the simple head types 
$\diamond^\altbot$, $\diamond^\natural$ and $\diamond^{\Join}$ 
are all $\natural$-stable (as well as the degenerate head type $\diamond^{\flat}$). 
The following result shows that $\natural$-stable algebras are not as
intractable as it might first appear.

\begin{proposition}\label{prop_head_type_5}
If $(H,\diamond)$ is a $\natural$-stable head type then any $(H,\diamond)$-algebra 
is $\natural$-stable. Conversely, each $\natural$-stable 
bottomed $\Lambda$-algebra is a $\diamond$-algebra for some $\natural$-stable simple head type 
$\diamond$.
\end{proposition}

\proof
Let $(H,\diamond)$ be a $\natural$-stable head type and $(Y,q)$ be a $(H,\diamond)$-algebra;
there thus exists a proper bottomed homomorphism $\pi : (Y,q) \to (H,\diamond)$
Let $v_1,\, v_2 \in \ass{\sdom{k}}{Y}$ with $q_k(v_1) \in \nonbot{Y}_{\scod{k}}$ and
such that $v_2(\eta) \in \nonbot{Y}_{\adom{k}{\eta}}$ for all $\eta \in \domsdom{k}$ with 
$v_1(\eta) = \nonbot{Y}_{\adom{k}{\eta}}$. Put 
$w_1 = \ass{\sdom{k}}{\pi}(v_1)$, $w_2 = \ass{\sdom{k}}{\pi}(v_2)$;
then $w_1,\, w_2 \in \ass{\sdom{k}}{H}$ with 
$\diamond_k(w_i) = \pi_{\scod{k}}(q_k(v_i))$ and
$w_i(\eta) = \pi_{\adom{k}{\eta}}(v_i(\eta))$ for all $\eta \in \domsdom{k}$, $i = 1,\,2$.
Therefore $\diamond_k(w_1) \in \nonbot{H}_{\scod{k}}$ and
$w_2(\eta) \in \nonbot{H}_{\adom{k}{\eta}}$ for all $\eta \in \domsdom{k}$ with 
$w_1(\eta) \in \nonbot{H}_{\adom{k}{\eta}}$ and, 
since $(H,\diamond)$ is $\natural$-stable, this implies that
$\pi_{\scod{k}}(q_k(v_2)) = \diamond_k(w_2) \in \nonbot{H}_{\scod{k}}$.
Hence $q_k(v_2) \in \nonbot{Y}_{\scod{k}}$ and thus  $(Y,q)$ is $\natural$-stable. 

Conversely, suppose $(Y,q)$ is $\natural$-stable, for each $b \in B$ let 
$H_b = \BOneptset$ and $\varepsilon_b : Y_b \to H_b$ be the mapping with
$\varepsilon_b(\bot_b) = \altbot$ and $\varepsilon_b(y) = \natural$ for each 
$y \in \nonbot{Y}_b$. Let $k \in K$; if $v_1,\,v_2\in \ass{\sdom{k}}{Y}$ with
$\ass{\sdom{k}}{\varepsilon}(v_1) = \ass{\sdom{k}}{\varepsilon}(v_2)$ then, since $(Y,q)$ 
is $\natural$-stable, it follows that 
$\varepsilon_{\scod{k}}(q_k(v_1)) = \varepsilon_{\scod{k}}(q_k(v_2))$. 
There thus exists a mapping $\diamond_k : \ass{\sdom{k}}{H} \to H_{\scod{k}}$
such that $\diamond_k(\ass{\sdom{k}}{\varepsilon}(v)) = \varepsilon_{\scod{k}}(q_k(v))$
for each $v \in \ass{\sdom{k}}{Y}$. Moreover, under the additional condition
that $\diamond_k(w) = \natural$ for each 
$w \in \ass{\sdom{k}}{H} \setminus \Im(\ass{\sdom{k}}{\varepsilon})$, this mapping
$\diamond_k$ is unique. (Note that the additional condition is needed if the mapping
$\ass{\sdom{k}}{\varepsilon}$ is not surjective, which is possible if $Y_b = \{\bot_b\}$
for some $b \in B$.) Then $\diamond$ is a simple head type and by construction
$(Y,q)$ is a $\diamond$-algebra. Furthermore $(H,\diamond)$ is $\natural$-stable:
This is more-or-less clear if the mapping $\varepsilon_b$ is surjective for each $b \in B$.
The general case is left for the reader.

\begin{proposition}\label{prop_head_type_6}
Let $(H,\diamond)$ be a $\natural$-stable head type and $V$ be an $A$-family of bottomed sets. 
Then any $V$-minimal regular $(H,\diamond)$-algebra $(X,p)$ is intrinsically free:
For each $(H,\diamond)$-algebra $(Y,q)$ and each family $\tau : V \to \rest{Y}{A}$ 
of bottomed mappings there exists a unique bottomed
homomorphism $\pi :(X,p) \to (Y,q)$ with $\rest{\pi}{A} = \tau$.
\end{proposition}

\proof 
As in the proof of Proposition~\ref{prop_reg_bot_algs_1}
let $(Z,r)$ be a $V$-initial bottomed $\Lambda$-algebra, let 
$\lambda$ be the unique bottomed homomorphism from $(Z,r)$ to $(X,p)$ fixing $V$
and let $\mu : (Z,r) \to (Y,q)$ be the unique bottomed homomorphism 
such that $\rest{\mu}{A} = \tau$. Moreover, again let $M,\, N \subset Z$ be the families defined by  
$M_b = \{ z \in Z_b : \lambda_b(z) = \bot_b \}$ and $N_b = \{ z \in Z_b : \mu_b(z) = \bot_b \}$
for each $b \in B$. Then by Lemma~\ref{lemma_reg_bot_algs_7}
it is enough to show that $M \subset N$. Moreover, the proof of
Lemma~\ref{lemma_reg_bot_algs_8} showed that $M = L$, where
$L_b = \{ z \in Z_b : \delta_b(z) = \bot_b \}$ for each $b \in B$ and where
$\delta : (Z,r) \to (H,\diamond)$ is the unique bottomed homomorphism such that 
$\delta_a(z) = \natural$ for all $z \in \nonbot{V_a}$, $a \in A$,

Now by assumption there exists a proper bottomed homomorphism $\alpha$ from $(Y,q)$ to $(H,\diamond)$.
Thus $\gamma = \alpha \fcomp \mu : (Z,r) \to (H,\diamond)$ is a bottomed homomorphism and $N = L'$, 
where $L'_b = \{ z \in Z_b : \gamma_b(z) = \bot_b \}$ for each $b \in B$ (since the mappings in
the family $\alpha$ are proper). For each $b \in B$ let
\[ D_b = \{ z \in Z_b : \gamma_b(z) = \bot_b
    \ \textrm{whenever}\ \delta_b(z) = \bot_b \}\;;\]
then the family $D$ is bottomed and $\rest{D}{A} = V$, since $L_a = \{ \bot_a\}$ for each $a \in A$.
Moreover, $D$ is invariant in $(Z,r)$: Let $k \in K$ and $v \in \ass{\sdom{k}}{D}$;
for each $\eta \in \domsdom{k}$ we then have
$\ass{\sdom{k}}{\gamma}(v)(\eta) =
\gamma_{\adom{k}{\eta}}(v(\eta)) = \bot_{\adom{k}{\eta}}$ whenever
$\ass{\sdom{k}}{\delta}(v)(\eta) =
\delta_{\adom{k}{\eta}}(v(\eta)) = \bot_{\adom{k}{\eta}}$ 
and thus $\gamma_{\scod{k}}(r_k(v)) =  \diamond_k(\ass{\sdom{k}}{\gamma}(v)) = \bot_{\scod{k}}$
whenever $\delta_{\scod{k}}(r_k(v)) =  \diamond_k(\ass{\sdom{k}}{\delta}(v)) = \bot_{\scod{k}}$,
since $(H,\diamond)$ is $\natural$-stable, i.e., $r_k(v) \in D_{\scod{k}}$.
Therefore $D = X$, since $(X,p)$ is $V$-minimal
(by Proposition~\ref{prop_bot_algs_homs_3}), and this implies that $M = L \subset L' = N$. \eop

As usual we end the section by considering the case when $\Lambda$ is the disjoint 
union of the signatures $\Lambda_i$, $i \in F$. Let $(H,\diamond)$ be a head type for
the signature $\Lambda$. Then by Lemma~\ref{lemma_algs_homs_1}
$(H,\diamond) = \oplus_{i\in F} (H^i,\diamond^i)$, where $H^i = \rest{H}{B_i}$ and
$\diamond^i = \rest{\diamond}{K_i}$ for each $i \in F$, and clearly 
$(H^i,\diamond^i)$ is a head type for the signature $\Lambda_i$.

\begin{proposition}\label{prop_head_type_7}
For each $i \in F$ let $(X^i,p^i)$ be an $(H^i,\diamond^i)$-algebra.
Then the sum $\oplus_{i\in F} (X^i,p^i)$ is an $(H,\diamond)$-algebra.
\end{proposition}

\proof This follows immediately from Lemma~\ref{lemma_algs_homs_2}. \eop

Note that if $(H,\diamond)$ is $\natural$-invariant (resp.\ $\natural$-stable)
then so is $(H^i,\diamond^i)$ for each $i \in F$. Moreover, the same holds for the property
of being a simple head type.


%% file: sbika4.tex
\startchapter{Partially ordered sets}
\label{domains}

In this chapter we present those basic facts about partially ordered sets which will be 
needed in the subsequent chapters. 
Section~\ref{posets} gives some elementary results about bottomed partially ordered sets 
(posets). Then in Section~\ref{comp_posets} the corresponding results for 
complete posets are presented. Section~\ref{init_compl} discusses the initial completion 
of posets and their relation to what are called algebraic posets.
The presentation in Section~\ref{init_compl} more-or-less follows that in
Wright, Wagner and Thatcher \cite{ADJ2}. Initial completions are really just ideal
completions, a concept that goes back to Birkhoff \cite{birkhoff67} (which was first 
published in 1940).

The use of complete posets as a tool for dealing with the denotational 
semantics of programming languages was originated by Dana Scott 
in the late sixties. The reader interested in finding out more about this topic
should consult Scott and Gunter \cite{scott_gunter}; an account of its origins can be 
found in Scott \cite{scott}.

\startsection{Bottomed partially ordered sets}
\label{posets}

A \definition{partial order}\index{partial order}
$\sqle$ on a set $X$ is a binary relation satisfying:
\begin{evlist}{25pt}{0.6ex}
\item[(1)] $x \sqle x$ for all $x \in X$.
\medskip
\item[(2)] If $x_1 \sqle x_2$ and $x_2 \sqle x_1$ then $x_1 = x_2$.
\medskip
\item[(3)] If $x_1 \sqle x_2$ and $x_2 \sqle x_3$ then $x_1 \sqle x_3$. 
\medskip\smallskip
\end{evlist}
If $\sqle$ is a partial order on a set $X$ then there can be at most one element 
$\bot_X \in X$ with $\bot_X \sqle x$ for all $x \in X$. If such an element $\bot_X$ exists 
then $\sqle$ is said to be a \definition{bottomed partial order}
\index{bottomed partial order}\index{partial order!bottomed}and
$(X,\sqle)$ is said to be a \definition{bottomed partially ordered set}
\index{bottomed partially ordered set}\index{set!bottomed partially ordered}with 
\definition{bottom element}\index{bottom element} $\bot_X$. Bottomed partially ordered 
sets will always be referred to simply as \definition{posets}\index{poset} in this study. 
The reader should be warned, however, that the existence of a bottom element is not 
usually taken to be part of the definition of a poset.

We mostly just write $X$ instead of $(X,\sqle)$ and assume $\sqle$ can 
be determined from the context. Something like `$X$ is a poset with partial order $\sqle$' 
or `$\sqle$ is the partial order on $X$' can be employed when it is necessary to refer to 
the partial order explicitly. It is useful to consider the set $\Oneptset$ as a poset 
(naturally with respect to the unique partial order on $\Oneptset$).

If $X_1$ and $X_2$ are posets then a mapping $h : X_1 \to X_2$ is said to be
\definition{monotone}\index{monotone mapping}\index{mapping!monotone} if 
$h(x) \sqle_2 h(x')$ for all $x ,\, x' \in X_1$ with $x \sqle_1 x'$
(where $\sqle_j$ is the partial order on $X_j$ for $j = 1,\, 2$ ). A monotone mapping
$h$ to be 
\definition{bottomed}\index{bottomed monotone mapping}\index{monotone mapping!bottomed}
(or \definition{strict})\index{strict monotone mapping}\index{monotone mapping!strict} 
if $h(\bot_1) = \bot_2$.

\begin{proposition}\label{prop_posets_1}
(1)\enskip For each poset $X$ the identity mapping $\id_X : X \to X$ is monotone.

(2)\enskip
If $X_1$, $X_2$, $X_3$ are posets and $g : X_1 \to X_2$ and $h : X_2 \to X_3$ are monotone 
mappings then the mapping $h\comp g : X_1 \to X_3$ is also monotone. 
\end{proposition}

\proof This is clear. \eop

Proposition~\ref{prop_posets_1} implies there is a category whose objects are posets and 
whose morphisms are monotone mappings. This category will be denoted (along with its 
objects) by $\mathsf{Posets}$. Note that the morphisms in $\mathsf{Posets}$ are not 
assumed to be bottomed.

A monotone mapping $h : X_1 \to X_2$ is said to be an 
\definition{order isomorphism}\index{order isomorphism} if it is an isomorphism in the 
category $\mathsf{Posets}$, i.e., if there exists a 
monotone mapping $g : X_2 \to X_1$ such that $g\comp h = \id_{X_1}$ and 
$h\comp g = \id_{X_2}$. It is clear that an order isomorphism is bottomed.
An order isomorphism is of course
a bijective mapping, but the set-theoretic inverse of a bijective monotone mapping
$h : X_1 \to X_2$ need not be monotone (even if $X_1$ and $X_2$ are sets with only
three elements). 

If $X$ is a poset with bottom element $\bot_X$ then the bottomed set
$(X,\bot_X)$ will be referred to as the \definition{underlying bottomed set}
\index{underlying bottomed set}\index{bottomed set!underlying}of $X$. This bottomed 
set will usually also be denoted just by $X$, the particular
usage of the symbol $X$ being determined by the context.
If $X$ is a bottomed set with bottom element $\bot_X$ then by a 
\definition{partial order}\index{partial order} $\sqle$ on $X$ is meant a partial order
on the set $X$ with $\bot_X \sqle x$ for all $x \in X$.

A poset $X$ is said to be \definition{flat}\index{flat poset}\index{poset!flat}
if $x_1 \sqle x_2$ if and only 
$x_1 \in \{\bot_X,x_2\}$. Such a poset is of course determined uniquely by its underlying
bottom set, and if $X$ is a bottomed set then the partial order defined in this way will
be referred to as the 
\definition{flat}\index{flat partial order}\index{partial order!flat} partial order on $X$.

If $X$ is an $S$-family of posets then, unless something explicit to the contrary is 
stated, the partial order on $X_s$ will always be denoted by $\sqle_s$ 
and the bottom element by $\bot_s$ for each $s \in S$.
If $X$ is a finite $S$-family of posets then there is a partial order $\sqle$ on the set 
$\utimes{X}$ defined by stipulating that 
$v \sqle v'$ if and only if $v(s) \sqle_s v'(s)$ for each $s \in S$. 
The element $\bot \in \utimes{X}$ with $\bot(s) = \bot_s$ for all $s \in S$ then satisfies
$\bot \sqle v$ for all $v \in \utimes{X}$, and so $(\utimes{X},\sqle)$ is a poset with
bottom element $\bot$. A reference to the poset $\utimes{X}$ always means with
this partial order.
Note that if $\grave{X}$ is the corresponding $S$-family of bottomed sets
(i.e., $\grave{X}_s$ is the underlying bottomed set of $X_s$ for each 
$s \in S$) then $\utimes{\grave{X}}$ is the underlying bottomed set of $\utimes{X}$.

Let $n \ge 2$ and for each $\oneto j n $ let $(X_j,\sqle_j)$ be a poset. Then the 
above construction produces the usual \definition{product partial order}
\index{product partial order}\index{partial order!product}$\sqle$ on 
$X_1 \times \cdots \times X_n$ in which $(\svector{x}{n}) \sqle (\svector{x'}{n})$ if 
and only if $x_j \sqle_j x'_j$ for each $j$.

As with the categories $\mathsf{Sets}$ and $\mathsf{BSets}$ the process of taking
finite products results in a mapping
$\utimes : \ftyped{\mathsf{Posets}} \to \mathsf{Posets}$. If $S$ is an arbitrary set, $X$ 
an $S$-family of posets and $\gamma$ a finite $S$-typing then the poset 
$\utimes (X \fcomp \gamma)$ will be denoted by $\ass{\gamma}{X}$.

Again let $X$ be a finite $S$-family of posets.

\begin{lemma}\label{lemma_posets_1}
For each $s \in S$ the projection mapping $\proj_s : \utimes{X} \to X_s$ 
defined by $\proj_s(v) = v(s)$ for each $v \in \utimes{X}$ is monotone. 
\end{lemma}

\proof This is clear. \eop

\begin{proposition}\label{prop_posets_2}
A mapping $h : Y \to \utimes{X}$ from a poset $Y$ to the poset $\utimes{X}$ is monotone 
if and only if for each $s \in S$ the mapping $\proj_s \comp h$ from $Y$ to 
$X_s$ is monotone. 
\end{proposition}

\proof Straightforward. \eop

\begin{proposition}\label{prop_posets_3}
Let $Y$ be a further $S$-family of posets and let $\varphi : X \to Y$ be an $S$-family of 
monotone mappings. Then the mapping $\utimes{{\varphi}} : \utimes{X} \to \utimes{Y}$ 
is monotone. 
\end{proposition}

\proof If $v,\,v' \in \utimes{X}$ with $v \sqle v'$ then for each $s \in S$ 
\[ \utimes{\varphi}(v)(s)  = \varphi_s(v(s)) \sqle_s  \varphi_s(v'(s))
 = \utimes{\varphi}(v')(s)\;,\]
since $\varphi_s$ is monotone, and hence $\utimes{\varphi}$ is monotone. \eop

The following is a special case of Proposition~\ref{prop_posets_3}: Let $n \ge 2$ and for 
$\oneto{j}{n}$ let 
$X_j$ and $X'_j$ be posets and $h_j : X_j \to X'_j$ be a monotone mapping. Then 
the mapping $h : X_1 \times \cdots \times X_n \to X'_1 \times \cdots \times X'_n$ defined 
by
\[ h(\svector{x}{n}) = (h_1(x_1),\ldots,h_n(x_n))\]
for each $(\svector{x}{n}) \in X_1 \times \cdots \times X_n$ is monotone.

\newpage

\startsection{Complete posets}
\label{comp_posets}

We now define what it means for a poset to be complete and what it means for a monotone 
mapping between complete posets to be continuous. We then show that the results presented 
in Section~\ref{posets} hold for complete posets with `monotone' replaced by `continuous'.
Recall that `poset' means a partially ordered set with a bottom element.

Let $X$ be a poset with partial order $\sqle$ and let $D$ be a non-empty subset of $X$. 
An element $x \in X$ is called an \definition{upper bound}\index{upper bound} of $D$ if 
$x' \sqle x$ for all $x' \in D$; $x$ is called the 
\definition{least upper bound}\index{least upper bound}\index{upper bound!least} of $D$ 
if $x$ is an 
upper bound of $D$ and $x \sqle x'$ for each upper bound $x'$ of $D$. (It is clear that 
there can be at most one element $x \in X$ having these properties.)  If the least 
upper bound of $D$ exists then it will be denoted by $\lub D$.

A subset $D$ of a poset $X$ is said to be 
\definition{directed}\index{directed set}\index{set!directed} if it is non-empty and 
for each $x_1,\, x_2 \in D$ there exists $x \in D$ such that $x_1 \sqle x$ and 
$x_2 \sqle x$. The set of directed subsets of $X$ will be denoted by $\directed{X}$. If 
$h : X_1 \to X_2$ is monotone then clearly $h(D) \in \directed{X_2}$ for each 
$D \in \directed{X_1}$. 

A poset $X$ is said to be \definition{complete}\index{complete poset}\index{poset!complete}
if the least upper bound $\lub D$ of $D$ 
exists for each $D \in \directed{X}$. If $X_1$ and $X_2$ are complete posets then a 
mapping $h : X_1 \to X_2$ is said to be 
\definition{continuous}\index{continuous mapping}\index{mapping!continuous} if $h$ is 
monotone and 
$h(\lub D) = \lub h(D)$ for each $D \in \directed{X_1}$. Note that if $h : X_1 \to X_2$ 
is monotone then $\lub h(D) \sqle_2 h(\lub D)$ always holds (with $\sqle_2$ the partial
order on $X_2$). Thus a monotone mapping $h$ is continuous if and only if 
$h(\lub D) \sqle_2 \lub h(D)$ for each $D \in \directed{X_1}$.

\begin{proposition}\label{prop_comp_posets_1}
(1)\enskip The identity mapping $\id_X : X \to X$ is continuous for each complete poset $X$.

(2)\enskip
If $X_1$, $X_2$ and $X_3$ are complete posets and $g : X_1 \to X_2$ and
$h : X_2 \to X_3$ are continuous mappings then the mapping $h\comp g : X_1 \to X_3$
is also continuous. 
\end{proposition}

\proof This is clear. \eop

Proposition~\ref{prop_comp_posets_1} implies there is a category whose objects are 
complete posets and whose 
morphisms are continuous mappings. This category will be denoted (along with its objects)
by $\mathsf{CPosets}$. Of course, $\mathsf{CPosets}$ is in fact a subcategory of 
$\mathsf{Posets}$.

\begin{lemma}\label{lemma_comp_posets_1}
Let $X$ be a complete poset and $Y$ a poset isomorphic to $X$; then $Y$ is also complete. 
Moreover, if $h : Y \to X$ is any order isomorphism then $h$ and the inverse mapping 
$h^{-1} : X \to Y$ are both automatically continuous. 
\end{lemma}

\proof Let $D \in \directed{Y}$; then $D' = h(D)$ is an element of $\directed{X}$, thus let 
$x = \lub D'$ and put $y = h^{-1}(x)$. Now $x'$ is an upper bound of $D'$ if and only if 
$h^{-1}(x')$ is an upper bound of $D$, and hence $y = \lub D$. This shows that $Y$ is 
complete and that $h$ is continuous (since  $h(\lub D) = h(y) = x = \lub D' = \lub h(D)$). 
The continuity of $h^{-1}$ follows by reversing the roles of $X$ and $Y$. \eop  

In what follows let $X$ be a finite $S$-family of complete posets.

\begin{lemma}\label{lemma_comp_posets_2}
Let $D \in \directed{\utimes{X}}$ and for each $s \in S$ put 
$D_s = \{ v(s) : v \in D \}$. Then $D_s \in \directed{X_s}$, and the least 
upper bound of $D$ is the assignment $v \in \utimes{X}$ defined by $v(s) = \lub D_s$
for each $s \in S$. 
\end{lemma}

\proof Straightforward. \eop

\begin{proposition}\label{prop_comp_posets_2}
The poset $\utimes{X}$ is complete. 
\end{proposition}

\proof This follows immediately from Lemma~\ref{lemma_comp_posets_2}. \eop

There is thus a mapping $\utimes : \ftyped{\mathsf{CPosets}} \to \mathsf{CPosets}$
obtained by taking finite products in $\mathsf{CPosets}$.
If $S$ is an arbitrary set, $X$ 
an $S$-family of complete posets and $\gamma$ a finite $S$-typing then the complete poset 
$\utimes (X \fcomp \gamma)$ will be denoted by $\ass{\gamma}{X}$.

\begin{lemma}\label{lemma_comp_posets_3}
For each $s \in S$ the projection mapping $\proj_s : \utimes{X} \to X_s$ 
defined by $\proj_s(v) = v(s)$ for each $v \in \utimes{X}$ is continuous. 
\end{lemma}

\proof This follows immediately from Lemma~\ref{lemma_comp_posets_2}. \eop

\begin{proposition}\label{prop_comp_posets_3}
A mapping $h : Y \to \utimes{X}$ from a complete poset $Y$ to the complete poset
$\utimes{X}$ is continuous if and only if for each $s \in S$ the mapping
$\proj_s \comp h$ from $Y$ to $X_s$ is continuous. 
\end{proposition}

\proof If $h$ is continuous then $\proj_s \comp h$ is also continuous for each 
$s \in S$ (since it is the composition of two continuous mappings). Conversely, suppose
that $\proj_s \comp h$ is continuous for each $s \in S$; in particular, 
(as in Proposition~\ref{prop_posets_2} $h$ is then monotone. Let $D \in \directed{Y}$; 
then by Lemma~\ref{lemma_comp_posets_2} it follows that for each $s \in S$
\begin{eqnarray*}
 \left( \lub h(D) \right)(s) &=& \lub \{ v(s) : v \in h(D) \}\\
 &=& \lub \proj_s(h(D))  = (\proj_s \comp h) \left(\lub D \right)
  = h\left(\lub D\right) (s) 
\end{eqnarray*}
and so $\lub h(D) = h(\lub D)$. This shows $h$ is continuous. \eop 

\begin{proposition}\label{prop_comp_posets_4}
Let $\varphi : X \to Y$ be a family of continuous mappings (with $Y$ a further $S$-family 
of complete posets). Then the mapping
$\utimes{\varphi} : \utimes{X} \to \utimes{Y}$ is also continuous. 
\end{proposition}

\proof By Proposition~\ref{prop_posets_3} the mapping $\utimes{\varphi}$ is monotone. Let 
$D \in \directed{\utimes{X}}$; then, with two applications of 
Lemma~\ref{lemma_comp_posets_2} and since $\varphi_s$ is continuous, it follows that
\begin{eqnarray*} 
\lefteqn{\utimes{\varphi}\left(\lub D\right)(s)
        = \varphi_s \left( \left(\lub D\right)(s)\right)
 = \varphi_s \left(\lub \{ v(s) : v \in D \}\right)}\hspace{30pt}\\
& =& \lub \{ \varphi_\eta(v(s)) : v \in D \}
  = \lub \{ \utimes{\varphi}(v)(s) : v \in D \}
  = \left(\lub \utimes{\varphi}(D)\right) (s)
\end{eqnarray*}
for each $s \in S$, and therefore $\utimes{\varphi}(\lub D) = \lub \utimes{\varphi}(D)$.
Thus $\utimes{\varphi}$ is continuous. \eop

Note the following special case of Proposition~\ref{prop_comp_posets_4}: Let $n \ge 2$ and 
for each $\oneto{j}{n}$
let $X_j$ and $X'_j$ be complete posets and $h_j : X_j \to X'_j$ be a continuous mapping. 
Then the mapping $h : X_1 \times \cdots \times X_n \to X'_1 \times \cdots \times X'_n$ 
defined by
\[ h(\svector{x}{n}) = (h_1(x_1),\ldots,h_n(x_n))\]
for each $(\svector{x}{n}) \in X_1 \times \cdots \times X_n$ is continuous.


\startsection{Initial completions and algebraic posets}
\label{init_compl}

A poset $Y_1$ is said to be a \definition{subposet}\index{subposet} of a poset $Y_2$ 
if $Y_1 \subset Y_2$, $Y_1$ and $Y_2$ have a common bottom element
and the partial order on $Y_1$ is obtained by restricting the partial order on $Y_2$. 
(This means that if $\sqle_j$ is the partial order on $Y_j$ for $j = 1,\, 2$ and  
$y,\, y' \in Y_1$ then $y \sqle_1 y'$ if and only if $y \sqle_2 y'$.) The  poset $Y_2$ is 
then said to be an 
\definition{extension}\index{extension of a poset}\index{poset!extension of} of $Y_1$.

A poset $X$ is said to be a 
\definition{completion}\index{completion of a poset}\index{poset!completion of} of a 
poset $Y$ if $X$ is a complete extension of $Y$ and each element of $X$ is the least 
upper bound (in $X$) of some element of $\directed{Y}$. 

\textit{Warning}\enspace Let $X$ be a complete extension of a poset $Y$ and let 
$D \in \directed{Y}$. Then it is possible that $D$ has a least upper bound in $Y$ which is 
not equal to the least upper bound of $D$ in $X$. (Note, however, that if the least upper 
bound of $D$ in $X$ is an element of $Y$ then in this case it must also be the least upper 
bound of $D$ in $Y$.) When speaking just of the least upper bound of $D$ or writing 
$\lub D$ then the least upper bound in $X$ is always meant. 

If $Y$ is a poset with partial order $\sqle$ and $X$ is an extension of $Y$ then, unless 
something explicit to the contrary is stated, the partial order on $X$ will also be 
denoted by $\sqle$.

\begin{proposition}\label{prop_init_compl_1}
Let $Y$ be a finite $S$-family of posets and for each $s \in S$ let 
$X_s$ be a completion of $Y_s$. Then $\utimes{X}$ is a completion of $\utimes{Y}$. 
\end{proposition}

\proof Denote the partial orders on $\utimes{Y}$ and $\utimes{X}$ by $\sqle$.
Let $v \in \utimes{X}$. Then for each $s \in S$ there exists a subset
$D_s \in \directed{Y_s}$ with $v(s) = \lub D_s$ (since $X_s$ is a completion of $Y_s$). Put
\[ D = \{ u \in \utimes{Y} : u(s) \in D_s \ \textrm{for each}\ s \in S \}\;;\]
then $D \in \directed{\utimes{Y}}$: If $v_1,\,v_2 \in D$ then for each $s \in S$ there 
exists $y_s \in D_s$ with $v_1(s) \sqle_s  y_s$ and
$v_2(s) \sqle_s y_s$; thus if $v \in \utimes{Y}$ is defined by letting 
$v(s) = y_s$ for each $s \in S$ then $v \in D$ and $v_1 \sqle v$, $v_2 \sqle v$. Moreover, 
$v = \lub D$:
If $u \in D$ then $u(s) \in D_s$ and so $u(s) \sqle_s \lub D_s = v(s)$ for each 
$s \in S$, i.e., $u \sqle v$. On the other hand, if 
$v' \in \utimes{X}$ is an upper 
bound of $D$ then $u(s) \sqle_s v'(s)$ for each $u \in D$ and this implies
that $y \sqle_s v'(s)$ for all $y \in D_s$ (since for each $y \in D_s$ there 
exists $u \in D$ with $u(s) = y$) and thus that 
$v(s) = \lub D_s \sqle_s v'(s)$ for each $s \in S$, i.e., 
$v \sqle v'$. This, together with Proposition~\ref{prop_comp_posets_3}, shows that 
$\utimes{X}$ is a completion of $\utimes{Y}$. \eop

\begin{lemma}\label{lemma_init_compl_1}
Let $X$ be a completion of a poset $Y$ and let $h : Y \to X'$ be a monotone mapping from 
$Y$ to a complete poset $X'$. Then there exists at most one extension of $h$ to a 
continuous mapping $h' : X \to X'$ (i.e., there exists at most one continuous mapping 
$h' : X \to X'$ with $h'(y) = h(y)$ for all $y \in Y$). 
\end{lemma}

\proof Let $x \in X$; then there exists $D \in \directed{Y}$ with $x = \lub D$
and it therefore follows that $h'(x) = \lub h(Y)$. \eop

\begin{lemma}\label{lemma_init_compl_2}
Let $X_1$ and $X_2$ be completions of a poset $Y$. Then there is at most one
continuous mapping $h : X_1 \to X_2$ such that $h(y) = y$ for all $y \in Y$.
Moreover, if such a mapping exists then it surjective.
\end{lemma}

\proof The first statement follows immediately from Lemma~\ref{lemma_init_compl_1}. Thus 
suppose that $h$ exists and let $x_2 \in X_2$; then there exists $D \in \directed{Y}$ with 
$x_2 = \lub D$ (the least upper bound of $D$ in $X_2$). Let $x_1 = \lub D$ (the least 
upper bound of $D$ in $X_1$); then $h(x_1) = h(\lub D) = \lub h(D) = \lub D = x_2$, since 
$h$ is continuous. Hence $h$ is surjective. \eop

Let $Y$ be a poset, which is considered to be fixed in what follows.
Then there is a category whose objects are completions of $Y$
and where if $X_1$ and $X_2$ are such completions then the morphisms in $\Hom(X_1,X_2)$
are the continuous mappings $h : X_1 \to X_2$ with $h(y) = y$ for each $y \in Y$.
Of course, Lemma~\ref{lemma_init_compl_2} implies that $\Hom(X_1,X_2)$ can contain at most 
one morphism 
for each $X_1,\, X_2$. A completion $X$ of $Y$ is said to be 
\definition{initial}\index{initial completion}\index{completion!initial} if it 
is an initial object in this category. Thus by the previous remark, a completion $X$ is 
initial if for each completion $X'$ of $Y$ there exists a continuous mapping 
$h : X \to X'$ with $h(y) = y$ for all $y \in Y$. 

Proposition~\ref{prop_init_compl_2} below states that $Y$ possesses an initial completion. 
For reasons to be explained at the end of the section, this completion is often also 
referred to as the 
\definition{ideal completion}\index{ideal completion}\index{completion!ideal} of $Y$.

A seemingly stronger requirement than being initial on a completion of $Y$ is the 
following: A completion $X$ is said to have the
\definition{continuous extension property}\index{continuous extension property}
 if whenever $h : Y \to X'$ is a monotone 
mapping from $Y$ to a complete poset $X'$ then $h$ can be uniquely extended to a continuous 
mapping $h' : X \to X'$ (i.e., there exists a unique continuous mapping $h' : X \to X'$ 
with $h'(y) = h(y)$ for all $y \in Y$). It is clear that a completion $X$ of $Y$ with the 
continuous extension property is initial, since if $X'$ is any completion of $Y$ then the 
mapping $i : Y \to X'$ with $i(y) = y$ for all $y \in Y$ is monotone. It turns out that
the converse is also true.

Let $D_1,\, D_2 \in \directed{Y}$; then $D_1$ is said to be 
\definition{cofinal}\index{cofinal} in $D_2$ 
if for each $y_1 \in D_1$ there exists $y_2 \in D_2$ with $y_1 \sqle y_2$. If $X$ is a 
complete extension of $Y$ and $D_1,\, D_2 \in \directed{Y}$ with $D_1$ cofinal in $D_2$ 
then clearly $\lub D_1 \sqle \lub D_2$.

\begin{proposition}\label{prop_init_compl_2}
There exists a initial completion of $Y$, and the following five statements are 
equivalent for a completion $X$ of $Y$:
\begin{evlist}{15pt}{0.8ex}
\item[(1)] $X$ is initial. 
\item[(2)] $X$ has the continuous extension property.
\item[(3)] $D_1$ is cofinal in $D_2$ whenever $D_1,\, D_2 \in \directed{Y}$ with 
$\lub D_1 \sqle \lub D_2$.
\item[(4)] $y \sqle y'$ for some $y' \in D$ whenever $y \in Y$ and $D \in \directed{Y}$ 
with $y \sqle \lub D$.
\item[(5)] 
$y \sqle x$ for some $x \in D$ whenever $y \in Y$ and $D \in \directed{X}$ with 
$y \sqle \lub D$.
\end{evlist}
\end{proposition}

\proof Let $\sqle$ be the partial order on $Y$. It will be shown first that there 
exists a completion $X$ of $Y$ for which statement (3) holds, i.e., such that
$D_1$ is cofinal in $D_2$ whenever $D_1,\, D_2 \in \directed{Y}$ with 
$\lub D_1 \sqle \lub D_2$.

Elements $D_1,\,D_2 \in \directed{Y}$
are said to be \definition{mutually cofinal}\index{mutually cofinal} 
if each is cofinal in the other. Mutual 
cofinality clearly defines an equivalence relation on the set $\directed{Y}$. Let $X$ 
denote the set of equivalence classes, and if $D \in \directed{Y}$ then denote by $[D]$ 
the equivalence class containing $D$. If $D_1$ is equivalent to $D'_1$ and $D_2$ 
equivalent to $D'_2$ then clearly $D_1$ is cofinal in $D_2$ if and only if $D'_1$ is 
cofinal in $D'_2$. Thus define a relation $\preceq$ on $X$ by letting $[D_1] \preceq [D_2]$
if $D_1$ is cofinal in $D_2$; it is immediate that $\preceq$ is a partial order. 

Now $\{y\} \in \directed{Y}$ for each $y \in Y$, which means a mapping $i : Y \to X$ can 
be defined by putting $i(y) = [\{y\}]$ for each $y \in Y$. Then $i$ is an 
\index{embedding}\definition{embedding}, i.e., $y_1 \sqle y_2$ holds if and only if 
$i(y_1) \preceq i(y_2)$ 
(since $y_1 \sqle y_2$ if and only if $\{y_1\}$ is cofinal in $\{y_2\}$), and $Y$ can thus 
be considered as a subposet of $X$ by identifying $Y$ with $i(Y)$. But this just means that 
$X$ is an extension of $Y$. 

\begin{lemma}\label{lemma_init_compl_3}
The poset $X$ (with the partial order $\preceq$) is a completion of $Y$ such that if 
$D_1,\, D_2 \in \directed{Y}$ with $\lub D_1 \sqle \lub D_2$ then $D_1$ is cofinal in $D_2$.
\end{lemma}

\proof Let $C \in \directed{X}$. For each $x \in C$ choose $D_x \in \directed{Y}$ with 
$[D_x] = x$ and put $D = \bigcup_{x \in C} D_x$. Then $D \in \directed{Y}$. (Let 
$y_1,\, y_2 \in D$; then there exist $x_1,\, x_2 \in C$ such that $y_1 \in D_{x_1}$ and 
$y_2 \in D_{x_2}$ and, since $C \in \directed{X}$, there exists $x \in C$ with 
$x_1 \preceq x$ and $x_2 \preceq x$. But this means that $y'_1,\, y'_2 \in D_x$ can be 
found with $y_1 \sqle y'_1$ and $y_2 \sqle y'_2$ and hence there exists 
$y \in D_x \subset D$ such that $y_1 \sqle y'_1 \sqle y$ and $y_2 \sqle y'_2 \sqle y$, 
because $D_x \in \directed{Y}$.) It is now enough to show that $\hat{x} = [D]$ is the 
least upper bound of $C$ in $X$. If $x \in C$ and $y \in D_x$ then $y \sqle y$ and 
$y \in D$; thus $D_x$ is cofinal in $D$. This implies that $x \preceq \hat{x}$, i.e., 
$\hat{x}$ is an upper bound of $C$. Now let $x' = [D']$ be any upper bound of $C$. If 
$y \in D$ then $y \in D_x$ for some $x \in C$ and $D_x$ is cofinal in $D'$, since 
$x \preceq x'$. There therefore exists $y' \in D'$ with $y \sqle y'$ and hence $D$ is 
cofinal in $D'$, i.e., $\hat{x} \preceq x'$. The poset $X$ is therefore complete.

Now if $x = [D] \in X$ and $C = \bigcup_{y \in D} \{y\}$ 
then $C$ is a directed subset of $Y \subset X$ and, as above, 
$x = \lub C$. This shows $X$ is a completion of $Y$.
Finally, if $D_1,\, D_2 \in \directed{Y}$ with $\lub D_1 \preceq \lub D_2$ then 
$[D_1] \preceq [D_2]$, since $\lub D = [D]$ for each $D \in \directed{Y}$, which by 
definition means that $D_1$ is cofinal in $D_2$. \eop

It will be shown next that any completion of $Y$ for which statement (3) holds
has the continuous extension property. We first establish the following fact:

\begin{lemma}\label{lemma_init_compl_4}
Let $X$ be a complete extension of $Y$ (although not necessarily a completion of $Y$) and 
suppose that whenever $D_1,\, D_2$ are elements of $\directed{Y}$ with 
$\lub D_1 \sqle \lub D_2$ then $D_1$ is cofinal in $D_2$. Let $X'$ denote the set of all 
elements of $X$ having the form $\lub D$ for some $D \in \directed{Y}$. Then for each 
$D' \in \directed{X'}$ there exists $D \in \directed{Y}$ with $D$ cofinal in $D'$ and 
$\lub D = \lub D'$. 
\end{lemma}

\proof Let $D' \in \directed{X'}$ and for each $x \in D'$ choose $D_x \in \directed{Y}$ 
with $x = \lub D_x$. Then $D = \bigcup_{x \in D'} D_x \in \directed{Y}$. To see this let 
$y_1,\, y_2 \in D$; then there exist $x_1,\, x_2 \in D'$ with $y_1 \in D_{x_1}$ and 
$y_2 \in D_{x_2}$ and, since $D' \in \directed{X'}$, there exists $x \in D'$ such that 
$x_1 \sqle x$ and $x_2 \sqle x$. But then by assumption $D_{x_1}$ is cofinal in $D_x$ and 
$D_{x_2}$ is cofinal in $D_x$. Hence $y'_1,\, y'_2 \in D_x$ can be found with 
$y_1 \sqle y'_1$ and $y_2 \sqle y'_2$ and, since $D_x \in \directed{Y}$, it follows that 
$y_1 \sqle y'_1 \sqle y$ and $y_2 \sqle y'_2 \sqle y$ for some $y \in D_x \subset D$. 
Now if $y \in D$ then $y \in D_x$ for some $x \in D'$, and so $y \sqle x $; this implies 
$D$ is cofinal in $D'$ and thus also that $\lub D \sqle \lub D'$. On the other hand, if 
$x \in D'$ then $x \sqle \lub D$, since $x = \lub D_x$ and $D_x \subset D$, and therefore 
$\lub D' \sqle \lub D$. Hence $\lub D = \lub D'$. \eop

\begin{lemma}\label{lemma_init_compl_5}
Let $X$ be a completion of $Y$ for which statement (3) holds. Then $X$ has the continuous 
extension property. 
\end{lemma}

\proof Let $h : Y \to X'$ be a monotone mapping of $Y$ to a complete poset $X'$. If 
$D_1,\, D_2 \in \directed{Y}$ with $\lub D_1 = \lub D_2$ then each of $D_1$ and $D_2$ is 
cofinal in the other and therefore $\lub h(D_1) = \lub h(D_2)$. Thus, since each element 
of $X$ has the form $\lub D$ for some $D \in \directed{Y}$, a mapping $h' : X \to X'$ can 
be defined by putting $h'(x) = \lub h(D)$, where $D$ is any element of $\directed{Y}$ 
with $x = \lub D$. Then $h'$ is monotone and an extension of $h$, because 
$\{y\} \in \directed{Y}$ and $\lub \{y\} = y$ for each $y \in Y$. 
Now in order to show the mapping $h'$ is continuous it is enough to show 
that $h'(\lub D') \sqle \lub h'(D')$ for each $D' \in \directed{X}$ (with $\sqle$ also 
denoting the partial order on $X'$). Let $D' \in \directed{X}$; then by 
Lemma~\ref{lemma_init_compl_4} there 
exists $D \in \directed{Y}$ with $D$ cofinal in $D'$ and $\lub D = \lub D'$. Hence
\[ h'\left(\lub D'\right) = h'\left(\lub D\right) = \lub h(D) = \lub h'(D)
 \sqle \lub h'(D')\;.\]
Finally, Lemma~\ref{lemma_init_compl_1} implies that this extension $h'$ of $h$ is 
unique. \eop

It was already noted that a completion having the continuous extension property is
initial. Together with Lemmas \ref{lemma_init_compl_3} and \ref{lemma_init_compl_5}
this implies that there exists an initial extension of $Y$.

\begin{lemma}\label{lemma_init_compl_6}
Statement (3) holds for any initial completion of $Y$.
\end{lemma}

\proof It has already been seen that there exists a completion for which statement (3) holds
and that any such completion is initial. Moreover, it is easily checked that statement 
(3) holds for any completion isomorphic to a completion for which this is the case. Thus 
by Proposition~\ref{prop_sets_2} statement (3) holds for any initial completion of $Y$. \eop

It has now been established that statements (1), (2) and (3) are equivalent. But it is 
clear that (3) and (4) are equivalent and that (5) implies (4); it thus remains to show 
that the last statement is implied by the others and this follows immediately from 
Lemma~\ref{lemma_init_compl_4}. This completes the proof of 
Proposition~\ref{prop_init_compl_2}. \eop

\textit{Warning}\enspace If $X$ is an initial completion of a poset $Y$ then in general 
$X$ is not an initial completion of itself. In fact, the class of complete posets which 
are initial completions of themselves is very special and will be characterised below in 
Proposition~\ref{prop_init_compl_4}.

\begin{proposition}\label{prop_init_compl_3}
Let $S$ be a finite set, let $Y$ be a family of posets, and for each $s \in S$ let $X_s$ 
be an initial completion of $Y_s$. Then $\utimes{X}$ is an initial completion of 
$\utimes{Y}$. 
\end{proposition}

\proof Denote the partial orders on $\utimes{Y}$ and $\utimes{X}$ by $\sqle$.
By Proposition~\ref{prop_init_compl_1} $\utimes{X}$ is a completion of $\utimes{Y}$. 
Let $D$ and $D'$ be elements of $\directed{\utimes{Y}}$ with $\lub D \sqle \lub D'$. For 
each $s \in S$ let $D_s = \{ u(s) : u \in D \}$ and 
$D'_s = \{ u(s) : u \in D' \}$. Then $D_s,\,D'_s \in \directed{Y_s}$ and 
$\lub D_s \sqle_s  \lub D'_s$ and thus, since $X_\eta$ is initial, $D_s$ is 
cofinal in $D'_s$ for each $s \in S$. From this it follows that $D$ is cofinal in 
$D'$: Let $v \in D$; then for each $s \in S$ there exists $v_s \in D'$ with
$v(s) \sqle_s  v_s(s)$, and therefore, since $S$ is finite and $D'$ is 
directed, there exists $v' \in D'$ with $v_s \sqle v'$ for all
$s \in S$. But then $v \sqle v'$. Hence by Proposition~\ref{prop_init_compl_2}
($(3)\Rightarrow (1)$) $\utimes{X}$ is an initial completion of $\utimes{Y}$. \eop

\begin{proposition}\label{prop_init_compl_4}
Let $X$ be a complete poset. Then $X$ is an initial completion of itself if and only if 
each directed subset of $X$ contains a maximum element (i.e., if and only if for each 
$D \in \directed{X}$ then there exists $x \in D$ with $x' \sqle x$ for all $x' \in D$). 
\end{proposition}

\proof Suppose first that $X$ is an initial completion of itself and let
$D \in \directed{X}$ with $\lub D = x$. Then $x \sqle \lub D$ and so by 
Proposition~\ref{prop_init_compl_2} ($(1)\Rightarrow (5)$)
$x \sqle x'$ for some $x' \in D$. But this is only possible if $x = x'$ and therefore 
$x \in D$, i.e., $D$ contains a maximum element (namely $x$). The converse follows 
directly from Proposition~\ref{prop_init_compl_2} ($(5)\Rightarrow(1)$). \eop 

If $Y$ is a poset then a sequence of elements $\{y_n\}_{n \ge 0}$ from $Y$ is called a 
\definition{chain}\index{chain} in $Y$ if $y_n \sqle y_{n+1}$ for each $n \ge 0$. A chain 
$\{y_n\}_{n \ge 0}$ is said to be 
\definition{finite}\index{finite chain}\index{chain!finite} if $y_m = y_n$ for all 
$m \ge n$ 
for some $n \ge 0$. Now let $X$ be a complete poset; if $X$ is an initial completion of 
itself then by Proposition~\ref{prop_init_compl_4} each chain in $X$ must be finite 
(since the elements in a 
chain form a directed set). The converse is in fact also true: If each chain in $X$ is 
finite then $X$ is an initial completion of itself. This follows from 
Proposition~\ref{prop_init_compl_4} together with a standard application of Zorn's lemma.

We now introduce a condition on posets, being \definition{algebraic},
which characterises those posets which arise as initial completions. Moreover, it turns 
out that an algebraic poset $X$ is the initial completion of exactly one poset $Y$, and 
this poset $Y$ is the set of what are called the \definition{compact} elements of $X$.

Let $X$ be a complete poset. An element $x \in X$ is said to be 
\definition{compact}\index{compact element} if whenever $D \in \directed{X}$ with 
$x \sqle \lub D$ then $x \sqle x'$ for some
$x' \in D$. Note that in particular the bottom element $\bot$ of $X$ is always compact.
Let $\compact(X)$ denote the set of compact elements of $X$ and for
each $x \in X$ put $\compact_x  = \{ y \in \compact(X) : y \sqle x \}$. 

A poset $X$ is said to be 
\definition{algebraic}\index{algebraic poset}\index{poset!algebraic} if it is complete 
and if for each $x \in X$ the set $\compact_x$ is directed with $\lub \compact_x = x$.

\begin{proposition}\label{prop_alg_posets_1}
A poset $X$ is algebraic if and only if it the initial completion
of some poset $Y$. Moreover, in this case $Y = \compact(X)$.
\end{proposition}

\proof This follows directly from Lemmas \ref{lemma_alg_posets_1}, 
\ref{lemma_alg_posets_2} and \ref{lemma_alg_posets_3}. \eop

\begin{lemma}\label{lemma_alg_posets_1}
If $X$ is an initial completion of a poset $Y$ then $Y = \compact(X)$.
\end{lemma}

\proof By Proposition~\ref{prop_init_compl_2} ($(1)\Rightarrow (5)$) 
$Y \subset \compact(X)$. Conversely, let $x \in \compact(X)$;
there thus exists $D \in \directed{Y}$ with $x = \lub D$, since $D$ is a completion of
$Y$. But then $x \sqle \lub D$ and so $x \sqle y$ for some $y \in  D$. This is only
possible if $x = y$ and hence in particular $x \in Y$, i.e., $\compact(X) \subset Y$. \eop

\begin{lemma}\label{lemma_alg_posets_2}
Let $X$ be an initial completion of a poset $Y$ and for each $x \in X$ let 
$D_x = \{ y \in Y : y \sqle x \}$. Then $D_x \in \directed{Y}$ and $\lub D_x = x$. 
\end{lemma}

\proof Let $x \in X$; then, since $X$ is a completion of $Y$, there exists 
$D \in \directed{Y}$ with $\lub D = x$. Thus $D \subset D_x$ and so in particular 
$D_x \ne \varnothing$. Let $y \in D_x$; then $y \sqle x = \lub D$ and hence by 
Proposition~\ref{prop_init_compl_2} ($(1) \Rightarrow (4)$)
there exists $y' \in D$ with $y \sqle y'$. Therefore if 
$y_1,\, y_2 \in D_x$ then there exist $y'_1,\, y'_2 \in D$ with $y_1 \sqle y'_1$ and 
$y_2 \sqle y'_2$. But $D \in \directed{Y}$ and so there exists $y' \in D$ with 
$y'_1 \sqle y'$ and $y'_1 \sqle y'$, i.e., there exists $y' \in D_x$ with $y_1 \sqle y'$ 
and $y_1 \sqle y'$. This shows that $D_x \in \directed{Y}$. It is now clear that
$\lub D_x = x$, since $x = \lub D \sqle \lub D_x$ and by definition $x$ is an upper bound 
of $D_x$. \eop

Lemmas \ref{lemma_alg_posets_1} and \ref{lemma_alg_posets_2} imply that the initial 
completion of a poset is algebraic.

\begin{lemma}\label{lemma_alg_posets_3}
Let $X$ be an algebraic poset. Then $X$ is an initial completion of $\compact(X)$.
\end{lemma}

\proof By definition $X$ is a completion of $\compact(X)$ and thus by
Proposition~\ref{prop_init_compl_2} ($(5) \Rightarrow (1)$) 
$X$ is an initial  completion of $\compact(X)$. \eop

\begin{proposition}\label{prop_alg_posets_2}
Let $X$ be a finite $S$-family of algebraic posets. Then $\utimes{X}$ is an algebraic
poset.
\end{proposition}

\proof For each $s \in S$ let $Y_s = \compact(X_s)$. By Lemma~\ref{lemma_alg_posets_3}
$X_s$ is an initial completion of $Y_s$ for each $s \in S$ and thus by 
Proposition~\ref{prop_init_compl_3} $\utimes{X}$ is an initial completion of $\utimes{Y}$. 
Therefore by Proposition~\ref{prop_alg_posets_1} $\utimes{X}$ is algebraic. \eop

The class of all algebraic posets will be denoted by $\mathsf{APosets}$.
Moreover, $\mathsf{APosets}$ will also be used to denote the corresponding full
subcategory of $\mathsf{CPosets}$. By Proposition~\ref{prop_alg_posets_2} there is
a mapping $\utimes : \ftyped{\mathsf{APosets}} \to \mathsf{APosets}$
obtained by taking finite products in $\mathsf{APosets}$.

The final topic of this section explains why the initial completion also goes under the 
name of the ideal completion. In what follows let $Y$ be a fixed poset. A non-empty 
subset $I$ of $Y$ is called an \definition{ideal of $Y$}\index{ideal} if $y \in I$ whenever 
$y \sqle y'$ for some $y' \in I$. An ideal $I$ is said to be 
\definition{directed}\index{directed ideal}\index{ideal!directed} if 
$I \in \directed{Y}$. In particular, the set
\[ I(y) = \{ y' \in Y : y' \sqle y \}\]
is a directed ideal for each $y \in Y$\index{principal ideal}\index{ideal!principal}
(\definition{the principal ideal generated by $y$}).
It is clear that if
$y_1,\,y_2 \in Y$ then $y_1 \sqle y_2$ if and only if $I(y_1) \subset I(y_2)$. 
Now denote by $\Idirected(Y)$ the set of directed ideals of $Y$, regarded as a 
poset with the inclusion ordering. By the above remark $\Idirected(Y)$ can be 
considered as an extension of $Y$ (by identifying the element $y$ with the principal 
ideal $I(y)$ for each $y \in Y$).

\begin{proposition}\label{prop_init_compl_5}
The poset $\Idirected(Y)$ is an initial completion (called 
\definition{the ideal completion})\index{ideal completion}\index{completion!ideal} of $Y$. 
\end{proposition}

\proof It is enough to show that $\Idirected(Y)$ is isomorphic to the initial completion 
constructed in the proof of Proposition~\ref{prop_init_compl_2}. For each non-empty 
subset $D$ of $Y$ let $I(D) = \{ y \in Y : y \sqle y'\ \textrm{for some}\ y' \in D \}$.

\begin{lemma}\label{lemma_init_compl_8}
(1)\enskip $I(D)$ is an ideal with $D \subset I(D)$. Moreover, if $I$ is any ideal with 
$D \subset I$ then $I(D) \subset I$, i.e., $I(D)$ is the smallest ideal containing $D$. 

(2)\enskip If $D \in \directed{Y}$ then $I(D)\in \directed{Y}$; moreover, $D$ and $I(D)$ 
are mutually cofinal.

(3)\enskip If $D_1,\,D_2 \in \directed{Y}$ then $I(D_1) \subset I(D_2)$ if and only if 
$D_1$ is cofinal in $D_2$. 
\end{lemma}

\proof This is straightforward. \eop

Now let $X$ be as in the proof of Proposition~\ref{prop_init_compl_2}.
Recall that mutual cofinality defines 
an equivalence relation on the set $\directed{Y}$ and that $X$ is the set of equivalence 
classes, considered as a poset with the partial order $\preceq$ defined by stipulating 
that $[D_1] \preceq [D_2]$ if $D_1$ is cofinal in $D_2$ (and where $[D]$ denotes the 
equivalence class containing $D$ for each $D \in \directed{Y}$).
If $D_1,\,D_2 \in \directed{Y}$ then by Lemma~\ref{lemma_init_compl_8}~(3) 
$I(D_1) = I(D_2)$ if and only if $D_1$ and $D_2$ are equivalent, thus by 
Lemma~\ref{lemma_init_compl_8}~(2) a mapping 
$h : X \to \Idirected(Y)$ can be defined by letting $h([D]) = I(D)$ for each 
$D \in \directed{Y}$. Moreover, by Lemma~\ref{lemma_init_compl_8}~(3) $h$ is an embedding.
In fact $h$ is also surjective, since by Lemma~\ref{lemma_init_compl_8}~(1) and (2) it 
follows that $D = I(D) = h([D])$ for each $D \in \Idirected(Y)$. Hence $h$ is an order 
isomorphism.
But $h(y) = y$ for each $y \in Y$ (because $y$ is identified with $[\{y\}]$ in $X$, with 
$I(y)$ in $\Idirected(Y)$, and $I(\{y\}) = I(y)$), and therefore
$X$ and $\Idirected(Y)$ are isomorphic extensions of $Y$. \eop


%% file: sbika5.tex
\startchapter{Ordered and continuous algebras}
\label{ord_cont_algs}

This chapter deals with the third and fourth steps in our programme of specifying data 
objects. The third step is based on the observation that if $(X,p)$ is a regular 
bottomed $\Lambda$-algebra satisfying some additional conditions
then for each $b \in B$ there is a unique partial order $\sqle_b$ on the set $X_b$ 
such that $x \sqle_b x'$ can reasonably be interpreted as meaning that $x$ is 
less-defined than $x'$. In particular, this should mean that $\bot_b \sqle x$ for each 
$x \in X_b$ and the mapping $p_k : \ass{\sdom{k}}{X} \to X_{\scod{k}}$ be  
monotone for each $k \in K$. 

Considering $X_b$ as a poset with this partial order $\sqle_b$ results in a $B$-family
of posets $X$ together with a $K$-family of monotone mappings $p$, and the
pair $(X,p)$ will then be called an ordered $\Lambda$-algebra.
The fourth step is to  complete the posets in the family
$X$, and the appropriate completion here 
is the initial completion presented in Section~\ref{init_compl}. After the completion
has been made and each $p_k$ has been extended to a continuous mapping we end up with a 
continuous algebra, i.e., a $\Lambda$-algebra $(Y,q)$ with a $B$-family of complete
posets $Y$ and a $K$-family of continuous mappings $q$.
The results in this chapter for $\diamond^\natural$-algebras
can be found in Goguen, Thatcher, Wagner and Wright \cite{ADJ1} and
in Courcelle and Nivat \cite{courcelle_nivat}.

\startsection{Ordered algebras}
\label{ord_algs}

An \definition{ordered $\Lambda$-algebra}\index{ordered algebra}\index{algebra!ordered}
is any pair $(X,p)$ consisting of a $B$-family 
of posets $X$ and a $K$-family of mappings $p$ such that $p_k$ is a monotone 
mapping from $\ass{\sdom{k}}{X}$ to $X_{\scod{k}}$ for each $k \in K$. Recall once again
that `poset' means a partially ordered set with a bottomed element and that
$\ass{\sdom{k}}{X}$ denotes the poset $\utimes(X \fcomp \sdom{k})$.
If $(X,p)$ is 
an ordered $\Lambda$-algebra and $X^o_b$ is the bottomed set underlying $X_b$ for each 
$b \in B$ then $(X^o,p)$ is a bottomed $\Lambda$-algebra, called the 
\definition{underlying bottomed $\Lambda$-algebra}\index{underlying bottomed algebra}
\index{bottomed algebra!underlying}of $(X,p)$. However, the 
underlying bottomed set $X^o_b$ will usually just be denoted by $X_b$, which means that 
the ordered and the underlying bottomed $\Lambda$-algebras are both denoted by $(X,p)$.
What a particular usage of $(X,p)$ refers to will always be clear from the context.  

If $(X,p)$ is an ordered $\Lambda$-algebra then, unless something to the contrary is 
stated, the partial order on $X_b$ will always be denoted by $\sqle_b$ and the bottom 
element by $\bot_b$. For each $k \in K$ the partial order on $\ass{\sdom{k}}{X}$ will be
denoted by $\ass{\sdom{k}}{\sqle}$, thus if $v_1,\,v_2 \in \ass{\sdom{k}}{X}$ then
$v_1 \opass{\sdom{k}}{\sqle} v_2$ if and only if 
$v_1(\eta) \sqle_{\adom{k}{\eta}} v_2(\eta)$ for each $\eta \in \domsdom{k}$.

An ordered $\Lambda$-algebra $(X,p)$ will be called 
\definition{intrinsic}\index{intrinsic ordered algebra}\index{ordered algebra!intrinsic} 
if the underlying bottomed $\Lambda$-algebra is regular
and if whenever $b \in B\setminus A$ and $x_1,\,x_2 \in \nonbot{X}_b$ then 
$x_1 \sqle_b x_2$ if and only if $x_1 = p_k(v_1)$ and $x_2 = p_k(v_2)$ with $k \in K_b$ 
and $v_2 \opass{\sdom{k}}{\sqle} v_2$. We consider intrinsic ordered algebras
as being those for which $x \sqle_b x'$  can really be interpreted as meaning
that $x$ is less-defined than $x'$.

\bigskip
\fbox{\begin{exframe}
\textit{Example \thesection.1} 
Let $(Y,q)$ be the initial bottomed $\Lambda$-algebra introduced in 
Example~\ref{algs_homs}.1
and for each $b \in B$ define a partial order $\sqle_b$ on $Y_b$ as follows:
\exparskip
\begin{eelist}{15pt}
\item If $a, \, a' \in Y_{\mathtt{bool}}$ then $a \sqle_{\mathtt{bool}} a'$ if and only if 
$a \in \{\bot_{\mathtt{bool}},a'\}$.
\exparskip
\item Let $n \in \Nat$ and $x \in Y_{\mathtt{nat}}$. Then $x \sqle_{\mathtt{nat}} n$ if 
and only if either $x = n$ or $x = m^\bot$ for some $m \le n$. Moreover, 
$x \sqle_{\mathtt{nat}} n^\bot$ if and only if $x = m^\bot$ for some $m \le n$. 
\exparskip
\item If $n, \, n' \in Y_{\mathtt{int}}$ then $n \sqle_{\mathtt{int}} n'$ if and only if 
$n \in \{\bot_{\mathtt{int}},n'\}$.
\exparskip
\item If $p, \, p' \in Y_{\mathtt{pair}}$ then $p \sqle_{\mathtt{pair}} p'$ if and only if 
either $p = \bot_{\mathtt{pair}}$ or $p = (x,y)$ and $p' = (x',y')$ with 
$x \sqle_{\mathtt{int}} x'$ and $y \sqle_{\mathtt{int}} y'$.
\exparskip
\item If $\ell,\,\ell' \in Y_{\mathtt{list}}$ then $\ell \sqle_{\mathtt{list}} \ell'$ 
      if and only if either
\begin{eelist}{10pt}
\item $\ell =  \llist{z}{m}$, 
      $\ell' = \llist{z'}{m}$ with $z_j \sqle_{\mathtt{int}} z'_j$ for each $j$, or 
\item $\ell =  (\llist{z}{m})^\bot$ and $\ell'$ either $\llist{z'}{n}$ or
      $(\llist{z'}{n})^\bot$, with $m \le n$ and $z_j \sqle_{\mathtt{int}} z'_j$ for each 
      $\oneto{j}{m}$. 
\end{eelist}
\end{eelist}
\exparskip
(Recall here that 
\begin{eelist}{15pt}
\item $Y_{\mathtt{bool}} = \Bool^\bot = \Bool \cup \{\bot_{\mathtt{bool}}\}$, 
\item $Y_{\mathtt{nat}} = \Nat \cup \Bot{\Nat}$ with $0^\bot = \bot_{\mathtt{nat}}$, 
\item $Y_{\mathtt{int}} = \Int \cup \{\bot_{\mathtt{int}}\}$, \quad
      $Y_{\mathtt{pair}} = Y_{\mathtt{int}}^2 \cup \{\bot_{\mathtt{pair}}\}$,
\item $Y_{\mathtt{list}} = Y_{\mathtt{int}}^* \cup \Bot{Y_{\mathtt{int}}^*}$
      with $\onept^\bot = \bot_{\mathtt{list}}$.)
\end{eelist}
\exparskip
It is then straightforward to check that with these partial orders $(Y,q)$ becomes
an intrinsic ordered $\Lambda$-algebra.
\end{exframe}}

\bigskip\bigskip
Ordered algebras will be obtained as ordered extensions of bottomed algebras: If $(Y,p)$ 
is a bottomed $\Lambda$-algebra then an ordered $\Lambda$-algebra $(X,p)$ is said to be an 
\definition{ordered extension}\index{ordered extension}\index{extension!ordered} of 
$(Y,p)$ if $(Y,p)$ is the underlying bottomed algebra of $(X,p)$. Defining 
an ordered extension of a given bottomed $\Lambda$-algebra $(Y,p)$ thus amounts to 
specifying an appropriate family of partial orders which turns $p$ into a family of 
monotone mappings. The main result of the section (Proposition~\ref{prop_ord_algs_3})
says that each regular bottomed $\Lambda$-algebra satisfying some additional conditions
has a unique intrinsic ordered extension.
 
The simplest example of an ordered $\Lambda$-algebra is obtained by starting with the
flat bottomed extension $(Y^\bot,q^\bot)$ of a $\Lambda$-algebra $(Y,q)$
and considering $Y^\bot_b$ as a poset with 
$\sqle_b$ the flat order on $Y^\bot_b$ (thus $y \sqle_b y'$ if and only if
$y \in \{\bot_b,y'\}$). Then each of the mappings $q^\bot_k$ is monotone 
(since if $v_1,\, v_2 \in \ass{\sdom{k}}{(Y^\bot)}$ with
$v_1 \opass{\sdom{k}}{\sqle} v_1$ and $q^\bot_k(v_1) \ne \bot_{\scod{k}}$ then
$v_1 \in \ass{\sdom{k}}{Y}$ and $v_1 = v_2$). This means that $(Y^\bot,q^\bot)$ is an 
ordered $\Lambda$-algebra which will be called the 
\definition{flat ordered extension of}\index{flat ordered extension} $(Y,q)$. If $(Y,p)$ 
is an initial $\Lambda$-algebra then, as was already
noted, the bottomed $\Lambda$-algebra $(Y^\bot,q^\bot)$ is regular, and in this
case the ordered $\Lambda$-algebra $(Y^\bot,q^\bot)$ is trivially intrinsic.

Recall that a bottomed $\Lambda$-algebra $(Y,q)$ is said to be $\natural$-stable if 
whenever $k \in K$ and $v_1,\,v_2 \in \ass{\sdom{k}}{Y}$ are such that 
$q_k(v_1) \in \nonbot{Y}_{\scod{k}}$ and $v_2(\eta) \in \nonbot{Y}_{\adom{k}{\eta}}$ for 
all $\eta \in \domsdom{k}$ with 
$v_1(\eta) \in \nonbot{Y}_{\adom{k}{\eta}}$ then also $q_k(v_2) \in \nonbot{Y}_{\scod{k}}$.
By Proposition~\ref{prop_head_type_5} a bottomed $\Lambda$-algebra is $\natural$-stable if 
and only if it is a $\diamond$-algebra for some $\natural$-stable simple head type 
$\diamond$. Finally, recall that
the simple head types $\diamond^\altbot$, $\diamond^\natural$ and 
$\diamond^{\Join}$, as well as the degenerate head type $\diamond^{\flat}$, are all 
$\natural$-stable.

\begin{lemma}\label{lemma_ord_algs_1}
A simple head type $\diamond$ is $\natural$-stable if and only if
$(H,\diamond)$ is an ordered $\Lambda$-algebra, naturally considering $\BOneptset$ as a 
poset with $\bot \sqle \natural$.
\end{lemma}

\proof This is clear. \eop

As usual, we consider a set-up including the situation typical for open signatures.
Let $V$ be an $A$-family of posets, which is considered to be fixed in what 
follows. An ordered $\Lambda$-algebra $(X,p)$ is said to be 
\definition{bound to $V$}\index{ordered algebra bound to a family} if 
$\rest{X}{A} = V$. Of course, if $\Lambda$ is closed (i.e., if $A = \varnothing$) then 
there is only one $A$-family of posets and any ordered $\Lambda$-algebra is 
bound to it. The underlying bottomed set of the poset $V_b$ will be denoted by $U_b$.
Thus $U$ also denotes the corresponding family of bottomed sets.

\begin{proposition}\label{prop_ord_algs_3}
Let $(Y,p)$ be a $U$-minimal regular $\natural$-stable $\Lambda$-algebra. Then there 
exists a unique intrinsic ordered extension $(X,p)$ of $(Y,p)$ which is bound to $V$. 
\end{proposition}

\proof This occupies the second half of the section. \eop

If $(X,p)$ and $(Y,q)$ are ordered $\Lambda$-algebras then an
\definition{ordered homomorphism}\index{ordered homomorphism}\index{homomorphism!ordered}
$\pi$ from $(X,p)$ to $(Y,q)$ is a bottomed 
homomorphism of the underlying bottomed algebras such that $\pi_b$ is a monotone mapping 
for each $b \in B$. 
If $(X,p)$ and $(X',p')$ are ordered $\Lambda$-algebras bound to $V$ 
then a an ordered homomorphism $\pi : (X,p) \to (X',p')$ is said to 
\definition{fix}\index{homomorphism fixing a family} $V$ 
if $\pi_a(x) = x$ for each $x \in V_a$, $a \in A$. Again, if $\Lambda$ is closed then
this imposes no requirement on an ordered homomorphism.

\begin{proposition}\label{prop_ord_algs_1}
(1)\enskip If $(X,p)$ is an ordered $\Lambda$-algebra bound to $V$ then
the $B$-family of identity mappings $\id : X \to X$ defines an ordered homomorphism 
from $(X,p)$ to itself fixing $V$.

(2)\enskip If $\pi : (X,p) \to (Y,q)$ and $\varrho : (Y,q) \to (Z,r)$ are ordered
homomorphisms fixing 
$V$ then the composition $\varrho\fcomp\pi$ is an ordered homomorphism from $(X,p)$ to 
$(Z,r)$ fixing $V$.
\end{proposition}

\proof This follows immediately from Propositions \ref{prop_algs_homs_1} and 
\ref{prop_posets_1}. \eop

Proposition~\ref{prop_ord_algs_1} implies there is a category whose objects are ordered
$\Lambda$-algebras bound to $V$ with morphisms ordered homomorphisms fixing $V$. 
There exist initial objects in this category but, as in Chapter~\ref{bot_algs}, they are 
really too special. What is required is the analogue of 
Proposition~\ref{prop_head_type_5}, and this is 
given in Proposition~\ref{prop_ord_algs_2} below.

An ordered  $\Lambda$-algebra $(X,p)$ is defined to be 
\definition{$V$-minimal}\index{minimal ordered algebra}\index{ordered algebra!minimal} 
if it is bound to $V$ and the underlying bottomed $\Lambda$-algebra is $U$-minimal.
Similarly, it is defined to be an
\definition{$(H,\diamond)$-algebra} if the underlying bottomed $\Lambda$-algebra is 
an $(H,\diamond)$-algebra.

\begin{proposition}\label{prop_ord_algs_2}
For each $\natural$-stable simple head type $\diamond$ there exists a $V$-minimal intrinsic
ordered $\diamond$-algebra. Moreover, any such $\diamond$-algebra $(X,p)$ is an 
initial object in the full subcategory of ordered $\diamond$-algebras bound to $V$. 
In fact, $(X,p)$ is intrinsically free: For each 
ordered $\diamond$-algebra $(Y,q)$ and each family of bottomed monotone mappings
$\tau : V \to \rest{Y}{A}$ there exists a unique ordered homomorphism
$\pi : (X,p) \to (Y,q)$ such that $\rest{\pi}{A} = \tau$. 
\end{proposition}

\proof Proposition~\ref{prop_head_type_1} implies there exists a 
$U$-minimal regular $\diamond$-algebra $(Y,p)$, and by Proposition~\ref{prop_head_type_5}
$(Y,q)$ is $\natural$-stable. Thus by Proposition~\ref{prop_ord_algs_3} there 
exists an intrinsic ordered extension $(X,p)$ of $(Y,p)$ bound to $V$. But this means 
that $(X,p)$ is a $V$-minimal intrinsic ordered $\diamond$-algebra.
To show that any such $\diamond$-algebra is intrinsically free the following fact is needed:

\begin{lemma}\label{lemma_ord_algs_2}
Let $(X,p)$ be a $V$-minimal intrinsic ordered $\Lambda$-algebra, let $(Y,q)$ be an
ordered $\Lambda$-algebra and
$\pi : (X,p) \to (Y,q)$ a bottomed homomorphism of the underlying bottomed 
$\Lambda$-algebras such that the mapping $\pi_a$ is monotone for each $a \in A$.
Then $\pi_b$ is monotone for each $b \in B$, i.e., $\pi$ is ordered. 
\end{lemma}

\proof For each $b \in B$ let $X'_b$ denote the set of those elements $x \in X_b$ such 
that $\pi_b(x) \sqle_b \pi_b(x')$ for all $x' \in X_b$ 
with $x \sqle_b x'$. Then $\rest{X'}{A} = V$, since by assumption $\pi_a$ is monotone for
each $a \in A$, and $\bot_b \in X'_b$ for each $b \in B$, since $\pi_b(\bot_b) = \bot_b$.
Moreover, the family $X'$ is invariant in $(X,p)$: Let $k \in K$ and 
$v \in \ass{\sdom{k}}{(X')} $, $x' \in X_{\scod{k}}$ with 
$p_k(v) \sqle_{\scod{k}} x'$. If $p_k(v) = \bot_{\scod{k}}$ then trivially 
$\pi_{\scod{k}}(p_k(v)) \sqle_{\scod{k}} \pi(x')$, 
hence suppose $p_k(v) \ne \bot_{\scod{k}}$. There then exists a unique 
$v' \in \ass{\sdom{k}}{X}$ such that
$x' = p_k(v')$, and $v \opass{\sdom{k}}{\sqle} v'$. But 
$v(\eta) \sqle_{\adom{k}{\eta}} v'(\eta)$ and
$v(\eta) \in X'_{\adom{k}{\eta}}$, and so 
$\pi_{\adom{k}{\eta}}(v(\eta)) 
\sqle_{\adom{k}{\eta}} \pi_{\adom{k}{\eta}}(v'(\eta))$ 
for each $\eta \in \domsdom{k}$, i.e., 
$\ass{\sdom{k}}{\pi}(v) \opass{\sdom{k}}{\sqle} \ass{\sdom{k}}{\pi}(v')$. 
Therefore
\[\pi_{\scod{k}}(p_k(v)) = q_k (\ass{\sdom{k}}{\pi}(v)) 
 \sqle_{\scod{k}} q_k(\ass{\sdom{k}}{\pi}(v')) = \pi_{\scod{k}}(p_k(v')) 
= \pi_{\scod{k}}(x')\;.\]
It follows that $X' = X$, since $(X,p)$ is $V$-minimal, which implies $\pi$ is a 
family of monotone mappings, i.e., $\pi$ is an ordered homomorphism. \eop

Now for the proof of Proposition~\ref{prop_ord_algs_3}. Let $(X,p)$ be a $V$-minimal intrinsic 
ordered $\diamond$-algebra, let $(Y,q)$ be an
ordered $\diamond$-algebra and let $\tau : V \to \rest{Y}{A}$ be a family of bottomed monotone
mappings. The underlying bottomed $\Lambda$-algebra of 
$(X,p)$ is then a $U$-minimal regular $\diamond$-algebra, and so by
Proposition~\ref{prop_head_type_6} there exists a unique bottomed homomorphism
$\pi : (X,p) \to (Y,q)$ such that $\rest{\pi}{A} = \tau$. Thus by Lemma~\ref{lemma_ord_algs_2}
$\pi$ is an ordered homomorphism
from $(X,p)$ to $(Y,q)$.  Finally, $\pi$ is unique, since any 
ordered 
homomorphism from $(X,p)$ to $(Y,q)$ is also a bottomed homomorphism of the underlying 
bottomed $\Lambda$-algebras, and by Proposition~\ref{prop_head_type_6} there is a unique 
such bottomed homomorphism $\pi$ with $\rest{\pi}{A} = \tau$. \eop

Before starting with the proof of Proposition~\ref{prop_ord_algs_3} we first look at
a couple of elementary properties of intrinsic ordered $\Lambda$-algebras.
A poset $X$ is said to be 
\definition{locally finite}\index{locally finite poset}\index{poset!locally finite} if
the principal ideal $\{ y \in X : y \sqle x \}$ is finite for each $x \in X$.
(For a partially ordered set with a bottom element this is equivalent to the usual
definition of being locally finite, which is that 
$\{ y \in X : x' \sqle y \sqle x \}$ should be finite for all $x,\,x' \in X$.)

\begin{proposition}\label{prop_ord_algs_6}
If $(X,p)$ is a $V$-minimal intrinsic ordered $\Lambda$-algebra and the posets in the 
$A$-family $V$ are all locally finite then $X$ is a $B$-family of locally finite posets. 
\end{proposition}

\proof For each $b \in B$ let $\fpi{X_b}$ denote the set of elements $x \in X_b$ for which 
the ideal $\{ y \in X : y \sqle x \}$ is finite. Then clearly $\bot_b \in \fpi{X_b}$ and 
it it follows more-or-less directly from the definition of being intrinsic that $\fpi{X}$ 
is an invariant family in the underlying bottomed $\Lambda$-algebra. Moreover, by 
assumption $\rest{\fpi{X}}{A} = V$, and hence $\fpi{X} = X$, since $(X,p)$ is $V$-minimal,
i.e., $X$ is a $B$-family of locally finite posets. \eop

Note the following special case of Proposition~\ref{prop_ord_algs_6}: If the signature
$\Lambda$ is closed (i.e., if $A = \varnothing$) and $(X,p)$ is a minimal intrinsic 
ordered $\Lambda$-algebra then $X$ is always a $B$-family of locally finite posets. 

If $X$ is a poset then $x \in X$ is said to be 
\definition{maximal}\index{maximal element in poset} if 
$\{ y \in X : x \sqle y \} = \{x\}$; the set of maximal elements will be denoted by
$\maximal{X}$. 

\begin{lemma}\label{lemma_ord_algs_6}
If $(X,p)$ is an intrinsic ordered $\Lambda$-algebra then
$\maximal{X}$ is an invariant family in the underlying
bottomed $\Lambda$-algebra.
\end{lemma}

\proof This follows more-or-less directly from the definition of being intrinsic. \eop

We now prepare for the proof of Proposition~\ref{prop_ord_algs_3}; this involves 
defining a family of partial orders having 
certain properties, and these properties form the basis for the following definition: Let 
$(Y,p)$ be a regular bottomed $\Lambda$-algebra, and for each $b \in B$ let $\sqle_b$ 
be a partial order on the set $Y_b$. Then the $B$-family of partial orders $\sqle$ will be 
called an \definition{ordering associated}\index{associated ordering} with $(Y,p)$ 
(or just an \definition{associated ordering}) if the following two conditions hold:
\begin{itemize}
\item[(1)] $(Y_b,\sqle_b)$ is a poset with bottom element $\bot_b$ for each
$b \in B$.
\item[(2)] If $b \in B\setminus A$ and $y_1,\,y_2 \in \nonbot{Y}_b$ then $y_1 \sqle_b y_2$ 
if and only if $y_1 = p_k(v_1)$ and $y_2 = p_k(v_2)$ with $k \in K_b$ and 
$v_1 \opass{\sdom{k}}{\sqle} v_2$. 
\end{itemize}
If $(X,p)$ is an intrinsic ordered extension of $(Y,p)$ and $\sqle_b$ is the 
partial order on $X_b$ for each $b \in B$ then $\sqle$ is an ordering associated 
with $(Y,p)$. Conversely, if $\sqle$ is an ordering associated with $(Y,p)$ and
$X_b = (Y_b,\sqle_b)$ for each $b \in B$ then $(X,p)$ will be an intrinsic ordered 
extension of $(Y,p)$, provided $(X,p)$ is an ordered $\Lambda$-algebra. But in 
general this is not the case, since it is possible that there exist
$v_1,\,v_2 \in \ass{\sdom{k}}{X}$ with $v_1 \opass{\sdom{k}}{\sqle} v_2$ and 
$p_k(v_1) \ne \bot_{\scod{k}}$ but with $p_k(v_2) = \bot_{\scod{k}}$ (see 
Example~\thesection.4 at the end of the section). However, if $(Y,q)$ is 
$\natural$-stable then this problem does not arise:

\begin{lemma}\label{lemma_ord_algs_3}
Let $(Y,p)$ be a $\natural$-stable regular bottomed $\Lambda$-algebra, let $\sqle$ 
be an ordering associated with $(Y,p)$ and put $X_b = (Y_b,\sqle_b)$ for each 
$b \in B$. Then $(X,p)$ is an ordered $\Lambda$-algebra (and thus an intrinsic ordered 
extension of $(Y,p)$).
\end{lemma}

\proof It must be shown that the mappings in the family $p$ are monotone. Thus consider 
$k \in K$ and let $v_1,\, v_2 \in \ass{\sdom{k}}{X}$ with
$v_1 \opass{\sdom{k}}{\sqle} v_2$; if $p_k(v_1) = \bot_{\scod{k}}$ then 
of course $p_k(v_1) \sqle_{\scod{k}} p_k(v_2)$ 
holds trivially, and so it can be assumed that $p_k(v_1) \in \nonbot{Y}_{\scod{k}}$. Now 
$v_2(\eta) \in \nonbot{Y}_{\adom{k}{\eta}}$ for all $\eta \in \domsdom{k}$ with
$v_1(\eta) \in \bot_{\adom{k}{\eta}}$, since
$v_1(\eta) \sqle_{\adom{k}{\eta}} v_2(\eta)$ for each $\eta \in \domsdom{k}$, and therefore
$p_k(v_2) \in \nonbot{Y}_{\scod{k}}$ (since $(Y,p)$ is $\natural$-stable). Hence by 
the definition of an associated ordering $p_k(v_1) \sqle_{\scod{k}} p_k(v_2)$, and this 
shows that $p_k$ is monotone. \eop

\bigskip
\fbox{\begin{exframe}
\textit{Example \thesection.2\enspace} Let $\Lambda = (B,K,\Theta)$ be the signature
in Example~\ref{algs_homs}.3, let $V$ be an
$A$-family of posets and let $(Y,q)$ be the bottomed $\Lambda$-algebra defined in 
Example~\ref{bot_algs_homs}.3 bound to the underlying sets in the family $V$.
For each $b \in B\setminus A$ define a partial order $\sqle_b$ on $Y_b$ as follows:
\exparskip
\begin{eelist}{15pt}
\item If $b \in \{\mathtt{bool},\mathtt{atom},\mathtt{int}\}$ and $a, \, a' \in Y_b$ 
      then $a \sqle_b a'$ if and only if $a \in \{\bot_b,a'\}$.
\exparskip
\item If $p, \, p' \in Y_{\mathtt{pair}}$ then $p \sqle_{\mathtt{pair}} p'$ if and only if 
      either $p = \bot_{\mathtt{pair}}$ or $p = (x,y)$ and $p' = (x',y')$ with 
      $x,\,x' \in \nonbot{V}_{\mathtt{x}}$, $y,\,y' \in \nonbot{V}_{\mathtt{y}}$, 
      $x \sqle_{\mathtt{x}} x'$ and $y \sqle_{\mathtt{y}} y'$.
\exparskip
\item If $\ell,\,\ell' \in Y_{\mathtt{list}}$ then $\ell \sqle_{\mathtt{list}} \ell'$ if 
      and only if either $\ell = \bot_{\mathtt{list}}$ or 
      $\ell,\,\ell' \in (\nonbot{Y}_{\mathtt{z}})^*$ with $\ell = \llist{z}{m}$, 
      $\ell' = \llist{z'}{m}$ and $z_k \sqle_{\mathtt{z}} z'_k$ for each $\oneto{k}{m}$.
\exparskip
\item If $y,\,y' \in Y_{\mathtt{lp}}$ then $y \sqle_{\mathtt{lp}} y'$ if and 
      only if either $y = \bot_{\mathtt{lp}}$ or $y,\,y' \in \nonbot{Y}_{\mathtt{pair}}$ 
      with $y \sqle_{\mathtt{pair}} y'$ or $y,\,y' \in \nonbot{Y}_{\mathtt{list}}$ with
      $y \sqle_{\mathtt{list}} y'$.
\end{eelist}
\exparskip
(Recall here that 
\begin{eelist}{15pt}
\item $Y_{\mathtt{bool}} = \Bool^\bot = \Bool \cup\{\bot_{\mathtt{bool}}\}$, \enskip 
      $Y_{\mathtt{atom}} = \Oneptset^\bot = \Oneptset \cup\{\bot_{\mathtt{atom}}\}$,
\item $Y_{\mathtt{int}} = \Int^\bot = \Int \cup \{\bot_{\mathtt{int}}\}$, \enskip
      $Y_{\mathtt{pair}} 
      = (\nonbot{V}_{\mathtt{x}} \times \nonbot{V}_{\mathtt{y}}) 
      \cup \{\bot_{\mathtt{pair}}\}$, 
\item $Y_{\mathtt{list}} = (\nonbot{V}_{\mathtt{z}})^* \cup \{\bot_{\mathtt{list}}\}$, 
      \enskip
      $Y_{\mathtt{lp}} = \nonbot{Y}_{\mathtt{pair}} \times \nonbot{Y}_{\mathtt{list}} 
      \cup \{\bot_{\mathtt{lp}}\}$.) 
\end{eelist}
\exparskip
The reader is left to check that with these partial orders $(Y,q)$ becomes
an intrinsic ordered $\Lambda$-algebra. Thus in fact $(Y,q)$ is a $V$-minimal 
intrinsic ordered $\diamond^\altbot$-algebra.
\end{exframe}}

\bigskip\bigskip

Lemma~\ref{lemma_ord_algs_3} and Proposition~\ref{prop_head_type_5} imply that
Proposition~\ref{prop_ord_algs_3} is an immediate corollary of the following result:

\begin{proposition}\label{prop_ord_algs_4}
Let $(Y,p)$ be a regular bottomed $\Lambda$-algebra bound to $U$ and let $\le$ be an 
$A$-family of bottomed partial orders on $U$, i.e., $\le_a$ is a bottomed partial order on 
$U_a$ for each $a \in A$. Then there exists an ordering $\sqle$ associated with $(Y,p)$ 
such that $\rest{\sqle}{A} = \le$. Moreover, if $(Y,p)$ is, in addition, $U$-minimal 
then this is the unique such associated ordering. 
\end{proposition}

\proof For the duration of this proof a $B$-family of partial orders $\sqle$ will be called 
an \definition{extension of $\le$} if $\rest{\sqle}{A} = \le$ and $\sqle_b$ is a bottomed 
partial order on $Y_b$ for each $b \in B\setminus A$. If $\sqle$ is an extension of 
$\le$ then a binary relation $\sqle'_b$ can be defined on $Y_b$ for each $b \in B$ as 
follows:
\begin{itemize}
\item[(1)] If $a \in A$ then $\sqle'_a = \le_a$. 
\item[(2)] If $b \in B\setminus A$ and $y_1,\,y_2 \in \nonbot{Y}_b$ then $y_1 \sqle'_b y_2$ 
holds if and only if $k_1 = k_2$ and $v_1 \opass{\sdom{k_1}}{\sqle} v_2$, where 
$k_1,\, k_2 \in K_b$ and $v_1 \in \ass{\sdom{k_1}}{X}$, $v_2 \in \ass{\sdom{k_2}}{X}$
are the unique elements such that
$y_1 = p_{k_1}(v_1)$ and $y_2 = p_{k_2}(v_2)$.
\item[(3)] If $b \in B\setminus A$ then $\bot_b \sqle'_b y$ holds for all $y \in Y_b$ but 
$y \sqle'_b \bot_b$ does not hold for any $y \in \nonbot{Y}_b$.
\end{itemize}

\begin{lemma}\label{lemma_ord_algs_4}
$\sqle'$ is an extension of $\le$. 
\end{lemma}

\proof It is enough to show that $\sqle'_b$ is a partial order for each
$b \in B \setminus A$, and it is clear that $\sqle'_b$ is reflexive. To show that
$\sqle'_b$ is anti-symmetric consider $y_1,\, y_2 \in Y_b$ with $y_1 \sqle'_b y_2$ and 
$y_2 \sqle'_b y_1$. Then by (3) either $y_1 = y_2 = \bot_b$, or $y_1$ and $y_2$ both lie 
in $\nonbot{Y}_b$, in which case $y_1 = y_2$ holds by (2) and the fact that 
$\opass{\sdom{k}}{\sqle}$ is 
anti-symmetric for each $k \in K_b$. A similar argument shows that $\sqle'_b$ is 
transitive. \eop

The family $\sqle'$ of partial orders given by Lemma~\ref{lemma_ord_algs_4} will be called 
the \definition{first refinement} of the family $\sqle$. Now in order to show the existence 
of an associated ordering it is useful to first introduce a somewhat weaker concept: A 
$B$-family $\sqle$ is called a \definition{weak ordering} if it is an extension of $\le$
and if whenever $b \in B\setminus A$ and $y_1,\,y_2 \in \nonbot{Y}_b$ are such that
$y_1 \sqle_b y_2$ then $y_1 = p_k(v_1)$ and $y_2 = p_k(v_2)$ with $k \in K_b$ and 
$v_1 \opass{\sdom{k}}{\sqle} v_2$. 

\begin{lemma}\label{lemma_ord_algs_5}
Suppose $\sqle$ is a weak ordering. Then the first refinement $\sqle'$ is also a weak 
ordering and $y_1 \sqle'_b y_2$ whenever $y_1 \sqle_b y_2$. 
\end{lemma}

\proof It will be shown first that if $y_1,\, y_2 \in Y_b$ with $y_1 \sqle_b y_2$ then 
$y_1 \sqle'_b y_2$. If $b \in A$ then this holds by definition, and if $b \in B\setminus A$
then the only non-trivial case is with $y_1,\, y_2 \in \nonbot{Y}_b$. Here let
$k_1,\,k_2 \in K_b$ and $v_1 \in \ass{\sdom{k_1}}{X}$, $v_2 \in \ass{\sdom{k_2}}{X}$ be 
the unique 
elements such that $y_1 = p_{k_1}(v_1)$ and $y_2 = p_{k_1}(v_2)$. Thus 
$p_{k_1}(v_1) \sqle_b p_{k_1}(v_2)$ and hence $k_1 = k_2$ and 
$v_1 \opass{\sdom{k_1}}{\sqle} v_2$ (since 
$\sqle$ is a weak ordering). But then $y_1 \sqle'_b y_2$ holds by the definition
of $\sqle'_b$.

It remains to show that $\sqle'$ is a weak ordering, and by Lemma~\ref{lemma_ord_algs_4}
$\sqle'$ is an extension of $\le$. Therefore consider $b \in B\setminus A$ and 
$y_1,\,y_2 \in \nonbot{Y}_b$ with $y_1 \sqle'_b y_2$ and let $k_1, \, k_2 \in K_b$ and 
$v_1 \in \ass{\sdom{k_1}}{X}$, $v_2 \in \ass{\sdom{k_2}}{X}$ be the unique elements such 
that $y_1 = p_{k_1}(v_1)$ and $y_2 = p_{k_2}(v_2)$. Then 
$p_{k_1}(v_1) \sqle'_b p_{k_2}(v_2)$ 
and so (by the definition of $\sqle'_b$) $k_1 = k_2$ and 
$v_1 \opass{\sdom{k_1}}{\sqle} v_2$; hence by 
the first part $v_1 \opass{\sdom{k_1}}{(\sqle')} v_2$. \eop

For each $b \in B\setminus A$ let $\sqle^0_b$ be the flat partial order on $Y_b$ and for 
each $a \in A$ let $\sqle^0_a = \le_a$ (so $\sqle^0$ can be thought of as the flat 
extension of $\le$). It is clear that $\sqle^0$ is a weak ordering, and therefore by 
Lemma~\ref{lemma_ord_algs_5} a sequence of weak orderings $\sqle^m$, $m \ge 0$, can be 
defined by letting $\sqle^{m+1}$ be the first refinement of $\sqle^m$ for each $m \ge 0$. 
Now define a partial order $\sqle_b$ on $Y_b$ for each $b \in B$ by stipulating that 
$y_1 \sqle_b y_2$ if and only if $y_1 \sqle^m_b y_2$ for some (and thus for all 
sufficiently large) $m \ge 0$. It is easy to see that $\sqle$ is then a weak ordering, 
and so in particular $\rest{\sqle}{A} = \le$.

In fact $\sqle$ is an associated ordering: 
Consider $v_1,\, v_2 \in \ass{\sdom{k}}{X}$ with $v_1 \opass{\sdom{k}}{\sqle} v_2$ and 
$p_k(v_1),\, p_k(v_2) \in \nonbot{Y}_b$. Then 
$v_1(\eta) \sqle_{\adom{k}{\eta}} v_2(\eta)$ for each 
$\eta \in \domsdom{k}$, and so for each $\eta$ there exists $m_\eta \ge 0$ such that
$v_1(\eta) \sqle^{m_\eta}_{\adom{k}{\eta}} v_2(\eta)$. Put 
$m = \max \{ m_\eta : \eta \in \domsdom{k} \}$; then 
$v_1(\eta) \sqle^m_{\adom{k}{\eta}} v_2(\eta)$ for each $\eta$, and this means that 
$v_1 \ass{\sdom{k}}{(\sqle^m)} v_2$. Therefore $p_k(v_1) \sqle^{m+1}_b p_k(v_2)$ by the 
definition of $\sqle^{m+1}_b$, and hence $p_k(v_1) \sqle_b p_k(v_2)$.

The uniqueness when $(Y,p)$ is $U$-minimal still has to be considered. Thus let 
$\sqle$ and $\sqle'$ be two orderings associated with $(Y,p)$ with 
$\rest{\sqle}{A} = \le = \rest{\sqle'}{A}$. For each $y \in Y_b$ define
$L_b(y) = \{ y' \in Y_b : y' \sqle_b y \}$ and 
$L'_b(y) = \{ y' \in Y_b : y' \sqle'_b y \}$; put 
$Y'_b = \{ y \in Y_b : L'_b(y) = L_b(y) \}$. Then $Y'$ is a bottomed family with 
$\rest{Y'}{A} = \rest{Y}{A}$ and it is straightforward to check that the family $Y'$ is 
invariant in $(Y,p)$. Hence if $(Y,p)$ is $U$-minimal then $Y' = Y$, i.e., 
$\sqle' = \sqle$. \eop

\textit{Proof of Proposition~\ref{prop_ord_algs_3}\enspace} 
This now follows immediately from Lemma~\ref{lemma_ord_algs_3} and Propositions 
\ref{prop_head_type_5} and \ref{prop_ord_algs_4}. \eop

Finally, consider the case when $\Lambda$ is the disjoint union of the 
signatures $\Lambda_i$, $i \in F$.

\begin{proposition}\label{prop_ord_algs_5}
For each $i \in F$ let $(X^i,p^i)$ be an intrinsic ordered $\Lambda_i$-algebra. Then
the sum $\oplus_{i\in F} (X^i,p^i)$ is an intrinsic ordered $\Lambda$-algebra.
\end{proposition}

\proof Straightforward. \eop

\bigskip
\fbox{\begin{exframe}
\textit{Example \thesection.3\enspace} Let $\Lambda = (B,K,\Theta)$ be the signature
in Example~\ref{algs_homs}.3, let $V$ be an
$A$-family of posets and let $(Y,q)$ be the bottomed $\Lambda$-algebra defined in 
Example~\ref{bot_algs_homs}.3 bound to the underlying sets in the family $V$.
For each $b \in B\setminus A$ define a partial order $\sqle_b$ on $Y_b$ as follows:
\exparskip
\begin{eelist}{15pt}
\item If $b \in \{\mathtt{bool},\mathtt{atom},\mathtt{int}\}$ and $a, \, a' \in Y_b$ 
      then $a \sqle_b a'$ if and only if $a \in \{\bot_b,a'\}$.
\exparskip
\item If $p, \, p' \in Y_{\mathtt{pair}}$ then $p \sqle_{\mathtt{pair}} p'$ if and only if 
      either $p = \bot_{\mathtt{pair}}$ or $p = (x,y)$ and $p' = (x',y')$ with 
      $x,\,x' \in V_{\mathtt{x}}$, $y,\,y' \in V_{\mathtt{y}}$, 
      $x \sqle_{\mathtt{x}} x'$ and $y \sqle_{\mathtt{y}} y'$.
\exparskip
\item If $\ell,\,\ell' \in Y_{\mathtt{list}}$ then $\ell \sqle_{\mathtt{list}} \ell'$ 
      if and only if either
\begin{eelist}{10pt}
\item $\ell =  \llist{z}{m}$, 
      $\ell' = \llist{z'}{m}$ with $z_j \sqle_{\mathtt{z}} z'_j$ for each $\oneto{j}{m}$, 
      or 
\item $\ell =  (\llist{z}{m})^\bot$ and $\ell'$ either $\llist{z'}{n}$ or
      $(\llist{z'}{n})^\bot$, with $m \le n$ and $z_j \sqle_{\mathtt{z}} z'_j$ for each 
      $\oneto{j}{m}$. 
\end{eelist}
\exparskip
\item If $y,\,y' \in Y_{\mathtt{lp}}$ then $y \sqle_{\mathtt{lp}} y'$ if and only if either 
      $y = \bot_{\mathtt{lp}}$ or $y,\,y' \in Y_{\mathtt{pair}}$ with 
      $y \sqle_{\mathtt{pair}} y'$ or $y,\,y' \in Y_{\mathtt{list}}$ with
      $y \sqle_{\mathtt{list}} y'$.
\end{eelist}
\exparskip
(Recall here that 
\begin{eelist}{15pt}
\item $Y_{\mathtt{bool}} = \Bool^\bot = \Bool \cup\{\bot_{\mathtt{bool}}\}$, \enskip 
      $Y_{\mathtt{atom}} = \Oneptset^\bot = \Oneptset \cup\{\bot_{\mathtt{atom}}\}$,
\item $Y_{\mathtt{int}} = \Int^\bot = \Int \cup \{\bot_{\mathtt{int}}\}$, \enskip
      $Y_{\mathtt{pair}} 
      = (V_{\mathtt{x}} \times V_{\mathtt{y}}) \cup \{\bot_{\mathtt{pair}}\}$, 
\item $Y_{\mathtt{list}} = V_{\mathtt{z}}^* \cup \Bot{V_{\mathtt{z}}^*}$
      with $\bot_{\mathtt{list}} = \onept^\bot$.
\item $Y_{\mathtt{lp}} = Y_{\mathtt{pair}} 
                  \cup Y_{\mathtt{list}} \cup \{\bot_{\mathtt{lp}}\}$.) 
\end{eelist}
\exparskip
The reader is left to check that with these partial orders $(Y,q)$ becomes
an intrinsic ordered $\Lambda$-algebra. Thus in fact $(Y,q)$ is a $V$-minimal 
intrinsic ordered $\diamond^\natural$-algebra.
\exparskip
With a minor modification the above discussion also deals with the bottomed 
$\Lambda$-algebra $(Y,q)$ defined in Example~\ref{head_type}.2. The only difference in the
bottomed $\Lambda$-algebras is that $Y_{\mathtt{pair}}$ is now taken to be
$V_{\mathtt{x}} \times V_{\mathtt{y}}$ (with 
$\bot_{\mathtt{pair}} = (\bot_{\mathtt{x}},\bot_{\mathtt{y}})$).
The order $\sqle_{\mathtt{pair}}$ on $Y_{\mathtt{pair}}$ is then defined by:
\exparskip
\begin{eelist}{15pt}
\item If $p, \, p' \in Y_{\mathtt{pair}}$ then $p \sqle_{\mathtt{pair}} p'$ if and only if 
      $p = (x,y)$ and $p' = (x',y')$ with $x \sqle_{\mathtt{x}} x'$ and 
      $y \sqle_{\mathtt{y}} y'$.
\end{eelist}
With these partial orders (the remaining partial orders being defined as above)
$(Y,q)$ becomes a $V$-minimal intrinsic ordered $\diamond^{\Join}$-algebra.
\end{exframe}}

\bigskip
\fbox{\begin{exframe}
\textit{Example~\thesection.4} Consider 
the $\Lambda$-algebra $(X,p)$ in Example~\ref{algs_homs}.1, and let
$(Y,q)$ be the $\Lambda$-algebra defined by
\begin{eelist}{20pt}
\item $Y_{\mathtt{nat}} = \Nat \cup \{\bot_{\mathtt{nat}},\bot^o_{\mathtt{nat}}\}$ with
$\bot^o_{\mathtt{nat}} \notin \Nat \cup \{\bot_{\mathtt{nat}}\}$,
\item $Y_\beta = X^\bot_b$\ for all $b \in B \setminus \{\mathtt{nat}\}$,
\item $q_{\mathtt{Zero}} : \Oneptset \to Y_{\mathtt{nat}}$ with
      $q_{\mathtt{Zero}}(\onept) = 0$,
\item $q_{\mathtt{Succ}} : Y_{\mathtt{nat}} \to Y_{\mathtt{nat}}$ with
$q_{\mathtt{Succ}}(n) = \left\{ \begin{array}{cl}
                           n + 1 &\ \textrm{if}\ n \in \Nat, \\
            \bot_{\mathtt{nat}} &\ \textrm{if}\ n = \bot^o_{\mathtt{nat}}, \\
               \bot^o_{\mathtt{nat}} &\ \textrm{if}\ n = \bot_{\mathtt{nat}}\, 
\end{array}\right.$
\exparskip
\item and with $q_\kappa = p^\bot_k$ for all 
$k \in K \setminus \{\mathtt{Zero},\mathtt{Succ}\}$,
\end{eelist}
where $(X^\bot,p^\bot)$ is the flat bottomed extension of $(X,p)$. Then 
it is easy to see that $(Y,q)$ is a minimal regular bottomed $\Lambda$-algebra, and 
so by Proposition~\ref{prop_ord_algs_4}
there is a unique ordering $\sqle$ associated with $(Y,q)$.
However, $q_{\mathtt{Succ}}$ is not monotone, since
$\bot_{\mathtt{nat}} \sqle_{\mathtt{nat}} \bot^o_{\mathtt{nat}}$
but $q_{\mathtt{Succ}}(\bot_{\mathtt{nat}}) = \bot^o_{\mathtt{nat}}$ and
$q_{\mathtt{Succ}}(\bot^o_{\mathtt{nat}}) = \bot_{\mathtt{nat}}$.
\end{exframe}}

\bigskip

\startsection{Continuous algebras}
\label{cont_algs}

A \definition{continuous $\Lambda$-algebra}\index{continuous algebra}\index{algebra!continuous}
is any pair $(X,p)$ in which $X$ is a 
$B$-family of complete posets and $p$ is a $K$-family of mappings such that 
$p_k$ is a continuous mapping from $\ass{\sdom{k}}{X}$ to $X_{\scod{k}}$ for each
$k \in K$, recalling that $\ass{\sdom{k}}{X}$ denotes the complete poset 
$\utimes(X \fcomp \sdom{k})$.
A continuous $\Lambda$-algebra is in particular an ordered $\Lambda$-algebra.

A continuous $\Lambda$-algebra is deemed to have a property (such as being intrinsic or 
being an $(H,\diamond)$-algebra) if it has this property as an ordered  $\Lambda$-algebra.

The main result here is Proposition~\ref{prop_cont_algs_2}, which is the result 
corresponding to Proposition~\ref{prop_ord_algs_2} for continuous algebras. 
The continuous $\Lambda$-algebras we will be dealing with are all initial completions
of ordered $\Lambda$-algebras, and so we must first say precisely what this means.

\begin{lemma}\label{lemma_cont_algs_1}
Let $(Y,q)$ be an ordered $\Lambda$-algebra and for each $b \in B$ let $X_b$ be an initial
completion of the poset $Y_b$. Then for each $k \in K$ the mapping 
$q_k : \ass{\sdom{k}}{Y} \to Y_{\scod{k}}$ extends uniquely to a continuous mapping 
$p_k : \ass{\sdom{k}}{X} \to X_{\scod{k}}$. 
\end{lemma}

\proof By assumption the mapping $q_k$ is monotone and so it is still monotone considered
as a mapping from $\ass{\sdom{k}}{Y}$ to $X_{\scod{k}}$; moreover, by 
Proposition~\ref{prop_init_compl_3}
$\ass{\sdom{k}}{X}$ is an initial completion of $\ass{\sdom{k}}{Y}$. The result
therefore follows from Proposition~\ref{prop_init_compl_2} ($(1)\Rightarrow(2)$). \eop 

Lemma~\ref{lemma_cont_algs_1} allows the following definition to be made: A continuous 
$\Lambda$-algebra $(X,p)$ is said to be an 
\definition{initial completion}\index{initial completion}\index{completion!initial} of 
an ordered $\Lambda$-algebra $(Y,q)$ if 
$X_b$ is an initial completion of the poset $Y_b$ for each $b \in B$ and $p_k$ is 
the unique continuous extension of $q_k$ for each $k \in K$. 
It follows immediately from Proposition~\ref{prop_init_compl_2} that there exists an initial
completion of $(Y,q)$.

\begin{lemma}\label{lemma_cont_algs_2}
Let $(Y,q)$ be an intrinsic ordered $\Lambda$-algebra. Then any
initial completion $(X,p)$ of $(Y,q)$ is also intrinsic.
\end{lemma}

\proof Let $b \in B \setminus A$ and $x \in \nonbot{X}_b$. Since $X_b$ is a
completion of $Y_b$ there exists $D \in \directed{Y_b}$ with $x = \lub D$, and then 
$\nonbot{D} = D \setminus \{\bot_b\}$ is also an element of $\directed{Y_b}$ with
$x = \lub \nonbot{D}$. If $y_1,\, y_2 \in D$ with $y_1 \sqle_b y_2$ then, since $(Y,q)$ 
is intrinsic, there exists a unique $k \in K_b$ and unique elements
$u_1,\,u_2 \in \ass{\sdom{k}}{Y}$ with $y_1 = q_k(u_1)$ and $y_2 = q_k(u_2)$, and then
$u_1 \opass{\sdom{k}}{\sqle} u_2$. Thus, since $\nonbot{D}$ is directed, there exists a 
unique 
$k \in K_b$ and for each $y \in \nonbot{D}$ a unique element $u \in \ass{\sdom{k}}{Y}$ with 
$y = q_k(u)$. 
Moreover, the set $C = q_k^{-1}(\nonbot{D})$ is an element 
of $\directed{\ass{\sdom{k}}{Y}}$ and $q_k(C) = \nonbot{D}$. Put $v = \lub C$; then, since 
$p$ is continuous, 
\[ p_k(v) = p_k\left(\lub C\right) = \lub p_k(C) = \lub q_k(C) = \lub \nonbot{D} = x\;.\]
This shows that for each $x \in \nonbot{X}_b$ there exists $k \in K_b$ and an element
$v \in \ass{\sdom{k}}{X}$ with $x = p_k(v)$. Now let $x_1,\,x_2 \in \nonbot{X}_b$ with 
$x_1 \sqle_b x_2$ and suppose $x_1 = p_{k_1}(v_1)$ and $x_2 = p_{k_2}(v_2)$ with 
$k_1,\,k_2 \in K_b$ and $v_1 \in \ass{\sdom{k_1}}{X}$, $v_2 \in \ass{\sdom{k_2}}{X}$. 
Then, since $\ass{\sdom{k_i}}{X}$ is a completion of $\ass{\sdom{k_i}}{Y}$ for $i = 1,\,2$, 
there exists $C_i \in \directed{\ass{\sdom{k_i}}{Y}}$ such that $v_i = \lub C_i$. 
But then 
\begin{eqnarray*}
 \lub q_{k_1}(C_1) &=& \lub p_{k_1}(C_1) = p_{k_1}\left(\lub C_1\right) 
= p_{k_1}(v_1) = x_1\\
& \sqle_b & x_2 = p_{k_2}(v_2) = p_{k_2}\left(\lub C_2\right) = \lub p_{k_2}(C_2) 
= \lub q_{k_2}(C_2)\;,
\end{eqnarray*}
which by Proposition~\ref{prop_init_compl_2} ($(1)\Rightarrow (3)$)
and Lemma~\ref{lemma_cont_algs_1} implies that $q_{k_1}(C_1)$ is cofinal in
$q_{k_2}(C_2)$. However, by the regularity of $(Y,q)$ this is only possible if $k_1 = k_2$ 
and $C_1$ is cofinal in $C_2$, and then also 
$v_1 = \lub C_1 \opass{\sdom{k_1}}{\sqle} \lub C_2 = v_2$.
In particular (with $x_1 = x_2 = x$) this shows, together with the first part of the proof,
that for each $x \in \nonbot{X}_b$ there exists a unique $k \in K_b$ and a unique  element
$v \in \ass{\sdom{k}}{X}$ with $x = p_k(v)$. Moreover, it then also clearly shows that 
$(X,p)$ is intrinsic. \eop

\begin{lemma}\label{lemma_cont_algs_3}
Let $\diamond$ be a simple head type and $(Y,q)$ be an ordered $\diamond$-algebra. Then 
any initial completion $(X,p)$ of $(Y,q)$ is also an $\diamond$-algebra.
\end{lemma}

\proof  Let $k \in K$ and $v \in \ass{\sdom{k}}{X}$. Then there 
exists $C \in \directed{\ass{\sdom{k}}{Y}}$ with $v = \lub C$, thus by 
Lemma~\ref{lemma_comp_posets_2}
$v(\eta) = \lub \{ u(\eta) : u \in C \}$ for each $\eta \in \domsdom{k}$ and 
$p_k(v) = \lub q_k(C)$.
Therefore, since $C$ is directed and $\domsdom{k}$ is finite, there exists $u \in C$ such 
that $p_k(v) \ne \bot_{\scod{k}}$ if and only if $q_k(u) \ne \bot_{\scod{k}}$ and such that 
$v(\eta) \ne \bot_{\adom{k}{\eta}}$ if and only if $u(\eta) \ne \bot_{\adom{k}{\eta}}$ 
for each $\eta\in \domsdom{k}$.
This implies that $\varepsilon_{\scod{k}}(p_k(v)) = \varepsilon_{\scod{k}}(p_k(u))$ and that
$\varepsilon_{\adom{k}{\eta}}(v(\eta)) = \varepsilon_{\adom{k}{\eta}}(u(\eta))$ for each 
$\eta \in \domsdom{k}$. Hence
\[ \varepsilon_{\scod{k}}(p_k(v)) = \varepsilon_{\scod{k}}(p_k(v')) 
 = \varepsilon_{\scod{k}}(q_k(v')) = \diamond_k(\ass{\sdom{k}}{\varepsilon}(v')) 
 = \diamond_k(\ass{\sdom{k}}{\varepsilon}(v))\;,\]
which shows that $(X,p)$ is also a $\diamond$-algebra. \eop 

Let $(X,p)$ and $(Y,q)$ be continuous $\Lambda$-algebras; then a 
\definition{continuous homomorphism}\index{continuous homomorphism}\index{homomorphism!continuous}
$\pi$ from $(X,p)$ to $(Y,q)$ is an 
ordered homomorphism such that $\pi_b$ is a continuous mapping for each $b \in B$. 

In what follows let $V$ be an $A$-family of complete posets.

\begin{proposition}\label{prop_cont_algs_1}
(1)\enskip If $(X,p)$ is a continuous $\Lambda$-algebra bound to $V$ then
the $B$-family of identity mappings $\id : X \to X$ defines a continuous homomorphism 
from $(X,p)$ to itself fixing $V$.

(2)\enskip If $\pi : (X,p) \to (Y,q)$ and $\varrho : (Y,q) \to (Z,r)$ are continuous
homomorphisms fixing 
$V$ then the composition $\varrho\fcomp\pi$ is a continuous homomorphism from $(X,p)$ to 
$(Z,r)$ fixing $V$.
\end{proposition}

\proof This follows immediately from Propositions \ref{prop_algs_homs_1}
and \ref{prop_comp_posets_1}. \eop

Proposition~\ref{prop_cont_algs_1} implies that there is a category whose objects are 
continuous $\Lambda$-algebras bound to $V$ with morphisms continuous homomorphisms fixing 
$V$. Lemma~\ref{lemma_comp_posets_1} implies that a morphism $\pi$ in this category is an 
isomorphism if and only if the mapping $\pi_b$ is an order isomorphism for each $b \in B$. 

Now let $\diamond$ be a $\natural$-stable simple head type and let $V$ be an $A$-family of 
algebraic posets. For $a \in A$ let $U_a = \compact(V_a)$ be the set of algebraic
elements $V_a$, considered as a subposet of $V_a$. By Proposition~\ref{prop_ord_algs_2}
there then exists a 
$U$-minimal intrinsic ordered $\diamond$-algebra $(Y,q)$.
Now by Proposition~\ref{prop_alg_posets_2}
$V_a$ is an initial completion of $U_a$ for each $a \in A$, and
hence there exists an initial completion $(X,p)$ of $(Y,q)$ with $X_a = V_a$ for each 
$a \in A$. Thus by Lemmas \ref{lemma_cont_algs_2} and \ref{lemma_cont_algs_3}
$(X,p)$ is an intrinsic
continuous $\diamond$-algebra, and by definition $(X,p)$ is bound to $V$.
Moreover, by Proposition~\ref{prop_alg_posets_2} $X$ is a family of algebraic posets.

\begin{proposition}\label{prop_cont_algs_2}
$(X,p)$ is an initial object in the full subcategory of continuous $\diamond$-algebras 
bound to $V$. In fact $(X,p)$ is intrinsically free:
For each continuous $\diamond$-algebra $(X',p')$ and each family
$\tau : V \to \rest{X'}{A}$ of bottomed continuous mappings there exists a unique 
continuous homomorphism $\pi : (X,p) \to (X',p')$ such that $\rest{\pi}{A} = \tau$.
\end{proposition}

\proof Let $\tau' : U \to \rest{X'}{A}$ be the restriction of $\tau$ to $U$. Then
$\tau'$ is a family of bottomed monotone mappings and hence by Proposition~\ref{prop_ord_algs_2} 
there exists a unique ordered homomorphism $\varrho : (Y,q) \to (X',p')$ 
such that $\rest{\varrho}{A} = \tau'$.
Now by Proposition~\ref{prop_init_compl_2} ($(1)\Rightarrow (2)$)
the monotone mapping $\varrho_b : Y_b \to X'_b$  extends uniquely
to a continuous mapping $\pi_b : X_b \to X'_b$ and $\pi$ is a homomorphism from
$(X,p)$ to $(X',p')$: Let $k \in K$; Proposition~\ref{prop_comp_posets_4} implies that 
$\ass{\sdom{k}}{\pi} : \ass{\sdom{k}}{X} \to \ass{\sdom{k}}{(X')}$ is continuous, and so 
$\pi_{\scod{k}} \comp p_k$ and $p'_k \comp \ass{\sdom{k}}{\pi}$ are both 
continuous mappings from $\ass{\sdom{k}}{X}$ to $X'_{\scod{k}}$ with
\[(\pi_{\scod{k}} \comp p_k)(v) = \pi_{\scod{k}}(p_k(v)) 
= \varrho_{\scod{k}}(q_k(v)) = q'_k (\ass{\sdom{k}}{\varrho})(v)
        = p'_k(\ass{\sdom{k}}{\pi})(v) = (p'_k \comp \ass{\sdom{k}}{\pi})(v) \]
for all $v \in \ass{\sdom{k}}{Y}$. Hence by Lemma~\ref{lemma_init_compl_1}
$\pi_{\scod{k}} \comp p_k = p'_k \comp \ass{\sdom{k}}{\pi}$, since $\ass{\sdom{k}}{X}$ is a
completion of $\ass{\sdom{k}}{Y}$, i.e., $\pi :(X,p) \to (X',p')$ is a continuous 
homomorphism, and clearly $\rest{\pi}{A} = \tau$, since $\tau_a$ is the unique continuous
extension of $\tau'_a$ for each $a \in A$.
Finally, if $\pi$ and $\pi'$ are continuous homomorphisms from $(X,p)$ to $(X',p')$ 
with $\rest{\pi}{A} = \tau = \rest{\pi'}{A}$
then by Lemma~\ref{lemma_bot_algs_homs_3} $\pi_b(y) = \pi'_b(y)$ for all $y \in Y_b$, 
$b \in B$, and so by Lemma~\ref{lemma_init_compl_1} $\pi = \pi'$.  \eop


%% file: sbika6.tex
\startchapter{Polymorphism}
\label{poly}

This chapter deals with polymorphism as it appears in all modern functional
programming languages. In order to make things simpler our approach is a bit more 
restrictive than that allowed in languages such as \textit{Haskell} or \textit{ML}.
However, this does not really impose any restrictions on what one (as programmer) can 
actually do.

\startsection{Algebras in categories}
\label{cat_algs}

As always let $\Lambda = (B,K,\Theta)$ be a signature with parameter set $A$. 
In Chapter~\ref{uni_alg} $\Lambda$-algebras were introduced as pairs $(X,p)$ with $X$ a 
$B$-family of sets and $p$ an appropriate $K$-family of mappings. Then in 
Chapter~\ref{bot_algs} bottomed $\Lambda$-algebras $(X,p)$ were considered; the family $X$ 
is then a $B$-family of bottomed sets. Finally, in Chapter~\ref{ord_cont_algs} ordered 
and continuous $\Lambda$-algebras were introduced with a $B$-family of posets 
(resp.\ complete posets) and a $K$-family of monotone (resp.\ continuous) mappings.

In order to emphasise what is common in all these cases it is worth formulating
what it means for an algebra to be defined in a category. The resulting formalism
is here little more than abstract nonsense, but it will play a useful role in the rest
of the chapter.

In what follows let $\mathsf{C}$ be a category (as defined in Section~\ref{sets}).
The objects of $\mathsf{C}$ will be denoted by $\mathcal{C}$ and the class of all 
morphisms by $\mathcal{M}$, i.e., $\mathcal{M}$ is the union of all the sets $\Hom(X,Y)$, 
$X,\,Y \in \mathcal{C}$.
 
If $S$ is a set and $\varphi : S \to \mathcal{M}$ is an $S$-family of morphisms
then $\varphi : X \to Y$ will be written to indicate that $X$ and $Y$ are the $S$-families 
of objects such that $\varphi_s \in \Hom(X_s,Y_s)$ for each $s \in S$. If $\psi : Y \to Z$
a further $S$-family of morphisms then the $S$-family of composed morphisms
will be denoted by $\psi\fcomp\varphi$, thus $\psi\fcomp\varphi : X \to Z$ is the
$S$-family of morphisms with $(\psi\fcomp\varphi)_s =  \psi_s\comp\varphi_s$ for each
$s \in S$.

If $\mathcal{S}$ is a class then the class of all finite families of elements from $\mathcal{S}$ will be 
denoted by $\ffam{\mathcal{S}}$. Note that if $S$ is an arbitrary set and $X : S \to \mathcal{S}$ 
an $S$-family of elements from $\mathcal{S}$ then $X \comp \gamma \in \ffam{\mathcal{S}}$ for each
finite $S$-typing $\gamma \in \ftyped{S}$.

The category $\mathsf{C}$ together with mappings 
$\utimes : \ffam{\mathcal{C}} \to \mathcal{C}$ and
$\utimes : \ffam{\mathcal{M}} \to \mathcal{M}$ will be called a 
\index{category}\definition{$\utimes$-category} if the following conditions hold:
\begin{evlist}{25pt}{0.8ex}
\item[---] 
$\utimes \varphi \in \Hom(\utimes X,\utimes Y)$ whenever
$\varphi \in \ffam{\mathcal{M}}$ with $\varphi : X \to Y$.
\item[---] 
$(\utimes \varrho) \comp (\utimes \pi) = \utimes(\varrho \fcomp \pi)$
whenever $\pi,\,\varrho \in \ffam{\mathcal{M}}$ with
$\pi : X \to Y$ and $\varrho : Y \to Z$.
\item[---] 
$\utimes \id = \id_{\utimes X}$ for each $X \in \ffam{\mathcal{C}}$, with
$\id : X \to X$ the family of identity morphisms.
\end{evlist}
The object $\utimes X$ should be thought of as the product of the objects in the family 
$X$ and the morphism $\utimes \varphi$ as the product of the morphisms in the family 
$\varphi$. However, in general no assumptions will placed on the mappings $\utimes$ to 
justify this interpretation.

The categories $\mathsf{Sets}$, $\mathsf{BSets}$, $\mathsf{Posets}$, $\mathsf{CPosets}$
and $\mathsf{APosets}$ are all $\utimes$-categories: In each case  
$\utimes : \ffam{\mathcal{C}} \to \mathcal{C}$ is defined as in Section~\ref{sets} or 
Chapter~\ref{domains} and if $\varphi : X \to Y$ is a finite $S$-family of morphisms
then $\utimes{\varphi} : \utimes{X} \to \utimes{Y}$ is defined by
\[\utimes{\varphi} (v)(s) = \varphi_s (v(s))\]
for each $v \in \utimes{X}$, $s \in S$, noting that in all of these cases a morphism is a 
mapping between the underlying sets. The conditions imposed on the mapping
$\utimes : \ffam{\mathcal{M}} \to \mathcal{M}$ thus follow from Lemma~2.1.

In what follows let $\mathsf{C}$ be a $\utimes$-category. If $S$ is a set and
$X : S \to \mathcal{C}$ an $S$-family of objects then for each finite $S$-typing 
$\gamma$ the object $\utimes (X \fcomp \gamma)$ will be denoted by $\ass{\gamma}{X}$. 
In the same way, if $\pi : S \to \mathcal{M}$ is an $S$-family of 
morphisms then the morphism $\utimes (\pi \fcomp \gamma)$ will be denoted by 
$\ass{\gamma}{\pi}$. Note that this notation agrees with that already being employed 
in the categories listed above.

\begin{lemma}\label{lemma_cat_algs_1}
(1)\enskip Let $X : S \to \mathcal{C}$ be an $S$-family of objects and 
$\id : X \to X$ the $S$-family of identity morphisms. Then 
$\ass{\gamma}{\id} : \ass{\gamma}{X} \to \ass{\gamma}{X}$ is 
also the identity morphism for each finite $S$-typing $\gamma$.

(2)\enskip Let $X,\,Y,\,Z : S \to \mathcal{C}$ be $S$-families of objects and let 
$\varphi : X \to Y$ and $\psi : Y \to Z$ be $S$-families of morphisms. Then 
$\ass{\gamma}{(\psi\fcomp\varphi)} = \ass{\gamma}{\psi}\comp\ass{\gamma}{\varphi}$
for each finite $S$-typing $\gamma$.

(3)\enskip Let $X,\,Y : S \to \mathcal{C}$ be $S$-families of objects and let 
$\varphi : X \to Y$ be an $S$-family of morphisms. If $\varphi_s \in \Hom(X_s,Y_s)$ 
is an isomorphism for each $s \in S$ then for each finite $S$-typing $\gamma$
the morphism $\ass{\gamma}{\varphi}$ is an isomorphism and
$(\ass{\gamma}{\varphi})^{-1} = \ass{\gamma}{(\varphi^{-1})}$, where $\varphi^{-1}$ is the
$S$-family of morphisms with $(\varphi^{-1})_s = (\varphi_s)^{-1}$ for each $s \in S$.
\end{lemma}

\proof (1)\enskip This follows since $\id\fcomp\gamma : \ass{\gamma}{X} \to \ass{\gamma}{X}$
is a family of identity morphisms and hence
$\ass{\gamma}{\id} = \utimes{(\id\fcomp\gamma)} = \id_{\ass{\gamma}{X}}$.

(2)\enskip It is easy to see that
$(\psi\fcomp\varphi)\comp\gamma = (\psi\comp\gamma)\fcomp(\varphi\comp\gamma)$
holds (exactly as in Lemma~\ref{lemma_sets_2}) and therefore
\[\ass{\gamma}{(\psi\fcomp\varphi)} = \utimes{((\psi\fcomp\varphi)\comp\gamma)} 
= \utimes{((\psi\comp\gamma)\fcomp(\varphi\comp\gamma))} 
= \utimes{(\psi\comp\gamma)} \comp \utimes{(\varphi\comp\gamma)} 
= \ass{\gamma}{\psi}\comp\ass{\gamma}{\varphi}\;.\]

(3)\enskip By (1) and (2)
$(\ass{\gamma}{\varphi}) \fcomp \ass{\gamma}{(\varphi^{-1})}$ and
$\ass{\gamma}{(\varphi^{-1})} \comp (\ass{\gamma}{\varphi})$ are identity morphisms, hence
the morphism $\ass{\gamma}{\varphi}$ is an isomorphism and
$(\ass{\gamma}{\varphi})^{-1} = \ass{\gamma}{(\varphi^{-1})}$. \eop

Now a pair $(X,p)$ will be called a 
\definition{$\Lambda(\mathsf{C})$-algebra}\index{algebra (in a category)} if 
$X : B \to \mathcal{C}$ is a $B$-family of objects and $p : K \to \mathcal{M}$ a 
$K$-family of morphisms such that $p_k \in \Hom(\ass{\sdom{k}}{X},X_{\scod{k}})$
for each $k \in K$. (The mapping $\utimes : \ffam{\mathcal{M}} \to \mathcal{M}$ is not 
involved in this definition; it is only needed in the definition of a homomorphism between
$\Lambda(\mathsf{C})$-algebras to be given below.)

In particular, a $\Lambda(\mathsf{BSets})$-algebra is thus nothing but a bottomed
$\Lambda$-algebra, a $\Lambda(\mathsf{Posets})$-algebra is an ordered
$\Lambda$-algebra and a $\Lambda(\mathsf{CPosets})$-algebra is just a continuous
$\Lambda$-algebra.

In order to deal with homomorphisms between $\Lambda(\mathsf{C})$-algebras it is necessary 
to introduce a restricted class of morphisms. This is because in all the cases being 
considered, with the exception of the original set-up in Chapter~\ref{uni_alg}, the 
mappings occurring in homomorphisms are required to be bottomed. Thus in what follows 
let $\mathsf{C}'$ be a subcategory of $\mathsf{C}$ having the same
objects as $\mathsf{C}$. For each $X,\,Y \in \mathcal{C}$ the set of morphisms from
$X$ to $Y$ in $\mathsf{C}'$ will be denoted by $\Hom'(X,Y)$, and the class of all such
morphisms will be denoted by $\mathcal{M}'$. In all the categories listed above, with 
the exception of $\mathsf{Sets}$, the morphisms in $\mathcal{M}'$ will be the bottomed 
morphisms in $\mathcal{M}$; in $\mathsf{Sets}$ there is no restriction, and so here 
$\mathcal{M}' = \mathcal{M}$, i.e., $\mathsf{C}' = \mathsf{C}$. 

Let $(X,p)$ and $(Y,q)$ be $\Lambda(\mathsf{C})$-algebras. A $B$-family 
$\pi : X \to Y$ of morphisms from $\mathcal{M}'$ will be called a
\definition{$\mathsf{C}'$-homomorphism}\index{homomorphism} from $(X,p)$ to $(Y,q)$ if
\[q_k \comp \ass{\sdom{k}}{\pi} = \pi_{\scod{k}} \comp p_k\]
for each $k \in K$. This fact will also be expressed by saying that
$\pi : (X,p) \to (Y,q)$ is a $\mathsf{C}'$-homomorphism.

\begin{proposition}\label{prop_cat_algs_1}
(1)\enskip The $B$-family of identity morphism $\id : X \to X$ defines a 
$\mathsf{C}'$-homomorphism from a $\Lambda(\mathsf{C})$-algebra $(X,p)$ to itself. 

(2)\enskip If $\pi : (X,p) \to (Y,q)$ and 
$\varrho : (Y,q) \to (Z,r)$ are $\mathsf{C}'$-homomorphisms then the composition 
$\varrho\fcomp\pi$ is a $\mathsf{C}'$-homomorphism from $(X,p)$ to $(Z,r)$.

(3)\enskip If $\pi : (X,p) \to (Y,q)$ is a $\mathsf{C}'$-homomorphism and 
$\pi_b \in \Hom'(X_b,Y_b)$ is an isomorphism for each $b \in B$ then the $B$-family
$\pi^{-1}$ of inverse morphisms is a $\mathsf{C}'$-homomorphism from $(Y,q)$ to $(X,p)$.
\end{proposition}

\proof (1)\enskip This follows immediately from Lemma~\ref{lemma_cat_algs_1}~(1),
since for each $k \in K$
\[ p_k \comp \ass{\sdom{k}}{\id} = p_k \comp \id_{\ass{\sdom{k}}{X}} 
= p_k = \id_{\scod{k}} \comp p_k\;.\]

(2)\enskip Let $k \in K$; then by Lemma~\ref{lemma_cat_algs_1}~(2)
\[ r_k \comp \ass{\sdom{k}}{(\varrho\fcomp\pi)} 
  = r_k \comp\ass{\sdom{k}}{\varrho}\comp \ass{\sdom{k}}{\pi}
  = \varrho_{\scod{k}} \comp q_k \comp\ass{\sdom{k}}{\pi} 
   = \varrho_{\scod{k}} \comp\pi_{\scod{k}} \comp p_k 
  = (\varrho\fcomp\pi)_{\scod{k}} \comp p_k\]
and hence $\varrho\fcomp\pi$ is a $\mathsf{C}'$-homomorphism from $(X,p)$ to $(Z,r)$. 

(3)\enskip Let $k \in K$. Then 
$q_k\comp\ass{\sdom{k}}{\pi} = \pi_{\scod{k}}\comp p_k$, and therefore by 
Lemma~\ref{lemma_cat_algs_1}~(3) it follows that 
$p_k\comp\ass{\sdom{k}}{(\pi^{-1})} = p_k\comp (\ass{\sdom{k}}{\pi})^{-1} 
 = \pi_{\scod{k}}^{-1}\comp q_k$, 
which implies that $\pi^{-1}$ is a $\mathcal{C}'$-homomorphism from $(Y,q)$ to $(X,p)$. \eop

Denote by $\Lambda(\mathsf{C})$ the class of all $\Lambda(\mathsf{C})$-algebras.
Proposition~\ref{prop_cat_algs_1} implies there is a category having
$\Lambda(\mathsf{C})$ as objects and $\mathsf{C}'$-homomorphisms between 
$\Lambda(\mathsf{C})$-algebras as morphisms; this category will be denoted by 
$\Lambda(\mathsf{C},\mathsf{C}')$. In the situations we have been considering there is 
then given some full subcategory $\mathsf{H}$ of $\Lambda(\mathsf{C},\mathsf{C}')$: 
For example, the main results in 
Chapters \ref{bot_algs} and \ref{ord_cont_algs} involve a $\natural$-stable simple
head type $\diamond$, and in 
this case the objects of $\mathsf{H}$ consist of the $\diamond$-algebras.

Now for an open signature the interest is mainly in algebras bound to a given $A$-family of 
objects: If $V : A \to \mathcal{C}$ is such an $A$-family then a 
$\Lambda(\mathsf{C})$-algebra $(X,p)$ is said to be 
\definition{bound to $V$}\index{algebra bound to a family} if 
$\rest{X}{A} = V$. If $(X,p)$ and $(Y,q)$ are $\Lambda(\mathsf{C})$-algebras bound to $V$ 
then a $\mathsf{C}'$-homomorphism $\pi : (X,p) \to (Y,q)$ is said to
\definition{fix}\index{homomorphism!fixing a family} $V$ if $\pi_a = \id_{X_a}$ for 
each $a \in A$. Clearly the family of identity morphisms fixes $V$ and the composition of 
two $\mathsf{C}'$-homomorphisms fixing $V$ is again a $\mathsf{C}'$-homomorphism fixing 
$V$. For each $A$-family $V : A \to \mathcal{C}$ there is thus a category whose objects 
are $\Lambda(\mathsf{C})$-algebras bound to $V$ and whose morphisms are 
$\mathsf{C}'$-homomorphisms fixing $V$. In this category there is again the full 
subcategory whose objects are, in addition, objects of $\mathsf{H}$. Note that the 
constructions in the previous chapters produce initial (and even intrinsically free) objects 
in these subcategories, 
and in each case it was possible to characterize what it means to be initial.

If $\mathsf{H}$ is a full subcategory of $\Lambda(\mathsf{C},\mathsf{C}')$ then
a $\Lambda(\mathsf{C})$-algebra is called an \definition{$\mathsf{H}$-algebra} 
if it is an object of $\mathsf{H}$. Moreover, $\mathsf{H}$ is said to
\definition{possess intrinsically free objects} if for each
$A$-family $V : A \to \mathcal{C}$ there exists an intrinsically free 
$\mathsf{H}$-algebra $(X,p)$ bound to $V$, this meaning that $(X,p)$ is an  $\mathsf{H}$-algebra 
bound to $V$ such that for each $\mathsf{H}$-algebra $(Y,q)$ and each
$A$-family $\tau : V \to \rest{Y}{A}$ of $\mathsf{C}'$-morphisms
there exists a unique $\mathsf{C}'$-homomorphism $\pi : (X,p) \to (Y,q)$ such that
$\rest{\pi}{A} = \tau$.

We have the following three examples of a full subcategory $\mathsf{H}$ which possesses intrinsically free
objects, in each case with $\diamond$ a $\natural$-stable simple head type and with
$\mathsf{C}'$ defined by requiring the morphisms to be bottomed: 

\newpage
\begin{evlist}{20pt}{0.8ex}
\item[---] 
$\mathsf{C} = \mathsf{BSets}$ with the objects of $\mathsf{H}$ the bottomed 
$\diamond$-algebras. 
\item[---] 
$\mathsf{C} = \mathsf{Posets}$ with the objects of $\mathsf{H}$ the ordered 
$\diamond$-algebras. 
\item[---] 
$\mathsf{C} = \mathsf{APosets}$ with the objects of $\mathsf{H}$ the continuous 
$\diamond$-algebras. 
\end{evlist}
(These statements follow from 
Propositions~\ref{prop_head_type_6}, \ref{prop_ord_algs_2} and \ref{prop_cont_algs_2}.)

Recall that if $F$ is a non-empty set and $\Lambda_i = (B_i,K_i,\Theta_i)$ is a signature 
for each $i \in F$  then $\Lambda$ is said to be the disjoint union of the signatures 
$\Lambda_i$, $i \in F$, if the following conditions hold:
\begin{itemize}
\item[(1)] $B_i \cap B_i = \varnothing$ and $K_i \cap K_i = \varnothing$
whenever $i \ne j$.
\item[(2)] $B = \bigcup_{i\in F} B_i$ and $K = \bigcup_{i\in F} K_i$.
\item[(3)] $\Theta_i(k) = \Theta(k)$ for all $k \in K_i$, $i \in F$.
\end{itemize}
In particular, if $A_i$ is the parameter set of $\Lambda_i$ for each $i \in F$ then 
$A = \bigcup_{i\in F} A_i$ is the parameter set of $\Lambda$.

In what follows let $\Lambda$ be the disjoint union of the signatures $\Lambda_i$, 
$i \in F$. If $(X^i,p^i)$ is a $\Lambda_i(\mathsf{C})$-algebra for each $i \in F$ then a 
$\Lambda(\mathsf{C})$-algebra $(X,p)$ 
can be defined by putting $X_b = X^i_b$ for each 
$b \in B_i$ and $p_k = p^i_k$ for each $k \in K_i$. $(X,p)$ will be
called the \definition{sum}\index{sum of algebras}\index{algebras!sum of}
of the $\Lambda_i(\mathsf{C})$-algebras $(X^i,p^i)$, $i \in F$, and will
be denoted by $\oplus_{i\in F} (X^i,p^i)$. The converse also holds:

\begin{lemma}\label{lemma_cat_algs_2}
Let $(X,p)$ be a $\Lambda(\mathsf{C})$-algebra and for each $i \in F$ put
$X^i = \rest{X}{B_i}$ and $p^i = \rest{p}{K_i}$. Then $(X^i,p^i)$ is a 
$\Lambda_i(\mathsf{C})$-algebra and $(X,p) = \oplus_{i\in F} (X^i,p^i)$.
\end{lemma}

\proof Straightforward. \eop

\begin{lemma}\label{lemma_cat_algs_3}
For each $i \in F$ let $(X^i,p^i)$, $(Y^i,q^i)$ be $\Lambda_i(\mathsf{C})$-algebras
and let $\pi^i : X^i \to Y^i$ be a family of morphisms. 
Let $\pi$ be the $B$-family of morphisms with $\pi_b = \pi^i_b$ for each $b \in B_i$.
Then $\pi^i : (X^i,p^i) \to (Y^i,q^i)$ is a $\mathsf{C}'$-homomorphism for each $i \in F$
if and only if $\pi : \oplus_{i\in F} (X^i,p^i) \to \oplus_{i\in F}(Y^i,q^i)$ is a 
$\mathsf{C}'$-homomorphism.
\end{lemma}

\proof Straightforward. \eop

For each $i \in F$ let $\mathsf{H}_i$ be a full subcategory of $\Lambda_i(\mathsf{C},\mathsf{C}')$.
Then by Lemma~\ref{lemma_cat_algs_2} a full subcategory $\mathsf{H}$ of 
$\Lambda(\mathsf{C},\mathsf{C}')$ can 
be defined by stipulating $(X,p) = \oplus_{i\in F} (X^i,p^i)$ to be an object of $\mathsf{H}$ if 
and only if $(X^i,p^i)$ is an $\mathsf{H}_i$-algebra for each $i \in F$.
$\mathsf{H}$ will be called the \definition{sum} of the subcategories
$\mathsf{H}_i$, $i\in F$, and will be denoted by $\oplus_{i\in F} \mathsf{H}_i$.

\begin{proposition}\label{prop_cat_algs_2}
Let $V : A \to \mathcal{C}$ be a family of objects and for each
$i \in F$ let $(X^i,p^i)$ be an intrinsically free $\mathsf{H}_i$-algebra
bound to $V^i$, with $V^i : A_i \to \mathcal{C}$ the restriction of $V$ to $A_i$. Then
$\oplus_{i\in F} (X^i,p^i)$ is an intrinsically free $\mathsf{H}$-algebra bound to $V$.
\end{proposition}

\proof This follows immediately from Lemma~\ref{lemma_cat_algs_3}. \eop

If $\mathsf{H}_i$ possesses intrinsically free objects for 
each $i \in F$ then Proposition~\ref{prop_cat_algs_2} implies in particular that 
$\oplus_{i\in F} \mathsf{H}_i$ also possesses intrinsically free  objects.

Consider for a moment a head type $(H,\diamond)$ for the signature $\Lambda$. By 
Lemma~\ref{lemma_algs_homs_1} $(H,\diamond) = \oplus_{i\in F} (H^i,\diamond^i)$, where
$H^i = \rest{H}{B_i}$ and $\diamond^i = \rest{\diamond}{K_i}$ for each $i \in F$, and 
clearly $(H^i,\diamond^i)$ is a head type for the signature $\Lambda_i$.
Moreover, $\diamond$ is a $\natural$-stable simple head type if and only if each
$\diamond^i$ is a $\natural$-stable simple head type.

\begin{lemma}\label{lemma_cat_algs_4}
Let $\diamond$ be a $\natural$-stable simple head type.

(1)\enskip
If $\mathsf{C} = \mathsf{BSets}$ and the objects of $\mathsf{H}_i$ are the bottomed 
$\diamond^i$-algebras for each $i \in F$ then the objects of
$\oplus_{i\in F} \mathsf{H}_i$ are the bottomed $\diamond$-algebras.

(2)\enskip If $\mathsf{C} = \mathsf{Posets}$ and the objects of $\mathsf{H}_i$ are the 
ordered $\diamond^i$-algebras for each $i \in F$ then the objects of
$\oplus_{i\in F} \mathsf{H}_i$ are the ordered $\diamond$-algebras. 

(3)\enskip
If $\mathsf{C} = \mathsf{APosets}$ and the objects of $\mathsf{H}_i$ are the continuous 
$\diamond^i$-algebras for each $i \in F$ then the objects of
$\oplus_{i\in F} \mathsf{H}_i$ are the  continuous $\diamond$-algebras. 
\end{lemma}

\proof These also follow immediately from Lemma~\ref{lemma_cat_algs_3}. \eop

For the rest of the chapter we work with the set-up introduced above.
This means that the following are given:
\begin{evlist}{20pt}{0.8ex}
\item[---] 
A $\otimes$-category $\mathsf{C}$ with objects $\mathcal{C}$ and morphisms $\mathcal{M}$.
\item[---] 
A subcategory $\mathsf{C}'$ of $\mathsf{C}$ having the same objects as $\mathsf{C}$.
\end{evlist}
There is then the category $\Lambda(\mathsf{C},\mathsf{C}')$ whose objects 
$\Lambda(\mathsf{C})$ are the $\Lambda(\mathsf{C})$-algebras and whose morphisms are
the $\mathsf{C}'$-homomorphisms between $\Lambda(\mathsf{C})$-algebras.

\newpage
\bigskip
\startsection{The polymorphic signature}
\label{poly_sig}

In what follows let $T$ be a fixed set (which could well be empty), whose elements will be referred 
to as \definition{type variables}. The aim of this section is to define a new signature
$\xx{\Lambda} = (\xx{B},\xx{K},\xx{\Theta})$ to be called 
the\index{polymorphic signature}\index{signature!polymorphic} 
\definition{polymorphic signature associated with $\Lambda$ and the type variables $T$}. 
The elements in $\xx{B}$ are called \definition{polymorphic types}, and those in
$\xx{K}$ \definition{polymorphic operator names}.

To get an idea of what this is all about consider the second representation of the 
signature $\Lambda$ in Example~\thesection.1. This corresponds more-or-less to that used 
in most functional programming languages in that each  type in $B \setminus A$ is 
augmented with the parameters (i.e., the elements of $A$) on which it depends. 
A bit more precisely, this means that each type in $B \setminus A$ must contain all of the 
parameters occurring on the `right-hand side' of the `equation' specifying it.

\bigskip
\fbox{\begin{exframe}
\textit{Example \thesection.1} Recall the signature $\Lambda = (B,K,\Theta)$ 
with parameter set $A = \{\mathtt{x},\mathtt{y},\mathtt{z}\}$ 
from Example~\ref{algs_homs}.3. This signature can be represented 
(using the conventions introduced in Example~\ref{algs_homs}.2) in the form
\begin{eelist}{100pt}
\item $\mathtt{bool\ ::=\ True\ |\ False}$
\item $\mathtt{atom\ ::=\ Atom}$
\item $\mathtt{int\ ::=\ } \cdots\, \mathtt{\ -2\ |\ -1\ |\ 0\ |\ 1\ |\ 2\ }\, \cdots$
\item $\mathtt{pair\ ::=\ Pair\ x\ y}$
\item $\mathtt{list\ ::=\ Nil\ |\ Cons\ z\ list}$
\item $\mathtt{lp\ ::=\ L\ list\ |\ P\ pair}$
\end{eelist}
\exparskip
An augmented form of this representation, in which the types $\mathtt{pair}$, 
$\mathtt{list}$ and $\mathtt{lp}$ are provided with parameters from the set $A$, is the 
following:
\begin{eelist}{100pt}
\item $\mathtt{bool\ ::=\ True\ |\ False}$
\item $\mathtt{atom\ ::=\ Atom}$
\item $\mathtt{int\ ::=\ } \cdots \mathtt{\ -2\ |\ -1\ |\ 0\ |\ 1\ |\ 2\ }\, \cdots$
\item $\mathtt{pair\ x\ y\ ::=\ Pair\ x\ y}$
\item $\mathtt{list\ z\ ::=\ Nil\ |\ Cons\ z\ (list\ z)}$
\item $\mathtt{lp\ x\ y\ z\ ::=\ L\ (list\ z)\ |\ P\ (pair\ x\ y)}$
\end{eelist}
\end{exframe}}

\bigskip

The process of adding parameters to the types in the general signature $\Lambda$ leads to 
the notion of a support. Denote by $\mathcal{P}_o(A)$ the set of all finite subsets of
$A$. A mapping $\lfloor \cdot \rfloor : B \to \mathcal{P}_o(A)$ 
is called a \definition{support}\index{support for a signature} for $\Lambda$ if
$\lfloor a \rfloor = \{a\}$ for each $a \in A$ and
$\lfloor \adom{k}{\eta} \rfloor \subset \lfloor \scod{k} \rfloor$ 
for all $\eta \in \domsdom{k}$ for each $k \in K$.
In general it is possible that no support exists. However, this problem does not
arise if $A$ is finite, since then
the mapping $\lfloor\cdot\rfloor^* : B \to \mathcal{P}_o(A)$ 
with $\lfloor b \rfloor^* = A$ for each $b \in B\setminus A$ 
and $\lfloor a \rfloor^* = \{a\}$ for each $a \in A$ is a support (and thus the maximal
support).

\begin{lemma}\label{lemma_poly_sig_1}
If there exists a support for $\Lambda$ then there exists a minimal support, i.e., a 
support $\lfloor\cdot\rfloor^o$ 
such that $\lfloor b \rfloor^o \subset \lfloor b \rfloor$ for all $b \in B$ for each 
support $\lfloor\,\cdot\,\rfloor$. 
\end{lemma}

\proof Define $\lfloor\cdot\rfloor^o : B \to \mathcal{P}_o(A)$ by for each $b \in B$ letting
\[\lfloor b\rfloor^o = \{ a \in A : a \in \lfloor b\rfloor \ \textrm{for every support}
                           \ \lfloor\cdot\rfloor \}\;.\]
Then $\lfloor\cdot\rfloor^o$ is clearly a support, and thus the minimal support for 
$\Lambda$. \eop

\bigskip
\fbox{\begin{exframe}
\textit{Example \thesection.2} 
For the signature $\Lambda$ in Example \thesection.1 the minimal support is the mapping 
$\lfloor \cdot \rfloor : B \to \mathcal{P}_o(A)$ defined by
\begin{eelist}{20pt}
\item $\lfloor\mathtt{pair}\rfloor = \{\mathtt{x},\mathtt{y}\}$,\enskip 
      $\lfloor\mathtt{list}\rfloor = \{\mathtt{z}\}$,\enskip
      $\lfloor\mathtt{lp}\rfloor = \{\mathtt{x},\mathtt{y},\mathtt{z}\}$,
\item $\lfloor\mathtt{x}\rfloor = \{\mathtt{x}\}$,\enskip
      $\lfloor\mathtt{y}\rfloor = \{\mathtt{y}\}$,\enskip
      $\lfloor\mathtt{z}\rfloor = \{\mathtt{z}\}$,
\item $\lfloor b\rfloor = \varnothing\,$ for all 
      $b \in B \setminus \{\mathtt{pair},\mathtt{list},\mathtt{lp},
      \mathtt{x},\mathtt{y},\mathtt{z}\}$. 
\end{eelist}
\end{exframe}}

\bigskip
In all practical applications $A$ will be finite and the minimal support is almost always 
the natural choice to make. However, it is convenient to work first with a general support.
Let $\lfloor\cdot\rfloor : B \to \mathcal{P}_o(A)$ be a support for $\Lambda$, which is 
considered to be fixed for the rest of the chapter. It should be emphasised that the 
constructions that follow all depend on the choice of  $\lfloor\cdot\rfloor$.

We start by giving a somewhat informal description of the polymorphic signature
$\xx{\Lambda} = (\xx{B},\xx{K},\xx{\Theta})$.
For this is it is convenient to write each element $b \in B\setminus A$ 
in the form $b\ a_1 \cdots a_n$, where $\lvector{a}{n}$ is some fixed enumeration of
the elements in the set $\lfloor b\rfloor$. With this device the sets $\xx{B}$ and 
$\xx{K}$ can be thought of as being defined by the following rules:
\begin{evlist}{25pt}{0.8ex}
\item[(1)] Each type variable $t \in T$ is an element of $\xx{B}$.
\item[(2)] If $b\ a_1\cdots a_n \in B\setminus A$ and $\lvector{\xx{b}}{n} \in \xx{B}$
then $b\ \xx{b}_1 \cdots \xx{b}_n$ is an element of $\xx{B}$.
\item[(3)] Each element of $\xx{B}$ can be obtained in a unique way using (1) and (2).
\item[(4)] If $k \in K$ and with $\scod{k} = b\ a_1\cdots a_m$ and 
$\lvector{\xx{b}}{m} \in \xx{B}$ then
$k\ \xx{b}_1 \cdots \xx{b}_m$ is an element of $\xx{K}$.
\item[(5)] Each element of $\xx{K}$ can be obtained in a unique way using (4).
\end{evlist}
The mapping $\xx{\Theta} : \xx{K} \to \ftyped{\xx{B}} \times \xx{B}$ is defined as
follows: Let $\xx{k} \in \xx{K}$; then $\xx{k}$ has a unique representation
of the form $k\ \xx{b}_1 \cdots \xx{b}_m$ with $k \in K$ and
$\lvector{\xx{b}}{m} \in \xx{B}$, where $m$ is the number of elements in the set
$\lfloor \scod{k}\rfloor$. Now define
$\scod{\xx{k}} = \scod{\xx{\Theta}}(\xx{k})$ to be the element
$\scod{k}\ \xx{b}_1 \cdots \xx{b}_m$ of $\xx{B}$ and define 
$\sdom{\xx{k}} = \sdom{\xx{\Theta}}(\xx{k})$ to be the typing with
$\domsdom{\xx{k}} = \domsdom{k}$ and with
$\sdom{\xx{k}} : \domsdom{k} \to \xx{B}$ given by
$\adom{\xx{k}}{\eta} = \adom{k}{\eta}\ \xx{b}_{\eta,1}\cdots\xx{b}_{\eta,n_\eta}$ for 
each $\eta \in \domsdom{k}$, where
$\xx{b}_{\eta,1}\cdots\xx{b}_{\eta,n_\eta}$ are those elements from
$\xx{b}_1 \cdots \xx{b}_m$ which are indexed by the elements of the subset 
$\lfloor \adom{k}{\eta}\rfloor$ of $\lfloor \scod{k}\rfloor$ (and note that a support
is defined so that $\lfloor \adom{k}{\eta}\rfloor \subset \lfloor \scod{k}\rfloor$ for 
each $\eta \in \domsdom{k}$). Example~\thesection.3 on the following page illustrates
theses definitions.

The definition of the signature $\xx{\Lambda}$ must now be made precise, and we first 
define $\xx{B}$. Recall that if $X$ and $Y$ are sets then $\total{X}{Y}$ denotes the set 
of all mappings from $X$ to $Y$, and if $\alpha : X \to Y$ is a mapping and $J$ a
set then $\alpha^J$ denotes the mapping from $X^J$ to $Y^J$ defined by
$\alpha^J(v) = \alpha \comp v$ for all $v \in X^J$.

Suppose $Y$ is a set which is going to be a 
candidate for the set of polymorphic types. Then for each $b \in B\setminus A$
a mapping $q_b : Y^{\lfloor b\rfloor} \to Y$ must also be given such that
$q_b(u)$ is the new type $b\ u(a_1) \cdots u(a_n)$ for each 
$u \in Y^{\lfloor b\rfloor}$. Such a pair $(Y,q)$ consisting of a set $Y$ and a 
corresponding $B{\setminus}A$-family of mappings $q$ will be called  
\definition{a set of polymorphic types}. 
If $(Y,q)$ and $(Y',q')$ are sets of polymorphic types then a mapping $\pi : Y \to Y'$ is 
called a \definition{homomorphism}\index{homomorphism} if 
$\pi\comp q_b = q'_b\comp\pi^{\lfloor b\rfloor}$
for each $b \in B\setminus A$. 
The identity mapping is clearly a homomorphism and the 
composition of two homomorphisms is again a homomorphism. A homomorphism $\pi : Y \to Y'$ 
is an \definition{isomorphism}\index{isomorphism} if there exists
a homomorphism $\pi' : Y' \to Y$ such that $\pi' \comp\pi = \id_Y$
and $\pi \comp\pi' = \id_{Y'}$. It is easy to see that a homomorphism is an 
isomorphism if and only if it is a bijection.

A set of polymorphic types $(Y,q)$ is said to be \definition{$T$-free}
\index{free set of polymorphic types}\index{set of polymorphic types!free}if $T \subset Y$ 
and for each set of polymorphic types $(Y',q')$ and for each mapping $w : T \to Y'$ there 
exists a unique homomorphism $\pi : (Y,q) \to (Y',q')$ such that $\pi(t) = w(t)$ for 
each $t \in T$.

\begin{lemma}\label{lemma_poly_sig_2}
There exists a $T$-free set of polymorphic types $(Y,q)$. Moreover, if $(Y',q')$ is a 
further $T$-free set of polymorphic types then there is a unique isomorphism
$\pi : (Y,q) \to (Y',q')$ such that $\pi(t) = t$ for each $t \in T$.
\end{lemma}

\proof Let $\Xi$ be the single-sorted signature $(B\setminus A,\lfloor\cdot\rfloor)$; then
there is obviously a one--to--one correspondence between sets of polymorphic types 
and $\Xi$-algebras. Moreover, a homomorphism between sets of polymorphic types
is the same as a $\Xi$-homomorphism. From these facts it follows that a 
$T$-free $\Xi$-algebra (with the set $T$ considered as a $\Oneptset$-family of sets
in the obvious way) corresponds exactly to a $T$-free set of polymorphic types, and 
therefore 
Proposition~\ref{prop_free_algs_2} implies there exists a $T$-free set of polymorphic 
types. The uniqueness holds trivially from the definition 
of being $T$-free. \eop

\bigskip
\fbox{\begin{exframe}
\textit{Example \thesection.3} Let $\Lambda = (B,K,\Theta)$ be the signature 
in Example~\thesection.1 considered the with the minimal support.
As in Example~\thesection.1 the types $\mathtt{pair}$, $\mathtt{list}$ and $\mathtt{lp}$
will be written as $\mathtt{pair\ x\ y}$, $\mathtt{list\ z}$ and $\mathtt{lp\ x\ y\ z}$.
The set of polymorphic types $\xx{B}$ is obtained by the following rules:
\begin{eelist}{30pt}
\item[(1)] Each type variable $t \in T$ is an element of $\xx{B}$,
\item[(2)] $\mathtt{bool}$, $\mathtt{atom}$ and $\mathtt{int}$ are elements of $\xx{B}$,
\item[(3)] $\mathtt{pair}\ \xx{b}_1\ \xx{b}_2$ is an element of $\xx{B}$ for all
           $\xx{b}_1,\, \xx{b}_2 \in \xx{B}$.
\item[(4)] $\mathtt{list}\ \xx{b}$ is an element of $\xx{B}$ for all $\xx{b} \in \xx{B}$.
\item[(5)] $\mathtt{lp}\ \xx{b}_1\ \xx{b}_2\ \xx{b}_3$ is an element of $\xx{B}$ for all
           $\xx{b}_1,\,\xx{b}_2,\,\xx{b}_3 \in \xx{B}$.
\item[(6)] Each element of $\xx{B}$ can be obtained in a unique way using (1), (2), (3), 
           (4) and (5).
\end{eelist}
\exsparskip
Let $\mathtt{v}$ be a type variable. Examples of elements of $\xx{B}$ are thus:
\begin{eelist}{20pt}
\item $\mathtt{list}\ \mathtt{v}$,\enskip 
      $\mathtt{pair}\ \mathtt{atom}\ \mathtt{int}$,\enskip 
      $\mathtt{list}\ (\mathtt{pair}\ \mathtt{v}\ \mathtt{bool})$,
\item $\mathtt{pair}\ \mathtt{atom}\ (\mathtt{list}\ \mathtt{int})$,\enskip
      $\mathtt{pair}\ (\mathtt{list}\ \mathtt{bool})\ (\mathtt{list}\ \mathtt{v})$.
\end{eelist}
\exsparskip
The set of polymorphic operator names $\xx{K}$ is obtained by the rules:
\begin{eelist}{30pt}
\item[(1)] $\mathtt{True}$, $\mathtt{False}$ and $\mathtt{Atom}$ are elements of $\xx{K}$.
\item[(2)] $\underline{n}$ is an element of $\xx{K}$ for each $n \in \Int$. 
\item[(3)] $\mathtt{Pair}\ \xx{b}_1\ \xx{b}_2$ is an element of $\xx{K}$ for all 
           $\xx{b}_1,\,\xx{b}_2 \in \xx{K}$.
\item[(4)] $\mathtt{Nil}\ \xx{b}$ and  $\mathtt{Cons}\ \xx{b}$ are elements of $\xx{K}$ 
           for all $\xx{b} \in \xx{K}$.
\item[(5)] $\mathtt{L}\ \xx{b}_1\ \xx{b}_2\ \xx{b}_3,\,
           \mathtt{P}\ \xx{b}_1\ \xx{b}_2\ \xx{b}_3 \in \xx{K}$ for all
           $\xx{b}_1,\,\xx{b}_2,\,\xx{b}_3 \in \xx{B}$.
\item[(6)] Each element of $\xx{K}$ can be obtained in a unique way using 
           (1), (2), (3), (4) and (5).
\end{eelist}
\exsparskip
Examples of elements of $\xx{K}$ are:
\begin{eelist}{20pt}
\item $\mathtt{Nil}\ (\mathtt{pair}\ \mathtt{v}\ \mathtt{v})$,\enskip 
      $\mathtt{Cons}\ (\mathtt{pair}\ \mathtt{v}\ \mathtt{bool})$,\enskip
      $\mathtt{Pair}\ (\mathtt{list}\ \mathtt{v})\ \mathtt{int}$, 
\item $\mathtt{Pair}\ \mathtt{atom}\ (\mathtt{list}\ \mathtt{int})$,\enskip 
      $\mathtt{L}\ (\mathtt{list}\ \mathtt{bool})\ (\mathtt{list}\ \mathtt{v})
     \ (\mathtt{pair}\ \mathtt{v}\ \mathtt{v})$.
\end{eelist}
\exsparskip
In the signature $\xx{\Lambda} = (\xx{B},\xx{K},\xx{\Theta})$ 
the types of the elements of $\xx{K}$ are: 
\begin{eelist}{20pt}
\item $\mathtt{True}$ and $\mathtt{False}$ have type $\onept \to \mathtt{bool}$,
      $\mathtt{Atom}$ has type $\onept \to \mathtt{atom}$,
\item $\underline{n}$ has type $\onept \to \mathtt{int}$ for each $n \in \Int$.
\item $\mathtt{Pair}\ \xx{b}_1\ \xx{b}_2$ has type 
      $\xx{b}_1\ \xx{b}_2 \to \mathtt{pair}\ \xx{b}_1\ \xx{b}_2$,
\item $\mathtt{Nil}\ \xx{b}$ has type $\onept \to \mathtt{list}\ \xx{b}$, 
\item $\mathtt{Cons}\ \xx{b}$ has type
      $\xx{b}\ (\mathtt{list}\ \xx{b}) \to \mathtt{list}\ \xx{b}$,
\item $\mathtt{L}\ \xx{b}_1\ \xx{b}_2\ \xx{b}_3$ has type
      $\mathtt{list}\ \xx{b}_3 \to \mathtt{lp}\ \xx{b}_1\ \xx{b}_2\ \xx{b}_3$,
\item $\mathtt{P}\ \xx{b}_1\ \xx{b}_2\ \xx{b}_3$ has type
      $\mathtt{pair}\ \xx{b}_1\ \xx{b}_2 \to \mathtt{lp}\ \xx{b}_1\ \xx{b}_2\ \xx{b}_3]$. 
\end{eelist}
\exsparskip
The element 
$\mathtt{P}\ \mathtt{atom}\ (\mathtt{list}\ \mathtt{v})
                   \ (\mathtt{pair}\ \mathtt{v}\ \mathtt{int})$ 
thus has the type
\begin{eelist}{20pt}
\item $\mathtt{pair}\ \mathtt{atom}\ (\mathtt{list}\ \mathtt{v})
 \to \mathtt{lp}\ \mathtt{atom}\ (\mathtt{list}\ \mathtt{v})
                   \ (\mathtt{pair}\ \mathtt{v}\ \mathtt{int})$.
\end{eelist}
\end{exframe}}

Let us fix a $T$-free set of polymorphic types $(\xx{B},\#)$. Because of the 
uniqueness (up to isomorphism) in Lemma~\ref{lemma_poly_sig_2} this will be 
referred to as \definition{the set of polymorphic types}.
The next lemma shows that $(\xx{B},\#)$ does have the properties corresponding to
the rules making up the informal definition.

\begin{lemma}\label{lemma_poly_sig_3}
The set of polymorphic types $(\xx{B},\#)$ has the following properties:

(1)\enskip $T \subset \xx{B}$ and
$\Im(\#_b) \subset \xx{B}\setminus T$ for each  $b \in B\setminus A$.

(2)\enskip
For each $\xx{b} \in \xx{B}\setminus T$ there exists a unique 
$b \in B\setminus A$ and a unique element $u \in \xx{B}^{\lfloor b\rfloor}$ such that 
$\xx{b} = \#_b(u)$. 

(3)\enskip The only subset of $\xx{B}$ containing $T$ and invariant under the mappings 
$\#_b$, $b \in B\setminus A$, is $\xx{B}$ itself. 
\end{lemma}

\proof This follows from Proposition~\ref{prop_free_algs_3} and the equivalence
of $T$-free sets of polymorphic types and $T$-free $\Xi$-algebras. \eop

The set $\xx{B}$ will certainly be non-empty if $T \ne \varnothing$ (since 
$T \subset \xx{B}$). Moreover, it will also be non-empty provided there is at least one 
type in $B \setminus A$ which does not depend on any parameters:

\begin{lemma}\label{lemma_poly_sig_4}
If there exists $b \in B \setminus A$ with $\lfloor b\rfloor = \varnothing$ 
then $\xx{B} \ne \varnothing$. 
\end{lemma}

\proof Here $\xx{B}^{\lfloor b \rfloor} = \xx{B}^\varnothing = \Oneptset$
and hence $\#_b(\onept) \in \xx{B}$. \eop

The hypothesis in Lemma~\ref{lemma_poly_sig_4} will be satisfied by any signature
containing the type $\mathtt{bool}$ (for any natural choice of the support).
Let us now assume that $\xx{B} \ne \varnothing$.

We next turn to the definition of $\xx{K}$. Suppose $Z$ is a set which is going to be a 
candidate for the set of polymorphic operator names. For each $k \in K$
a mapping $r_k : \xx{B}^{\lfloor \scod{k}\rfloor} \to Z$ must then also be given such that
$r_k(u)$ is the new operator name $k\ u(a_1) \cdots u(a_n)$ for each 
$u \in \xx{B}^{\lfloor b\rfloor}$. Such a pair $(Z,r)$ consisting of a set $Z$ and a 
corresponding $K$-family of mappings $r$ will be called  
\definition{a set of polymorphic operator names}. 
If $(Z,r)$ and $(Z',r')$ are sets of polymorphic operator names then a mapping
$\varrho : Z \to Z'$ is called a \definition{homomorphism}\index{homomorphism} if
$ \varrho\comp r_k = r'_k$ for all $k \in K$. It is clear that the identity mapping
$\id_Z : Z \to Z$ is a homomorphism and that the composition of two homomorphisms is again
a homomorphism.
A homomorphism $\varrho : Z \to Z'$ is an \definition{isomorphism}\index{isomorphism} 
if there exists
a homomorphism $\varrho' : Z' \to Z$ such that $\varrho' \comp\varrho = \id_Z$
and $\varrho \comp\varrho' = \id_{Z'}$. It is easy to see that a homomorphism is an 
isomorphism if and only if it is a bijection.

A set of polymorphic operator names $(Z,r)$ is said to be \definition{initial}
if for each set of polymorphic operator names $(Z',r')$ there is a unique
homomorphism from $(Z,r)$ to $(Z',r')$. 

\begin{lemma}\label{lemma_poly_sig_5}
There exists an initial set of polymorphic operator names.
\end{lemma}

\proof An initial set of polymorphic operator names $(Z,r)$ can be constructed explicitly 
by just taking $Z$ to be the disjoint union of the sets $\xx{B}^{\lfloor \scod{k}\rfloor}$,
$k \in K$, and letting $r_k : \xx{B}^{\lfloor \scod{k}\rfloor} \to Z$ be the inclusion 
mapping for each $k \in K$. It is then easy to check that $(Z,r)$ is initial.
(The reader is left to show that, alternatively, an initial set of polymorphic operator
names can be obtained via a $\xx{B}$-free algebra for an appropriate signature 
involving two types.) \eop

Proposition~\ref{prop_sets_2} implies that any two initial sets of polymorphic operator
names are isomorphic.

Now fix an initial set of polymorphic operator names $(\xx{K},\#')$. Because of the 
uniqueness (up to isomorphism) this will be referred to as 
\definition{the set of polymorphic operator names}.
The next lemma shows that $(\xx{K},\#')$ does have the properties corresponding to
the rules making up the informal definition.

\begin{lemma}\label{lemma_poly_sig_6}
For each $\xx{k} \in \xx{K}$ there exists a unique $k \in K$ and a unique 
element $u \in \xx{B}^{\lfloor \scod{k}\rfloor}$ such that $\xx{k} = \#'_k(u)$. 
\end{lemma}

\proof 
This clearly holds for the particular initial set of polymorphic operator
names given in the proof of Lemma~\ref{lemma_poly_sig_5}, and so it must also hold for 
any initial set of polymorphic operator names. \eop

It remains to define the mapping $\xx{\Theta} : \xx{K} \to \ftyped{\xx{B}} \times \xx{B}$.
Before doing this, however, it is convenient
to introduce two new families of mappings which reorganise the information contained in the families
$\#$ and $\#'$. If $C \subset D \subset A$ and $f : \xx{B}^C \to Z$ is a mapping then,
in order to increase the legibility, we just write $f(u)$ instead of $f(\rest{u}{C})$
for each $u \in \xx{B}^D$. For each $u \in \xx{B}^A$ let $\ipar_u : B \setminus A
\to \xx{B}$ be the mapping given by $\ipar_u(b) = \#_b(u)$ for each $b \in B \setminus A$ and 
$\ipar'_u : K \to \xx{K}$ the mapping given by
$\ipar'_u(k) = \#'_k(u)$ for each $k \in K$ (and of course
$\#_b(u)$ here really means $\#_b$ applied to the restriction of $u$ to 
$\lfloor b \rfloor$ and $\#'_k(u)$ means $\#'_k$ applied to the restriction of $u$ to 
$\lfloor \scod{k} \rfloor$).

\begin{lemma}\label{lemma_poly_sig_7}
(1)\enskip 
If $b,\, b' \in B \setminus A$ and $u,\,v \in \xx{B}^A$ then
$\ipar_u(b) = \ipar_v(b')$ if and only if
$b = b'$ and $\rest{u}{\lfloor b \rfloor} = \rest{v}{\lfloor b \rfloor}$; moreover, 
$\bigcup_{u \in \xx{B}^A} \Im(\ipar_u) = \xx{B} \setminus T$.

(2)\enskip 
If $k,\, k' \in K$ and $u,\,v \in \xx{B}^A$ then
$\ipar'_u(k) = \ipar'_v(k')$ if and only if
$k = k'$ and $\rest{u}{\lfloor \scod{k} \rfloor} = \rest{v}{\lfloor \scod{k} \rfloor}$;
moreover, $\bigcup_{u \in \xx{B}^A} \Im(\ipar'_u) = \xx{K}$.
\end{lemma}

\proof This follows immediately from Lemmas \ref{lemma_poly_sig_3} and \ref{lemma_poly_sig_6}. \eop

In particular, Lemma~\ref{lemma_poly_sig_7} implies that
$\ipar_u : B \setminus A \to \xx{B}$ and $\ipar'_u : K \to \xx{K}$ are both injective mappings for 
each $u \in \xx{B}^A$.

For what follows it is necessary to extend the mapping $\ipar_u : B\setminus A \to \xx{B}$ to a mapping
$\ipar_u : B \to \xx{B}$ by putting $\ipar_u(a) = u(a)$ for each $a \in A$ (but note that the statements
in Lemma~\ref{lemma_poly_sig_7}~(1) no longer hold for this extended mapping).

Let $\mathcal{B}$ and $\mathcal{C}$ be classes; a mapping $\varphi : \mathcal{B}^A \to \mathcal{C}^B$ is 
said to be \definition{$B$-compatible} with $\lfloor \cdot \rfloor$ if
$\varphi(u)(b) = \varphi(v)(b)$ whenever $u(a) = v(a)$ for all $a \in \lfloor b \rfloor$.
Similarly, a mapping $\psi : \mathcal{B}^A \to \mathcal{C}^K$ is 
said to be \definition{$K$-compatible} with $\lfloor \cdot \rfloor$ if
$\psi(u)(k) = \psi(v)(k)$ whenever $u(a) = v(a)$ for all $a \in \lfloor \scod{k} \rfloor$.
The following result really just restates the fact that 
$(\xx{B},\#)$ is an initial set of polymorphic types and $(\xx{K},\#')$ is an initial set of
polymorphic operator names, but using the mappings $\ipar_u,\,\ipar'_u$, $u \in \xx{B}^A$, instead of the 
families $\#$ and $\#'$.

\begin{proposition}\label{prop_poly_sig_1}
Let $\mathcal{C}$ be a class.

(1)\enskip Let $\varphi : \mathcal{C}^A \to \mathcal{C}^B$ be a mapping which is $B$-compatible with 
$\lfloor \cdot \rfloor$ 
and such that $\varphi(U)(a) = U(a)$ for all $U \in \mathcal{C}^A$, $a \in A$; let 
$\beta : T \to \mathcal{C}$.
Then there exists a unique mapping $\zeta : \xx{B} \to \mathcal{C}$ such that
$\rest{\zeta}{T} = \beta$ and
$\zeta \comp \ipar_u  = \varphi(\zeta \comp u)$ for all $u \in \xx{B}^A$.

(2)\enskip Let $\psi : \xx{B}^A \to \mathcal{C}^K$ be a mapping which is $K$-compatible with 
$\lfloor \cdot \rfloor$; then 
there exists a unique mapping $\zeta : \xx{K} \to \mathcal{C}$ such that
$\zeta \comp \ipar'_u  = \psi(u)$ for all $u \in \xx{B}^A$.
\end{proposition}

\proof 
(1)\enskip For each $b \in B \setminus A$ define 
$\alpha_b : \mathcal{C}^{\lfloor b \rfloor} \to \mathcal{C}$
by putting $\alpha_b(U) = \varphi(U')(b)$, where $U' : A \to \mathcal{C}$ is any extension of the 
mapping $U : \lfloor b \rfloor \to \mathcal{C}$ to $A$ (and  $\alpha_b(U)$ does not depend on which
extension is used because of the $B$-compatibility with $\lfloor \cdot \rfloor$).
Then, except that $\mathcal{C}$ is a class, $(\mathcal{C},\alpha)$ would be a $\Xi$-algebra, with 
$\Xi$ the single-sorted signature $(B\setminus A,\lfloor\cdot\rfloor)$, and
as noted in the proof of Lemma~\ref{lemma_poly_sig_2}, $(\xx{B},\#)$
is a $T$-free $\Xi$-algebra. Therefore by Proposition~\ref{prop_free_algs_5}
there exists a unique $\Xi$-homomorphism $\zeta : (\xx{B},\#) \to (\mathcal{C},\alpha)$
with $\rest{\zeta}{T} = \beta$. But $\zeta$ being a $\Xi$-homomorphism means that 
$\zeta\comp\#_b = \alpha_b\comp \zeta^{\lfloor b \rfloor}$
for all $b \in B\setminus A$. Now let $b \in B \setminus A$, $u \in \xx{B}^A$ and put
$u' = \rest{u}{\lfloor b \rfloor}$. Then $\zeta \comp u = \zeta^A(u)$ is an extension of 
$\zeta^{\lfloor b \rfloor}(u')$ to $A$ and so
$\zeta(\ipar_u(b)) = \zeta(\#_b(u')) = \alpha_b(\zeta^{\lfloor b\rfloor}(u')) = \varphi(\zeta\comp u)(b)$.
On the other hand, if $a \in A$ then
$\zeta(\ipar_u(a)) = \zeta(u(a)) = (\zeta\comp u)(a)
= \varphi(\zeta \comp u)(a)$ and hence
$\zeta \comp \ipar_u  = \varphi(\zeta \comp u)$ for all $u \in \xx{B}^A$.
Conversely, if $\xi : \xx{B} \to \mathcal{C}$ is any mapping with
$\xi \comp \ipar_u  = \varphi(\xi \comp u)$ for all $u \in \xx{B}^A$ then it immediately
follows that $\xi\comp\#_b = \alpha_b\comp \xi^{\lfloor b \rfloor}$
for all $b \in B\setminus A$. This implies that $\zeta$ is the unique mapping with
$\rest{\zeta}{T} = \beta$ and
$\zeta \comp \ipar_u  = \varphi(\zeta \comp u)$ for all $u \in \xx{B}^A$.

(2)\enskip For each $k \in K$ define $\alpha_k : \xx{B}^{\lfloor \scod{k} \rfloor} \to \mathcal{C}$
by putting $\alpha_k(u) = \psi(u')(k)$, where $u' : A \to \xx{B}$ is any extension of the 
mapping $u : \lfloor \scod{k} \rfloor \to \xx{B}$ to $A$ (and $\alpha_k(u)$ does not depend on which
extension is used because of the $K$-compatibility with $\lfloor \cdot \rfloor$).
Then $(\mathcal{C},\alpha)$ would  
be a set of polymorphic operator names if $\mathcal{C}$ were a set (and not a class), 
and so there exists a unique mapping $\zeta : \xx{K} \to \mathcal{C}$ 
such that $\zeta \comp \#'_k = \alpha_k$ for each $k \in K$. 
($\mathcal{C}$ being a class is clearly not a problem here, since it is not a problem if the initial 
set of polymorphic operator names given in the proof of Lemma~\ref{lemma_poly_sig_5} is used.)
It follows that $\zeta(\ipar'_u(k)) = \zeta(\#'_k(u')) = \alpha_k(u') = \psi(u)(k)$
for all $k \in K$, $u \in \xx{B}^A$, where $u' = \rest{u}{\lfloor \scod{k}\rfloor}$, and hence 
$\zeta \comp \ipar'_u  = \psi(u)$ for all $u \in \xx{B}^A$. The uniqueness follows from the fact that
$\bigcup_{u \in \xx{B}^A} \Im(\ipar'_u) = \xx{K}$. \eop

For $u \in \xx{B}^A$ let $\ipar^*_u : \ftyped{B} \to \ftyped{\xx{B}}$ be the induced mapping given by
$\ipar^*_u(\gamma) = \ipar_u \comp \gamma$ for each $\gamma \in \ftyped{B}$.

\begin{proposition}\label{prop_poly_sig_2}
There exists a unique mapping $\sdom{\xx{\Theta}} : \xx{K} \to \ftyped{\xx{B}}$ and a 
unique mapping $\scod{\xx{\Theta}} : \xx{K} \to \xx{B}$ such that
$\sdom{\xx{\Theta}} \comp \ipar'_u = \ipar^*_u \comp \sdom{\Theta}$ and 
$\scod{\xx{\Theta}} \comp \ipar'_u = \ipar_u \comp \scod{\Theta}$ for all $u \in \xx{B}^A$.
\end{proposition}

\proof 
These are both special cases of Proposition~\ref{prop_poly_sig_1}~(2).

To obtain the mapping $\sdom{\xx{\Theta}} : \xx{K} \to \ftyped{\xx{B}}$ 
consider the mapping $\psi : \xx{B}^A \to \ftyped{\xx{B}}^K$ given by
$\psi(u)(k) = \ipar_u \comp \sdom{k}$ for all $u \in \xx{B}^A$, $k \in K$. 
Let $k \in K$ and $u,\,v \in \xx{B}^A$ with $u(a) = v(a)$ for all 
$a \in \lfloor \scod{k} \rfloor$. If $\eta \in \domsdom{k}$ with
$\adom{k}{\eta} \in B\setminus A$ then by Lemma~\ref{lemma_poly_sig_7}~(1)
$\psi(u)(k)(\eta) = \ipar_u(\adom{k}{\eta}) = \ipar_v(\adom{k}{\eta}) = \psi(v)(k)(\eta)$, since
$\lfloor \adom{k}{\eta} \rfloor \subset \lfloor \scod{k} \rfloor$.  
On the other hand, if $\adom{k}{\eta} \in A$ then 
\[\psi(u)(k)(\eta) = \ipar_u(\adom{k}{\eta}) = u(\adom{k}{\eta}) = v(\adom{k}{\eta}) 
= \ipar_v(\adom{k}{\eta}) = \psi(v)(k)(\eta)\;,\]
since 
$\adom{k}{\eta} \in \{\adom{k}{\eta}\} =
\lfloor \adom{k}{\eta} \rfloor \subset \lfloor \scod{k} \rfloor$.  
This shows that $\psi$ is $K$-compatible with $\lfloor \cdot \rfloor$ and thus by 
Proposition~\ref{prop_poly_sig_1}~(2) there exists a unique mapping
$\sdom{\xx{\Theta}} : \xx{K} \to \ftyped{\xx{B}}$ such that
$\sdom{\xx{\Theta}} \comp \ipar'_u = \psi(u) = \ipar^*_u \comp \sdom{\Theta}$ 
for all $u \in \xx{B}^A$. 

To obtain the mapping $\scod{\xx{\Theta}} : \xx{K} \to \xx{B}$ define $\psi : \xx{B}^A \to \xx{B}^K$ 
by $\psi(u)(k) = \ipar_u(\scod{k})$ for all $u \in \xx{B}^A$, $k \in K$. By Lemma~\ref{lemma_poly_sig_7}~(1)
$\psi$ is $K$-compatible with $\lfloor \cdot \rfloor$ and thus by 
Proposition~\ref{prop_poly_sig_1}~(2) there exists a unique mapping
$\scod{\xx{\Theta}} : \xx{K} \to \xx{B}$ such that
$\scod{\xx{\Theta}} \comp \ipar'_u = \psi(u) = \ipar_u \comp \scod{\Theta}$ 
for all $u \in \xx{B}^A$. \eop

Now put $\xx{\Theta} = (\sdom{\xx{\Theta}},\scod{\xx{\Theta}})$. 
This completes the formal definition of the polymorphic signature
$\xx{\Lambda} = (\xx{B},\xx{K},\xx{\Theta})$. Note that if
$u \in \xx{B}^A$, $k \in K$ and $\xx{k} = \ipar'_u(k)$ then by 
definition $\scod{\xx{k}} = \ipar_u(\scod{k})$
and $\sdom{\xx{k}} = \ipar^*_u(\sdom{k}) = \ipar_u \comp \sdom{k}$.

\begin{proposition}\label{prop_poly_sig_3}
(1)\enskip 
Let $b \in B\setminus A$ and $u \in \xx{B}^A$ and put $\xx{b} = \ipar_u(b)$.
Then $\xx{K}_{\xx{b}} = \{ \ipar'_u(k) : k \in K_b \}$ 
(and so in particular $\xx{K}_{\xx{b}} \ne \varnothing$, 
since $K_b \ne \varnothing$).

(2)\enskip $\Im(\scod{\xx{\Theta}}) = \xx{B} \setminus T$, and thus $T$ is the parameter 
set of $\xx{\Lambda}$. 
\end{proposition}

\proof 
(1)\enskip 
Consider $\xx{k} \in \xx{K}_{\xx{b}}$, so
$\scod{\xx{k}} = \xx{b}$. By Lemma~\ref{lemma_poly_sig_7}~(2) there exist $k \in K$ and
$v \in \xx{B}^A$ with $\xx{k} = \ipar'_v(k)$ and then
$\ipar_v(\scod{k}) = \scod{\xx{k}} = \xx{b} = \ipar_u(b)$. Hence by Lemma~\ref{lemma_poly_sig_7}~(1)
$\scod{k} = b$ and $\rest{v}{\lfloor \scod{k}\rfloor} = \rest{u}{\lfloor \scod{k}\rfloor}$.
Therefore $k \in K_b$ and by
Lemma~\ref{lemma_poly_sig_7}~(2) $\xx{k} = \ipar'_v(k)= \ipar'_u(k)$.
Conversely, let $k \in K_b$ and put $\xx{k} = \ipar'_u(k)$; then $\scod{k} = b$ and thus
$\scod{\xx{k}} = \ipar_u(\scod{k}) = \ipar_u(b) = \xx{b}$, i.e., $\xx{k} \in \xx{K}_{\xx{b}}$.
This shows that $\xx{K}_{\xx{b}} = \{ \ipar'_u(k) : k \in K_b \}$. 

(2)\enskip Let $\xx{k} \in \xx{K}$; then by
Lemma~\ref{lemma_poly_sig_7}~(2) there exists $k \in K$
and $u \in \xx{B}^A$ such that $\xx{k} = \ipar'_u(k)$ and thus
\[ \scod{\xx{\Theta}}(\xx{k}) = \scod{\xx{\Theta}}(\ipar'_u(k))
 = \ipar_u(\scod{\Theta}(k)) \in \Im(\ipar_u) \subset \xx{B} \setminus T\;.\]
Conversely, if $\xx{b} \in \xx{B}\setminus T$ then by (1)
$\xx{K}_{\xx{b}} \ne \varnothing$, and hence $\xx{b} \in \Im(\scod{\xx{\Theta}})$. \eop

\begin{proposition}\label{prop_poly_sig_4}
Let $\xx{X} : \xx{B} \to \mathcal{C}$ be a $\xx{B}$-family of objects and
$\xx{p} : \xx{K} \to \mathcal{M}$ a $\xx{K}$-family of morphisms. Then the pair $(\xx{X},\xx{p})$ is a
$\xx{\Lambda}(\mathsf{C})$-algebra if and only if 
$(\xx{X} \comp \ipar_u,\xx{p}\comp \ipar'_u)$ is a $\Lambda(\mathsf{C})$-algebra for each $u \in \xx{B}^A$.
\end{proposition}

\proof 
Let $k \in K$, $u \in \xx{B}^A$ and put $\xx{k} = \ipar'_u(k)$.
Then $\scod{\xx{k}} = \ipar_u(\scod{k})$, $\sdom{\xx{k}} = \ipar_u \comp \sdom{k}$
and thus $\xx{X}_{\scod{\xx{k}}} = (\xx{X} \comp \ipar_u)_{\scod{k}}$ and
$\ass{\sdom{\xx{k}}}{\xx{X}} = \utimes (\xx{X} \comp \sdom{\xx{k}}) =
\utimes (\xx{X} \comp \ipar_u \comp \sdom{k}) = \ass{\sdom{k}}{(\xx{X}\comp \ipar_u)}$.
Hence $\xx{p}_{\xx{k}} \in \Hom( \ass{\sdom{\xx{k}}}{\xx{X}},\xx{X}_{\scod{\xx{k}}})$ if and only if 
$(\xx{p} \comp \ipar'_u)_{k} 
\in \Hom( \ass{\sdom{k}}{(\xx{X} \comp \ipar_u)},(\xx{X}\comp \ipar_u)_{\scod{k}})$ and this,
together with the fact that
$\bigcup_{u \in \xx{B}^A} \Im(\ipar'_u) = \xx{K}$, gives the result.
\eop

Let $(\xx{X},\xx{p})$ be a $\xx{\Lambda}(\mathsf{C})$-algebra. Note then that for each $u \in \xx{B}^A$
the $\Lambda(\mathsf{C})$-algebra $(\xx{X} \comp \ipar_u,\xx{p}\comp \ipar'_u)$ is bound to the $A$-family 
of objects $\xx{X} \comp u$.

\begin{proposition}\label{prop_poly_sig_5}
Let $(\xx{X},\xx{p})$ and $(\xx{Y},\xx{q})$ be $\xx{\Lambda}(\mathsf{C})$-algebras and let 
$\xx{\pi} : \xx{X} \to \xx{Y}$ be a $\xx{B}$-family of morphisms. Then $\xx{\pi}$ is a 
$\mathsf{C}'$-homomorphism 
from $(\xx{X},\xx{p})$ to $(\xx{Y},\xx{q})$ if and only if $\xx{\pi} \comp \ipar_u$ is a 
$\mathsf{C}'$-homomorphism 
from $(\xx{X}\comp \ipar_u,\xx{p} \comp \ipar'_u)$ to $(\xx{Y} \comp \ipar_u,\xx{q} \comp \ipar'_u)$ 
for each $u \in \xx{B}^A$.
\end{proposition}

\proof 
This is the same as the proof of Proposition~\ref{prop_poly_sig_4}. Let $k \in K$, $u \in \xx{B}^A$ and 
put $\xx{k} = \ipar'_u(k)$.
Then $\scod{\xx{k}} = \ipar_u(\scod{k})$, $\sdom{\xx{k}} = \ipar_u \comp \sdom{k}$
and thus $\xx{\pi}_{\scod{\xx{k}}} = (\xx{\pi} \comp \ipar_u)_{\scod{k}}$ and
$\ass{\sdom{\xx{k}}}{\xx{\pi}} = \utimes (\xx{\pi} \comp \sdom{\xx{k}}) =
\utimes (\xx{\pi} \comp \ipar_u \comp \sdom{k}) = \ass{\sdom{k}}{(\xx{\pi} \comp \ipar_u)}$.
Hence $\xx{q}_{\xx{k}} \comp \ass{\sdom{\xx{k}}}{\xx{\pi}} = \xx{\pi}_{\scod{\xx{k}}} \comp \xx{p}_{\xx{k}}$
if and only if
$(\xx{q} \comp \ipar'_u)_k \comp \ass{\sdom{k}}{(\xx{\pi} \comp \ipar_u)} 
= (\xx{\pi} \comp \ipar_u)_{\scod{k}} \comp (\xx{p} \comp \ipar'_u)_{k}$
and this, together with the fact that
$\bigcup_{u \in \xx{B}^A} \Im(\ipar'_u) = \xx{K}$, gives the result.
\eop

We end the section by explicitly constructing an initial term 
$\xx{\Lambda}$-algebra. By Proposition~\ref{prop_term_algs_5} this just amounts to 
choosing a suitable 
locally injective term algebra specifier $\Gamma' : \xx{K} \to \Omega$ and a suitable 
family of enumerations $i'_{\xx{K}}$. In fact there is a 
surprisingly simple way of making these choices, which will now be explained.

This starts out as if the interest were in the signature $\Lambda$ (and not 
$\xx{\Lambda}$): For each $k \in K$ choose a bijective mapping $i_k$ from $[n_k]$ to the 
set $\domsdom{k}$, where of course $n_k$ is the cardinality of $\domsdom{k}$.
Also fix a mapping $\Gamma : K \to \Omega$ (which could also be thought of as a term 
algebra specifier).

By Lemma~\ref{lemma_poly_sig_7}~(2) there is a unique mapping $\omega : \xx{K} \to K$ 
with $\omega \comp \ipar'_u = \id_K$ for each $u \in \xx{B}^A$. Moreover, $\omega$ is surjective
and for each $\xx{k} \in \xx{K}$ there exists $u \in \xx{B}^A$ such that 
$\xx{k} = \ipar'_u(\omega(\xx{k}))$.
In particular $\domsdom{\xx{k}} = \domsdom{\omega(\xx{k})}$ for each $\xx{k} \in \xx{K}$ 
(since if $u \in \xx{B}^A$ is such that $\xx{k} = \ipar'_u(\omega(\xx{k}))$ then
$\sdom{\xx{k}} = \ipar'_u \comp \sdom{\omega(\xx{k})}$).
Let $\xx{\Gamma} : \xx{K} \to \Omega$ be given by $\xx{\Gamma} = \Gamma \comp \omega$.
Then, since $\domsdom{\xx{k}} = \domsdom{\omega(\xx{k})}$,
a family of enumerations $i'_{\xx{K}}$ can be defined by letting
$i'_{\xx{k}} = i_{\omega(\xx{k})}$ for each $\xx{k} \in \xx{K}$.

Now let $(E_{\xx{B}},\blob_{\xx{K}})$ be the term $\xx{\Lambda}$-algebra specified by 
$\xx{\Gamma}$ and the family $i'_{\xx{K}}$. This means that $E_{\xx{b}} \subset \Omega^*$ for 
each $\xx{b} \in \xx{B}$ and 
the mapping $\blob_{\xx{k}} : \ass{\sdom{\xx{k}}}{E} \to E_{\scod{\xx{k}}}$ is given for 
each $\xx{k} \in \xx{K}$ by
\[ \blob_{\xx{k}}(u) = \Gamma(\omega(\xx{k})) \ u(i'_{\xx{k}}(1))
             \ \cdots\ u(i'_{\xx{k}}(n_{\xx{k}}))\]
for each $u \in \ass{\sdom{\xx{k}}}{E}$. Moreover, the family $E_{\xx{B}}$ can be 
regarded as being defined by the following rules:
\begin{itemize}
\item[(1)] If $\xx{k} \in \xx{K}$ with $\sdom{\xx{k}} = \onept$ then the 
           list consisting of the single component $\Gamma(\omega(\xx{k}))$ is an 
           element of $E_{\scod{\xx{k}}}$.
\item[(2)] If $\xx{k} \in \xx{K}$ with $\sdom{\xx{k}} \ne \onept$ 
           and $e_j \in E_{\xx{b}_j}$ for $\oneto{j} {n_{\xx{k}}}$, where 
           $\xx{b}_j = \adom{\xx{k}}{i'_{\xx{k}}(j)}$, then 
           $\Gamma(\omega(\xx{k}))\ e_1\ \cdots \ e_{n_{\xx{k}}}$ is an element of 
           $E_{\scod{\xx{k}}}$.
\item[(3)] The only elements in $E_{\xx{b}}$ are those which can be obtained using (1) 
           and (2). 
\end{itemize}

\begin{lemma}\label{lemma_poly_sig_8}
If $\Gamma$ is locally injective then so is $\xx{\Gamma}$. 
\end{lemma}

\proof Since $\xx{K}_t = \varnothing$ for each $t \in T$ it is enough to show that
the restriction of $\xx{\Gamma}$ to $\xx{K}_{\xx{b}}$ is injective for each
$\xx{b} \in \xx{B} \setminus T$. Thus consider $\xx{b} \in \xx{B} \setminus T$ 
and $\xx{k},\, \xx{k}' \in \xx{K}_{\xx{b}}$ with $\xx{\Gamma}(\xx{k}) = \xx{\Gamma}(\xx{k}')$.
Now by Lemma~\ref{lemma_poly_sig_7}~(1) there exists $b \in B\setminus A$ 
and $u \in \xx{B}^A$ such that $\xx{b} = \ipar_u(b)$, and then by
Proposition~\ref{prop_poly_sig_3}~(1) there exist $k,\,k' \in K_b$ with $\xx{k} = \ipar'_u(k)$ and 
$\xx{k}' = \ipar'_u(k')$. Therefore
\[ \Gamma(k) = \Gamma(\omega(\ipar'_u(k))) = \xx{\Gamma}(\xx{k}) = \xx{\Gamma}(\xx{k}')
 = \Gamma(\omega(\ipar'_u(k'))) = \Gamma(k')\]
and so $k = k'$, since the restriction of $\Gamma$ to $K_b$ is injective. But this implies 
that $\xx{k} = \xx{k}'$, i.e., the restriction of $\xx{\Gamma}$ to $\xx{K}_{\xx{b}}$ is 
injective. \eop

\begin{proposition}\label{prop_poly_sig_33}
If $\Gamma$ is locally injective then $(E_{\xx{B}},\blob_{\xx{K}})$ is an initial 
$\xx{\Lambda}$-algebra. 
\end{proposition}

\proof This follows immediately from Proposition~\ref{prop_term_algs_5} and 
Lemma~\ref{lemma_poly_sig_8}. \eop

\newpage

\startsection{Parameterised algebras}
\label{param_algs}

A mapping $\Phi : \mathcal{C}^A \to \Lambda(\mathsf{C})$ is said to be a
\definition{parameterised $\Lambda(\mathsf{C})$-algebra}
\index{parameterised algebra}\index{algebra!parameterised}if $\Phi_V$ is a 
$\Lambda(\mathsf{C})$-algebra bound to $V$ to each family $V : A \to \mathcal{C}$
(and note that we write $\Phi_V$ here instead of $\Phi(V)$).

Of course, if $\Lambda$ is closed (i.e., if $A = \varnothing$) then
$\mathcal{C}^A = \Oneptset$ and hence a parameterised $\Lambda(\mathsf{C})$-algebra 
is then nothing but a single $\Lambda(\mathsf{C})$-algebra. The constructions in this
section are thus really meant for open signatures.

Let $\mathsf{H}$ be a full subcategory of $\Lambda(\mathsf{C},\mathsf{C}')$. A parameterised 
$\Lambda(\mathsf{C})$-algebra $\Phi$ 
is said to be  a \definition{parameterised $\mathsf{H}$-algebra} if
$\Phi_V$ is an $\mathsf{H}$-algebra for each $V \in \mathcal{C}^A$ and
then a parameterised $\mathsf{H}$-algebra is said to be
\definition{intrinsically free}\index{intrinsically free algebra}\index{algebra!intrinsically free}
 if $\Phi_V$ is an intrinsically
free $\mathsf{H}$-algebra for each $V \in \mathcal{C}^A$. Thus by
definition there exists such a parameterised $\mathsf{H}$-algebra if and only 
if $\mathsf{H}$ possesses intrinsically free objects,
and in all the cases we are interested in this requirement is met. However, this is 
not the end of the story since there is an additional condition which needs to be 
satisfied and which has to do with how $\Phi_V$ behaves as a 
function of $V$. How this condition arises and the problem of constructing parameterised 
algebras satisfying it is the topic to be discussed in the present section.

Let $\Phi$ be a parameterised $\Lambda(\mathsf{C})$-algebra with $\Phi = (X,p)$; thus
$X : \mathcal{C}^A \to \mathcal{C}^B$ and 
$p : \mathcal{C}^A \to \mathcal{M}^K$, and we write
$(\param{X}{V},\param{p}{V})$ instead of $(X_V,p_V)$ for the
$\Lambda(\mathsf{C})$-algebra $\Phi_V$.
Let $\lfloor \cdot\rfloor$ be a support for $\Lambda$. A 
parameterised $\Lambda(\mathsf{C})$-algebra $\Phi : \mathcal{C}^A \to \Lambda(\mathsf{C})$  
is said to be \definition{compatible}\index{algebra!compatible with support}
with $\lfloor\,\cdot\,\rfloor$ if:
\begin{evlist}{25pt}{0.8ex}
\item[(1)] $\parax{X}{V}{b} = \parax{X}{W}{b}$
whenever $V_a = W_a$ for all $a \in \lfloor b \rfloor$.
\item[(2)] $\parax{p}{V}{k} = \parax{p}{W}{k}$
whenever $V_a = W_a$ for all $a \in \lfloor \scod{k} \rfloor$.
\end{evlist}
Note that if a parameterised algebra is compatible with the minimal support then it is 
compatible with any support. 

\bigskip
\fbox{\begin{exframe}
\textit{Example \thesection.1} Let $\Lambda = (B,K,\Theta)$ be the signature 
in Example~\thesection.1 considered the with the minimal support $\lfloor\cdot\rfloor$.
\exparskip
For each $V : A \to \mathsf{BSets}$ denote by $\Phi_V$
the bottomed $\Lambda$-algebra defined in Example~\ref{bot_algs_homs}.3 (and denoted there 
by $(Y,q)$); thus $\Phi_V$ is bound to $V$. This gives a parameterised 
$\Lambda(\mathsf{BSets})$-algebra $\Phi$ which is compatible with $\lfloor\cdot\rfloor$.
\end{exframe}}

\bigskip

In what follows we will need parameterised algebras which are compatible with a given 
support $\lfloor\cdot\rfloor$, and in general it is not even clear whether such an algebra
exists. In particular, there seems little hope of obtaining them using something like 
Proposition~\ref{prop_head_type_1}, Proposition~\ref{prop_ord_algs_2} or 
Proposition~\ref{prop_cont_algs_2}: The fact that a $\Lambda(\mathsf{C})$-algebra
$\Phi_V$ bound to $V$ is unique up to an appropriate isomorphism
for each $V$ does not say anything about how $\Phi_V$ behaves as a function of $V$.
This suggests that, in order to obtain parameterised algebras compatible with 
$\lfloor\cdot \rfloor$, it is necessary to give an absolutely explicit method of 
constructing certain classes of algebras. No doubt this is possible, but it would be 
rather involved. We thus prefer to take an easy way out,
which at first sight seems somewhat restrictive, but turns out not to be so.

The starting point for our approach is contained in the following trivial fact:

\begin{lemma}\label{lemma_param_algs_2}
If $A$ is finite then any parameterised $\Lambda(\mathsf{C})$-algebra 
$\Phi : \mathcal{C}^A \to \Lambda(\mathsf{C})$ 
is compatible with the maximal support $\lfloor\cdot\rfloor^* : B \to \mathcal{P}_o(A)$ 
(given by $\lfloor b \rfloor^* = A$ for each $b \in B\setminus A$ 
and $\lfloor a \rfloor^* = \{a\}$ for each $a \in A$). 
\end{lemma}

\proof Let $\Phi = (X,p)$. Let $b \in B$ and $V,\,W \in \mathcal{C}^A$ with $V_a = W_a$ for all 
$a \in \lfloor b \rfloor^*$. If $b \in B\setminus A$ then $\lfloor b \rfloor^* = A$, thus 
$V = W$ and hence $\parax{X}{V}{b} = \parax{X}{W}{b}$. 
On the other hand, 
if $b \in A$ then $\lfloor b \rfloor^* = \{b\}$, thus $V_b = W_b$
and again $\parax{X}{V}{b} = V_b = W_b = \parax{X}{W}{b}$. 
This shows that (1) holds, and (2) follows in the same way, since if
$k \in K$ then $\scod{k} \in B\setminus A$ and so
$\lfloor\scod{k}\rfloor^* = A$. \eop

The signature $\Lambda$ will be called 
\definition{simple}\index{simple signature}\index{signature!simple} 
if $A$ is finite and there
is only one support for $\Lambda$ (i.e., if the maximal support is also the minimal 
support). If $\Lambda$ is simple then Lemma~\ref{lemma_param_algs_2} says that
any parameterised $\Lambda(\mathsf{C})$-algebra is compatible with the minimal support 
(and thus is compatible with any support). Of course, $\Lambda$ will rarely be simple
and so this fact cannot be applied directly. However, it can be applied indirectly 
whenever $\Lambda$ is the disjoint union of simple signatures.

In what follows suppose $\Lambda$ is the disjoint union of the signatures 
$\Lambda_i = (B_i,K_i,\Theta_i)$, $i \in F$; thus $A = \bigcup_{i \in F} A_i$, with
$A_i$ the parameter set of $\Lambda_i$. For each $i \in F$ let 
$\Phi^i : \mathcal{C}^{A_i} \to \Lambda_i(\mathsf{C})$ be a 
parameterised $\Lambda_i(\mathsf{C})$-algebra, and 
define $\Phi : \mathcal{C}^A \to \Lambda(\mathsf{C})$ by putting
$\Phi_V = \oplus_{i \in F} \Phi^i_{V_i}$ for all $V \in \mathcal{C}^A$, where
$V_i$ is the restriction of $V$ to $A_i$. Thus $\Phi$ is a 
parameterised $\Lambda(\mathsf{C})$-algebra, which will be called the 
\definition{sum}\index{sum of parameterised algebras}\index{parameterised algebras!sum of}
of the parameterised algebras $\Phi^i$, $i \in F$, and will be denoted by
$\oplus_{i \in F} \Phi^i$.

\begin{proposition}\label{prop_param_algs_1}
If each of the signatures $\Lambda_i$, $i \in F$, is simple then any sum 
$\oplus_{i\in F}\Phi^i$ of parameterised $\Lambda_i(\mathsf{C})$-algebras
is compatible with the minimal support for $\Lambda$.
\end{proposition}

\proof 
For each $i\in F$ let $\Phi^i = (X^i,p^i)$, let $\lfloor\cdot\rfloor_i$ be the unique 
support for $\Lambda_i$ and define a mapping $\lfloor \cdot\rfloor : B \to \mathcal{P}_o(A)$
by letting $\lfloor b \rfloor = \lfloor b\rfloor_i$ for each $b \in B_i$; then
$\lfloor\cdot\rfloor$ is a support for $\Lambda$ and it is clear that 
$\lfloor\cdot\rfloor$  is in fact the minimal support.
Consider $b \in B$ and $V,\, W \in \mathcal{C}^A$ with  
$V_a = W_a$ for all $a \in \lfloor b \rfloor$; let $i$ be such that
$b \in B_i$. Then $\lfloor b \rfloor_i = \lfloor b \rfloor$ and $(V_i)_a = V_a$,
$(W_i)_a = W_a$ for all $a \in \lfloor b \rfloor_i$ (since
$\lfloor b \rfloor_i \subset A_i$). Thus
$\parax{X}{V}{b} = \parax{(X^i)}{{V_i}}{b} = \parax{(X^i)}{{W_i}}{b}
= \parax{X}{W}{b}$,
because by Lemma~\ref{lemma_param_algs_2} $\Phi^i$ is compatible with 
$\lfloor\cdot\rfloor_i$. In the same way 
$\parax{p}{V}{k} = \parax{p}{W}{k}$ whenever 
$V_a = W_a$ for all $a \in \lfloor \scod{k} \rfloor$, and therefore the 
sum  $\oplus_{i\in F}\Phi^i$ is compatible with $\lfloor\cdot\rfloor$. \eop

\begin{proposition}\label{prop_param_algs_2}
Let $\mathsf{H}_i$ be a full subcategory of $\Lambda_i(\mathsf{C},\mathsf{C}')$
which possesses intrinsically free objects for each $i \in F$ and let
$\mathsf{H} = \oplus_{i\in F} \mathsf{H}_i$. If each $\Lambda_i$ is simple 
then there exists an intrinsically free
parameterised $\mathsf{H}$-algebra 
compatible with the minimal support for $\Lambda$.
\end{proposition}

\proof There exists an intrinsically free  parameterised 
$\mathsf{H}_i$-algebra for each $i \in F$ and by
Proposition~\ref{prop_param_algs_1} their sum is a parameterised 
$\Lambda(\mathsf{C})$-algebra which is compatible with the minimal support for $\Lambda$. 
Moreover, by Proposition~\ref{prop_cat_algs_2} this sum is also $\mathsf{H}$-intrinsically free. \eop

Let us now say that the signature $\Lambda$ is 
\definition{semi-simple}\index{semi-simple signature}\index{signature!semi-simple} if 
it can be obtained as the disjoint union of simple signatures $\Lambda_i$, $i \in F$.

\begin{proposition}\label{prop_param_algs_3}
Suppose the signature $\Lambda$ is semi-simple. Then there exists an intrinsically
free parameterised $\mathsf{H}$-algebra 
compatible with the minimal support for $\Lambda$ in each of the following three cases,
in each case with $\diamond$ a $\natural$-stable simple head type and with
$\mathsf{C}'$ the subcategory defined by requiring the morphisms
to be bottomed:

(1)\enskip $\mathsf{C} = \mathsf{BSets}$ with the objects of $\mathsf{H}$ the bottomed 
$\diamond$-algebras.

(2)\enskip $\mathsf{C} = \mathsf{Posets}$ with the objects of $\mathsf{H}$ the ordered 
$\diamond$-algebras.

(3)\enskip
$\mathsf{C} = \mathsf{APosets}$ with the objects of $\mathsf{H}$ the continuous 
$\diamond$-algebras.
\end{proposition}

\proof These follow immediately from Proposition~\ref{prop_param_algs_2} and
Lemma~\ref{lemma_cat_algs_4}. \eop

If $\Lambda$ is semi-simple then the results of this section imply that in the
cases we are interested in there exist initial parameterised $\mathsf{H}$-algebras 
compatible with the minimal support for $\Lambda$. The problem is now, of course, that a 
typical open signature need not be semi-simple, and in fact this is the case with the 
signature in Example~\ref{algs_homs}.3. However, it 
is always possible to replace a given signature with a semi-simple signature (with a larger 
parameter set) which is, in a certain sense, more general.
This is illustrated in Example~\thesection.2, where a semi-simple replacement for
the signature in Example~\ref{algs_homs}.3 is presented. (It is left for the reader to
to work out how to make this replacement in the general case.)

\bigskip
\fbox{\begin{exframe}
\textit{Example \thesection.2} Consider the signature $\Lambda = (B,K,\Theta)$ 
with parameter set $A = \{\mathtt{x},\mathtt{y},\mathtt{z},\mathtt{u},\mathtt{v}\}$ 
given by
\begin{eelist}{20pt}
\item $B = \{\mathtt{bool},\mathtt{atom},\mathtt{int},\mathtt{pair},\mathtt{list},
      \mathtt{lp}, \mathtt{x},\mathtt{y},\mathtt{z},\mathtt{u},\mathtt{v}\}$,
\item $K = \{\mathtt{True},\mathtt{False},\mathtt{Atom},\mathtt{Pair},
      \mathtt{Nil},\mathtt{Cons},\mathtt{L},\mathtt{P}\} \cup \SynInt$,
\end{eelist}
and with $\Theta : K \to B^* \times B$ defined by
\begin{eelist}{20pt}
\item $\Theta(\mathtt{True}) = \Theta(\mathtt{False}) = (\onept,\mathtt{bool})$,
\item $\Theta(\mathtt{Atom}) = (\onept,\mathtt{atom})$, \enskip
      $\Theta(\mathtt{Pair}) = (\mathtt{x}\ \mathtt{y},\mathtt{pair})$, 
\item $\Theta(\mathtt{Nil}) = (\onept,\mathtt{list})$, \enskip
      $\Theta(\mathtt{Cons}) = (\mathtt{z}\ \mathtt{list},\mathtt{list})$,
\item $\Theta(\mathtt{L}) = (\mathtt{u},\mathtt{lp})$, \enskip
      $\Theta(\mathtt{P}) = (\mathtt{v},\mathtt{lp})$,
\item $\Theta(\underline{n}) = (\onept,\mathtt{int})$ for each $n \in \Int$,
\end{eelist}
and which can be represented in the (augmented) form
\begin{eelist}{90pt}
\item $\mathtt{bool\ ::=\ True\ |\ False}$
\item $\mathtt{atom\ ::=\ Atom}$
\item $\mathtt{int\ ::=\ } \cdots \mathtt{\ -2\ |\ -1\ |\ 0\ |\ 1\ |\ 2\ }\, \cdots$
\item $\mathtt{(pair\ x\ y)\ ::=\ Pair\ x\ y}$
\item $\mathtt{(list\ z)\ ::=\ Nil\ |\ Cons\ z\ (list\ z)}$
\item $\mathtt{(lp\ u\ v)\ ::=\ L\ u\ |\ P\ v}$
\end{eelist}
This signature $\Lambda$ is easily seen to be semi-simple. 
\end{exframe}}


\startsection{Polymorphic algebras}
\label{poly_algs}

Recall once again we have fixed a support $\lfloor\cdot\rfloor : B \to \mathcal{P}_o(A)$ 
for $\Lambda$ and a set of type variables $T$. A $T$-free set of polymorphic types
$(\xx{B},\#)$ and an initial set of polymorphic operator names $(\xx{K},\#')$ were then 
fixed. We are assuming that $\xx{B} \ne \varnothing$ and so these objects give rise to the 
polymorphic signature $\xx{\Lambda} = (\xx{B},\xx{K},\xx{\Theta})$.

If $\xx{\Phi} = (\xx{X},\xx{p}) : \mathcal{C}^T \to \xx{\Lambda}(\mathsf{C})$ is a
parameterised $\xx{\Lambda}(\mathsf{C})$-algebra then for each
$U \in \mathcal{C}^T$, $u \in \xx{B}^A$, the $\Lambda(\mathsf{C})$-algebra
$(\param{\xx{X}}{U}\comp \ipar_u ,\param{\xx{p}}{U}\comp \ipar'_u)$
will be denoted by $\xx{\Phi}_U \comp \ipar^o_u$.

\begin{proposition}\label{prop_poly_algs_1}
Let $\Phi : \mathcal{C}^A \to \Lambda(\mathsf{C})$ be a parameterised 
$\Lambda(\mathsf{C})$-algebra which is compatible with $\lfloor\cdot\rfloor$. Then there exists a unique 
parameterised $\xx{\Lambda}(\mathsf{C})$-algebra
$\xx{\Phi} = (\xx{X},\xx{p}) : \mathcal{C}^T \to \xx{\Lambda}(\mathsf{C})$ such that 
\[ \xx{\Phi}_U \comp \ipar^o_u = \Phi_V\]
with $V = \param{\xx{X}}{U} \comp u$ for all $U \in \mathcal{C}^T$, $u \in \xx{B}^A$. 
\end{proposition}

\proof 
Define a mapping $\varphi : \mathcal{C}^A \to \mathcal{C}^B$ by 
$\varphi(V)(b) = \parax{X}{V}{b}$ for all $V \in \mathcal{C}^A$, $b \in B$, where
$\Phi = (X,p)$; thus in particular
$\varphi(V)(b) = \parax{X}{V}{a} = V(a)$ for all $V \in \mathcal{C}^A$, $a \in A$.
Moreover, $\varphi$ is $B$-compatible with $\lfloor\cdot\rfloor$, since
$\Phi$ is compatible with $\lfloor\cdot\rfloor$. 
By Proposition~\ref{prop_poly_sig_3}~(1) there thus exists for each $U \in \mathcal{C}^T$ 
a unique mapping $\xx{X}^U : \xx{B} \to \mathcal{C}$ with $\rest{{\xx{X}^U}}{T} = U$ such that
$\xx{X}^U \comp \ipar_u  = \varphi(\xx{X}^U \comp u)$ for all $u \in \xx{B}^A$. It then follows that
$\xx{X}^U \comp \ipar_u  = \param{X}{V}$ with $V = \xx{X}^U \comp u$ for each $u \in \xx{B}^A$, because
$\varphi(\xx{X}^U \comp u)(b) = \parax{X}{V}{b}$ for each $b \in B$.

Now for each $U \in \mathcal{C}^T$ define $\psi^U : \xx{B}^A \to \mathcal{M}^K$ by
$\psi^U(u)(k) = \parax{p}{V}{k}$ for all $u \in \xx{B}^A$, $k \in K$, where
$V = \param{\xx{X}}{U} \comp u$. Then $\psi^U$ is $K$-compatible with 
$\lfloor\cdot\rfloor$, since $\Phi$ is compatible with $\lfloor\cdot\rfloor$. 
By Proposition~\ref{prop_poly_sig_3}~(2) there thus exists a unique mapping 
$\xx{p}^U : \xx{K} \to \mathcal{M}$ such that
$\xx{p}^U \comp \ipar'_u  = \psi^U(u)$ for all $u \in \xx{B}^A$. 
Therefore $\xx{p}^U \comp \ipar'_u  = \param{p}{V}$ with 
$V = \param{\xx{X}}{U} \comp u$ for all $u \in \xx{B}^A$.

By Proposition~\ref{prop_poly_sig_4} $(\param{\xx{X}}{U},\param{\xx{p}}{U})$ is a
$\xx{\Lambda}(\mathsf{C})$-algebra for each $U \in \mathcal{C}^T$ and
$(\param{\xx{X}}{U},\param{\xx{p}}{U})$ is bound to $U$, since 
by definition $\rest{{\xx{X}^U}}{T} = U$.
A parameterised $\xx{\Lambda}(\mathsf{C})$-algebra
$\xx{\Phi} : \mathcal{C}^T \to \xx{\Lambda}(\mathsf{C})$ can thus be defined by letting
$\xx{\Phi}_U = (\param{\xx{X}}{U},\param{\xx{p}}{U})$ for each $U \in \mathcal{C}^T$ and by construction
$\xx{\Phi}_U \comp \ipar^o_u = \Phi_V$ with $V = \param{\xx{X}}{U} \comp u$
for all $U \in \mathcal{C}^T$, $u \in \xx{B}^A$.
The uniqueness follows from the uniqueness in Proposition~\ref{prop_poly_sig_1}. \eop

The parameterised $\xx{\Lambda}(\mathsf{C})$-algebra $\xx{\Phi}$ given
in Proposition~\ref{prop_poly_algs_1} will be called the 
\definition{polymorphic $\xx{\Lambda}(\mathsf{C})$-algebra}\index{polymorphic algebra}
\index{algebra!polymorphic}corresponding to the 
parameterised $\Lambda(\mathsf{C})$-algebra $\Phi$.


The final task is to  show that the polymorphic algebra $\xx{\Phi}$ inherits the properties of 
$\Phi$ which are needed in applications. In particular, it will be established that 
the $\xx{\Lambda}(\mathsf{C})$-algebra $\xx{\Phi}_U$ is regular for each $U \in \mathcal{C}^T$
if the $\Lambda(\mathsf{C})$-algebra $\Phi_V$ is regular for each $V \in \mathcal{C}^A$.
We will employ a set-up here (using what we call a reduction) which allows things to be formulated
not just for a particular category such as $\mathsf{BSets}$.

Let $\mathsf{D}$ be a further category. A mapping
$\Delta : \mathcal{C}_{\mathsf{C}} \to \mathcal{C}_{\mathsf{D}}$ will be called
a \definition{reduction}\index{reduction} from $\mathsf{C}$ to $\mathsf{D}$ if 
\begin{evlist}{25pt}{0.8ex}
\item[---] 
$\Hom_{\mathsf{C}}(X,Y) \subset \Hom_{\mathsf{D}}(\Delta(X),\Delta(Y))$ for all objects
$X,\, Y \in \mathcal{C}_{\mathsf{C}}$, 
\item[---] 
$g \comp_{\mathsf{C}} f = g \comp_{\mathsf{D}} f$ for all 
$f \in \Hom_{\mathsf{C}}(X,Y)$, $g \in \Hom_{\mathsf{C}}(Y,Z)$
for all objects $X,\,Y,\,Z \in \mathcal{C}_{\mathsf{C}}$,
\item[---] 
the identity morphism in $\Hom_{\mathsf{C}}(X,X)$ is equal to the identity morphism in
$\Hom_{\mathsf{C'}}(\Delta(X),\Delta(X))$ for each object $X \in \mathcal{C}_{\mathsf{C}}$,
\end{evlist}
(where the subscripts $\mathsf{C}$ and $\mathsf{D}$ indicate which category is involved). 
If $\mathsf{D}$ (as well as $\mathsf{C}$) is a $\utimes$-category then a reduction
$\Delta : \mathcal{C}_{\mathsf{C}} \to \mathcal{C}_{\mathsf{D}}$ is called
a \definition{$\utimes$-reduction} if $\Delta(\utimes X) = \utimes (\Delta \comp X)$ for all
$X \in \ffam{\mathcal{C}_{\mathsf{C}}}$. 

A reduction should be thought of as as an operation which forgets some of the structure.
There are obvious $\utimes$-reductions 
from $\mathsf{APosets}$ and $\mathsf{CPosets}$ to $\mathsf{Posets}$, from $\mathsf{Posets}$
to $\mathsf{BSets}$ and from $\mathsf{BSets}$ to $\mathsf{Sets}$. These will be referred 
to as the
\index{standard reduction}\index{reduction!standard}\definition{standard $\utimes$-reductions}.

The category $\mathsf{C}$ together with a reduction from $\mathsf{C}$ to $\mathsf{Sets}$
is usually called a 
\index{concrete category}\index{category!concrete}\definition{concrete category}.

\begin{lemma}\label{lemma_poly_algs_1}
Suppose $\mathsf{D}$ is a $\utimes$-category and 
$\Delta$ is a $\utimes$-reduction from $\mathsf{C}$ to $\mathsf{D}$. Then 
$(\Delta \comp X,p)$ is a $\Lambda(\mathsf{D})$-algebra 
for each $\Lambda(\mathsf{C})$-algebra $(X,p)$.
Moreover, if $(X,p)$ is bound to the family $V : A \to \mathcal{C}_{\mathsf{C}}$ then 
$(\Delta \comp X,p)$ is bound to the family $\Delta \comp V : A \to \mathcal{C}_{\mathsf{D}}$.
\end{lemma}

\proof Let $k \in K$; then 
$p_k \in \Hom_{\mathsf{C}}(\ass{\sdom{k}}{X},X_{\scod{k}})
\subset \Hom_{\mathsf{D}}(\Delta(\ass{\sdom{k}}{X}),\Delta(X_{\scod{k}}))$ and thus
$p_k \in \Hom_{\mathsf{D}}(\ass{\sdom{k}}{(\Delta\comp X)},(\Delta \comp X)_{\scod{k}})$,
since
\[\ass{\sdom{k}}{(\Delta\comp X)} = \utimes(\Delta \comp X \comp \sdom{k})
= \Delta(\utimes(X \comp \sdom{k})) = \Delta(\ass{\sdom{k}}{X})\;.\]
This implies 
$(\Delta \comp X,p)$ is a $\Lambda(\mathsf{D})$-algebra. The final statement is
clear. \eop

The $\Lambda(\mathsf{D})$-algebra $(\Delta \comp X,p)$ in 
Lemma~\ref{lemma_poly_algs_1} will be denoted by $\Delta \fcomp (X,p)$.
Lemma~\ref{lemma_poly_algs_1} allows some of the definitions introduced in the previous 
chapters to be reformulated in the following style:

Let $\mathsf{C}$ be one of the categories $\mathsf{APosets}$, $\mathsf{CPosets}$ and 
$\mathsf{Posets}$ and let $\Delta$ be the standard reduction from  $\mathsf{C}$ to 
$\mathsf{BSets}$. If $(H,\diamond)$ is a head type then a $\Lambda(\mathsf{C})$-algebra 
$(X,p)$ was defined to be an $(H,\diamond)$-algebra if the bottomed $\Lambda$-algebra 
$\Delta\fcomp (X,p)$ is an $(H,\diamond)$-algebra. Moreover, if $(X,p)$ is bound to the 
family $V : A \to \mathcal{C}$ then $(X,p)$ was defined to be $V$-minimal if 
$\Delta\fcomp (X,p)$ is $\Delta\comp V$-minimal.

In the same way, let $\mathsf{C}$ be one of the categories $\mathsf{APosets}$ and 
$\mathsf{CPosets}$ and let $\Delta$ be the standard reduction from  $\mathsf{C}$ to 
$\mathsf{Posets}$ (so in fact $\Delta$ is just the inclusion mapping).
Then a $\Lambda(\mathsf{C})$-algebra $(X,p)$ was defined to be intrinsic
if the ordered $\Lambda$-algebra $\Delta\fcomp (X,p)$ is intrinsic.

In what follows let $\Phi = (X,p)$ be a parameterised $\Lambda(\mathsf{C})$-algebra 
compatible with $\lfloor\cdot\rfloor$ and let $\xx{\Phi} = (\mathbf{X},\mathbf{p})$ be the 
corresponding polymorphic $\mathbf{\Lambda}(\mathsf{C})$-algebra.

\begin{proposition}\label{prop_poly_algs_3}
Let $\Delta$ be a $\utimes$-reduction from $\mathsf{C}$ to $\mathsf{BSets}$ and 
suppose that $\Delta \comp \Phi_V$ is a regular bottomed 
$\Lambda$-algebra for each $V$. Then $\Delta\comp \mathbf{\Phi}_U$ is a regular bottomed 
$\xx{\Lambda}$-algebra for each $U$.
\end{proposition}

\proof 
Put $\Delta \comp \param{X}{V} = \param{\breve{X}}{V}$ for each $V \in \mathcal{C}^A$ and
$\Delta \comp \param{\mathbf{X}}{U} = \param{\breve{\mathbf{X}}}{U}$
for each $U \in \mathcal{C}^T$.
Let $U \in \mathcal{C}^T$, $\xx{b} \in \xx{B} \setminus T$ and 
$x \in \nonbot{(\parax{\breve{\mathbf{X}}}{U}{\xx{b}})}$; it must then be shown 
that there exists a unique $\xx{k} \in \xx{K}_{\xx{b}}$ and a unique 
$v \in \dom(\parax{\mathbf{p}}{U}{\xx{k}})$ such that $x = \parax{\mathbf{p}}{U}{\xx{k}}(v)$. 
Now by Lemma~\ref{lemma_poly_sig_7}~(2) there exists $b \in B\setminus A$ 
and $u \in \xx{B}^{\lfloor b\rfloor}$ with $\xx{b} = \ipar_u(b)$, and then by
Proposition~\ref{lemma_poly_sig_3}~(1) 
$\xx{K}_{\xx{b}} = \{ \ipar'_u(k) : k \in K_b \}$. Let $V = \param{\mathbf{X}}{U} \comp u$. 
Then $\parax{\mathbf{X}}{U}{\xx{b}} = \parax{X}{V}{b}$, which means  also that 
$\parax{\mathbf{\breve{X}}}{U}{\xx{b}} = \parax{\breve{X}}{V}{b}$, and 
$(\param{\mathbf{p}}{U} \comp \ipar'_u)_k = \parax{p}{V}{k}$ for 
each $k \in K_b$. But $(\param{\breve{X}}{V},\param{p}{V})$ is a regular bottomed 
$\Lambda$-algebra, $b \in B\setminus A$ and 
$x \in \nonbot{(\parax{\breve{X}}{V}{b})}$. There thus exists a unique
$k \in K_b$ and a unique $v \in \dom(\parax{p}{V}{k})$ such that $x = \parax{p}{V}{k}(v)$.
Hence, putting $\xx{k} = \ipar'_u(k)$, it follows that $\xx{k} \in \xx{K}_{\xx{b}}$ and
$x = \parax{\mathbf{p}}{U}{\xx{k}}(v)$. Conversely, if $x = \parax{\mathbf{p}}{U}{\xx{k}'}(v')$ with 
$\xx{k}' \in \xx{K}_{\xx{b}}$ and $v' \in \dom(\parax{\mathbf{p}}{U}{\xx{k}'})$ then 
$\xx{k}' = \ipar'_u(k')$ for some $k' \in K_b$, and so
$\parax{p}{V}{k}(v) = x = \parax{\mathbf{p}}{U}{\xx{k}'}(v') = \parax{p}{V}{k'}(v')$, which implies 
that $k' = k$, thus also that $\xx{k}' = \xx{k}$, and that $v' = v$  (since 
$(\param{\breve{X}}{V},\param{p}{V})$ 
is regular). This shows $\Delta \fcomp \mathbf{\Phi}_U$ is a regular bottomed 
$\xx{\Lambda}$-algebra. \eop

Let $(H,\diamond)$ be a simple head type for $\Lambda$ (so
$H_b = \BOneptset$ for each $b \in B$); then a simple head
type $(\xx{H},\xxx{\diamond})$ for the signature $\xx{\Lambda}$ (so again
$\xx{H}_{\xx{b}} = \BOneptset$ for each $\xx{b} \in \xx{B}$) can be defined as follows: 
Recall that there is a unique mapping $\omega : \xx{K} \to K$ 
with $\omega \comp \ipar'_u = \id_K$ for each $u \in \xx{B}^A$,
and that $\domsdom{\xx{k}} = \domsdom{\omega(\xx{k})}$ for each $\xx{k} \in \xx{K}$. 
Consider $\xx{k} \in \xx{K}$ and put $k = \omega(\xx{k})$;
Then
$\ass{\sdom{\xx{k}}}{\xx{H}} 
= \total{\domsdom{\xx{k}}}{\BOneptset} = \total{\domsdom{k}}{\BOneptset}
= \ass{\sdom{k}}{H}$ and $\xx{H}_{\scod{\xx{k}}} = \BOneptset = H_{\scod{k}}$, 
and so a mapping 
$\xxx{\diamond}_{\xx{k}} : \ass{\sdom{\xx{k}}}{\xx{H}} \to \xx{H}_{\xx{b}}$ 
can be defined by putting 
$\xxx{\diamond}_{\xx{k}}(v) = \diamond_k(v)$ for all 
$v \in \ass{\sdom{\xx{k}}}{\xx{H}}$. Thus in fact
$\xxx{\diamond} = \diamond \circ \omega$.
If $\diamond$ is $\natural$-invariant (resp.\ $\natural$-stable) then so is 
$\xxx{\diamond}$.

\begin{proposition}\label{prop_poly_algs_4}
Let $\Delta$ be a $\utimes$-reduction from $\mathsf{C}$ to $\mathsf{BSets}$ and 
suppose that $\Delta \fcomp \Phi_V$ is a  $\diamond$-algebra
for each $V$. Then $\Delta\fcomp \mathbf{\Phi}_U$ is a  
$\xxx{\diamond}$-algebra for each $U$.
\end{proposition}

\proof
Put $\Delta \comp \param{X}{V} = \param{\breve{X}}{V}$ for each $V \in \mathcal{C}^A$ and
$\Delta \comp \param{\mathbf{X}}{U} = \param{\breve{\mathbf{X}}}{U}$
for each $U \in \mathcal{C}^T$.
Let $U \in \mathcal{C}^T$ and $\xx{k} \in \xx{K}$; by Lemma~\ref{lemma_poly_sig_7}~(2) there exists 
$k \in K$ and $u \in \xx{B}^A$ such that 
$\xx{k} = \ipar'_u(k)$, and  $\domsdom{\xx{k}} = \domsdom{k}$.
Let $V = \param{\mathbf{X}}{U} \comp u$. Then
$\parax{\mathbf{X}}{U}{\xx{b}} = \parax{X}{V}{b}$,
$\ass{\sdom{\xx{k}}}{(\param{\mathbf{X}}{U})} = \ass{\sdom{k}}{(\param{X}{V})}$, 
hence also 
$\parax{\breve{\mathbf{X}}}{U}{\xx{b}} = \parax{\breve{X}}{V}{b}$,
$\ass{\sdom{\xx{k}}}{(\param{\breve{\mathbf{X}}}{U})} = \ass{\sdom{k}}{(\param{\breve{X}}{V})}$, 
and $\parax{\mathbf{p}}{U}{\xx{k}} = \parax{p}{V}{k}$. Therefore 
$\varepsilon_{\scod{\xx{k}}} = \varepsilon_{\scod{k}}$ 
(as mappings from $\parax{\breve{\mathbf{X}}}{U}{\scod{\xx{k}}} = \parax{\breve{X}}{V}{\scod{k}}$ to 
$\xx{H}_{\scod{\xx{k}}} = H_{\scod{k}}$) 
and $\ass{\sdom{\xx{k}}}{\varepsilon} = \ass{\sdom{k}}{\varepsilon}$ (as mappings 
from 
$\ass {\sdom{\xx{k}}}{(\param{\breve{\mathbf{X}}}{U})} = \ass{\sdom{k}}{(\param{\breve{X}}{V})}$
to $\ass{\sdom{\xx{k}}}{\xx{H}} = \ass{\sdom{k}}{H}$). Moreover, 
$\xxx{\diamond}_{\xx{k}} = \diamond_k$ holds by definition. But
$\varepsilon_{\scod{k}}\comp \parax{p}{V}{k} 
= \diamond_k\comp \ass{\sdom{k}}{\varepsilon}$, 
because $(\param{\breve{X}}{V},\param{p}{V})$ is a $\diamond$-algebra, and so
$\varepsilon_{\scod{\xx{k}}}\comp \parax{\xx{p}}{U}{\xx{k}} 
 = \xxx{\diamond}_{\xx{k}}\comp \ass{\sdom{\xx{k}}}{\varepsilon}$,
which implies that $\Delta \fcomp \mathbf{\Phi}_U$
is a $\xxx{\diamond}$-algebra. \eop

\begin{proposition}\label{prop_poly_algs_5}
Let $\Delta$ be a $\utimes$-reduction from $\mathsf{C}$ to $\mathsf{Posets}$ and 
suppose that $\Delta \fcomp \Phi_V$ is an intrinsic ordered
$\Lambda$-algebra for each $V$. Then $\Delta\fcomp \mathbf{\Phi}_U$ is an intrinsic ordered 
$\xx{\Lambda}$-algebra for each $U$.
\end{proposition}

\proof This is very similar to the proof of Proposition~\ref{prop_poly_algs_3} and is 
left for the reader. \eop


%% file: sbikabib.tex
\bigskip
\bigskip

{\sc Fakult\"at f\"ur Mathematik, Universit\"at Bielefeld}\\
{\sc Postfach 100131, 33501 Bielefeld, Germany}\\
\textit{E-mail address:} \texttt{preston@math.uni-bielefeld.de}\\
\textit{URL:} \texttt{http://www.math.uni-bielefeld.de/\symbol{126}preston}
